\documentclass{MSc_Thesis}

\begin{document}

\maketitle

\thispagestyle{empty}
\vspace{2cm}
\begin{center}
    {\Large \textbf{Abstract}}
    \vspace{0.5cm}
\end{center}

This dissertation addresses the reconstruction of Equations of State (EoSs) governing the internal structure of compact stars, by employing modern machine learning and deep learning methods. The reconstruction pipeline is based on data derived from mass–radius (M–R) curves, computed by solving the Tolman–Oppenheimer–Volkoff (TOV) equations for a wide variety of physically admissible EoSs. The manuscript is structured into seven chapters, split into a Theoretical Part (Chapters 1–4) and a Computational Part (Chapters 5–7). Chapter 1 explores the astrophysical properties of neutron stars, including matter stratification (crust–core), nuclear interactions under extreme densities, and the importance of equations of state in the description of hydrostatic equilibrium. The physical constraints governing viable hadronic EoS models are also analyzed. Chapter 2 extends the discussion to quark stars, starting from Quantum Chromodynamics (QCD) and the principle of asymptotic freedom, and focusing on effective models, such as MIT bag and CFL, under the stability conditions of strange quark matter and self-bound compact objects. Chapter 3 introduces the regression-based machine learning algorithms used in this study (Decision Tree, Random Forest, Gradient Boosting, XGBoost), with emphasis on multi-output regression, error metrics (MSE and MSLE), and model optimization techniques, such as cross-validation and grid search. In Chapter 4, emphasis is placed on the methodology for developing and fitting a complete neural network with multiple hidden layers, with the use of ReLU activations and backpropagation, and the incorporation of regularization techniques, such as batch normalization and dropout. Chapter 5 introduces the methodology for generating artificial EoSs: multimodal equations for hadronic stars, with any corrections to avoid causality violation, and parametric equations for quark stars. In addition, the methodology for numerically solving the TOV equations for each model is described. The EoS models are constructed in a way to ensure dense and detailed coverage of the M-R region, within which such compact objects are expected to be detected. In Chapter 6, the preparation steps of data for use in regression models are presented, including sampling and shuffling, to avoid bias and exclude data leakage. Furthermore, the construction of hyperparameter grids for the machine learning algorithms, as well as the architecture of the neural network used are fully documented. Chapter 7, includes the training and evaluation results for each regression model, with comparisons of MSE and MSLE metrics and training times. For the neural network, learning curves are additionally presented and the stability of the training procedure is assessed. Then the reconstruction of the 21 "main" equations of state for hadronic stars and 20 equations of state for quark stars (10 MIT bag, 10 CFL) is demonstrated, with the reconstruction accuracy evaluated based on comparison with the actual $\epsilon$-P curves. The methodology for producing the TOV equations and all links to directories, of the source code developed for this work, are included in separate appendices. This dissertation aims to provide a reusable and scalable framework for future EoS reconstruction studies, promoting the connection between theoretical astrophysics and computational science.
\clearpage
\clearpage

\thispagestyle{empty}
\selectlanguage{greek} 
\vspace{2cm}
\begin{center}
    {\Large \textbf{Περίληψη}}
    \vspace{0.5cm}
\end{center}

Η παρούσα διπλωματική εργασία ασχολείται με την ανακατασκευή των καταστατικών εξισώσεων \textlatin{(Equations of State – EoSs)} που περιγράφουν την εσωτερική δομή και φυσική των συμπαγών αστέρων, κάνοντας χρήση προηγμένων τεχνικών μηχανικής και βαθιάς μάθησης. Η προσέγγιση βασίζεται στη συστηματική αξιοποίηση και ανάλυση δεδομένων από καμπύλες μάζας–ακτίνας \textlatin{(M–R)}, παραγόμενες από την επίλυση των εξισώσεων \textlatin{TOV} για μεγάλο αριθμό φυσικά αποδεκτών καταστατικών εξισώσεων. Η εργασία δομείται σε επτά κεφάλαια, χωρισμένα σε δύο μέρη: το Θεωρητικό Μέρος (Κεφάλαια 1–4) και το Υπολογιστικό Μέρος (Κεφάλαια 5–7). Στο Κεφάλαιο 1, παρουσιάζονται οι αστροφυσικές ιδιότητες των αστέρων νετρονίων, με έμφαση στη διαστρωμάτωση της ύλης (κρούστα–πυρήνας), τις πυρηνικές αλληλεπιδράσεις σε υπερπυκνές καταστάσεις, και τη σημασία των καταστατικών εξισώσεων για την περιγραφή της υδροστατικής ισορροπίας. Αναλύονται υπάρχοντα μοντέλα για την ύλη νετρονίων, καθώς και οι φυσικοί περιορισμοί που πρέπει να ικανοποιούν. Στο Κεφάλαιο 2, η μελέτη επεκτείνεται στους κουάρκ αστέρες, με αφετηρία τη θεωρία Κβαντικής Χρωμοδυναμικής \textlatin{(QCD)} και την έννοια της ασυμπτωτικής ελευθερίας, ενώ εξετάζονται τα μοντέλα \textlatin{MIT bag} και \textlatin{CFL}, μαζί με τις συνθήκες σταθερότητας που επιβάλλονται στους αυτοδεσμευμένους αστέρες που συγκροτούνται από παράξενη κουάρκ ύλη. Στο Κεφάλαιο 3, εισάγονται οι βασικοί αλγόριθμοι μηχανικής μάθησης για παλινδρόμηση που χρησιμοποιούνται στην μελέτη \textlatin{(Decision Tree, Random Forest, Gradient Boosting, XGBoost)}, η μετάβαση από παλινδρόμηση μίας εξόδου σε παλινδρόμηση πολλαπλών εξόδων, και οι βασικές μετρικές απόδοσης \textlatin{(MSE, MSLE)}. Παρουσιάζονται επίσης μέθοδοι βελτιστοποίησης και αξιολόγησης μοντέλων, όπως η σταυρωτή επικύρωση \textlatin{(cross-validation)} και η αναζήτηση υπερπαραμέτρων \textlatin{(grid search)}. Στο Κεφάλαιο 4, δίνεται έμφαση στην μεθοδολογία ανάπτυξης και εκμάθησης ενός πλήρους νευρωνικού δικτύου με πολλαπλά κρυφά στρώματα, με την χρήση \textlatin{ReLU} ενεργοποιήσεων και οπισθοδιάδοσης \textlatin{(backpropagation)}, και την ενσωμάτωση τεχνικών κανονικοποίησης \textlatin{(regularization)}, όπως η κανονικοποίηση παρτίδας \textlatin{(batch normalization)} και η εγκατάλειψη \textlatin{(dropout)}. Το Κεφάλαιο 5 εισάγει τη μεθοδολογία παραγωγής τεχνητών καταστατικών εξισώσεων: πολυτροπικές εξισώσεις για αδρονικούς αστέρες, με τυχόν διορθώσεις ώστε να μην παραβιάζεται η αιτιότητα \textlatin{(causality)}, και παραμετρικές εξισώσεις για κουάρκ αστέρες. Επιπλέον, περιγράφεται η μεθοδολογία αριθμητικής επίλυσης των εξισώσεων \textlatin{TOV} για κάθε μοντέλο. Η κατασκευή των μοντέλων καταστατικών γίνεται με τέτοιο τρόπο, ώστε να εξασφαλίζεται πυκνή και λεπτομερής κάλυψη της περιοχής \textlatin{M–R}, εντός της οποίας αναμένεται να εντοπίζονται τέτοια συμπαγή αντικείμενα. Στο κεφάλαιο 6, παρουσιάζονται τα στάδια προετοιμασίας των δεδομένων για χρήση σε μοντέλα παλινδρόμησης, συμπεριλαμβανομένης της δειγματοληψίας \textlatin{(sampling)} και της αναδιάταξης \textlatin{(shuffling)}, για την αποφυγή μεροληψίας \textlatin{(bias)} και τον αποκλεισμό της διαρροής δεδομένων \textlatin{(data leakage)}. Επιπλέον, τεκμηριώνεται πλήρως η κατασκευή των πλεγμάτων υπερπαραμέτρων για τους αλγορίθμους μηχανικής μάθησης, καθώς και η αρχιτεκτονική του νευρωνικού δικτύου που χρησιμοποιείται. Το Κεφάλαιο 7, περιλαμβάνει τα αποτελέσματα εκπαίδευσης και αξιολόγησης για κάθε μοντέλο παλινδρόμησης, με συγκρίσεις των μετρικών \textlatin{MSE} και \textlatin{MSLE} και των χρόνων εκπαίδευσης. Για το νευρωνικό δίκτυο, παρουσιάζονται επιπλέον οι καμπύλες εκμάθησης \textlatin{(learning curves)} και αξιολογείται η σταθερότητα της εκπαίδευσης. Ακολουθεί η ανακατασκευή των 21 <<κύριων>> καταστατικών εξισώσεων για αδρονικούς αστέρες και 20 καταστατικών εξισώσεων για κουάρκ αστέρες (10 \textlatin{MIT bag}, 10 \textlatin{CFL}), με την ακρίβεια ανακατασκευής να αξιολογείται βάσει σύγκρισης με τις πραγματικές $\epsilon$-\textlatin{P} καμπύλες. Η μεθοδολογία παραγωγής των εξισώσεων \textlatin{TOV} και οι σύνδεσμοι προς το σύνολο του κώδικα, που αναπτύχθηκε στο πλαίσιο της εργασίας, περιλαμβάνονται σε ξεχωριστά παραρτήματα. Η παρούσα διατριβή φιλοδοξεί να αποτελέσει ένα επαναχρησιμοποιήσιμο και επεκτάσιμο πλαίσιο για μελλοντικές μελέτες ανακατασκευής καταστατικών εξισώσεων, προάγοντας τη σύνδεση μεταξύ θεωρητικής αστροφυσικής και υπολογιστικής επιστήμης.

\selectlanguage{english} 
\clearpage
\clearpage

\thispagestyle{empty}
\vspace{2cm}
\begin{center}
    {\Large \textbf{Acknowledgments}}
    \vspace{0.5cm}
\end{center}

This dissertation closes the cycle of my master's degree in MSc Computational Physics, at Physics Department AUTh. As this postgraduate journey comes to an end, I feel grateful for everything I have learned and experienced along the way. I would like to thank my family and friends for their support and encouragement throughout this time. Their help gave me strength to keep going, even during the most challenging moments. Most of all, I want to express my sincere gratitude to my supervising professors, Dr. Charalampos Moustakidis and Dr. Theodoros Diakonidis. Their valuable guidance, trust, and continuous support played a key role in the development of this thesis. I truly appreciated their willingness to share their knowledge, offer feedback, and work together on something meaningful.
\clearpage
\clearpage

\thispagestyle{empty}
\vspace*{5cm}
\large
\begin{center}
    \textit{"Coding might be difficult, but every line of code gets us closer to unlocking the secrets of the universe"} \\
    \vspace{0.5cm}
\end{center}
\normalsize
\clearpage
\clearpage

\tableofcontents
\clearpage
\listoffigures
\clearpage
\listoftables
\clearpage


\chaptertitleformat

\hypersetup{
    colorlinks=true,
    linkcolor=black,
    urlcolor=blue,
    citecolor=blue
}

\chapter*{Introduction}\label{Intro}
\addcontentsline{toc}{part}{Introduction}
The TOV equations can be considered as a process, which correlates an equation of state to a $M-R$ diagram. The latter reveals a relation between mass and radius for compact stars. The forward process involves establishing the relation between energy density and pressure, through an EoS model, and then proceed to solve TOV equations and generate a $M-R$ diagram. On the contrary, a backward (inverse) process, that allows one to obtain the EoS from a $M-R$ curve, has proved extremely difficult to find. In this dissertation, we aim to develop such a machine (or deep) learning model, that returns data of the equation of state, when data from the respective $M-R$ graph are given, successfully posing as this inverse process.

Yuki Fujimoto, Kenji Fukushima, and Koichi Murase in their papar of 2018: "Methodology study of machine learning
for the neutron star equation of state" \cite{fujimoto2018methodology}
present a methodology for addressing this kind of problems, focusing on the reconstruction of equations of state for Neutron Stars. They start by producing a large amount of mock hadronic EoSs, using the piecewise polytropes method. \cite{raithel2016neutron}.
Then, they filter the resulting EoSs, keeping only those EoSs, the $M-R$ curves of which reach approximately $2M_\odot$ ($M_\odot$ denotes the solar mass). They sample data from these $M-R$ curves and add artificial noise on them, in order to align with real observations of Neutron Stars. The analysis part comes next, where they build a Deep Neural Network with 3 hidden layers and fit it to predict the values of average sound velocity $c_s^2$ at each of five polytropic segments. They assess the performance of their model via learning curves, using MSLE as their primary loss function, or via direct comparison between the reconstructed and the original EoSs, as well as the reconstructed and original $M-R$ curves. Throughout the chapters of this dissertation, we follow a similar methodology for Neutron Stars' EoSs and expand the study, exploring additionally the reconstruction of Quark Stars' EoSs and employing several regression models.

\part{Theoretical Part}

\chapter{Neutron Stars}\label{NS Theory}
\section{Composition of matter in Neutron Stars}\label{NS struct}
Neutron stars exhibit a rather fascinating and complex structure, consisting of several layers. More specifically, a neutron star is considered to be composed of five major distinct layers: the atmosphere, the outer crust (or envelope), the inner crust (or just crust), and the outer and inner cores, each with its own special properties (see Fig.\ref{fig:NS_structure}). In the following, we present a brief overview of the most important features of each region \cite{schaffner2020compact}:

\begin{figure}[htb]
    \centering
    \includegraphics[width=0.45\linewidth]{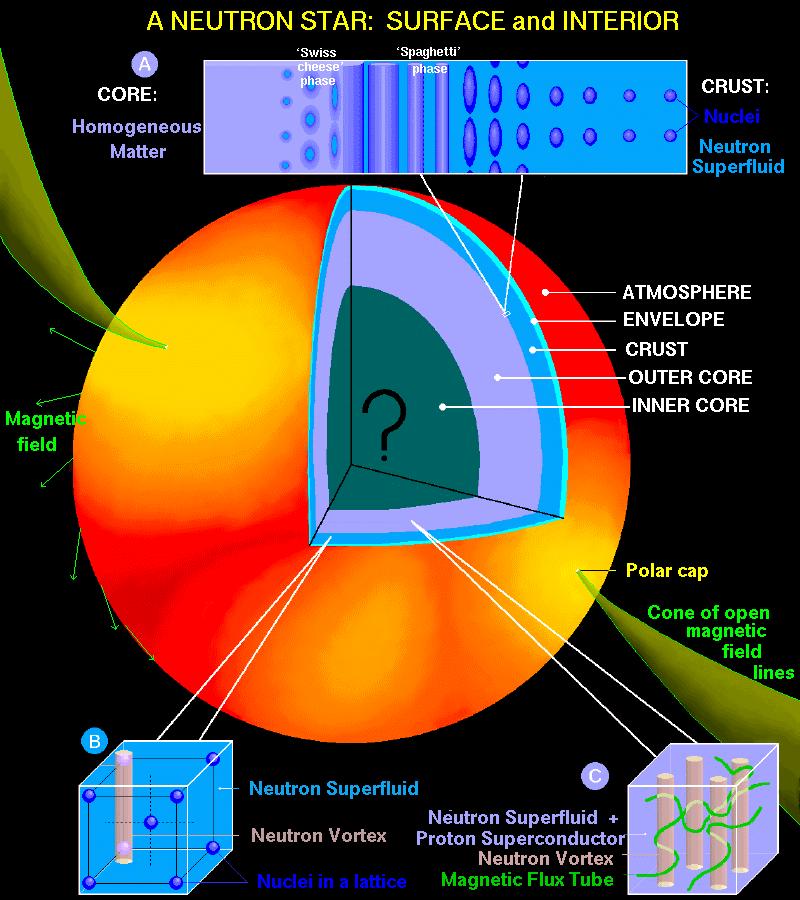}
    \caption{Schematic illustration for the composition of matter in a Neutron Star. \textit{Top}: the geometric transitions that might take place, from uniform matter (high densities) to spherical nuclei (low densities) are displayed. \textit{Center}: a colored representation of the layers of a Neutron Star is shown. \textit{Bottom}: aspects of the superfluidity and superconductivity of the crust and core are presented. Figure adapted from \href{https://www.astro.umd.edu/~miller/nstar.html}{www.astro.umd.edu}.}
    \label{fig:NS_structure}
\end{figure}

\begin{itemize}
    \item \textbf{Atmosphere}: being the outermost layer and having a thickness that varies from $0.3mm$ (for cold neutron stars) to $10cm$ (for hot neutron stars), the atmosphere of a neutron star contains atoms and has a negligible contribution to its total mass and radius. However, its composition, which differs from one neutron star to another, plays a crucial role when it comes to the spectral properties of the star, as it shapes the spectrum of the emitted photons. Thus, a detailed detection and examination of the thermoelectromagnetic radiation of a neutron star can provide important information about the production mechanism and the surface of the star, as well as the macroscopic features of the star, like the mass and the radius.

    \item \textbf{Envelope} (outer crust): the envelope also has a negligible contribution to the neutron star's mass, but it significantly affects the transport and release of thermal energy from the surface of the star. Energy densities above the critical value of $1.42\cdot10^4g$ $cm^{-3}$ are found in this region, allowing the electrons to form a free Fermi gas and causing the nuclei to be arranged in a lattice format to minimize their energy. As we go deeper towards the center of the star, the mass density $\rho$ and Fermi energy of the electrons increase further, and for $\rho \geq 2.1\cdot10^6g$ $cm^{-3}$, the electrons are becoming degenerate and relativistic. Therefore, the pressure is due to the degeneracy pressure of the electrons ideal gas, while the energy density is determined by the mass density of the nuclei. Additionally, these conditions favor the transformation of protons and electrons into neutrons: $p^{+}+e^{-}\rightarrow n_0 + \nu_{e}$, resulting in the enrichment of the nuclei with neutrons and the change of the chemical composition of the layer.

    \item \textbf{Crust} (inner crust): the crust starts when the mass density reaches the so-called neutron drip density, $\rho = \rho_{ND} \simeq 4\cdot10^{11}g$ $cm^{-3}$ and extends from $1km$ to $2km$, approximately, below the surface (see Fig.\ref{fig:NS_structure_2}). Above this density, the chemical potential of the neutron becomes zero, and neutrons begin to leak from the neutron-rich nuclei, forming another fluid, apart from the electron gas. At higher densities in the inner crust, the majority of neutrons are located in the fluid rather than in the interior of the nuclei. This formation of a solid-state physics system (the lattice of nuclei) immersed in a quantum-mechanical fluid (of neutrons and electrons) is responsible for some exotic phenomena. Neutron stars can rotate rapidly, causing the creation of vortices in the crust, which are responsible for angular momentum transfer and pulsar glitches. On the other hand, when pairs of bound neutrons break and recombine as pairs, neutrino-antineutrino pairs are emitted, making them an important cooling factor, especially in young neutron stars. Lastly, new phases, such as the various pasta phases, can also appear in this region of the neutron star.

    \begin{figure}[htb]
    \centering
    \includegraphics[width=0.5\linewidth]{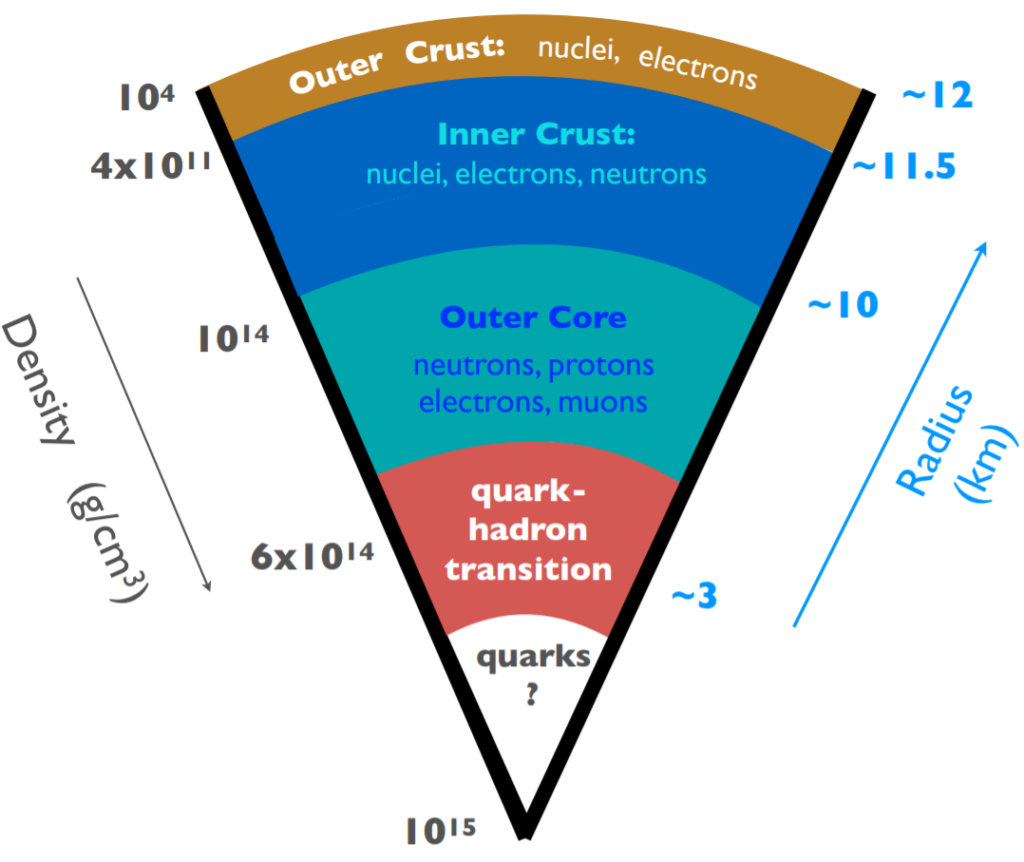}
    \caption{Schematic illustration for the layers in a Neutron Star. Characteristic values for the density and the thickness of each layer are included, along with speculations about the composition of matter in the core of the star. Figure adapted from \href{https://n3as.berkeley.edu/p/can-gravitational-waves-reveal-phase-transitions-in-the-cores-of-neutron-stars/}{n3as.berkeley.edu}.}
    \label{fig:NS_structure_2}
\end{figure}

    \item \textbf{Outer core}: the transition point from the crust to the core of the neutron star, is the nuclear saturation density: $n_0\simeq0.16fm^{-3}$ or $\rho_0=m_N\cdot n_0\simeq2.7\cdot 10^{14}$ $g\cdot cm^{-3}$ (with $m_N=939MeV$ the average nucleon mass). As this point is approached, the lattice structure of the nuclei, as well as the nuclei themselves, collapse and their components, namely neutrons and protons, are released. Pretty much, almost all of the star's mass is concentrated in the core. As for the outer part of the core, its density varies from $\simeq0.5\rho_0$ up to $2\rho_0$, while its thickness can reach several $km$. Inside this region, a composition of infinite-scale (or bulk) matter is developing: a soup of nucleons (mainly neutrons: giving the star its name), electrons and muons. The electrons and muons form a nearly ideal Fermi gas. In contrast, neutrons and protons interact with each other via strong nuclear forces, forming a strongly interactive Fermi superfluid, with potential superconducting properties. 
    
    \item \textbf{Inner core}: The mass density in this deepest layer of the neutron star can reach extremely high values, up to $10\rho_0-15\rho_0$, providing a fertile ground for the birth of exotic particles. Boson condensates, such as pions or kaons, and/or even strangeness-bearing hyperons, could be found inside the inner core. The transition to a mixed phase of hadronic and deconfined quark matter (see Fig.\ref{fig:NS_structure_2}), beyond a certain mass density is, also, a possible scenario. As of today, the exact composition of matter in the core, is still under research.
    
\end{itemize}

\section{Thermodynamics and the Equation of State (EoS)}
In Appendix \ref{Hydrostatic Equilibrium}, we showed how the hydrostatic equilibrium requirement in massive stars led from Einstein's equations to three independent differential equations and subsequently to the TOV equations. In these, four independent quantities are present: the two metric functions $\alpha(r)$ and $\beta(r)$ (or $m(r)$), the energy density $\epsilon(r)$ and the pressure $P(r)$. Therefore a fourth independent equation has to be added to complete the system of equations (see Eqs. \ref{ind_diff_eq1}, \ref{ind_diff_eq2} and \ref{ind_diff_eq3}). The missing equation is found in the relation between the pressure and energy density \cite{schaffner2020compact}:
\begin{equation}\label{EOS_gen_form}
    P=P(\epsilon)
\end{equation}
the so-called \textbf{equation of state} (or just \textbf{EoS}). This type of EoS is called \textit{barotropic} equation of state.

In principle, the EoS correlates all independent thermodynamical quantities of the system \cite{kanakis2019constraints,kourmpetis2024nuclear}: pressure $P$, energy density $\epsilon$, temperature $T$ and number of particles $N$. The existence of cold neutron matter inside the star, with $T=0$, eliminates the dependency on temperature. Assuming the conservation of baryon number, the dependency on the number of particles is also eliminated. Thus, for a fixed $N$, an equation of state that incorporates baryonic matter, results in having two independent variables in the star's interior: the pressure $P$ and the energy density $\epsilon$.

In this dissertation, the matter of the neutron star can be well described by a perfect fluid in equilibrium. As the \textit{First Law of Thermodynamics} states, for a fluid element constituting of $N$ baryons, the total energy $E$, including the rest-mass energy of the fluid element, is \cite{lackey2012neutron}:
\begin{equation}\label{1st_thermo_law_diff}
    dE=-PdV+TdS+\mu dN
\end{equation}
where $P$ is the pressure, $V$ the volume, $T$ the temperature, $S$ the entropy and $\mu$ the baryon chemical potential. The baryon chemical potential is defined as the increase in energy when
a baryon is added to the fluid element, and this includes the energy needed to, for example, add other particles to conserve charge.

The last term in Eq. \ref{1st_thermo_law_diff} can be eliminated, if we introduce the \textit{Gibbs} free energy: $G=E+PV-TS$. Then Eq. \ref{1st_thermo_law_diff} gives:
\begin{equation}\label{Gibbs_free_enrg}
    G=\mu N
\end{equation}
and the last term of Eq. \ref{1st_thermo_law_diff}, can be written, in terms of the rest mass of the fluid element $M_0$ and the specific Gibbs free energy $g=G/M_0=\mu/m_B$, as:
\begin{equation}\label{chem_pot_term}
    \mu dN = \frac{\mu}{m_B}dM_0=gdM_0
\end{equation}
where $m_B = 1.66\cdot10^{-24} g$ is the baryon rest mass. Treating the number of baryons $N$ as a constant, as mentioned before, means that the rest mass $M_0$ is also conserved, and its term therefore becomes equal to zero.

The first law can be rewritten in terms of only intensive quantities \cite{lackey2012neutron}:
\begin{equation}\label{1st_thermo_law_intens}
    d\frac{\epsilon}{\rho}=-Pd\frac{1}{\rho}+Tds
\end{equation}
where $n=N/V$ is the baryon number density, $\rho=M_0/V=m_Bn$ is the rest mass density, $\epsilon=E/V$ the energy density and $s=S/M_0$ the specific entropy. Equivalently we have:
\begin{equation}\label{1st_thermo_law_intens2}
    d\epsilon = hd\rho+\rho Tds
\end{equation}
where $h$ the specific enthalpy:
\begin{equation}\label{spec_enthalpy}
    h=\frac{E+PV}{M_0} = \frac{\epsilon+P}{\rho}
\end{equation}

Since we considered temperatures far below the Fermi temperature, the EoS is one-parameter dependent:
\begin{equation}\label{EOS_gen_form2}
\begin{matrix}
P=P(\rho), &
\epsilon=\epsilon({\rho})
\end{matrix}
\end{equation}
The expressions in Eq. \ref{EOS_gen_form2} are not independent, because the quantities $P,\epsilon,\rho$ are correlated to each other through the first law (Eq.\ref{1st_thermo_law_intens2}), with $ds=0$. Hence, when a relation is specified between the two of the three quantities, the third one is obtained from one of the following relations \cite{lackey2012neutron}:
\begin{equation*}
    P=\rho^2\frac{d(\epsilon/\rho)}{d\rho}
\end{equation*}
\begin{equation}\label{1st_thermo_law_relations}
   \frac{\epsilon}{\rho}=\frac{\epsilon_0}{\rho_0}+\int\limits^\rho_{\rho_0}\frac{P}{\rho^{'2}}d\rho^{'} 
\end{equation}
\begin{equation*}
\rho=\rho_0\cdot exp\left(\int\limits^{\epsilon}_{\epsilon_0}\frac{d\epsilon^{'}}{\epsilon^{'}+P}\right)
\end{equation*}
where $\epsilon_0$ is the energy density for a rest mass density $\rho_0$. At the surface of the star we have $P\rightarrow{0}$, $\epsilon_0\rightarrow{0}$ and $\rho_0\rightarrow{0}$ for a standard EoS and the ratio $\epsilon_0/\rho_0\rightarrow{1}$.

\section{Starting from the surface: Crust EoSs}\label{Crust_EOSs}
In our study, the equations of state we employ are in the form:
\begin{equation}\label{EOS_gen_form3}
    \epsilon = \epsilon(P)
\end{equation}
where the energy density $\epsilon$ and the pressure $P$ are in $[MeV\cdot fm^{-3}]$ units. We also take into account the existence of crust in the neutron star. Inside this region, the equation of state is modified with respect to pressure. In particular four sub-regions are formed, each of which is governed by its own equation of state as follows \cite{kanakis2019constraints}:
\begin{itemize}
    \item $P \in (9.34375\cdot 10^{-5}, P_{crust-core}]$:
    \begin{equation}\label{eos_crust_1}
        \epsilon_{crust\_1}=\epsilon(P)=0.00873+103.17338\left(1-e^{-\frac{P}{0.38527}}\right)+7.34979\left(1-e^{-\frac{P}{0.01211}}\right) 
    \end{equation}

    \item $P \in (4.1725\cdot 10^{-8},9.34375\cdot 10^{-5}]$:
    \begin{equation}\label{eos_crust_2}
        \epsilon_{crust\_2}=\epsilon(P)=0.00015 + 0.00203\left(1-e^{-344827.5\cdot P}\right)+0.10851\left(1-e^{-7692.3076\cdot P}\right)
    \end{equation}

    \item $P \in (1.44875\cdot10^{-11},4.1725\cdot 10^{-8}]$:
    \begin{equation}\label{eos_crust_3}
        \epsilon_{crust\_3}=\epsilon(P)=0.0000051\left(1-e^{-0.2373\cdot10^{10}\cdot P}\right)+0.00014\left(1-e^{-0.4020\cdot10^8\cdot P}\right)
    \end{equation}

    \item $P\leq1.44875\cdot10^{-11}$:
    \begin{equation}\label{eos_crust_4}
        \epsilon_{crust\_4}=\epsilon(P)=10^{c_0+c_1+c_2+c_3+c_4+c_5}
    \end{equation}
    with:
    
    $c_0=31.93753$, 
    
    $c_1=10.82611\cdot\log_{10}(P)$, 
    
    $c_2=1.29312\cdot[\log_{10}(P)]^2$, 
    
    $c_3=0.08014\cdot[\log_{10}(P)]^3$, 
    
    $c_4=0.00242\cdot[\log_{10}(P)]^4$,
    
    $c_5=0.000028\cdot[\log_{10}(P)]^5$
\end{itemize}
The pressure $P_{crust-core}$ marks the transition between the crust and the core of the neutron star. Its value depends on the EoS used for the matter in the core. It is $0.696$ $MeV\cdot fm^{-3}$ for the PS equation of state and $0.184$ $MeV\cdot fm^{-3}$ for the rest of the equations we present in the following section (\ref{Core_EOSs}).

\section{Going deeper: Core EoSs}\label{Core_EOSs}
As we mentioned in section \ref{NS struct}, the composition of matter in the core of a neutron star remains unknown. This leaves the field open for creating and testing different models as equations of state. Below we present the equations of state for the core, we employ in our study, predicting neutron stars with masses subject to observational constraints \cite{demorest2010two,antoniadis2013massive}. We have:
\begin{itemize}
    \item Akmal-Pandharipande-Ravenhall (APR) model  \cite{akmal1998equation}:
    \begin{equation}\label{APR-1_EOS}
        \text{APR-1}: \epsilon(P)=0.000719964\cdot P^{1.85898}+108.975\cdot P^{0.340074}
    \end{equation}

    \item Bowers-Gleeson-Pedigo (BGP) model  \cite{bowers1975relativistic}:
    \begin{equation}\label{BGP_EOS}
        \text{BGP}: \epsilon(P)= 0.0112475\cdot P^{1.59689}+102.302\cdot P^{0.335526}
    \end{equation}

    \item Bombaci-Logoteta (BL) models \cite{bombaci2018equation}:
    \begin{equation}\label{BL-1_EOS}
        \text{BL-1}: \epsilon(P)=0.488686\cdot P^{1.01457}+102.26\cdot P^{0.355095}\text{ }
    \end{equation}
    \begin{equation}\label{BL-2_EOS}
        \text{BL-2}: \epsilon(P)=1.34241\cdot P^{0.910079}+100.756\cdot P^{0.354129}
    \end{equation}

    \item Douchin-Haensel (DH) model \cite{douchin2001unified}:
    \begin{equation}\label{DH_EOS}
        \text{DH}: \epsilon(P)=39.5021\cdot P^{0.541485}+96.0528\cdot P^{0.00401285}
    \end{equation}

    \item Heiselberg and Hjorth-Jensen (HHJ) models  \cite{heiselberg2000phases}:
    \begin{equation}\label{HHJ-1_EOS}
        \text{HHJ-1}: \epsilon(P)= 1.78429\cdot P^{0.93761}+106.93652\cdot P^{0.31715}
    \end{equation}
    \begin{equation}\label{HHJ-2_EOS}
        \text{HHJ-2}: \epsilon(P)=1.18961\cdot P^{0.96539}+108.40302\cdot P^{0.31264}
    \end{equation}

    \item Hebeler-Lattimer-Pethick-Schwenk (HLPS) models \cite{hebeler2013equation}:
    \begin{equation}\label{HLPS-2_EOS}
        \text{HLPS-2}: \epsilon(P)=161.553+172.858\left(1-e^{-\frac{P}{22.8644}}\right)+2777.75\left(1-e^{-\frac{P}{1909.97}}\right)
    \end{equation}
    \begin{equation}\label{HLPS-3_EOS}
        \text{HLPS-3}: \epsilon(P)=81.5682+131.811\left(1-e^{-\frac{P}{4.41577}}\right)+924.143\left(1-e^{-\frac{P}{523.736}}\right)
    \end{equation}

    \item Momentum-Dependent Interaction (MDI) models \cite{prakash1997composition,moustakidis2009equation}:
    \begin{equation}\label{MDI-1_EOS}
        \text{MDI-1}: \epsilon(P)=4.1844\cdot P^{0.81449} + 95.00135\cdot P^{0.31736}\text{ }\text{ }\text{ }
    \end{equation}
    \begin{equation}\label{MDI-2_EOS}
        \text{MDI-2}: \epsilon(P)=5.97365\cdot P^{0.77374} + 89.24\cdot P^{0.30993}\text{ }\text{ }\text{ }\text{ }\text{ }\text{ }
    \end{equation}
    \begin{equation}\label{MDI-3_EOS}
        \text{MDI-3}: \epsilon(P)=15.55\cdot P^{0.666}+76.71\cdot P^{0.247}\text{ }\text{ }\text{ }\text{ }\text{ }\text{ }\text{ }\text{ }\text{ }\text{ }\text{ }\text{ }\text{ }
    \end{equation}
    \begin{equation}\label{MDI-4_EOS}
        \text{MDI-4}: \epsilon(P)=25.99587\cdot P^{0.61209}+65.62193\cdot P^{0.15512}
    \end{equation}

    \item Non Linear Derivative (NLD) model \cite{gaitanos2013momentum,gaitanos2015toward}:
    \begin{equation}\label{NLD_EOS}
        \text{NLD}: \epsilon(P)=119.05736+304.80445\left(1-e^{-\frac{P}{48.61465}}\right)+33722.34448\left(1-e^{-\frac{P}{17499.47411}}\right)
    \end{equation}

    \item Pethick-Schwenk (PS) model:
    \begin{equation}\label{PS_EOS}
        \text{PS}: \epsilon(P)= 1.69483+9805.95\left(1-e^{-0.000193624\cdot P}\right)+212.072\left(1-e^{-0.401508\cdot P}\right)
    \end{equation}

    \item Sharma-Centelles-Viñas-Baldo-Burgio (SCVBB) model \cite{sharma2015unified}:
    \begin{equation}\label{SCVBB_EOS}
        \text{SCVBB}: \epsilon(P)=0.371414\cdot P^{1.08004}+109.258\cdot P^{0.351019}
    \end{equation}

    \item Skyrme (Sk) models \cite{chabanat1997skyrme,farine1997nuclear}:
    \begin{equation}\label{Ska_EOS}
        \text{Ska}: \epsilon(P)=0.53928\cdot P^{1.01394}+94.31452\cdot P^{0.35135}
    \end{equation}
    \begin{equation}\label{SkI4_EOS}
        \text{SkI4}: \epsilon(P)=4.75668\cdot P^{0.76537}+105.722\cdot P^{0.2745}\text{ }
    \end{equation}

    \item Walecka (W) model \cite{walecka1974theory}:
    \begin{equation}\label{W_EOS}
        \text{W}: \epsilon(P)=0.261822\cdot P^{1.16851}+92.4893\cdot P^{0.307728}
    \end{equation}

    \item Wiringa-Fiks-Fabrocini (WFF) models \cite{wiringa1988equation}:
    \begin{equation}\label{WFF-1_EOS}
        \text{WFF-1}: \epsilon(P)=0.00127717\cdot P^{1.69617}+135.233\cdot P^{0.331471}
    \end{equation}
    \begin{equation}\label{WFF-2_EOS}
        \text{WFF-2}: \epsilon(P)=0.00244523\cdot P^{1.62692}+122.076\cdot P^{0.340401}
    \end{equation}
\end{itemize}

that is $21$ different EoS models. From now on we will refer to them as 'main' EoSs.

\section{Artificial generation of equations: polytropic and linear EoSs}\label{PolyLin_EOSs}
Employing the $21$ 'main' EoSs, of the relations \ref{APR-1_EOS}-\ref{WFF-2_EOS}, would suffice in a study of observational constraints \cite{kourmpetis2024nuclear} or tidal deformability \cite{kanakis2019constraints} in neutron stars. Indeed, these EoSs are quite different from each other. However, in our analysis, beyond the diversity of the equations of state, the dense coverage of a large area in the M-R plane, is also a key point, as it secures the production of a large amount of data, as well as the generality of the regression models we will develop. To meet these requirements, one has to follow a systematic way, artificially generating a sufficiently large number of different equations of state. 

In this dissertation, we resort in the use of polytropic equations of state. A region of mass density, between two boundaries $\rho_{min}$ and $\rho_{max}$, needs to be chosen and divided into $n$ segments. The EoS is then parametrized in terms of $n$ piecewise polytropes. Setting the values of mass density and pressure at polytropic segments' bounds as $\rho_i$ and $P_i$, respectively, each segment is given by \cite{raithel2016neutron}:
\begin{equation}\label{press_poly}
\begin{matrix}
    P=K_i\rho^{\Gamma_i} & (\rho_{i-1}\leq\rho\leq\rho_i)
\end{matrix}
\end{equation}
where the value of the constant $K_i$, is determined from the pressure and mass density at the previous fiducial point as follows:
\begin{equation}\label{K_poly}
    K_i=\frac{P_{i-1}}{\rho^{\Gamma_i}_{i-1}}=\frac{P_i}{\rho^{\Gamma_i}_i}
\end{equation}
and the polytropic index of the segment $\Gamma_i$, is given by:
\begin{equation}\label{Gamma_poly}
    \Gamma_i=\frac{\log_{10}(P_i/P_{i-1})}{\log_{10}(\rho_i/\rho_{i-1})}
\end{equation}
The value of $\Gamma_i$ at each segment is usually arbitrarily chosen. Furthermore, for a given number $l$ of possible choices for $\Gamma_i$ and a certain number of $n$ polytropic segments, one can produce:
\begin{equation}\label{number_of_polyEoSs}
    f=l^n
\end{equation}
differently parametrized EoSs.

The formulas of these EoSs are obtained by integrating Eq. \ref{1st_thermo_law_intens}, for $ds=0$ and $\Gamma_i\neq1$ to \cite{raithel2016neutron}:
\begin{equation}\label{poly_EoS_Gn1_form}
    \epsilon(\rho)=(1+a)\rho c^2+\frac{K}{\Gamma_i-1}\rho^{\Gamma_i}
\end{equation}
where $a$ is an integration constant. The value of $a$ is determined by requiring the continuity of the EoS along any mass density section at either endpoint and Eq. \ref{poly_EoS_Gn1_form} becomes:
\begin{equation}\label{poly_EoS_Gn1_gen_form}
    \epsilon(\rho)=\left[\frac{\epsilon(\rho_{i-1})}{\rho_{i-1}}-\frac{P_{i-1}}{\rho_{i-1}(\Gamma_i -1)}\right]\rho + \frac{K_i}{\Gamma_i-1}\rho^{\Gamma_i},\text{ }\text{ } (\rho_{i-1}\leq\rho\leq\rho_i)
\end{equation}
where $K_i$ and $\Gamma_i$ are calculated as in Eq. \ref{K_poly} and Eq. \ref{Gamma_poly}, respectively. In the same way, integrating Eq. \ref{1st_thermo_law_intens}, for $ds=0$ and $\Gamma_i=1$, gives \cite{raithel2016neutron}:
\begin{equation}\label{poly_EoS_G1_gen_form}
    \epsilon(\rho)=\frac{\epsilon(\rho_{i-1})}{\rho_{i-1}}\rho + K_i\ln\left(\frac{1}{\rho_{i-1}}\right)\rho-K_i\left(\frac{1}{\rho}\right)\rho, \text{ }\text{ } (\rho_{i-1}\leq\rho\leq\rho_i)
\end{equation}

We can rewrite Eqs. \ref{poly_EoS_Gn1_gen_form} and \ref{poly_EoS_G1_gen_form}, with the energy density being a function of pressure $\epsilon(P)$, to align with the expressions in Eqs. \ref{eos_crust_1}-\ref{WFF-2_EOS}. To do so, we use the polytropic relation between the pressure and mass density from Eq. \ref{press_poly}. For $\Gamma_i\neq1$, we have:
\begin{equation}\label{poly_EoS_Gn1_gen_form2}
    \epsilon(P)=\left[\frac{\epsilon(\rho_{i-1})}{\rho_{i-1}}-\frac{P_{i-1}}{\rho_{i-1}(\Gamma_i -1)}\right]\left(\frac{P}{K_i}\right)^{\Gamma_i^{-1}}+\frac{P}{\Gamma_i-1}, \text{ }\text{ } (P_{i-1}\leq P\leq P_i)
\end{equation}
and for $\Gamma_i=1$
\begin{equation}\label{poly_EoS_G1_gen_form2}
    \epsilon(P)=\frac{\epsilon(\rho_{i-1})}{\rho_{i-1}}\frac{P}{K_i}+\ln\left(\frac{1}{\rho_{i-1}}\right)P-\ln\left(\frac{K_i}{P}\right)P, \text{ }\text{ } (P_{i-1}\leq P\leq P_i)
\end{equation}.

Now, the equations of state are subject to the condition of causality. Derived from special relativity, causality dictates that the pressure gradient of any EoS has an upper bound of \cite{raithel2016neutron}:
\begin{equation}\label{caus_limit}
    \frac{dP}{d\epsilon}\equiv\left(\frac{c_s}{c}\right)^2\leq1
\end{equation}
known as the causality limit. That is, the local speed of sound $c_s$ should not exceed the speed of light $c$.

It is quite possible, though, that some of the mock polytropic EoSs we create, violate the condition of Eq. \ref{caus_limit}, after a certain value of mass density $\rho_{tr}$ (or equivalently pressure $P_{tr}$). In this cases, we assume the transition from the polytropic parametrized EoS to an EoS with linear behavior, at pressure $P_{tr}$ and beyond. A \textit{Maxwell} construction is well-suited in describing this kind of transitions \cite{laskos2025speed}:
\begin{equation}\label{Maxwell_construct}
    \epsilon(P) = \begin{cases}
    \epsilon_{Hadron}(P),&P\leq P_{tr} \\
    \epsilon(P_{tr}) +\Delta\epsilon+(c_s/c)^{-2}(P-P_{tr}), &P>P_{tr}
    \end{cases}
\end{equation}
where $\epsilon_{Hadronic}(P)$ in the first line of Eq. \ref{Maxwell_construct} stands for the hadronic phase before the transition, governed by a continuous EoS (polytropic or other), and the second line refers to the maximally stiff high density phase. Notice, that there is no mixed phase region (as in \textit{Gibbs} construction) and that the two phases co-exist only at the phase transition pressure $P_{tr}$. Lastly, the term $\Delta\epsilon$ establishes the discontinuity of the energy density and hence the discontinuity of the total EoS $\epsilon(P)$.

\chapter{Quark Stars}\label{QS Theory}
\section{Quantum Chromodynamics (QCD)}
In contrast to the complexity in neutron stars structure, consisting of different hadronic phases, nucleons and (perhaps) hyperons (see section \ref{NS struct}), quark stars configurations contain quark matter at high densities. To this day, quarks, as well as leptons, are in principle elementary particles, being the building blocks of heavier and more complex particles. The baryons, like nucleons and hyperons (baryons containing a strange quark), feature three quark, while the mesons, like pions and kaons, feature a quark-antiquark pair. Both baryons and
mesons belong to the general category of hadrons, compositions of particles that interact via the strong force. Therefore, the study of the properties of quark stars should start from the fundamental principles of quantum field theory, underlying strong interactions, which are summarized as quantum chromodynamics (or simply QCD) \cite{schaffner2020compact}.

Quarks interact with each other by exchanging gluons. Both quarks and gluons are carriers of color charge. As QCD states, gluons also interact with each other, as they possess a color charge, suggesting that there are nonlinear interactions in the theory. From this perspective, QCD is rendered as a non-abelian theory, resulting in intriguing phenomena, specifically the one of confinement: the quarks are confined within the boundaries of the hadrons, notably the baryons and mesons, in such a way as to ensure overall color neutrality. That is, the overall color of the components - quarks and gluons - cancels out, allowing only color-neutral hadronic states to be observed from an external viewpoint.

Moreover, QCD interactions turn out to be short-range, with an intrinsic scale that corresponds to the size of hadrons, measured at approximately $1 fm$. Applying the uncertainty principle, this
represents an energy scale of about $200 MeV$, the scale that is commonly associated with QCD. Lastly, we denote that, gluons are massless, whereas quarks have nonzero masses. Their respective charges and masses are presented in Table \ref{tab:quark-properties} below.
\begin{table}[h]
\centering

a) Light Quarks

\begin{tabular}{|c|c|c|c|c|c|c|}
\hline
\hline
Quark Flavor & Up (u) & Down (d) & Strange (s) \\
[0.1cm]
\hline
Charge & $+2/3$ & $-1/3$ & $-1/3$ \\
[0.15cm]
Mass & $2.2^{+0.5}_{-0.4}$ $MeV$ & $4.7^{+0.5}_{-0.3}$ $MeV$ &
$95^{+9}_{-3}$ $MeV$\\
\hline
\hline
\end{tabular}

\vspace{0.1cm}
b) Heavy Quarks

\begin{tabular}{|c|c|c|c|c|c|c|}
\hline
\hline
Quark Flavor & Charm (c) & Bottom (b) & Top (t) \\
[0.1cm]
\hline
Charge & $+2/3$ & $-1/3$ & $+2/3$ \\
[0.2cm]
Mass & $1.275^{+0.025}_{-0.035}$ $GeV$ & $4.18^{+0.04}_{-0.03}$ $GeV$ & $173.0^{+0.4}_{-0.4}$ $GeV$\\
\hline
\hline
\end{tabular}

\caption{Charges and masses of a) light quarks and b) heavy quarks from the standard model \cite{tanabashi2018review}.}
\label{tab:quark-properties}
\end{table}

\section{Condensates of Quarks and Gluons}\label{condens_quark_gluon}
Although a large number of conducted experiments have validated the predictions of QCD, the phenomenon of confinement remains a mystery in search of interpretation. Another important feature of QCD is that the total mass of a nucleon, is not derived simply as the sum of the individual masses of the quarks, of which it is composed. Indeed, the typical mass of a nucleon is roughly $1GeV$, while the masses of three up/down quarks add up at a total mass of several $MeV$’s! A similar case, is observed in hyperons, as well. Thus, the origins of the nucleon and hyperon masses, need to be traced to the gluonic contribution. This finding raises two
important speculations \cite{schaffner2020compact}:
\begin{itemize}
    \item At a first lowest order of approximation, one can assume the masses of light quarks to be negligible. The QCD Lagrangian without a quark mass term features a symmetry, called chiral symmetry\footnote{Chirality is an abstract quantum property that relates closely to helicity. The behavior of the latter differs, depending on wether the particle has a nonvanishing mass or no. A massless particle, traveling at the speed of light, exhibits no frame in which is at rest ("rest frame"), so no one can catch up to it. This leads to a fixed value for its helicity, in all reference frames. On the contrary, a particle with mass has no fixed value for its helicity, since observations done from different reference frames, can result in different values for the helicity (left- or right-helicity). Chirality is an inherent property of the particle, equivalent to helicity in the massless limit and features a specific value for each particle in all valid reference frames \cite{kourmpetis2024nuclear}.}: ”\textit{Left and right-handed quarks cannot be distinguished in a massless theory}”.

    \item Being the dominant contributors to the masses of hadrons (nucleons and hyperons), gluons form a condensate that permeates space-time and acts as a background to the energy density of the hadron. Since gravity is coupled to energy, the energy density contributed by the gluon condensate, appears as the hadron's mass to an external observer.
\end{itemize}

A second condensate related to the quarks is also present: the quark condensate. This condensate occurs from the nonvanishing vacuum expectation value of the quark fields and quarks inside hadrons are coupled to it, achieving a constituent
quark mass, which is about one-third of the nucleon mass, that is roughly $300MeV$, for the light quarks. As it seems, the quark condensate acts as a mass term for the quarks inside hadrons. Due to the existence of this condensate, the interactions between left-handed and right-handed quarks are getting energetically favored over maintaining the chiral symmetry. Consequently, left-handed and
right-handed quarks couple to each other, breaking spontaneously the chiral symmetry of QCD. The \textit{Goldstone} theorem reads, that the breaking of a continuous symmetry is followed by the emission of \textit{Goldstone} bosons, which in principle are massless. In this case, though, the chiral symmetry breaks also explicitly due to the small masses of the light quarks, leading to the emission of bosons with small nonzero masses. Hence, these bosons are called \textit{pseudo-Goldstone} bosons and are associated with pions.

Now, solid-state physics dictates the melting of condensates beyond a critical temperature. Both gluon and quarks condensates do not deviate from this rule. The melting of the gluon condensate can be linked to the transition from confined situations
of the hadrons to a deconfined state of quarks and gluons, the \textbf{quark-gluon plasma}. On the contrary, the melting of the quark condensate plays a significant role to the restoration of chiral symmetry. The solution of the QCD Lagrangian performed on supercomputers, using the lattice gauge, estimates the critical temperature of QCD to be around $T_c \approx 150MeV$. In principle,
the condensates should also melt at high densities, but the value of such a critical density is still under research.

\section{Asymptotic Freedom}\label{asymp_free}
The main feature of QCD at high temperature, high density or high energy is the asymptotic freedom \cite{schaffner2020compact}. As it turns out, the physical processes involved, significantly affect the interacting strength between quarks and gluons. The continuous increase in energy, leads to the reduction and ultimately the elimination of the coupling constant, at infinitely high energies. Having reached this point, the quarks cease to interact with each other and become asymptotically free. The information about the strength of the interactions in QCD is commonly incorporated using the formalism of the QCD fine-structure constant:
\begin{equation}\label{QCD_constant}
    a_s=\frac{g^2}{4\pi}
\end{equation}
where $g$ is the coupling constant between quarks and gluons\footnote{Notice the similarity with the formalism of fine-structure constant in quantum electrodynamics: $a_e=\frac{e^2}{4\pi}$, where the coupling constant $e$ is the electromagnetic charge.}.

\begin{figure}[htb]
    \centering
    \includegraphics[width=0.7\linewidth]{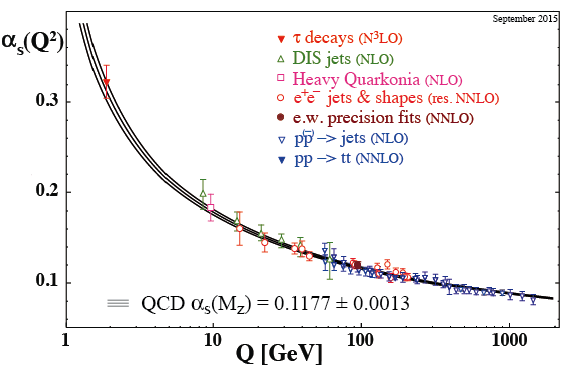}
    \caption{Summary of measurements of $a_s$ as a function of the energy scale $Q$. The respective degree of QCD perturbation theory used in the extraction of $a_s$ is indicated in brackets (NLO: next-to-leading order; NNLO: next-to-next-to leading order; res. NNLO: NNLO matched with resummed next-to-leading logs; N$^3$LO: next-to-NNLO). Figured adapted from \cite{d2015high} and \href{https://www.researchgate.net/figure/Summary-of-measurements-of-a-s-as-a-function-of-the-energy-scale-Q-The-respective-degree_fig2_287249926}{www.researchgate.net}.}
    \label{fig:QS_interact_strength}
\end{figure}

In Fig. \ref{fig:QS_interact_strength}, the measured values
of $a_s$ are depicted as a function of the energy scale of the experimental reaction probing QCD. The drawn line
that passes through the experimental data points, is the prediction of QCD, which was obtained by a single measured value of $a_s$ at a specific energy scale. It is proven, then, that the coupling strength decreases at a significant rate, as the energy level increases. At the lowest energy, the decay of the $\tau$-lepton gives the value of $a_s(1.8GeV)=0.3$, while at the highest energy, $a_s$ decreases to about $a_s(100GeV)=0.1$. For direct comparison, the fine-structure constant of quantum
electrodynamics is about $a_s \approx 1/137$, that is two orders of magnitude smaller. Thus, QCD interactions are much stronger than electromagnetic ones.

\section{The MIT Bag Model}\label{MIT_bag_model}
In the 1970s, a group at Massachusetts Institute for Technology (MIT) in Boston, USA, came up with a different perspective, as for the illustration of hadrons in terms of quarks and gluons. In this new approach, one considers a collection of quarks, carrying all the relevant quantum numbers of the hadron, namely the baryon number, charge, isospin, quark flavor (strangeness, charm, beauty, etc.), as well as spin and parity.

\begin{wrapfigure}{r}{0.5\textwidth}
  \centering
  \includegraphics[width=0.45\textwidth]{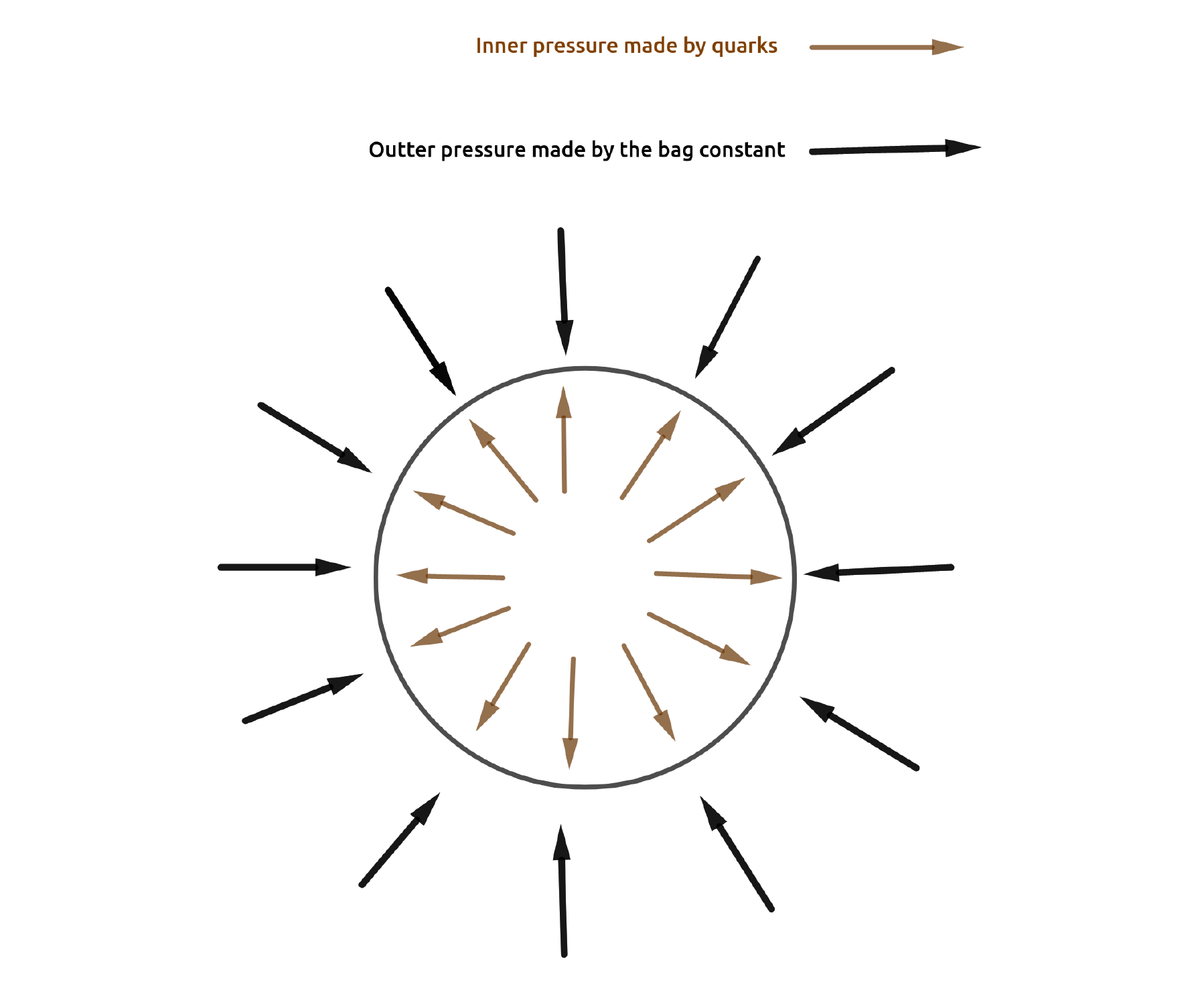}
  \caption{Schematic illustration of the MIT bag model. Figure adapted from \cite{cesario2023vector}.}
    \label{MIT_bag_model_fig}
\end{wrapfigure}

A boundary condition was applied, to ensure the coherence of the quarks, preventing them to escape. This boundary condition, was successful in simulating confinement: the quarks are confined within a sphere of determined radius. To ensure the stability of this construction, the research group incorporated an equilibrium between the pressure arising from the kinetic energy of the quarks and an external pressure applied to the surface of the hadron. This external pressure is inwardly directed, as shown in Fig. \ref{MIT_bag_model_fig}, leading the hadron to experience a negative pressure from the vacuum surrounding it. Eventually, the total pressure vanishes at the boundary of the hadron. This formation of confined quarks was called the 'bag', while the external pressure, being a crucial parameter of the model, is known as the bag parameter ($B$). The typical values of this parameter are of the order of $B^{1/4}$ $\sim150MeV$ or $B\sim66$ $MeV\cdot fm^{-3}$. In terms of the energy density
of nuclear matter at saturation density (see section \ref{NS struct}), this translates to $0.4-2.8\epsilon_0$. It should not be unexpected, that these scales align with those of nuclear matter, since QCD serves as the fundamental theory, dictating both the behavior of of the nuclear interaction and the hadron mass spectrum.

At a first glance, one might wonder about the presence of negative pressure in the MIT bag model. Is there any relation with the energy density of the gluon condensate, we mentioned at section \ref{condens_quark_gluon}? In fact, as thermodynamic consistency demands, a constant vacuum energy density should be associated with a constant negative vacuum pressure of equal magnitude. Let's just write the \textit{First Law of Thermodynamics}:
\begin{equation}\label{1st_thermo_law_integr}
    \epsilon = -P+T\cdot s + \mu \cdot n
\end{equation}
and consider the vacuum case of vanishing temperature $T$ and number density $n$. This gives:
\begin{equation}\label{1st_thermo_law_vacuum}
    \epsilon_{vac}=-P_{vac}=B
\end{equation}
Therefore, the bag parameter $B$ relates to the energy density of the gluon condensate in vacuum. Considering Eq. \ref{1st_thermo_law_vacuum}, the total pressure reads:
\begin{equation}\label{total_press_quark}
    P = P_{quarks}+P_{vac}=P_{quarks}-\epsilon_{vac}
\end{equation}
Now, the missing part is a connection between the pressure and the energy density of quarks. Assuming non-interacting quarks with vanishing mass, this relation would resemble that of a relativistic gas ($P=\epsilon/3$). Hence, the total energy density can be written as:
\begin{equation}\label{total_enrg_quark}
    \epsilon = \epsilon_{quarks}+\epsilon_{vac}=3P_{quarks}+\epsilon_{vac}
\end{equation}
Substituting Eq. \ref{total_enrg_quark} to Eq. \ref{total_press_quark} we get:
\begin{equation}\label{MIT_EOS_press}
    P=\frac{1}{3}(\epsilon-4\epsilon_{vac})=\frac{1}{3}(\epsilon-4B)
\end{equation}
The relation in Eq. \ref{MIT_EOS_press} is the EoS of quark matter derived from the MIT bag model.

\section{Selfbound stars and constraints}\label{Selfbound Stars}
Getting a closer look at the EoS of the MIT bag model, reveals an intriguing property: the pressure vanishes at a nonvanishing value of the energy density\footnote{Notice that neutron star matter has a vanishing energy density at zero pressure}. This brings us on the verge of the discovery of a new class of stars, selfbound stars \cite{schaffner2020compact}:
\begin{itemize}
    \item \textbf{Selfbound Stars}: "\textit{Compact star configurations with an EoS where the energy
density has a nonvanishing value at zero pressure}".
\end{itemize}

If bulk strange quark matter is selfbound, then spheres comprising strange quark matter are bound by virtue of the vanishing total pressure. The gravitational attraction is not needed for hydrostatic equilibrium. In fact, gravity only defines a
limit on the maximum mass of the quark matter sphere. The energy density in the interior of these bound spheres of strange quark matter is constant, obtained by the vacuum energy density. Therefore, selfbound stars feature a mass–radius relationship of a sphere with fixed energy density. That is:
\begin{equation}\label{mass_radius_relate}
    M\propto R^3
\end{equation}
We denote that these spheres can be arbitrarily tiny in size, thus the mass–radius relationship for selfbound stars originates from zero. In other words, its possible for selfbound stars to possess extremely small radii. This characteristic, is what distinguishes
them from neutron stars for
small star masses. At the limit of small masses, the radius of neutron stars increases, while the radius of selfbound stars gets smaller\footnote{However, for strange stars with crust, the total radius gets bigger with decreasing mass, like in neutron stars.}.
The main differences between selfbound stars and neutron stars are summarized in Table \ref{tab:QS_vs_NS} below:
\begin{table}[h]
\centering

\begin{tabular}{|p{5cm}|p{5.5cm}|p{5.5cm}|}
\hline
\hline
 & \textbf{Selfbound Stars} & \textbf{Neutron Stars} \\
[0.1cm]
\hline
EOS & Vanishing pressure at nonzero
energy density & Vanishing pressure only at
vanishing energy density \\
[0.15cm]
Stability & Bound by interactions & Bound by gravity \\
[0.15cm]
Mass–radius relation & Starts at the origin (without
a crust) & Starts at large radii for small
masses \\
[0.15cm]
Minimum mass & Arbitrarily small masses and
radii possible & $M\approx 0.1 M_\odot$ at $R=200$ $km$ \\
\hline
\hline
\end{tabular}

\caption{Comparison of the properties of selfbound stars and neutron stars \cite{schaffner2020compact}}
\label{tab:QS_vs_NS}
\end{table}

The microscopic quark spheres made of strange quark matter are commonly referred to as 'strangelets'. Their existence is
hypothetical, since their stability depends on the binding energy of strange quark matter compared to nuclear matter. Strange quark matter poses as the best candidate for absolutely stable matter,
implying that nuclear matter would not be the ground state of matter. One might wonder why we don’t observe totally
stable matter. The answer lies in the fact that the timescale of the decay of nuclear matter is actually much longer than the age of the universe. So, strange matter needs to satisfy two conditions:
\begin{enumerate}
    \item The absence of strange quarks from droplets of quark matter leads to configurations that are less stable than nuclei, so nuclei cannot decay to quark droplets via strong interactions.
    \item Quark matter involving strange quarks is more stable than nuclear matter, preventing its decay to nuclei.
\end{enumerate}

The first of the conditions can be expressed in a form, such that the binding energy per baryon number (or mass number) of two-flavor quark matter droplets of up and down
quarks is higher than the total energy of the most stable nucleus, $^{56}Fe$. We also include, a correction in energy of $4MeV$, derived from the surface term for the quark matter blob  \cite{schaffner2020compact}, resulting in the following two-flavor constraint for bulk matter:
\begin{equation}\label{quark_constraint1}
    \frac{E_{bulk}}{A}\bigg|_{N_f=2} >m_N-\frac{E_b}{A}\bigg|_{^{56}Fe}
    + \frac{E_{surf}}{A}\bigg|_{N_f=2}\approx 934\text{ }MeV
\end{equation}
As for the second condition, it indicates that three-flavor quark matter in bulk is more stable than $^{56}Fe$, so the constraint arising for three-flavor quark matter reads:
\begin{equation}\label{quark_constraint2}
    \frac{E_{bulk}}{A}\bigg|_{N_f=3}<m_N - \frac{E_b}{A}\bigg|_{^{56}Fe}\approx 930\text{ }MeV
\end{equation}

Now, we can incorporate the constraints of Eqs. \ref{quark_constraint1} and \ref{quark_constraint2} into the MIT bag model for three massless quark flavors. The \textit{Hugenholtz–van Hove theorem}\footnote{The Hugenholtz–van Hove theorem states that the Fermi energy $E_F$ of a Fermi gas
with interactions at $T=0$ equals to $E_F=d\epsilon/dn_B=(\epsilon+P)/n_B$, that is the average energy $\epsilon$ plus the pressure $P$
per particle. Using the thermodynamic relation $\epsilon = -P +\mu\cdot n$, this theorem suggests that the chemical potential
equals to the Fermi Energy, even at the presence of interactions. At saturation density $n_0$, the pressure vanishes, hence the theorem equates the binding energy in bulk with the chemical potential (see \cite{schaffner2020compact} p. 176).} 
suggests that the binding energy in bulk can be set equal to the baryon chemical potential at zero pressure:
\begin{equation}\label{HvanHove_theorem}
    \frac{E_{bulk}}{A}\bigg|_{N_f=2}=\mu_B\bigg|_{N_f=2}>934\text{ }MeV
\end{equation}
For two-flavor quark matter (up and down, $N_f=2$), the baryon chemical potential is:
\begin{equation}\label{baryon_chem_pot}
    \mu_B= \mu_u+2\mu_d
\end{equation}
while the charge neutrality condition dictates that:
\begin{equation}\label{charge_neutrality}
    q_un_u=q_dn_d\implies n_d=2n_u
\end{equation}
where $q$ and $n$ are the charge and number density of each quark flavor (it is $q_u=+2/3$ and $q_d=-1/3$, from Table \ref{tab:quark-properties}). Equivalently, we can rewrite Eq. \ref{charge_neutrality} as:
\begin{equation}\label{charge_neutrality_equiv}
    \mu_d=2^{1/3}\mu_u
\end{equation}
since $n_i=\mu^3_i/3\pi^2$, with $i$ indicating the different quark flavor. Substituting Eq. \ref{charge_neutrality_equiv} to Eq. \ref{baryon_chem_pot} gives:
\begin{equation}\label{baryon_chem_pot2}
    \mu_B=(1+2^{4/3})\mu_u
\end{equation}
When equilibrium is achieved, the pressure of quarks equals to the bag constant:
\begin{equation}\label{press_equil}
    P_{quarks} = B = \frac{1}{4\pi^2}(\mu^4_u+\mu^4_d)=\frac{1}{4\pi^2}(1+2^{4/3})\mu^4_u=\frac{1}{4\pi^2}(1+2^{4/3})^{-3}\mu^4_B
\end{equation}
Solving Eq. \ref{press_equil} for $\mu_B$ and substituting the result to Eq. \ref{HvanHove_theorem}, results in the following relation:
\begin{equation}
    \frac{E_{bulk}}{A}\bigg|_{N_f=2} = \mu_B\bigg|_{N_f=2}=\left((1+2^{4/3})^34\pi^2B\right)^{1/4}>934\text{ }MeV
\end{equation}
which yields a lower bound for the bag constant:
\begin{equation}\label{bag_low_constraint}
    B^{1/4}>145\text{ }MeV \text{ }\text{ } or \text{ }\text{ } B>57\text{ }MeV\cdot fm^{-3}
\end{equation}
Notice that we used the relation:
\begin{equation}\label{bag_units}
    B=\frac{(B^{1/4})^4}{(\hbar c)^3}
\end{equation}
for unit conversion, with $\hbar  c= 197.327$ $MeV\cdot fm$.

\section{Color Superconductivity}\label{Superconduct}

Inside color superconductive matter the color charge can be transported without any resistance, like the electric charge can travel losslessly in traditional superconductors. At the asymptotic limit of high energies and low temperatures, the strength of quark interactions reduces significantly (see section \ref{asymp_free}), leading to a Fermi surface that consists of nearly free quarks. In 1957, Bardeen, Cooper, and Schrieffer \cite{bardeen1957microscopic,bardeen1957theory}, discovered that, at the presence of an attractive interaction channel near the Fermi surface, a state with lower free energy
than that of a simple Fermi surface is formed. This state involves quarks near the Fermi surface, pairing up into the so-called 'Cooper pairs' and breaking the color gauge symmetry. This phase clearly differs from the normal phase of quark matter, which is just a Fermi gas of
weakly interacting quarks.

\section{The Color-Flavor Locked (CFL) phase of matter}\label{CFL matter}
Now, at asymptotically high densities, the
masses of quarks are negligibly small compared to the quark
chemical potential. As it is widely accepted, under these conditions, strange quark matter is most favored to enter a superfluid phase, where quarks of all three flavors and colors form Cooper pairs and have the same Fermi momentum. This results in breaking the chiral symmetry and also, prevents the presence of electrons \cite{oikonomou2023color,flores2017constraining}. This phase is known as Color-Flavor-Locked (CFL) phase. Color-flavor locking affects significantly many features of quark matter, for example transport properties.

More importantly yet, it introduces corrections in the equation of state (EoS), of order $(\Delta/\mu^2)$, which is around a few percent for typical values of the
color superconducting gap ($\Delta\sim0-150\text{ }MeV$) and the baryon chemical potential ($\mu\sim300-400\text{ }MeV$). The effect, though, is proportionally very large in the low pressure regime that affects the absolute stability of quark matter. Therefore, self-bound stars that consist entirely of quark matter, from the center up to the
stellar surface, known as \textit{strange stars}, may occur for a wide range of parameters of the MIT bag model EOS \cite{lugones2002color}. Furthermore, researches upon the
structure of these objects reveal that color superconductivity
affects significantly the mass-radius relationship of strange
stars, allowing for very large maximum masses \cite{lugones2003high,horvath2004self}.

Now, we will study the thermodynamics of the CFL phase. The equation of state for CFL quark matter can be derived in the general framework of the MIT bag model. The pressure and energy density are given, to order of $\Delta^2$ and $m_s^2$ (with $m_s$ the mass of strange quark), as follows \cite{flores2017constraining}:
\begin{equation}\label{CFL_EOS_press}
    P = \frac{3\mu^4}{4\pi^2}+\frac{9\alpha\mu^2}{2\pi^2}-B
\end{equation}
\begin{equation}\label{CFL_EOS_enrg}
    \epsilon = \frac{9\mu^4}{4\pi^2}+\frac{9\alpha\mu^2}{2\pi^2}+B
\end{equation}
where
\begin{equation}\label{alpha_cfl}
    \alpha=-\frac{m_s^2}{6}+\frac{2\Delta^2}{3}
\end{equation}
and $\Delta$ is the gap parameter, representing the contribution of color superconductivity \cite{oikonomou2023color}. Combining the equations above, we can results in an analytic expression for both $P=P(\epsilon)$ and $\epsilon=\epsilon(P)$:
\begin{equation}\label{CFL_EOS_enrg2}
    \epsilon = 3P+4B-\frac{9\alpha\mu^2}{(\hbar c)^3\pi^2}, \text{ }\text{ }\text{ }\text{ }\text{ }\text{ }\text{ }
    \mu^2=-3\alpha+\left[\frac{4}{3}\pi^2(B+P)(\hbar c)^3+9\alpha^2\right]^{1/2}  
\end{equation}
\begin{equation}\label{CFL_EOS_press2}
   P=\frac{\epsilon}{3}-\frac{4B}{3}+\frac{3\alpha\mu^2}{(\hbar c)^3\pi^2},\text{ }\text{ }\text{ }\text{ }\text{ }\text{ }\text{ }
    \mu^2=-\alpha+\left[\alpha^2+\frac{4}{9}\pi^2(\epsilon-B)(\hbar c)^3\right]^{1/2}  
\end{equation}
where we used the product $\hbar c$ (see section \ref{Selfbound Stars}), once again for unit conversion.

The absolute stability of CFL quark matter, requires the energy per baryon to be less than the neutron mass $m_n$ at vanishing pressure ($P=0$) and temperature ($T=0$). Thus, the following condition must be satisfied \cite{lugones2002color}:
\begin{equation}\label{CFL_stable_baryonenrg}
    \frac{\epsilon}{n_B}\bigg|_{P=0}=3\mu\leq m_n = 939 \text{ }MeV
\end{equation}
This result is derived directly from the shared Fermi momentum among the three quark flavors in CFL matter and is valid at $T=0$ without any approximation. Since this condition must be fulfilled at vanishing pressure points, using the second relation from Eq. \ref{CFL_EOS_enrg2}, we get \cite{flores2017constraining}:
\begin{equation}\label{bag_high_constraint}
    B<-\frac{1}{(\hbar c)^3}\left(\frac{m_s^2m_n^2}{12\pi^2}+\frac{\Delta^2m_n^2}{3\pi^2}+\frac{m_n^4}{108\pi^2}\right)
\end{equation}
The last equation defines a region in the $m_s-B$ plane on which
the energy per baryon is smaller than $m_n$ for a given $\Delta$. Equivalently, a stability window is defined at the plane $B-\Delta$, for a given mass of strange quark $m_s$. We will make use of both Eq. \ref{bag_low_constraint} and Eq. \ref{bag_high_constraint} later on chapter \ref{TOV Solve}, when we will discuss the methodology for solving the TOV equations for Quark Star EoSs.

\chapter{Machine Learning Regression}\label{ML Theory}
The reconstruction issue we address in this dissertation, requires matching multiple numerical data from the mass-radius curves with multiple numerical data from the corresponding EoSs. That is, the issue is of the multivariate multiple regression problem type. We could consider a possible solution, using classical regression techniques such as the general least-squares method or other linear methods. However, we will exploit more modern and more advanced techniques of machine learning (ML), focusing on their ability to solve such problems. In particular, we will explore the capabilities and limitations of the following four algorithms: the \textit{Decision Tree}, the \textit{Random Forest}, the \textit{Gradient Boosting} and the \textit{Extreme Gradient Boosting} (or \textit{XGBoost} for short). In this chapter, we briefly present the properties and operation of each of them.

\section{Decision Trees}\label{dtree_reg}
Decision trees for regression are called \textit{Regression Trees} and fall under the general category of tree-based algorithms. Their operation is based on the division of the data into subsets (branches, nodes, and leaves), in such a way as to reduce the dispersion of target values within each subset. Geometrically speaking, the feature space is partitioned into a set of rectangles, and then a simple model (i.e. a constant) is used to fit the data inside each one of them. Predictions are typically
made from the mean of the target values in the final leaves.

\begin{figure}
    \centering
    \includegraphics[width=\linewidth]{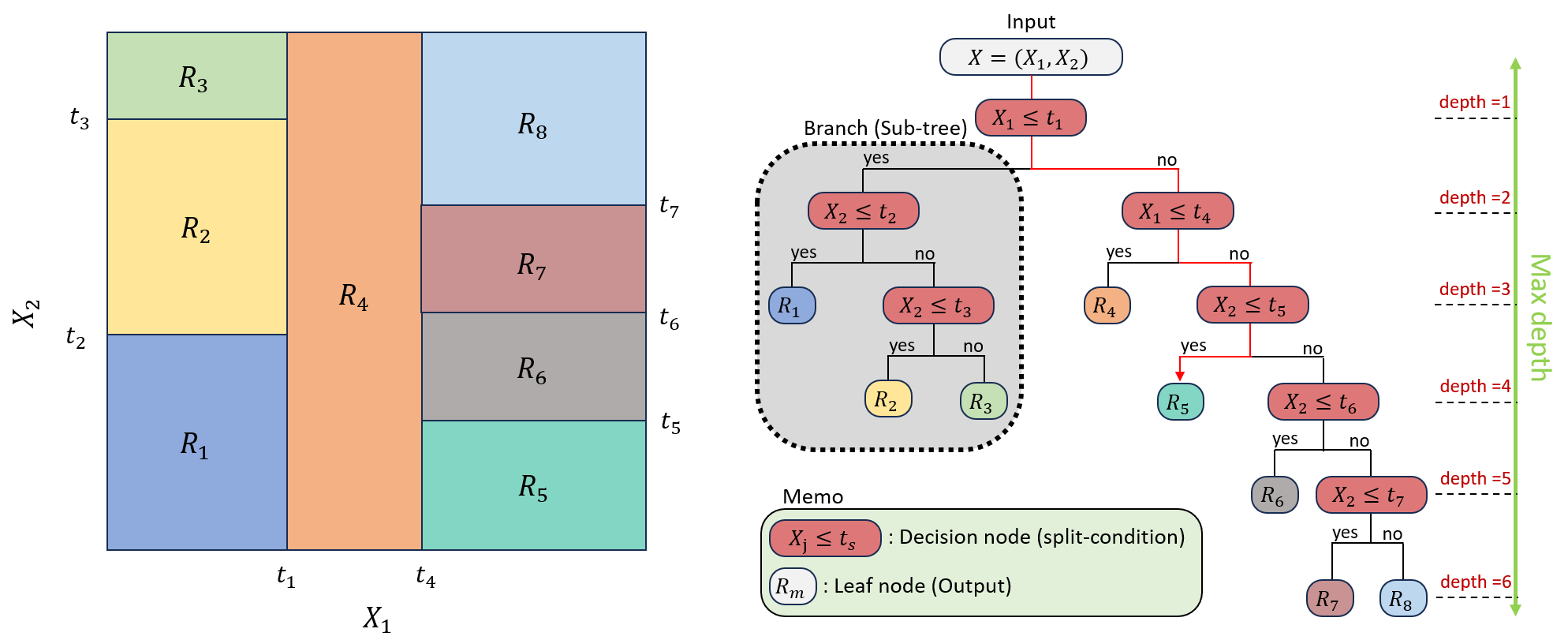}
    \caption{\textit{Left}: a partition of the
two-dimensional feature space of our mock regression problem, obtained by recursive binary splitting. \textit{Right}: the binary Decision Tree corresponding to the partition of the feature space in left.}
    \label{fig:dtree_flow}
\end{figure}

Let us assume a regression problem with a continuous response variable $Y$ and two feature variables $X_1$ and $X_2$, having the same scale of values (for simplicity we could consider both taking values in the unit interval). In the left graph of Fig. \ref{fig:dtree_flow}, a possible partition of the two-dimensional feature space is displayed, and drawing lines that are parallel to the coordinate axes. Notice, that the formed rectangles do not overlap with each other, to keep things simple. We accomplish that by using recursive binary splitting. The space is initially divided into two regions, according to the condition $X_1\leq t_1$. The region for $X_1\leq t_1$, is then divided at two points $X_2=t_2$ and $X_2=t_3$. Moreover, the region for $X_1>t_1$ is split at $X_1=t_4$. For the region $t_1<X_1\leq t_4$ ($R_4$), no further partition is applied. On the contrary, the region for $X_1>t_3$ is split at $X_2=t_5$, $X_2=t_6$ and $X_2=t_7$. This process results in a partition into eight regions $R_1,R_2,\dots,R_8$, as shown in Fig. \ref{fig:dtree_flow}. The corresponding model is fitted to predict the response by the mean of training $Y$ values in each region. We write \cite{hastie2009elements}:
\begin{equation}\label{dtree_predict_example}
    \hat{f}(X)=\sum\limits_{m=1}^8 c_mI\{(X_1,X_2)\in R_m\}
\end{equation}
where the constant $c_m$ is the prediction of $Y$.

The binary tree in the right graph of Fig. \ref{fig:dtree_flow} represents the exact same model. At the top of the tree the entire dataset in given as input. The observations that satisfy the condition at each junction (red rounded rectangles) are appointed to the left branch, while the rest to the right branch. At the terminal nodes or leaves of the tree, the eight regions $R_1,R_2,\dots,R_8$ are obtained. This is a significant advantage of the recursive binary tree, its interpretability. The
feature space partition is described in total by a single tree. In problems with more than two inputs, the visualization of the feature space becomes very difficult, since diagrams like the left one in Fig. \ref{fig:dtree_flow} are hard to adapt in higher dimensions. However, the illustration with a binary tree can be implemented exactly the same way, providing an overview of the algorithm workflow. 

Another key advantage of \textit{Decision Trees}, is that they can model complex, nonlinear relationships between features and the target variables. In the general case of single-output multivariate regression, the data involve one target variable $Y$ and $p$ feature variables $X_j$. That is, for a dataset with $N$ samples (rows), each observation is described by a vector $(y_i,x_i)$, with $i=1,2,\dots,N$ and $x_i=(x_{i1},x_{i2},\dots,x_{ip})$. The algorithm should automatically select the splitting variables and the split points, as well as the structure (topology) of the tree. For a partition of the feature space into $M$ regions $R_1,R_2,\dots,R_M$ and a modeling of the response as a constant $c_m$, Eq. \ref{dtree_predict_example} generalizes as \cite{hastie2009elements}:
\begin{equation}\label{dtree_predict_general}
    f_{tree}(X)=\sum\limits_{m=1}^M c_mI\{x\in R_m\}
\end{equation}
The tree model belongs to the space known as CART \cite{breiman2017classification}. Adopting the minimization of the sum $\sum(y_i-f(x_i))^2$ as our criterion, leads to the average of $y_i$ in region $R_m$ as the best prediction $\hat{c}_m$:
\begin{equation}\label{dtree_best_predict}
    \hat{c}_m=avg(y_i|x_i\in R_m)
\end{equation}
Finding the best binary partition, requires a careful approach to ensure computational feasibility. We start with the whole dataset and consider a splitting variable $X_j$, that splits the feature space into a pair of half-planes at point $t_s$:
\begin{equation}\label{dtree_binary_split}
    R_1(j,s)=\{X|X_j\leq t_s\} \text{ and }  R_2(j,s)=\{X|X_j> t_s\}
\end{equation}
Then we search for the splitting variable $X_j$ and split point $t_s$ that satisfy the condition \cite{hastie2009elements}:
\begin{equation}\label{dtree_binary_split_conditions}
    \min_{j,s}\left[\min_{c_1}\left( \sum\limits_{x_i\in R_1(j,s)}(y_i-c_1)^2 \right) + \min_{c_2}\left( \sum\limits_{x_i\in R_2(j,s)}(y_i-c_2)^2 \right)\right]
\end{equation}
According to Eq. \ref{dtree_best_predict}, for any choice of $X_j$ and $t_s$, the inner minimizations of Eq. \ref{dtree_binary_split_conditions} are solved by:
\begin{equation}\label{dtree_inner_min_cond}
    \hat{c}_1=avg(y_i|x_i\in R_1(j,s)) \text{ and } \hat{c}_2=avg(y_i|x_i\in R_2(j,s))
\end{equation}
For every splitting variable, the determination of the split point $s$ can be done very quickly, thus making the determination of the best pair $(j, s)$ a feasible task, by scanning through all of the inputs. After founding the best split, we move on with the partition the data inside the two resulting regions of Eq. \ref{dtree_binary_split} and repeat the splitting process on each of them. Then, this process is repeated recursively on all of the resulting regions.

This makes the \textit{Regression Trees} a versatile algorithm. However, it also reveals their main weakness. One might wonder, how large a tree has to be grown? On the one hand, a very large tree might result in overfitting (like having the $4$ regions for $X_1>t_4$ in left of Fig. \ref{fig:dtree_flow}). On the other hand, a small tree might not capture the important structure of the data. As it seems, the tree size is a tuning parameter dictating the model’s complexity, and the best tree size should be adaptively chosen from the data.

The optimal strategy is to develop a large tree $T_0$, stopping the splitting process, only when a minimum node size (say 4) is achieved. Then this large tree is pruned using \textit{cost-complexity pruning}. As sub-tree (or branch) $T$, with $T\subset T_0$, is defined any tree that can be derived by pruning $T_0$: collapsing any number of its internal (non-terminal) nodes. In right graph of Fig. \ref{fig:dtree_flow} we have marked a sub-tree, as a gray area with dashed black borders. We saw previously that index $m$ reflects the regions $R_m$. Additionally, we use $|T|$ to express the number of terminal nodes in $T$. Defining:
\begin{equation}\label{dtree_complexity_preliminaries}
\begin{aligned}
N_m &= \#\{x_i\in R_m\}, \\
\hat{c}_m &= \frac{1}{N_m}\sum\limits_{x_i\in R_m} y_i, \\
Q_m(T) &= \frac{1}{N_m}\sum\limits_{x_i\in R_m} (y_i-\hat{c}_m)^2
\end{aligned}
\end{equation}
we can obtain the cost complexity criterion \cite{hastie2009elements}:
\begin{equation}\label{dtree_complexity_criterion}
    C_a(T) = \sum\limits_{m=1}^{|T|}N_mQ_m(T) + a|T|
\end{equation}
The basic idea revolves around determining the sub-tree $T_a\subseteq T_0$, that minimizes $C_a(T)$, for every $\alpha$. The tuning parameter $a\geq0$ regulates the trade-off between the size of the tree and its goodness of fit
to the data. Large values of $a$ give smaller trees $T_a$, and small values of $a$ might result in large overfitted trees $T_a$. Finally, the notation in Eq. \ref{dtree_complexity_criterion} points that for $\alpha=0$ the solution is the full tree $T_0$. For more details about cost-complexity pruning see \cite{breiman2017classification}.

\section{Random Forest}\label{rf_reg}

As flexible and computationally light as they can be, \textit{Decision Trees}, suffer from \textit{high variance}. That is, splitting the training data into two parts randomly, and fitting a decision tree to both halves, would yield results that might be quite different. On the contrary, a process with \textit{low variance} will give similar results, if used repeatedly to distinct datasets.  An ingenious technique for treating high variance of statistical learning routines is \textit{Bootstrap aggregation} or \textit{Bagging} \cite{james2013introduction}.

For a set of $n$ independent observations $Z_1,Z_2,\dots,Z_n$, each having variance $\sigma^2$, the variance of the mean $\bar{Z}$ of the observations is given by $\sigma^2/n$. That is, \textit{averaging a set of observations reduces variance}. This property can be expanded and applied to a  statistical learning method in order to minimize its variance and increase the test dataset accuracy. One could consider $B$ different training sets from population, get the method's predictions $\hat{f}^1(x),\hat{f}^2(x),\dots,\hat{f}^B(x)$ and average them
to get a single low-variance result \cite{james2013introduction}:
\begin{equation}\label{average_predict}
\hat{f}_{avg}(x) = \frac{1}{B}\sum\limits_{b=1}^B \hat{f}_{tree}^b(x)
\end{equation}
In practice, access to a lot of training datasets may not be feasible. In this case, we can bootstrap, by considering repeated samples from the (single) training dataset, we already possess, and generate $B$ separate bootstrapped training datasets. Now, training our method on the $b$-th bootstrapped training set gives the prediction $\hat{f}^{*b}(x)$. Averaging over all predictions results in \cite{james2013introduction}:
\begin{equation}\label{bag_predict}
\hat{f}_{bag}(x) = \frac{1}{B}\sum\limits_{b=1}^B \hat{f}_{tree}^{*b}(x)
\end{equation}

\begin{figure}
    \centering
    \includegraphics[width=0.75\linewidth]{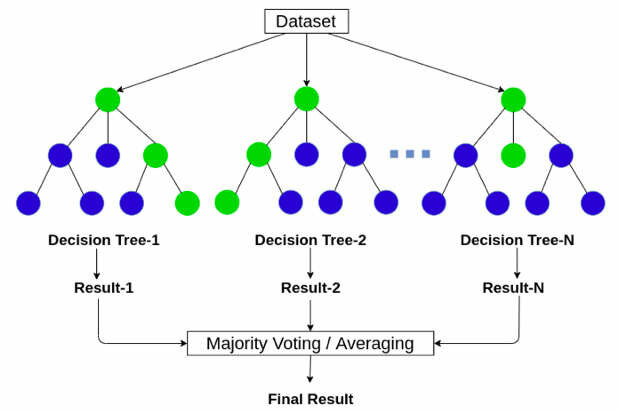}
    \caption{Structure of the Random Forest regressor algorithm. The \textit{Majority Voting} step is applied to classification problems. On the contrary, the \textit{Averaging} step is applied to regression problems, like the one we are studying in this dissertation. Figure adapted from \cite{ha2023experimental} and \href{https://www.mdpi.com/2076-3417/13/13/7660}{www.mdpi.com}.} 
    \label{fig:rf_flow}
\end{figure}

This is the \textit{Bagging} technique. Essentially, it can be a generalization of \textit{Decision Trees}, significantly improving their accuracy. One can simply develop $B$ regression trees, using $B$ bootstrapped training
sets, and calculate the average of the resulting predictions. These trees are grown deep, and they are not pruned. Each individual tree features high variance, but has low bias. Averaging over the $B$ trees reduces the variance. A single bagging process might involve the combination of hundreds or even thousands of trees, yielding some impressive improvements in accuracy and stability of predictions. However, the algorithm becomes computationally heavy.

The \textit{Random Forest} algorithm features a small tweak to further improve the concept of bagged trees, by \textit{decorrelating} the trees. An illustration of its basic structure is presented in Fig. \ref{fig:rf_flow}. As shown, the main concept of building a number of decision trees on bootstrapped training datasets remains the same. However, the separation of the full dataset into training subsets (\textit{predictors}) and the use of these to create trees, is done randomly. In other words, from a full dataset of $N$ available predictors, only a random sample of $m$ predictors is selected as split candidates, whenever a split in a tree is going to happen. The split is allowed to use only one of those $m$ predictors. At each new split a fresh sample of $m$ predictors is considered. Typically, for regression we select $m\approx N/3$ (see section 15.3 in \cite{hastie2009elements}). That is, the algorithm is not even allowed to utilize a majority of the available $N$ predictors \cite{james2013introduction}.

At a first glance, this might seem odd to the reader, but it underlines a clever rationale. Let's assume the existence of a very strong predictor in the dataset, along with a number of other less strong but good predictors. Then the final collection of bagged trees, will contain mostly or even entirely trees, that use this strong predictor in the top split. Consequently, all of the bagged trees will share a similar structure and the predictions from them will be highly correlated. Averaging many highly correlated quantities does not result in as large of a reduction in variance as averaging many uncorrelated quantities. The bagging trees perform like a single tree in this case.

\textit{Random Forests}, in general, overcome this problem by forcing each split to choose only from a subset of the predictors. Therefore, on average $(N - m)/N$ of the splits will not even include the strong predictor as a potential candidate, and so other predictors will have more of a chance to be selected. One can think of this process as decorrelating the trees, thus making the average of the resulting trees less variable and hence more reliable. Obviously, the predictor subset size $m$ is a tuning parameter, allowing one to experiment over training time and accuracy level. The final prediction of the \textit{Random Forest} for an input $x_i=(x_{i1},x_{i2},\dots,x_{ip})$, of a feature vector $\textbf{X}$, and for a use of $k$ regression trees is:
\begin{equation}\label{rf_predict}
\hat{f}_{RF}(x_i,\Theta_k) = \frac{1}{k}\sum\limits_{b=1}^k \hat{f}_{tree}^{*b}(x_i)
\end{equation}
with $\Theta$ the vector of random selected trees. Regarding the generalization prediction error ($PE^{*}$) of the forest, the latter reads:
\begin{equation}\label{rf_error}
    PE^{*}_{RF} = E_{\textbf{X},Y}[E_\Theta(Y-\hat{f}_{RF}(\textbf{X},\Theta))]^2
\end{equation}
where $E$ stands for mean value. For more details see \cite{breiman2001random}.

\section{Gradient Boosting}\label{gradboost_reg}

Except being involved in averaging models, like \textit{Random Forests}, \textit{Regression Trees} can be used as \textit{base estimators} in additive and boosting models. According to Eq. \ref{dtree_predict_general} and \cite{hastie2009elements} the structure of the tree can formally expressed as:
\begin{equation}\label{dtree_formalism}
    T(x;\Theta)=\sum\limits_{j=1}^J \gamma_jI(x\in R_j)
\end{equation}
with parameters $\Theta=\{R_j,\gamma_j\}_1^J$ and $J$ being usually a meta-parameter. The parameters are determined by minimizing the empirical risk $L$ \cite{hastie2009elements}:
\begin{equation}\label{empirical_risk_min}
    \hat{\Theta} = \arg\min_\Theta \sum\limits_{j=1}^J \sum\limits_{x_i\in R_j} L(y_i,\gamma_j)
\end{equation}

We discussed in section \ref{dtree_reg}, that this is an unfeasible computational problem and approximations are usually developed to address it. It is optimal to split this problem into two parts:
\begin{itemize}
    \item \textbf{Finding $R_j$}: This is the difficult part. Typically a greedy strategy is used, incorporating a top-down recursive partitioning
    algorithm for the determination of $R_j$. Additionally, Eq. \ref{empirical_risk_min} is often being approximated by a smoother and more convenient criterion for the optimization of $R_j$:
    \begin{equation}\label{empirical_risk_approx}
        \tilde{\Theta} = \arg\min_\Theta  \sum\limits_{i=1}^N \tilde{L}(y_i,T(x_i,\Theta))
    \end{equation}

    \item \textbf{Finding $\gamma_j$ from given $R_j$}: Having an estimation for $R_j$: $\hat{R}_j=\tilde{R}_j$, estimating $\gamma_j$ is typically trivial and the original criterion of Eq. \ref{empirical_risk_min} can be used, for more accurate results. Often, $\hat{\gamma}_j=\bar{y}_j$, that is the mean of the $y_i$ falling in region $R_j$, as we discussed in section \ref{dtree_reg}.
\end{itemize}

\begin{figure}[h]
    \centering
    \includegraphics[width=\linewidth]{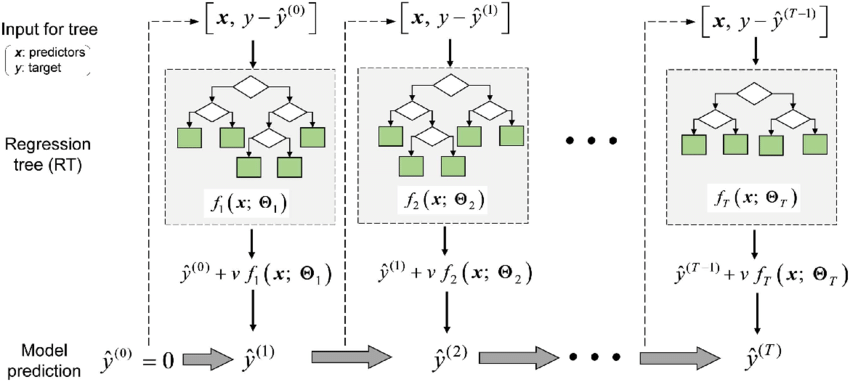}
    \caption{Schematic diagram of the \textit{Gradient Boosted} regression tree. Figure adapted from \cite{wang2020ss} and \href{https://www.researchgate.net/publication/342270212_SS-XGBoost_A_Machine_Learning_Framework_for_Predicting_Newmark_Sliding_Displacements_of_Slopes}{www.researchgate.net}.} 
    \label{fig:gradboost_flow}
\end{figure}

A boosted tree model is the sum of such trees $T(x;\Theta)$ of Eq. \ref{dtree_formalism}:
\begin{equation}\label{boost_tree_predict}
    f_M(x) = \sum\limits_{m=1}^M T(x;\Theta_m)
\end{equation}
induced in a forward stagewise manner (see Algorithm 10.2 in \cite{hastie2009elements}). Each step in the forward stagewise procedure involves the solution of the following equation:
\begin{equation}\label{dtree_regions_optimize}
    \hat{\Theta}_m = \arg \min_{\Theta_m} \sum\limits_{i=1}^N L(y_i,f_{m-1}(x_i)+T(x_i;\Theta_m))
\end{equation}
to obtain the region set and constants $\Theta_m=\{R_{jm},\gamma_{jm}\}_1^{J_m}$ of the next tree, given the current tree model $f_{m-1}(x)$. Knowing the regions $R_{jm}$, one can proceed to find the optimal values for the constants $\gamma_{jm}$ in each region. This is a straightforward process, solving the equation:
\begin{equation}\label{dtree_constants_optimize}
    \hat{\gamma}_{jm} = \arg \min_{\gamma_{jm}} \sum\limits_{x_i\in R_{jm}} L(y_i,f_{m-1}(x_i)+\gamma_{jm})
\end{equation}
Obtaining the regions $R_{jm}$ from Eq. \ref{dtree_regions_optimize} is a difficult task, even more difficult than for a single tree. However, the problem becomes quite simple, for some special cases. We will denote the case of squared-error loss, which we use later in chapter \ref{Build models Comp}, on our implementation of the regression techniques. For this type of loss function, the solution of Eq.\ref{dtree_regions_optimize} is as hard as for a single tree. It basically reflects the regression tree that best predicts the current residuals: $y_i-f_{m-1}(x_i)$, and $\hat{\gamma}_{jm}$ is the mean of these residuals in each corresponding region. 

Now, figure \ref{fig:gradboost_flow} shows the structure of a \textit{Gradient Boosted Regression Tree} model. To avoid confusion $T$ in figure corresponds to $M$ in Eq. \ref{boost_tree_predict}. But what does gradient has to do with boosting? In general, the loss in using a model $f(x)$ to predict $y$ on the training data is:
\begin{equation}\label{loss_gen_form}
    L(f) = \sum\limits_{i=1}^N L(y_i,f(x_i))
\end{equation}
The goal is the minimization of $L(f)$ with respect to $f$. This task can be viewed as a numerical optimization problem \cite{hastie2009elements}:
\begin{equation}\label{num_opt_matrix_form}
\hat{\textbf{f}} = \arg \min_{\textbf{f}} L(\textbf{f})
\end{equation}
where the 'parameters' $\textbf{f}\in \mathbb{R}^N$ are the values of the approximating function $f(x_i)$ at each of the $N$ data points $x_i$:
\begin{equation}\label{predict_vector}
    \textbf{f} = \{f(x_1),f(x_2),\dots,f(x_N)\}^T
\end{equation}
Numerical optimization procedures return the solution of Eq. \ref{num_opt_matrix_form} as a sum of component vectors \cite{hastie2009elements}:
\begin{equation}\label{boost_predict}
    \textbf{f}_M = \sum\limits_{m=0}^M \textbf{h}_m, \text{ }\text{ }\text{}\text{  }\text{ }\text{ } \textbf{h}_m\in \mathbb{R}^N
\end{equation}
Setting $\textbf{f}_0=\textbf{h}_0$, as the initial guess, each successive $\textbf{f}_m$ is induced based on the current parameter vector $\textbf{f}_{m-1}$, that is the sum of the previously
induced updates. Notice that, different numerical optimization methods suggest different prescriptions for computing each increment vector $\textbf{h}_m$ ('step').

A commonly used method, is the classical \textit{Steepest Descent} optimization procedure. This procedure is built on the principal of consecutive improvements along the direction of the gradient of the loss function \cite{natekin2013gradient}. The algorithm chooses $\textbf{h}_m=-\rho_m\textbf{g}_m$, where $\rho_m$ is a scalar and $\textbf{g}_m\in \mathbb{R}^N$ is the gradient of $L(\textbf{f})$, evaluated at $\textbf{f}=\textbf{f}_{m-1}$. The components of the gradient vector $\textbf{g}_m$ are \cite{hastie2009elements}:
\begin{equation}\label{loss_gradient}
    g_{im} = \left[\frac{\partial L(y_i,f(x_i))}{\partial f(x_i)}\right]_{f(x_i)=f_{m-1}(x_i)}
\end{equation}
The \textit{step length} $\rho_m$ is obtained from the solution of the "line search" equation \cite{hastie2009elements,friedman2001greedy}:
\begin{equation}\label{loss_step}
    \rho_m = \arg \min_\rho L(\textbf{f}_{m-1}-\rho\textbf{g}_m)
\end{equation}
Then the current solution is updated:
\begin{equation}\label{SteepDesc_predict_update}
    \textbf{f}_m = \textbf{f}_{m-1} - \rho_m\textbf{g}_m
\end{equation}
and this process is repeated at each iteration. \textit{Steepest Descent} can actually be considered a very greedy strategy, since $-\textbf{g}_m$ points to the local direction in $\mathbb{R}^N$ in which $L(\textbf{f})$ is most rapidly decreasing at $\textbf{f}=\textbf{f}_{m-1}$.

The forward stagewise boosting, implemented in Eq. \ref{dtree_regions_optimize} for a boosted tree, is also a very greedy strategy. The solution at each step, is the tree for which Eq. \ref{dtree_regions_optimize} is minimized, given the current model $f_{m-1}$ and its fits $f_{m-1}(x_i)$. Hence, the solution tree predictions $T(x_i;\Theta_m)$ are analogous to the components of the negative gradient in Eq. \ref{loss_gradient}. The main difference between the two, is that the tree components: $\textbf{t}_m=\{T(x_1;\Theta_m),\dots,T(x_N;\Theta_m)\}^N$, are in principle not independent. They are constrained to be the predictions of a decision tree with $J_m$ terminal nodes (leaves). In contrast, the negative gradient is the unconstrained maximal descent direction. Besides, another analogy is found between the line search equations Eq. \ref{dtree_constants_optimize} (\textit{Stagewise Boosted Tree}) and Eq. \ref{loss_step} (\textit{Steepest Descent}). The difference is that Eq. \ref{dtree_constants_optimize} incorporates a separate line search for those components of $\textbf{t}_m$ that correspond to each separate terminal region $\{T(x_i;\Theta_m)\}_{x_i \in R_{jm}}$.

So, which approach is better, \textit{Forward Stagewise Boosting} or \textit{Steepest Descent}, and what can the analogy between them offer? For any differentiable loss function $L(y,f(x))$ the calculation  of its gradient (see Eq. \ref{loss_gradient}) is a trivial task. On the contrary, the solution of Eq. \ref{dtree_regions_optimize} is difficult to be adjusted to robust criteria (see section 10.6 in \cite{hastie2009elements}). Therefore, in a problem of loss minimization, \textit{Steepest Descent} would be the optimal strategy. However, this strategy has a crucial weakness. The gradient from Eq. \ref{loss_gradient} is defined only at the training data points $x_i$. Thus, $f_M(x)$ can not be generalized to new foreign data, not included in the training dataset.

A possible resolution to this dilemma might occur from the combination of the two strategies: a \textit{Gradient Boosting} technique. Indeed, at $m$-th iteration one could induce a tree $T(x;\Theta_m)$, whose predictions are as close as possible to the negative gradient. Applying squared error as the metric criterion for closeness, this results in \cite{hastie2009elements}:
\begin{equation}\label{gradboost_regions_optimize}
    \tilde{\Theta}_m = \arg \min_\Theta \sum\limits_{i=1}^N (-g_{im}-T(x_i;\Theta))^2
\end{equation}
That is, the tree $T$ is fitted to the negative gradient values from Eq.. \ref{loss_gradient} by least squares. Of course, the solution regions $\tilde{R}_{jm}$ to Eq. \ref{gradboost_regions_optimize} might (slightly) differ from the regions $R_{jm}$ that solve Eq. \ref{dtree_regions_optimize}. They are similar enough, though, to serve the same purpose. Besides, the forward stagewise boosting procedure, and top-down decision tree induction, are already approximation procedures. After determining and  constructing the tree using Eq. \ref{gradboost_regions_optimize}, the corresponding constants in each region are obtained by \ref{dtree_constants_optimize}. Finally, the use of squared-error loss makes the negative gradient just the ordinary residual: $-g_{im}=y_i-f_{m-1}(x_i)$, so that Eq. \ref{gradboost_regions_optimize} is on its own equivalent to standard least-squares boosting.

\textit{Gradient Boosting} is one of the most powerful and accurate machine learning techniques but prone to overfitting. Indeed, except the size of the constituent trees $J$, the number of boosting iterations $M$ is also a meta-parameter of the gradient boosting algorithm. With each iteration the training risk $L(f_M)$ should decrease, so that for $M$ large enough this risk can take arbitrarily small values. Unfortunately, though, fitting the training data too well can result in overfitting, which degrades the risk on future predictions. Hence, an optimal number $M^*$ has to be found, which minimizes future risk. This value is application dependent and can be estimated using a validation sample (similar to neural networks, see chapter \ref{DL Theory}). 

Other regularization strategies can be applied in addition to the control of the $M$ value. \textit{Shrinkage} is a known technique, in which the contribution of each tree is scaled by a factor $0<\nu<1$, when this tree is added to the current approximation. The update of the boosted model approximation at each iteration, then, reads:
\begin{equation}\label{gradboost_shrink_predict_update}
    f_m(x) = f_{m-1}(x) + \nu\cdot\sum\limits_{j=1}^J \gamma_{jm}I(x\in R_{jm})
\end{equation}
The parameter $\nu$ can be viewed as a controlling parameter for the learning rate of the boosting procedure. It is not entirely independent from the value of $M$, though. For example, smaller values of $\nu$ result in larger values of M for the same training risk, so that there is a tradeoff between them. In practice, it is found (see \cite{friedman2001greedy}) that for smaller values of $\nu$ the algorithm performs better in test data, i.e. the test risk is smaller, requiring correspondingly larger values of $M$. The optimal strategy is to set tiny values for $\nu$ ($\nu<0.1$) and then determine $M$ by early stopping \cite{hastie2009elements}. This approach yields impressive improvements, over regression models with no shrinkage ($\nu=1$). Of course, these improvements come with a computational price: smaller values of $\nu$ favor larger values of $M$ to be used, and computation is proportional to the latter. However, with small trees, featuring no pruning, being induced at each step, many iterations are generally computationally feasible, even on very large datasets.

At last, inspired by the \textit{bagging} concept and its sampling procedures, we discussed in section \ref{rf_reg}, one could consider such a device in gradient boosting, to achieve better performance and computational efficiency. At each iteration, a fraction $\eta$ of the training observations is sampled (without replacement), and then this subsample is used to grow the next tree, instead of the whole training dataset. The rest of the algorithm remains the same. A typical value for $\eta$ is $\frac{1}{2}$, but for large number of observations $N$, can be much smaller. Gradient boosting with subsampling is called \textit{Stochastic Gradient Boosting} \cite{friedman2002stochastic}. In practice, the \textit{Subsampling} reduces the computing time by the same
fraction $\eta$ and in many cases might actually build a more accurate model. However, it seems that subsampling without shrinkage does poorly \cite{hastie2009elements}.

\section{Extreme Gradient Boosting (XGBoost)}\label{xgboost_reg}

The \textit{XGBoost} name stands for \textit{Extreme Gradient Boosting}.  It starts with a simple prediction, usually the target mean. It computes residuals, the errors from this prediction. The first tree learns to correct these errors. New trees are added to fix the remaining mistakes. This continues until a stopping rule is met. The algorithm basically, encapsulates the main idea of the \textit{Gradient Boosting} algorithm with additional improvements in overfitting, scalability and computational efficiency. In 2016, Chen and Geustrin \cite{chen2016xgboost}, suggested design ideas of such a model, backed up by a theoretical basis. Here, we will present briefly the basic remarks of their work. 

We will start with the minimization of the following regularized objective, for a gradient boosted tree prediction $\phi$:
\begin{equation}\label{xgboost_regular_gen}
\begin{aligned}
    &L(\phi) = \sum\limits_i l(y_i,\hat{y}_i) + \sum\limits_k \Omega(f_k) \\
    &\text{where }\text{    } \Omega(f) = \gamma T + \frac{1}{2}\lambda||w||^2
\end{aligned}    
\end{equation}
with $f_k$ the prediction of the $k$-th single tree , $T$ the leaves of the tree and $w$ the leaves weights. The first term in Eq. \ref{xgboost_regular_gen} involves a differentiable convex loss function $l$, that measures the closeness between the prediction $\hat{y}_i$ and the target $y_i$. The second term $\Omega$ penalizes the complexity of the model. This additional regularization term helps to avoid overfitting, by smoothing the final learnt weights. Now, we saw earlier that a boosted tree makes predictions in an additive manner, that is at each iteration the existed prediction $\hat{y}_i$ is updated with an extra term, obtained from the newly created tree (see Eqs. \ref{SteepDesc_predict_update} and \ref{gradboost_shrink_predict_update}). Hence, based on Eq. \ref{xgboost_regular_gen}, at the $t$-th iteration, letting the prediction of the $i$-th instance be $\hat{y}_i^{(t)}$, we will have to add $f_t$ and minimize the objective below:
\begin{equation}\label{xgboost_regular_iter}
    L^{(t)} = \sum_{i=1}^Nl(y_i,\hat{y}_i^{(t-1)}+f_t(\textbf{x}_i)) + \Omega(f_t)
\end{equation}
where $\textbf{x}_i \in \mathbb{R}^m$ is the vector of the feature variables values of the $i$-th observation. Equation \ref{xgboost_regular_iter} underlines a greedy strategy: the addition of $f_k$ that most improves the model. The objective can be optimized using second-order approximations:
\begin{equation}\label{xgboost_regular_2nd_approx}
    L^{(t)} \simeq\sum\limits_{i=1}^N[l(y_i,\hat{y}^{(t-1)})+g_if_t(\textbf{x}_i)+\frac{1}{2}h_if_t^2(\textbf{x}_i)]+\Omega(f_t)
\end{equation}
where $g_i=\partial_{\hat{y}^{(t-1)}}l(y_i,\hat{y}^{(t-1)})$ and $h_i=\partial^2_{\hat{y}^{(t-1)}}l(y_i,\hat{y}^{(t-1)})$  are first and second order gradient statistics on the loss function. Removing the constant terms leads to the following simplified objective at step $t$:
\begin{equation}\label{xgboost_regular_2nd_approx_simpl}
    \tilde{L}^{(t)}=\sum\limits_{i=1}^N[g_if_t(\textbf{x}_i)+\frac{1}{2}h_if_t^2(\textbf{x}_i)]+\Omega(f_t)
\end{equation}

Assuming $I_j=\{i|q(\textbf{x}_i)=j\}$ is the instance set of leaf $j$, Eq. \ref{xgboost_regular_2nd_approx_simpl}, can be re-written by expanding $\Omega$:
\begin{equation}\label{xgboost_regular_2nd_approx_expand}
    \tilde{L}^{(t)}=\sum\limits_{i=1}^N[g_if_t(\textbf{x}_i)+\frac{1}{2}h_if_t^2(\textbf{x}_i)]+\gamma T+\frac{1}{2}\lambda\sum\limits_{j=1}^Tw^2_j=\sum\limits_{j=1}^T[(\sum\limits_{i\in I_j}g_i)w_j+\frac{1}{2}(\sum\limits_{i\in I_j}h_i+\lambda)w_j^2]+\gamma T
\end{equation}
For a fixed tree structure $q(\textbf{x})$, one can compute the optimal weight $w^{*}_j$ of leaf $j$ by \cite{chen2016xgboost}:
\begin{equation}\label{xgboost_leaf_weight_opt}
    w^{*}_j=-\frac{\sum_{i\in I_j}g_i}{\sum_{i\in I_j}h_i+\lambda}
\end{equation}
and calculate the corresponding optimal loss value by:
\begin{equation}\label{xgboost_loss_val_opt}
    \tilde{L}^{(t)}(q)=-\frac{1}{2}\sum\limits_{j=1}^T \frac{(\sum_{i\in I_j}g_i)^2}{\sum_{i\in I_j}h_i+\lambda}+\gamma T
\end{equation}
Hence, Eq. \ref{xgboost_loss_val_opt} can be viewed and utilized as a scoring function, for measuring the quality of a tree structure. This score is analogous to the impurity score for assessing decision trees, but now obtained for a wider range of objective functions. Of course, it is impossible to scan all the possible tree structures $q$. Instead, a greedy algorithm, that starts from a single leaf and adds branches to the tree in a repetitive manner, is used. Considering $I_L$ and $I_R$ are the instance sets of left
and right nodes after the split, and letting $I=I_L \cup I_R$, the loss reduction after the split reads \cite{chen2016xgboost}:
\begin{equation}\label{xgboost_loss_split_opt}
    L_{split}=\frac{1}{2}\left[ \frac{(\sum_{i\in I_L}g_i)^2}{\sum_{i\in I_L}h_i+\lambda}+
    \frac{(\sum_{i\in I_R}g_i)^2}{\sum_{i\in I_R}h_i+\lambda}-
    \frac{(\sum_{i\in I}g_i)^2}{\sum_{i\in I}h_i+\lambda}\right]-\gamma
\end{equation}
In practice, this formula is used to assess the split candidates.

Besides, regularization, the two other techniques, \textit{Shrinkage} and \textit{Subsampling}, we mentioned in section \ref{gradboost_reg}, are incorporated to further prevent overfitting. Shrinkage reduces the contribution of each individual tree and allows the construction of future trees to improve the model. In contrast, subsampling (especially column subsampling) speeds up computations, when used as part of a parallel \textit{XGBoost} algorithm.

Under the constraints of effective enumeration over possible splitting points and distributed setting, an approximate algorithm is necessary. This algorithm should propose candidate splitting points according to percentiles of feature distribution. Then, the algorithm would map the continuous features into buckets split by these candidate points, aggregating the statistics and determining the best solution among proposals based on these aggregated statistics. 

In this framework, Chen and Geustrin \cite{chen2016xgboost} followed two approaches in their implementations: the global variant and the local variant. In the first approach, all the candidate splits are proposed during the initial phase of tree construction, and the same proposals are used for split finding at all levels. The global method used less proposal steps than the local method, but usually requires more candidate points, since candidates are not refined after each split. On the contrary, the local variant re-proposes and refines the candidates after each split. This way it can potentially be more suitable for deeper trees. In practice, the global proposal might be as accurate as the local one given enough candidates.

Chen and Geustrin \cite{chen2016xgboost} made further suggestions, regarding the criteria for the proposal of splitting candidates, the awareness of the algorithm to sparse data and the computational effectiveness of the learning procedure. For the criteria, they developed a so-called \textit{weighted quantile sketch} algorithm, capable to handle large datasets with weighted data. As of the sparsity, involved in many datasets, due to missing or frequent zero data, or strangely encoded artifacts, they proposed a configuration with a default direction in each tree node. Their algorithm sets the optimal default directions based on the given data, by treating the non-presence as a missing value and learning the best direction to handle missing values. \textit{Sparsity awareness} is proven to be a significant property of an algorithm, resulting in accelerations of about $50$ times, compared to naive versions (without sparsity awareness). Finally, for the effectiveness of the algorithm, they proposed the storage of data in in-memory units called \textit{block}, where the data are stored in
compressed column (CSC) format and  each column is sorted
by the corresponding feature value. This block configuration can be adjusted and used differently, in order to achieve better and more effective usage of available hardware resources, such as the CPU, the cache-memory and the disk read speed, by even making the \textit{XGBoost} algorithm to run in parallel.

\section{From single to multiple multivariate regression}

Every algorithm we presented in the previous sections, considers the existence of multiple features and makes predictions for a single-output, performing single multivariate regression. The escalation to multi-output multivariate regression, to serve the needs of our study, is actually non-trivial for all techniques. For example, the general \textit{Least Squares} method can be easily expanded to predict multiple responses. The single-output model for $n$ observations and $q$ features, has the following matrix notation \cite{rencher2012methods}:
\begin{equation}\label{single_least_squares_matrix_form}
\begin{pmatrix}
    y_1 \\ y_2 \\ \vdots \\ y_n
\end{pmatrix} =
\begin{pmatrix}
    1 & x_{11} & x_{12} & \cdots & x_{1q} \\
    1 & x_{21} & x_{22} & \cdots & x_{2q} \\
    \vdots & \vdots & \vdots & \ddots & \vdots\\
    1 & x_{n1} & x_{n2} & \cdots & x_{nq} \\
\end{pmatrix}
\begin{pmatrix}
    \beta_0 \\ \beta_1 \\ \vdots \\ \beta_q
\end{pmatrix}
+
\begin{pmatrix}
    \epsilon_1 \\ \epsilon_2 \\ \vdots \\ \epsilon_n
\end{pmatrix}
\end{equation}
or in compact form
\begin{equation}\label{single_least_squares_compact}
    \textbf{y} = \textbf{X}\textbf{b}+\textbf{e}
\end{equation}
where $\textbf{y}$ is the $n\times1$ column-vector of the single response variable, $\textbf{X}$ the $n\times(q+1)$ features matrix, $\textbf{b}$ the $(q+1)\times1$ column-vector containing the regression parameters and $\textbf{e}$ the $n\times1$ column-vector of the prediction errors. The parameters $b_j$ are estimated by minimizing the sum of squares of deviations. The vector $\hat{\textbf{b}}=(\hat{\beta}_0,\hat{\beta}_1,\dots,\hat{\beta}_q)^T$ is given by \cite{rencher2012methods}:
\begin{equation}\label{single_least_squres_param_estimation}
    \hat{\textbf{b}} = (\textbf{X}^T\textbf{X})^{-1}\textbf{X}^T\textbf{y}
\end{equation}
For multi-output regression with $n$ observations, $q$ features and $p$ responses, Eq. \ref{single_least_squares_matrix_form} generalizes as \cite{rencher2012methods}:
\begin{equation}\label{multi_least_squares_matrix_form}
\begin{pmatrix}
    y_{11} & y_{12} & \cdots & y_{1p} \\
    y_{21} & y_{22} & \cdots & y_{2p} \\ 
    \vdots & \vdots & \ddots & \vdots \\ 
    y_{n1} & y_{n2} & \cdots & y_{np}
\end{pmatrix} =
\begin{pmatrix}
    1 & x_{11} & x_{12} & \cdots & x_{1q} \\
    1 & x_{21} & x_{22} & \cdots & x_{2q} \\
    \vdots & \vdots & \vdots & \ddots & \vdots\\
    1 & x_{n1} & x_{n2} & \cdots & x_{nq} \\
\end{pmatrix}
\begin{pmatrix}
    \beta_{01} & \beta_{02} & \cdots & \beta_{0p} \\ 
    \beta_{11} & \beta_{12} & \cdots & \beta_{1p} \\ 
    \vdots & \vdots & \ddots & \vdots \\ 
    \beta_{q1} & \beta_{q2} & \cdots & \beta_{qp} \\ 
\end{pmatrix}
+
\begin{pmatrix}
    \epsilon_{11} & \epsilon_{12} & \cdots & \epsilon_{1p} \\
    \epsilon_{21} & \epsilon_{22} & \cdots & \epsilon_{2p} \\ 
    \vdots & \vdots & \ddots & \vdots \\ 
    \epsilon_{n1} & \epsilon_{n2} & \cdots & \epsilon_{np}
\end{pmatrix}
\end{equation}
or in compact form:
\begin{equation}\label{multi_least_squares_compact}
    \textbf{Y} = \textbf{X}\textbf{B}+\textbf{E}
\end{equation}
That is, we have now the $n\times p$ matrix $\textbf{Y}$ for the responses, the same $n\times (q+1)$ matrix \textbf{X} for the features, the $(q+1)\times p $ matrix $\textbf{B}$ containing the regression parameters and the $n\times p$ matrix $\textbf{E}$ for the prediction errors. By analogy with the univariate case in Eq. \ref{single_least_squres_param_estimation}, the elements of matrix $\textbf{B}$ are estimated by:
\begin{equation}\label{multi_least_squres_param_estimation}
    \hat{\textbf{B}} = (\textbf{X}^T\textbf{X})^{-1}\textbf{X}^T\textbf{Y}
\end{equation}

A similar generalization cannot be applied globally to the machine learning models we described earlier, due to their different structure and loss function calculation. Studies \cite{de2002multivariate,gonzalez2016applying,dapogny2017multi} have shown that single or bagged tree techniques, i.e. \textit{Decision Trees} and \textit{Random Forest} in our study, can be expanded to natively support multiple outputs. At the leaf nodes a vector of values is obtained, instead of a single value. This allows for joint training of the model. That is, only one model is fitted, by optimizing the loss on all responses. There are various metrics, that can be used to score the accuracy, like the \textit{Mean Squared Error} (MSE) or the \textit{Mean Squared Log Error} (MSLE). 

In contrast, the boosted tree structure of \textit{Gradient Boosting} and \textit{XGBoost} makes the escalation much more complicated. Later studies \cite{iosipoi2022sketchboost,zhang2020gbdt} have made progress in the direction of a joint training procedure, for these techniques. However, this progress has not yet been incorporated into the \texttt{sckit-learn} framework (see \href{https://scikit-learn.org/stable/}{documentation}) we use in our study. Hence, a different model is trained separately and independently for each response variable, and there is no direct joint calculation of the loss. In this case, a wrapper is needed, to summarize the trained models and make the final predictions for the full set of responses. The wrapper we use is \texttt{MultiOutputRegressor} (see \href{https://scikit-learn.org/stable/modules/generated/sklearn.multioutput.MultiOutputRegressor.html}{documentation}). An overview of the algorithms we use is presented in Table \ref{tab:multioutput-tree-methods}, below.

\vspace{1cm}
\begin{table}[h]
\centering
\begin{tabular}{|p{3.2cm}|p{6.5cm}|p{6.5cm}|}
\hline
\hline
\textbf{Algorithm} & \textbf{Advantages in Multi-Output Regression} & \textbf{Limitations in Multi-Output Regression} \\
\hline
\hline

\textbf{Decision Trees} & 
- Naturally support multi-output regression by predicting a vector of outputs at each leaf node.

- Very fast to train and easy to interpret.

- Simple to implement and understand. &
- Low predictive power compared to ensembles.

- Prone to overfitting, especially in small datasets.

- Cannot capture correlations between target variables.

\\ \hline

\textbf{Random Forests} & 
- Inherit native multi-output support from decision trees.

- Reduce overfitting via averaging across trees.

- Perform well on structured data with many outputs. &

- Still treats outputs independently in splits, so correlations are not explicitly modeled.

- Less interpretable due to ensemble nature.

- Higher computational cost compared to a single tree.

\\ \hline

\textbf{Gradient Boosting} & 
- Highly accurate when trained individually per target.

- Effective for non-linear patterns and small-to-medium datasets.

- Allows fine-tuned control via hyperparameters. &

- Does not natively support multi-output regression (wrapper needed).

- Requires separate model for each output, increasing training time.

- Unable to capture cross-output dependencies. 

- Slow training for large datasets

\\ \hline

\textbf{XGBoost} & 
- Very efficient and accurate implementation of gradient boosting.

- Includes experimental support for multi-output since v1.6+.

- Built-in regularization improves generalization.
&
- Experimental multi-output support is limited and not well documented.

- In most cases, still requires one model per output.

- High model complexity and tuning effort. 

\\ \hline \hline
\end{tabular}
\caption{Advantages and limitations of the four tree-based algorithms we presented in sections \ref{dtree_reg}, \ref{rf_reg}, \ref{gradboost_reg} and \ref{xgboost_reg}, when applied to multi-output multivariate regression tasks.}
\label{tab:multioutput-tree-methods}
\end{table}

\section{Cross-validation and Grid search}\label{CrossVal_GridSrch}
With the application of \texttt{MultiOutputRegressor} on the algorithms \textit{Gradient Boosting} and \textit{XGBoost}, all four algorithms in Table \ref{tab:multioutput-tree-methods} can now be evaluated, based on their performance in multi-output regression. We choose to use both the \textit{Mean Squared Error} (MSE) and the \textit{Mean Squared Log Error} (MSLE) loss functions for this evaluation. A summary of their differences is presented in Table \ref{tab:mse-vs-msle}. MSE specializes in errors on raw units and is used to reduce large absolute errors, but is greatly affected from outliers. In contrast, MSLE specializes in percentage errors and is used to maintain the same level of accuracy among all responses of a multiple output. However, only non-negative responses can be assessed, due to the logarithm involved in the calculation of the loss function.

\begin{table}[h]
\centering
\small
\begin{tabular}{|p{4.5cm}|p{4.5cm}|p{6.5cm}|}
\hline
\hline
\textbf{Criterion} & \textbf{Mean Squared Error (MSE)} & \textbf{Mean Squared Logarithmic Error (MSLE)} \\
\hline
\hline

\textbf{Definition} & 
Measures the average of the squared differences between actual and predicted values. & 
Measures the average of the squared differences between the logarithms of actual and predicted values.

\\ \hline

\textbf{Formula} & 
$\frac{1}{n} \sum_{i=1}^{n} (y_i - \hat{y}_i)^2$ & 
$\frac{1}{n} \sum_{i=1}^{n} \left(\log(1 + y_i) - \log(1 + \hat{y}_i)\right)^2$

\\ \hline

\textbf{Error Emphasis} & 
Penalizes large absolute errors. & 
Penalizes large relative errors (e.g., overestimation of small values).

\\ \hline

\textbf{Best Used For} & 
When prediction accuracy in raw units is important. & 
When we care more about percentage errors or underestimations.

\\ \hline

\textbf{Sensitivity to Outliers} & 
Very sensitive to outliers due to squaring large errors. & 
Less sensitive to outliers; compresses the effect of large values.

\\ \hline

\textbf{Target Requirements} & 
No restrictions (can be used with any real number). & 
Requires non-negative target and prediction values.

\\ \hline

\textbf{Interpretation} & 
Provides average squared deviation in original units. & 
Provides average squared deviation on a logarithmic scale.

\\ \hline

\textbf{Example Use Cases} & 
Price, weight, and temperature prediction. & 
Forecasting exponential growth, such as web traffic or revenue.

\\ \hline \hline
\end{tabular}
\caption{Comparison of Mean Squared Error (MSE) and Mean Squared Logarithmic Error (MSLE) loss functions for regression tasks.}
\label{tab:mse-vs-msle}
\end{table}

\begin{figure}[h]
    \centering
    \includegraphics[width=0.8\linewidth]{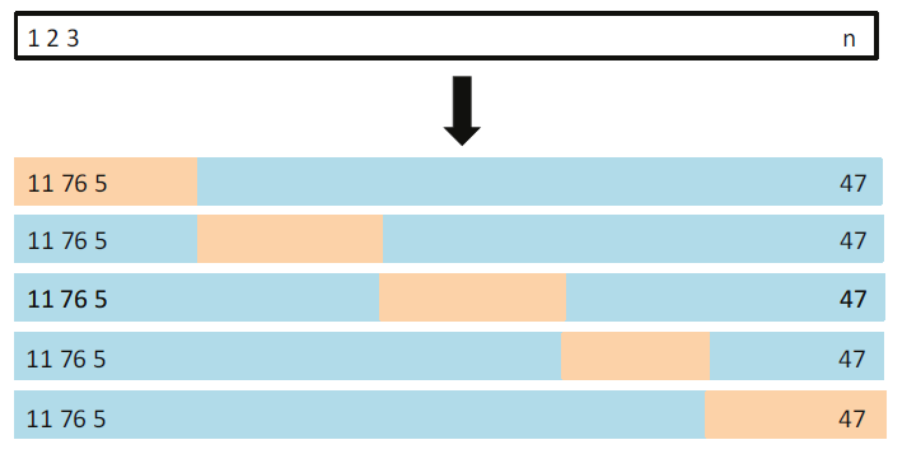}
    \caption{Schematic display of 5-fold CV. A set of $n$ observations is randomly split into five non-overlapping groups. Each of these fifths acts as a validation set (shown in beige), and the remainder as a training set (shown in blue). The test error is estimated by averaging the five resulting loss estimates. Figure adapted from \cite{james2013introduction}.}
    \label{fig:5fold_CV}
\end{figure}

So far, the assessment of each model is done only once, giving as input the test set after the end of training. This evaluation, though, might not be representative of the actual performance of the model in foreign data. We want to include an evaluation process during the training, as well. Cross-validation (CV) is the answer and, more specifically \textbf{k-fold cross-validation}. When this procedure is applied, the training dataset is divided into $k$ groups (folds) of approximately equal size. The first fold is considered a validation dataset and the method is trained on the remaining $k-1$ folds. Then, the selected loss function $L$ is computed on the observations in the left-out fold. This process is
repeated $k$ times: each time, a different group of observations is treated as a validation set. Hence, we get $k$ estimates of the test error; $L_1,L_2,\dots,L_k$. In Fig. \ref{fig:5fold_CV}, an example of the splitting during a 5-fold cross-validation process is presented. The total k-fold CV estimate is computed by averaging the values $L_i$ \cite{james2013introduction}:
\begin{equation}\label{kfoldCV_estimation}
    CV_{(k)} = \frac{1}{k}\sum\limits_{i=1}^k L_i
\end{equation}
Averaging reduces the variance of the loss estimation and provides a clearer picture of the generalization of the model and its performance on independent data, since the model was evaluated multiple times during training and on different validation data each time. However, large execution times might occur, if many folds are selected, rendering the cross-validation process, a computationally infeasible one.

\begin{table}[h]
    \centering
    \begin{tabular}{|p{6cm}|p{6cm}|}
    \hline
    \hline
    \textbf{Hyperparameters}  &  \textbf{Values}\\
    \hline
    \hline
    \vspace{0.005cm} max\_depth   & \vspace{0.005cm} [None, 5, 10, 20] \\
    \hline
    \vspace{0.005cm} min\_samples\_split  & \vspace{0.005cm} [2, 5, 10]\\
    \hline
    \vspace{0.005cm} min\_samples\_leaf   & \vspace{0.005cm} [1, 2, 5] \\
    \hline
    \vspace{0.005cm} max\_features   & \vspace{0.005cm} [None, 'sqrt', 'log2'] \\
    \hline
    \vspace{0.005cm} criterion  & \vspace{0.005cm} ['squared\_error', 'friedman\_mse'] \\
    \hline 
    \hline
    \end{tabular}
    \caption{Example of a grid of hyperparameters for the tuning of a Decision Tree model. The Grid Search algorithm will evaluate the model on a total of $4\times3\times3\times3\times2 = 216$ different combinations, in order to find the best one.}
    \label{tab:grid_example_dtree}
\end{table}

Now, a model features a variety of hyperparameters, like the depth of the tree (see section \ref{dtree_reg}) or the predictor size in the \textit{Random Forest} model (see section \ref{rf_reg}). The values of the latter are determined in advance and remain fixed throughout the training process. Thus, one can assess the accuracy of the model, with or without cross-validation, based on its performance with one-time defined hyperparameters. This in some ways limits the capabilities of the model, overlooking some combinations of hyperparameters for which the model would potentially perform better. 

We want to investigate the accuracy of the model, when different combinations of hyperparameters are applied, and choose the combination, for which the model performs the best. Of course, scanning through all combinations of hyperparameters, is computationally infeasible. Instead, we follow a greedy strategy called \textbf{Grid Search} \cite{liashchynskyi2019grid}. A subset of the hyperparameters space is given and the algorithm makes a complete search over it. In other words, we provide a set of values for each of the hyperparameters we want to tune and the algorithm explores all the possible combinations, to determine the one that leads to the minimum loss. An example is presented in Table \ref{tab:grid_example_dtree}. 

By definition, Grid Search suffers from high dimensional spaces. However, it often can easily be parallelized, since the hyperparameter values, that the algorithm works with, are usually independent of each other. Combined with cross-validation, grid search offers a powerful tool in training robust and reliable regression models. The final model is both fine-tuned and generalized to address foreign data.

\chapter{Deep Learning Regression}\label{DL Theory}

In the previous chapter \ref{ML Theory}, we saw that machine learning offers a variety of techniques to deal with our regression issue. Regression analysis, however, could also be implemented using artificial neural networks (ANN), or in our complex scenario, through deep neural networks (DNN). Deep learning (DL) is the most powerful tool to achieve learning complex relationships of data, especially if a sufficiently large dataset is available to ensure proper training. In this chapter, we present the basics in building and properly fitting a deep neural network model, in order to get as much as possible from deep learning regression.

\section{Building a Neural Network}\label{NNs_Building}
\subsection{Visible and Hidden layers}\label{NNs_layers}

\begin{figure}
    \centering
    \includegraphics[width=\linewidth,height=9cm]{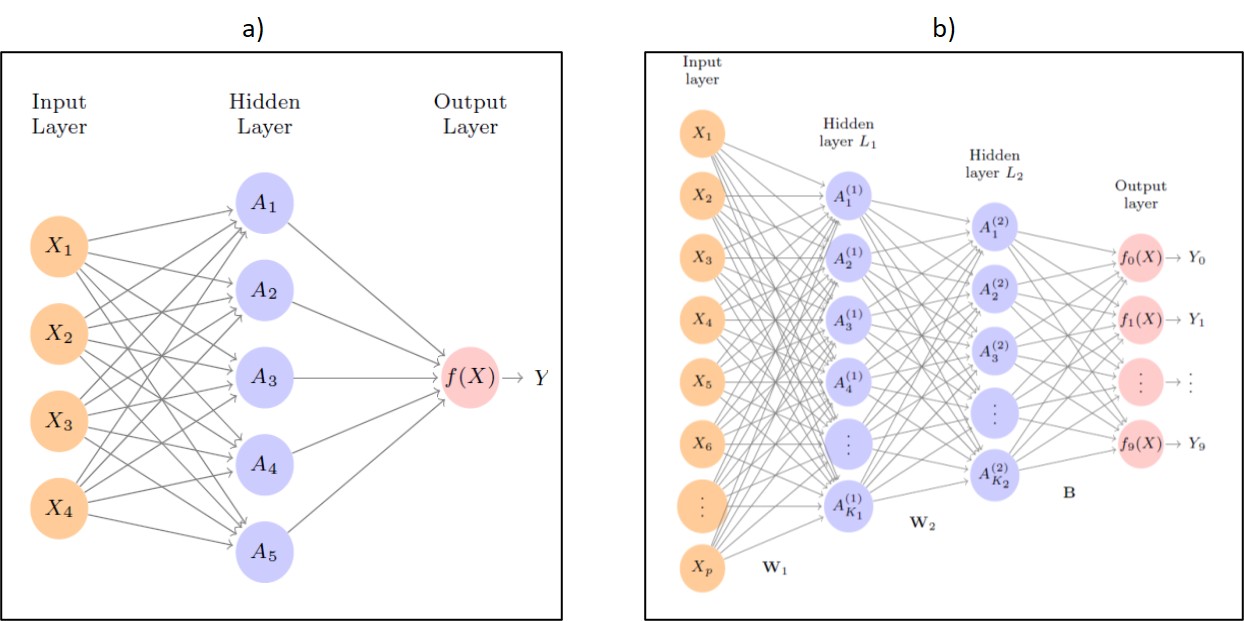}
    \caption{A neural network with a) a single hidden layer (ANN) and b) two hidden layers (DNN). Figures adapted from \cite{james2013introduction}.}
    \label{fig:ANNs_structure}
\end{figure}

The simplest neural network consists of two visible layers (\textit{input} and \textit{output}) and at least one \textit{middle-hidden layer}, as shown in a) of Fig. \ref{fig:ANNs_structure}. A vector of $p$ variables: $\textbf{X}=(X_1,X_2,\dots,X_p)$ is given as input. The response $Y$ is predicted by building a nonlinear function $f(\textbf{X})$. The neural network differs from the nonlinear machine learning predictors due to its particular \textit{feed-forward} structure. The four features $X_1,X_2,X_3,X_4$ in a) of Fig. \ref{fig:ANNs_structure} form the four units (\textit{neurons}) of the input layer. The arrows indicate that information from every \textit{input unit} feeds into each of the $K$ \textit{hidden units}. The value of $K$ is selected arbitrarily (here $K=5$). This neural network model has the following form:
\begin{equation}\label{ANN_predict}
    f(\textbf{X}) = \beta_0 + \sum\limits_{k=1}^K \beta_kh_k(\textbf{X}) = \beta_0 + \sum\limits_{k=1}^K\beta_kg(w_{k0}+\sum\limits_{j=1}^pw_{kj}X_j)
\end{equation}
and is built up in two steps. First, the $K$ \textit{activations} $A_k$, $k=1,\dots,K$ in the hidden layer are calculated as functions of the input features $X_1,\dots,X_p$:
\begin{equation}\label{ANN_activations}
    A_k = h_k(\textbf{X})=g(w_{k0}+\sum\limits_{j=1}^pw_{kj}X_j)
\end{equation}
Each $A_k$ can be considered as a different transformation $h_k(\textbf{X})$ of the original features. Then, these $K$ activations from the hidden layer feed into the output layer, leading to:
\begin{equation}\label{ANN_output}
    f(\textbf{X}) = \beta_0+\sum\limits_{k=1}^K \beta_kA_k
\end{equation}
a linear regression model in the $K=5$ activations. All the \textit{parameters} $\beta_0,\dots,\beta_K$ and the \textit{weights} $w_{10},\dots,w_{Kp}$ are estimated from the data.

The same pattern can be expanded to describe the operation of more complex neural networks, with more hidden layers and more output variables (more units in the output layer). Consider the neural network of b) in Fig. \ref{fig:ANNs_structure}. This network has an input layer with $p$ units (input vector $\textbf{X}$ of $p$ variables like before), a first hidden layer $L_1$ with $K_1$ units, a second hidden layer $L_2$ with $K_2$ units and an output layer with $10$ units ($10$ output variables). The activations $A_k^{(1)}$ of the first hidden layer $L_1$ are computed by \cite{james2013introduction}:
\begin{equation}\label{DNN_activ_L1}
    A_k^{(1)} = h_k^{(1)}(\textbf{X}) = g(w_{k0}^{(1)}+\sum\limits_{j=1}^pw_{kj}^{(1)}X_j)
\end{equation}
with $k=1,\dots,K_1$. The second hidden layer $L_2$ treats the activations $A_k^{(1)}$ of the first hidden layer $L_1$ as inputs and calculates new activations $A_l^{(2)}$:
\begin{equation}\label{DNN_activ_L2}
    A_l^{(2)} = h_l^{(2)}(\textbf{X}) = g(w_{l0}^{(2)}+\sum\limits_{k=1}^{K_1}w_{lk}^{(2)} A_k^{(1)})
\end{equation}
with $l=1,\dots,K_2$. Notice that each of the activations of the second layer is expressed as a function of the input vector $\textbf{X}$: $A_l^{(2)} = h_l^{(2)}(\textbf{X})$. By definition, $A_l^{(2)}$ are explicitly a function of the activations $A_k^{(1)}$ from the layer $L_1$. In turn, the activations $A_k^{(1)}$ are functions of $\textbf{X}$. This would also be the case with more
hidden layers. Thus, through a chain of transformations, the network manages to build up quite complex transformations of $\textbf{X}$, which eventually feed into the output layer as features.

As for the notation, the superscript indicates to which layer the activations and weights (coefficients) belong. For example, the activation $A_5^{(1)}$ belongs to the hidden layer $L_1$, while the weight $w_{25}^{(2)}$ belongs to the hidden layer $L_2$. The notation $\textbf{W}_1$ refers to the entire matrix of weights, that feed from the input layer to the first hidden layer $L_1$. This matrix will have $(p+1)\times K_1$ elements: there are $p+1$ rather than $p$ because we must account for the \textit{intercept} or \textit{bias} term $w_{k0}^{(1)}$. Similarly, each element $A_k^{(1)}$ feeds to the second hidden layer $L_2$ via the matrix of weights $\textbf{W}_1$, of dimensions $(K_1+1)\times K_2$, including the biases $w_{l0}^{(2)}$.

Getting to the output layer, the Deep Neural Network has to compute ten different linear models, one per each of the ten responses. We have \cite{james2013introduction}:
\begin{equation}\label{DNN_output}
    Z_m = \beta_{m0} +\sum\limits_{l=1}^{K_2}\beta_{ml}h_l^{(2)}(\textbf{X}) = \beta_{m0} +\sum\limits_{l=1}^{K_2}\beta_{ml}A_l^{(2)}
\end{equation}
for $m=0,1,\dots,9$. The matrix $\textbf{B}$ stores all $(K_2+1)\times 10$ of these $\beta_{ml}$ weights and $\beta_{m0}$ biases. If ten separate quantitative responses is the case, like the twelve responses of our multi-output regression problem (see subsection \ref{Data_sampling}), then it is:
\begin{equation}\label{DNN_predict}
    f_{m}(\textbf{X}) = Z_m
\end{equation}
for the prediction of the Deep Neural Network.

\subsection{Activation functions}\label{NNs_activ_func}
\begin{figure}[h]
    \centering
    \includegraphics[width=\linewidth,height=7cm]{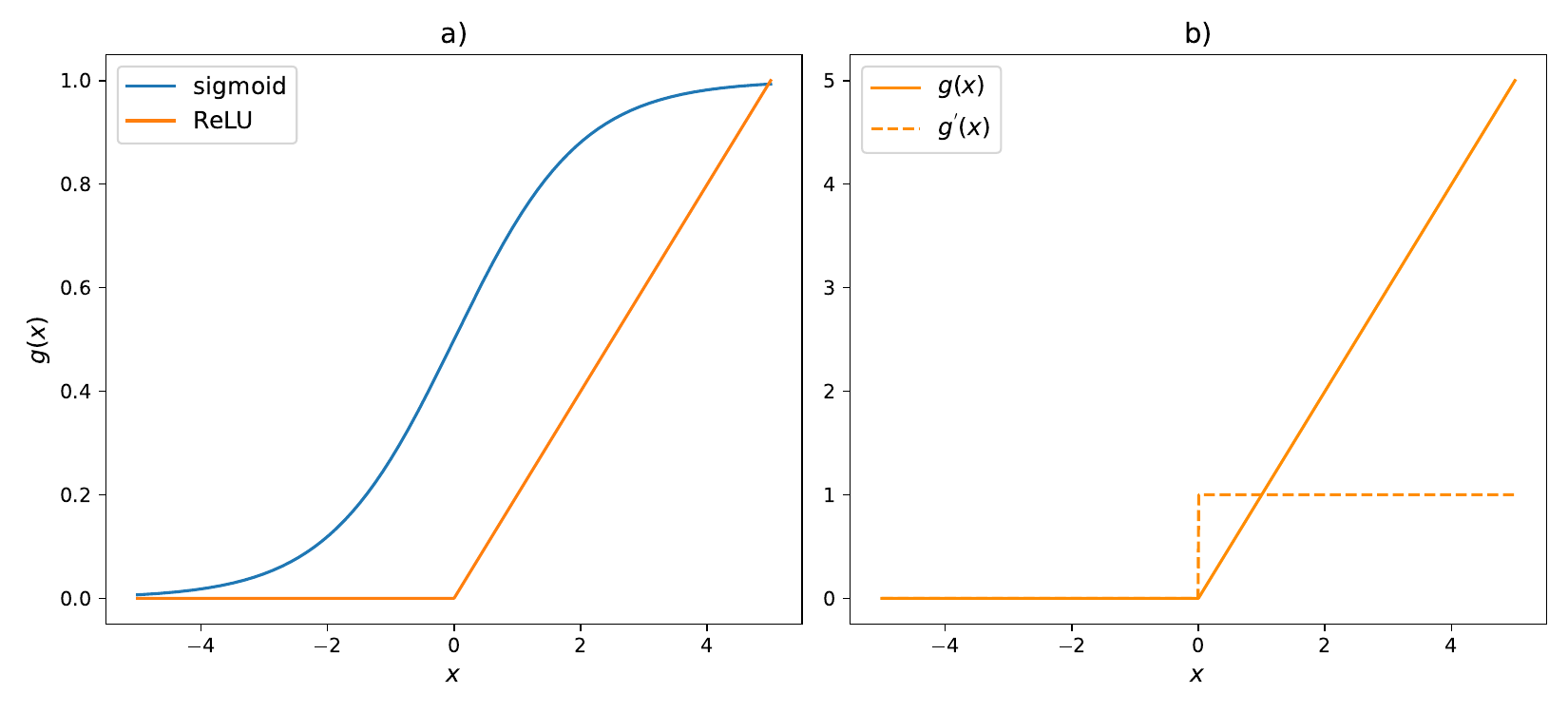}
    \caption{Activation functions. a) Comparison between the \textit{sigmoid} and the \textit{ReLU} activation functions. The \textit{ReLU} function has been scaled down by a factor of five for ease of comparison. b) The \textit{ReLU} function and its first derivative. See \texttt{activation\_functions.ipynb} in Table \ref{tab:train_test_DL}.}
    \label{fig:ANN_activ_func}
\end{figure}

Note the participation of the function $g$ in the equations. \ref{ANN_predict}, \ref{ANN_activations}, \ref{DNN_activ_L1} and \ref{DNN_activ_L2}, where the activations of the hidden layers are calculated. This function is called \textit{activation function} and conducts the transformation of the (input) information between the layers of the neural network. There are many choices, regarding the formula of the activation function. In the early years of deep learning and neural networks, the \textit{sigmoid} activation function was favored \cite{james2013introduction,ding2018activation}:
\begin{equation}\label{sigmoid_activ_func}
    g_{sigmoid}(x) = \frac{1}{1+e^{-x}}
\end{equation}
in which $x\in (-\infty,+\infty)$ and $g\in (0,1)$, as shown in a) of Fig. \ref{fig:ANN_activ_func}. This is the same function used in logistic regression to convert a linear function into probabilities between zero and one.

In modern neural networks, the
preferred choice is the \textit{ReLU} (rectified linear ReLU unit) activation function, which has the following form \cite{james2013introduction,ding2018activation}:
\begin{equation}\label{ReLU_activ_func}
    g_{ReLU}(x) = (x)_+ = 
    \begin{cases}
    x, & x\geq0 \\
    0, & x<0
    \end{cases}
\end{equation}
So, $ReLU$ is a piecewise function and its first derivative reads:
\begin{equation}\label{ReLU_derivative}
    g^{'}_{ReLU}(x) = 
    \begin{cases}
    1, & x\geq0 \\
    0, & x<0
    \end{cases}
\end{equation}
that is, a step function, as shown in b) of  Fig. \ref{fig:ANN_activ_func}. The value of $g^{'}_{ReLU}(x)$ is a constant for $x>0$. This gives some significant advantages on the \textit{ReLU} compared to other activation functions. In particular \cite{ding2018activation}:
\begin{itemize}
    \item The lack of exponential terms in \textit{ReLU} makes computations in the Neural Network that incorporates \textit{ReLU} much cheaper, than using the activation functions \textit{sigmoid} and \textit{hyperbolic tangent}, where exponential terms are present.

    \item The neural networks that incorporate \textit{ReLU} activations functions exhibit faster convergence, versus those networks with saturating activation functions, in terms of training time with gradient descent.

    \item The \textit{ReLU} function to the network the ability to easily obtain sparse representation. For $x<0$, the output is zero, providing the sparsity in the activation of neuron units and improving the efficiency of data learning. For $x\geq0$, the features of the data can be retained largely.

    \item The derivative of \textit{ReLU} function are fixed as the constant $1$. This helps in avoiding trapping into the local optimization and resolves the vanishing gradient effect,  occurred in \textit{sigmoid} and \textit{hyperbolic tangent} activation functions.

    \item Deep Neural Networks with \textit{ReLU} activation function, are able to reach their best performance without needing any unsupervised pre-training on purely supervised tasks with large labeled datasets.
\end{itemize}
These are the main reasons why we use the \textit{ReLU} activation function in our study. However, \textit{ReLU} function comes also with some disadvantages. The two main drawbacks are \cite{ding2018activation}:
\begin{itemize}
    \item The derivatives $g^{'}(x)$ vanish when $x<0$, so the \textit{ReLU} activation function is left-hard-saturating. Subsequently, the relative weights might not be updated any more and that results in the death of some neuron units, meaning that these neuron units will never be activated.

    \item The average of the units' outputs is identically positive, which will lead to a bias shift for units in the next layer. 
\end{itemize}
These two attributes both have negative impact on the convergence of the respective Deep Neural Networks.

\section{Fitting a Neural Network}\label{NNs_Fitting}
When it comes to fitting a neural network, the procedure is quite complex and we present a brief overview. Like in machine learning, the goal is the minimization of a loss function. Consider the simple neural network of subsection \ref{NNs_layers}, with a single hidden layer, weights $w_k=(w_{k0},w_{k1},\dots,w_{kp})$, $k=1,\dots,K$, and parameters $\beta=(\beta_0,\beta_1,\dots,\beta_K)$. Given observations $(\textbf{x}_i,y_i)$, $i=1,\dots,n$, one could fit the neural network model by solving a nonlinear least squares problem \cite{james2013introduction}:
\begin{equation}\label{NN_loss}
    \min_{\{w_k\}_1^K,\beta} \frac{1}{2}\sum\limits_{i=1}^n (y_i-f(x_i))^2
\end{equation}
where
\begin{equation}\label{ANN_predict_single}
    f(\textbf{x}_i) = \beta_0 + \sum\limits_{k=1}^K \beta_kh_k(\textbf{X}) = \beta_0 + \sum\limits_{k=1}^K\beta_kg(w_{k0}+\sum\limits_{j=1}^pw_{kj}x_{ij})
\end{equation}

\begin{figure}[h]
    \centering
    \includegraphics[width=0.9\linewidth]{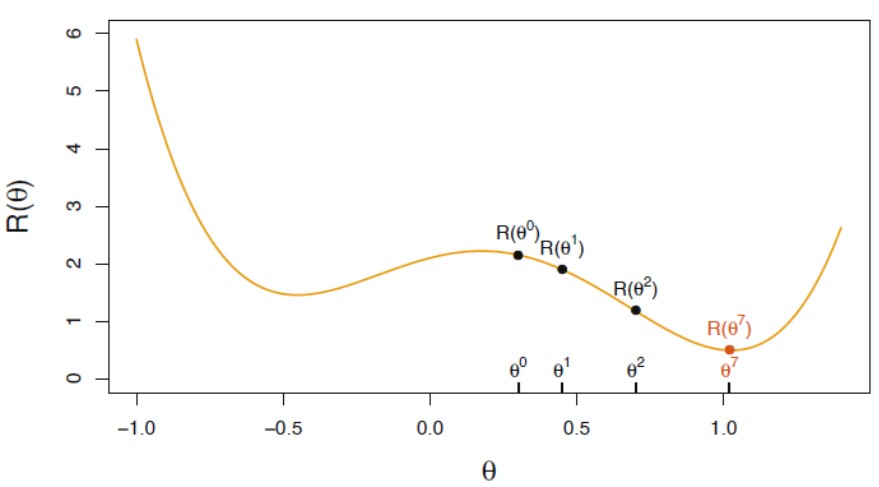}
    \caption{Illustration of gradient descent for one-dimensional $\theta$. The objective function $R(\theta)$ is not convex, and has \textit{two minima}, one at $\theta = -0.46$ (local), the other at $\theta = 1.02$ (global). Starting at some value $\theta_0$ (typically randomly chosen), each step in $\theta$ moves downhill - against the gradient - until it cannot go down any further. Here gradient descent reached the global minimum in $7$ steps. Figure adapted from \cite{james2013introduction}.}
    \label{fig:NNs_gradient_descent}
\end{figure}

The objective in Eq. \ref{NN_loss} might seem simple at first glance, but the nested configuration of the parameters and the symmetry of the hidden units, make the minimization procedure quite complicated. The problem is nonconvex in the parameters space and thus, there are multiple solutions. For example, assume the simple nonconvex function of a single variable $\theta$, depicted in Fig. \ref{fig:NNs_gradient_descent}. There are two solutions for the minima of this function: one is a \textit{local minimum} and the other is a \textit{global minimum}. Besides, the network we considered is the very simplest of neural networks and in this chapter we have shown the generalization to much more complex networks, where these problems are getting even more complex. To address these issues and avoid overfitting, two general strategies are employed when fitting neural networks:
\begin{itemize}
    \item \textit{Slow Learning Rate}: the model is trained in a slow iterative fashion, using \textit{gradient descent}. The fitting process is terminated when overfitting is detected.

    \item \textit{Regularization}: penalties are imposed on the parameters, like \textit{ridge} or \textit{lasso}(see section 6.2 in \cite{james2013introduction})
\end{itemize}

Assume all the parameters are represented with one long vector $\theta$. Then, the objective in Eq. \ref{NN_loss} can be rewritten as:
\begin{equation}\label{NN_loss_vector}
    R(\theta) = \frac{1}{2}\sum\limits_{i=1}^n (y_i-f_{\theta}(x_i))^2
\end{equation}
where the dependence of $f$ on the parameters is declared explicitly. The concept of \textit{gradient descent} is pretty simple:
\begin{enumerate}
    \item Start with a initial guess $\theta^0$ for all parameters in $\theta$ and set the steps' counter at $t=0$.

    \item Repeat the following steps until the objective in Eq. \ref{NN_loss_vector} fails to reduce:
    \begin{enumerate}
        \item Find a vector $\delta$ that corresponds to a small change in $\theta$, such that $R(\theta^{t+1})<R(\theta^t)$, i.e. the objective decreases, with $\theta^{t+1}=\theta^t+\delta$.
        
        \item Update the steps' counter: $t\xleftarrow{}t+1$.
    \end{enumerate}
\end{enumerate}
An illustration of this process is presented in Fig. \ref{fig:NNs_gradient_descent}. We can imagine being in a mountainous terrain, and the goal is to reach the bottom through a series of steps. As long as we move downhill, step by step, we must eventually reach the bottom. The example in Fig. \ref{fig:NNs_gradient_descent} reflects the lucky case: the initial guess $\theta^0$ ultimately led to the global minimum. In general, we can hope the procedure finishes at a (good) local minimum.

\subsection{Backpropagation}\label{NNs_backpropag}
One might wonder how we determine the vector $\delta$, i.e. the directions in which $\theta$ must move, so that the objective $R(\theta)$ in Eq. \ref{NN_loss_vector} reduces. The \textit{gradient} of $R(\theta)$, calculated at some current value $\theta=\theta^m$, is the vector of partial derivatives at that point \cite{james2013introduction}:
\begin{equation}\label{NN_gradient_gen}
    \nabla R(\theta^m) = \frac{\partial R(\theta)}{\partial \theta} \bigg|_{\theta=\theta^m}
\end{equation}
The subscript $\theta=\theta^m$ indicates that after computing the vector of derivatives, we calculate it at the current guess $\theta^m$. This returns the direction in $\theta$-space in which $R(\theta)$ exhibits the fastest \textit{increase}. The concept of gradient descent is to change $\theta$ slightly to the \textit{opposite} direction (since we aim to go downhill) \cite{james2013introduction}:
\begin{equation}\label{NN_theta_update}
    \theta^{m+1}\leftarrow \theta^m - \rho \nabla R(\theta^m)
\end{equation}
For a fairly small value of the \textit{learning rate} $\rho$, this step will result in reducing the objective $R(\theta)$: $R(\theta^{m+1})\leq R(\theta^m)$. If the gradient vector is found zero, then we may have reach a minimum of the objective.

The calculation of $\nabla R(\theta^m)$ in Eq. \ref{NN_gradient_gen} proves to be pretty simple, and retains its simplicity even for much more complex networks, due to the \textit{chain rule} of differentiation. Since $R(\theta)$ is a sum: $R(\theta)=\frac{1}{2}\sum_{i=1}^n R_i(\theta)=\frac{1}{2}\sum_{i=1}^n(y_i-f_\theta(x_i))^2$, its gradient is also a sum over the $n$ observations. Thus, we will only look at one of these terms:
\begin{equation}\label{NN_loss_vector_term}
    R_i(\theta) = \frac{1}{2}\left(y_i- \beta_0 - \sum\limits_{k=1}^K\beta_kg(w_{k0}+\sum\limits_{j=1}^pw_{kj}x_{ij})\right)^2
\end{equation}
For simplicity, we write $z_{ik}=w_{k0}+\sum\limits_{j=1}^pw_{kj}x_{ij}$. First, we calculate the derivative with respect to $\beta_k$ \cite{james2013introduction}:
\begin{equation}\label{NN_loss_deriv_bk}
    \frac{\partial R_i(\theta)}{\partial \beta_k} = \frac{\partial R_i(\theta)}{\partial f_\theta(x_i)} \cdot \frac{\partial f_\theta(x_i)}{\partial \beta_k} = -(y_i-f_\theta(x_i))\cdot g(z_{ik})
\end{equation}
Then, we calculate the derivative with respect to $w_{kj}$ \cite{james2013introduction}:
\begin{equation}\label{NN_loss_deriv_wkj}
    \frac{\partial R_i(\theta)}{\partial w_{kj}} = \frac{\partial R_i(\theta)}{\partial f_\theta(x_i)}\cdot \frac{\partial f_\theta(x_i)}{\partial g(z_{ik})}\cdot \frac{\partial g(z_{ik})}{\partial z_{ik}}\cdot \frac{\partial z_{ik}}{\partial w_{kj}} = 
    -(y_i-f_\theta(x_i))\cdot \beta_k\cdot g^{'}(z_{ik})\cdot x_{ij}
\end{equation}

Notice that both Eqs. \ref{NN_loss_deriv_bk} and \ref{NN_loss_deriv_wkj} include the residual $y_i-f_\theta(x_i)$. In Eq. \ref{NN_loss_deriv_bk}, a fraction of that residual gets assigned to each of the hidden units according to the value of $g(z_{ik})$. Similarly, in Eq. \ref{NN_loss_deriv_wkj}, a fraction of the same residual is assigned to input $j$ via hidden unit $k$. Hence, the act of differentiation distributes a fraction of the residual to each of the parameters via the chain rule - a process known as \textit{backpropagation} in the neural network literature.

\subsection{Regularization and Dropout learning}\label{DNN_regulirize}
Gradient descent usually arrives at a local minimum after many steps. In practice, there are various approaches to speed up the process. Moreover, when $n$ is large, instead of using all $n$ observations in Eqs. \ref{NN_loss_deriv_bk} and \ref{NN_loss_deriv_wkj}, we can sample a small fraction or \textit{mini-batch} of them each time we evaluate a gradient step. This process is known as stochastic gradient minibatch
descent (SGD) and is the state of the art for learning deep neural networks. Other techniques, include \textit{ridge} or \textit{lasso} regularization, or even \textit{early stopping} \cite{james2013introduction}, but these procedures will not concern us in our analysis. 

Instead, we apply \textit{batch normalization} between some of the hidden layers. This technique works by normalizing the data contained in each mini-batch. This means it computes the mean and variance of data in a batch and then adjusts the values so that they have similar range. After that it scales and shifts the values so that model learns effectively (see \href{https://www.geeksforgeeks.org/deep-learning/what-is-batch-normalization-in-deep-learning/}{documentation}).

\begin{figure}[h]
    \centering
    \includegraphics[width=0.8\linewidth]{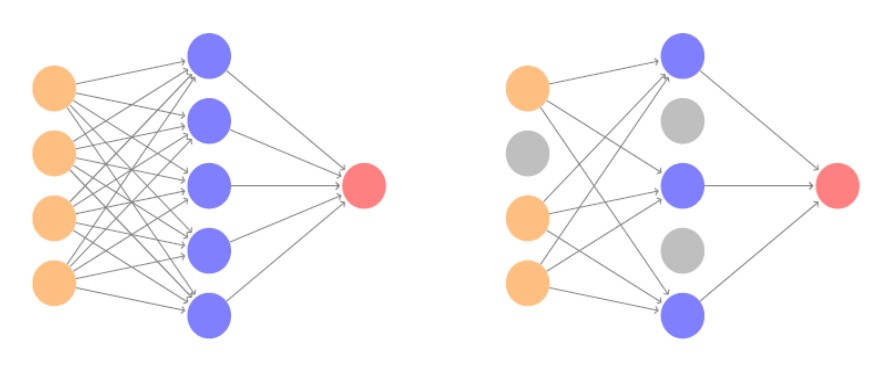}
    \caption{Dropout Learning. \textit{Left}: a fully connected network. \textit{Right}: network with dropout in the input and hidden layer. The nodes in gray are selected at random, and neglected in an instance of training. Figure adapted from \cite{james2013introduction}.}
    \label{fig:ANN_dropouts}
\end{figure}

We also, apply a fairly new and efficient form of regularization called \textit{dropout}. This form, is similar to ridge regularization in some aspects \cite{james2013introduction}. Getting inspiration from \textit{Random Forests} (see section \ref{rf_reg}), the basic idea of this technique lies in randomly
neglect a fraction $\phi$ of the units in a layer when fitting the model. A visualization of this procedure is presented in Fig. \ref{fig:ANN_dropouts}. This is done independently each time a training observation is processed. The remaining units stand in for those missing, and their weights are scaled up by a factor of $1/(1-\phi)$ to counterbalance the elimination. This prevents nodes from becoming over-specialized, and can be treated as a form of regularization. In practice, dropout is carried out by randomly zeroing the activations for the “dropped out” units, while keeping the model's architecture intact. 

Fortunately, the \texttt{TensorFlow} software we use in our study, offers a good \href{https://www.tensorflow.org/}{documentation} on how to set up and fit deep learning models to data. So, most of the technicalities are hidden from the user.

\subsection{Tuning}

Every neural network requires a number of choices, which all have an effect on the performance \cite{james2013introduction}:

\begin{itemize}
    \item \textit{The number of hidden layers} and \textit{the number of units per layer}. The trend nowadays, is that the number of units per hidden layer can be large and overfitting can be monitored through various types of regularization.

    \item \textit{Regularization tuning parameters}. In these belong the \textit{dropout rate} $\phi$ and the \textit{strength} $\lambda$ of lasso and ridge regularization, and are typically set independently at each layer.

    \item \textit{Elements of stochastic gradient descent}. In these belong the \textit{batch size}, the \textit{number of epochs}, and if used, details of \textit{data augmentation} (see section 10.3.4. in \cite{james2013introduction})
\end{itemize}

Choices and adjustments like the aforementioned can make a significant difference. More detailed tuning and fitting of a similar network could lead to errors orders of magnitude smaller. However, the calibration process can be tedious and can lead to overfitting, if done carelessly.

\part{Computational Part}

\chapter{Solving the TOV equations}\label{TOV Solve}
In this chapter, we will discuss the procedure for solving TOV equations (see appendix \ref{TOV Theory}). We will point out the basic preparation steps that need to be taken. In addition, we will present the solution results, for all the equations of state of Neutron and Quark stars mentioned in the previous chapters.

\section{Scaling and preliminaries}\label{TOV_scaling_n_prelim}
The first step, is the scaling of the TOV equations, bringing them to a form that is suitable for numerical integration. Their new form reads \cite{chatzisavvas2009complexity,kanakis2019constraints,kourmpetis2024nuclear}:
\begin{equation}\label{TOV_scaled}
\begin{aligned}
  \frac{d\bar{P}(r)}{dr} &= -1.474\frac{\bar{\epsilon}(r)\bar{m}_r(r)}{r^2}
  \left(1+\frac{\bar{P}(r)}{\bar{\epsilon}(r)}\right)
  \left(1+11.2\cdot10^{-6}r^3\frac{\bar{P}(r)}{\bar{m}_r(r)}\right)
  \left(1-2.948\frac{\bar{m}_r(r)}{r}\right)^{-1} \\
  \frac{d\bar{m}_r(r)}{dr} &= 11.2\cdot10^{-6}r^2\bar{\epsilon}(r)
\end{aligned}
\end{equation}

where
\begin{equation}\label{scales}
\begin{aligned}
m_r(r)&=\bar{m}_r(r)M_\odot \\
\epsilon(r)&=\bar{\epsilon}(r)\epsilon_0 \\
P(r)&=\bar{P}(r)\epsilon_0 \\
\epsilon_0&=1\text{ }MeV\cdot fm^{-3} \\
\frac{GM_\odot}{c^2} &= 1.474\text{ }km \\
\frac{4\pi}{M_\odot c^2} &= 0.7\cdot 10^{-4}\text{ }s^2\cdot kg^{-1}\cdot km^{-2}
\end{aligned}
\end{equation}

In Eqs. \ref{TOV_scaled} and \ref{scales}, the quantities $\bar{P}(r)$, $\bar{\epsilon}(r)$ and $\bar{m}_r(r)$ are dimensionless. The radius $r$ is measured in $km$. From the second equation in Eqs. \ref{TOV_scaled}, we get \cite{chatzisavvas2009complexity}:
\begin{equation}\label{TOV_total_mass}
    M=\bar{m}_r(R)=11.2\cdot10^{-6}\int\limits^R_0 r^2\bar{\epsilon}(r)dr=b_0\int \bar{\epsilon}(r)d\textbf{r}
\end{equation}
with $M$ the total mass of the star.

During the solution of Eqs. \ref{TOV_scaled}, the compact star is scanned from the inside out, i.e. from the center to the surface. To avoid division by zero ($r_{center}\rightarrow{0}$ and $m_{center}\rightarrow{0}$), we set the first radius interval of the numerical integration to be: $r\in[10^{-9},10^{-2}]$ $km$ and assume a microscopic mass at the center of the star: $M_0=10^{-12}M_
\odot$. The integration step is set to: $r_{step}=10^{-3}$ $km$. Finally, we consider that the surface of the star has been reached, when the pressure becomes less than $\bar{P}_{surface}=10^{-12}$ (or $P_{surface}=10^{-12}$ $MeV \cdot fm^{-3}$). The final values of $\bar{m}_r$ and $r$, represent the total mass $M$ and the radius $R$ of the star, for a given pressure $\bar{P}_c$ at the center of the star.

\section{Methodology and solutions for Neutron Stars EoSs}\label{TOV_solved_NS}
\subsection{Main EoSs}\label{TOV_solved_mainNS}
The simplest representation of the EoS of a Neutron Star, involves the $4$ crust EoSs (see section \ref{Crust_EOSs}) and a 'main' EoS as the core EoS (see section \ref{Core_EOSs}). The total EoS reads then:
\begin{equation}\label{total_EOS_mainNS}
    \epsilon(P) = 
    \begin{cases}
    \epsilon_{crust}(P), & P<P_{crust-core} \\
    \epsilon_{core}(P), & P\geq P_{crust_core}
    \end{cases}
\end{equation}
where
\begin{equation}\label{total_crustEOS}
    \epsilon_{crust}(P) =
    \begin{cases}
        \epsilon_{crust\_4}(P), & P\leq 1.44875\cdot10^{-11}\text{ }[MeV\cdot fm^{-3}] \\
        \epsilon_{crust\_3}(P), & 1.44875\cdot10^{-11} < P\leq 4.1725\cdot10^{-8}\text{ }[MeV\cdot fm^{-3}] \\
         \epsilon_{crust\_2}(P), & 4.1725\cdot10^{-8} < P\leq 9.34375\cdot10^{-5}\text{ }[MeV\cdot fm^{-3}] \\
         \epsilon_{crust\_1}(P), & 9.34375\cdot10^{-5}\text{ }[MeV\cdot fm^{-3}] < P \leq P_{crust-core}
    \end{cases}
\end{equation}
and
\begin{equation}\label{total_coreEOS_mainNS}
   \epsilon_{core}(P) = \epsilon_{main}(P) 
\end{equation} 
We denote, once more, that the value of $P_{crust-core}$ is $0.696$ $MeV\cdot fm^{-3}$ for the PS EoS and $0.184$ $MeV\cdot fm^{-3}$ for the rest of the 'main' EoSs.

\begin{figure}[h!]
    \centering
    \includegraphics[width=\linewidth]{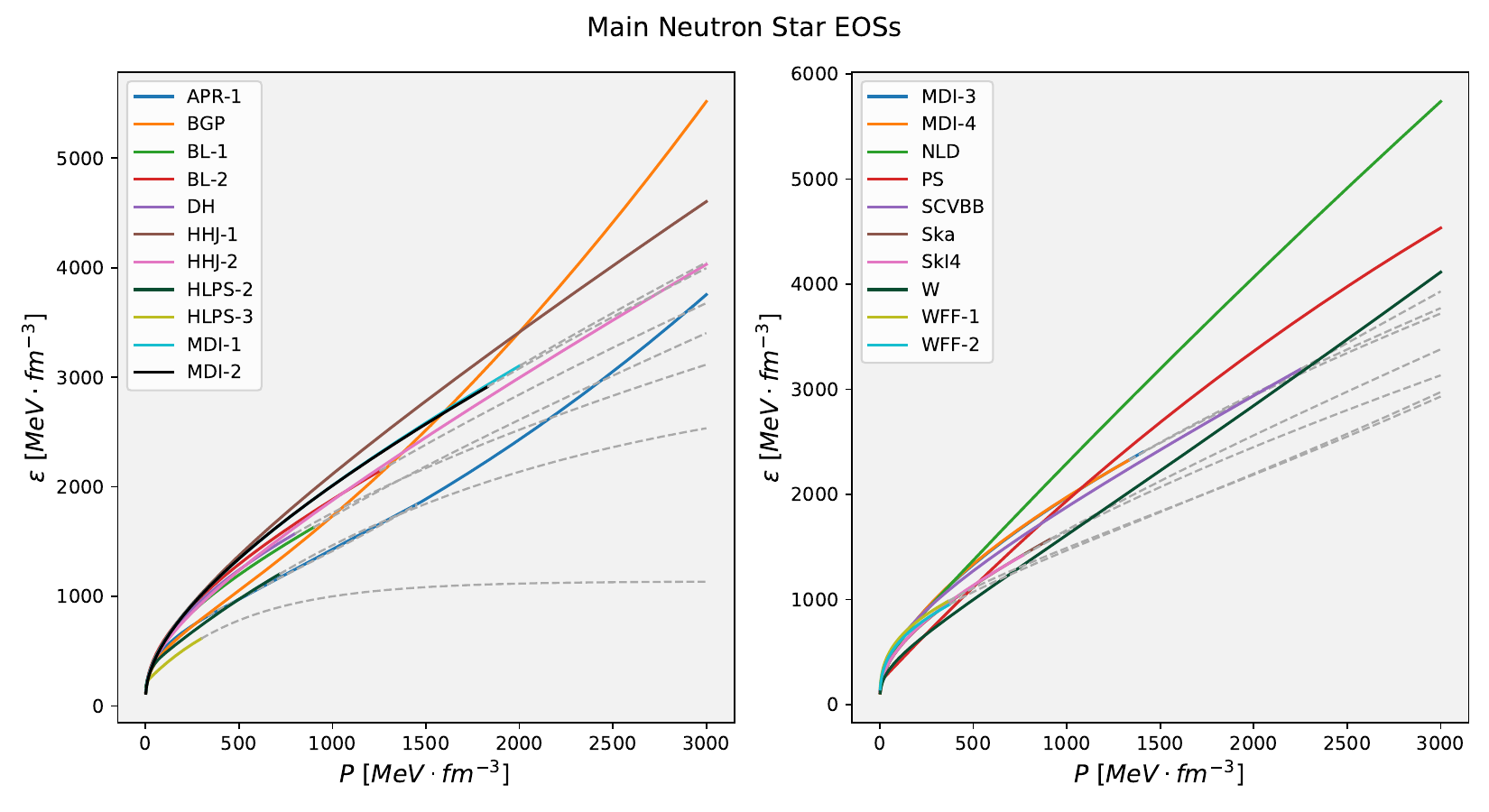}
    \caption{The 'main' EoSs of Neutron Stars (see section \ref{Core_EOSs}). The colored solid line parts of the curves correspond to points that do not violate causality ($\frac{d\epsilon}{dP}\geq1$), while the gray dashed line parts reflect to parts that violate causality ($\frac{d\epsilon}{dP}<1$)). See \texttt{ExoticStarsResults\_1.ipynb} in Table \ref{tab:TOV_handle_data}.}, 
    \label{fig:mainNS_EoSs}
\end{figure}

\begin{figure}[h!]
    \centering
    \includegraphics[width=\linewidth]{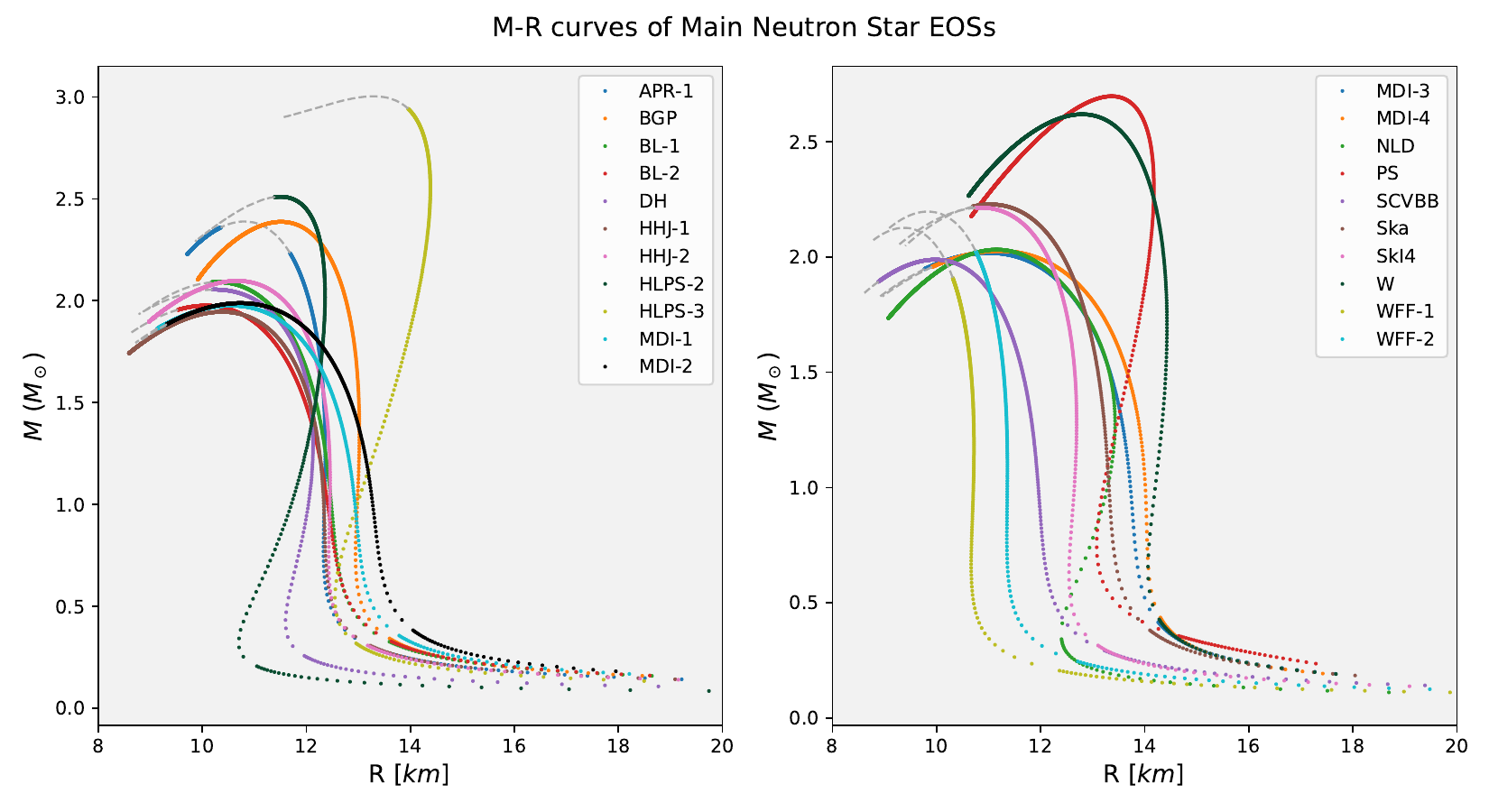}
    \caption{The $M-R$ curves that correspond to the 'main' EoSs of Neutron Stars (see section \ref{Core_EOSs}). The colored parts of the curves correspond to points that do not violate causality ($\frac{d\epsilon}{dP}\geq1$), while the gray dashed line parts reflect to points that violate causality ($\frac{d\epsilon}{dP}<1$)). Notice that all curves include the prediction for the potential maximum mass of the star. See \texttt{ExoticStarsResults\_1.ipynb} in Table \ref{tab:TOV_handle_data}.}, 
    \label{fig:mainNS_MR}
\end{figure}

In Fig. \ref{fig:mainNS_EoSs} the $\epsilon - P$ curves that correspond to the $21$ "main" EoSs are depicted. Notice that lots of them violate causality after a value of pressure $P$. The violation though, happens, in most cases, at rather high pressures (or equivalently high mass densities): around $1000$ $MeV\cdot fm^{-3}$ or higher. Of course, there are exceptions, like the HLPS-3, the WFF-1 and WFF-2 ones, where the causality violation happens at pressure less than $500$ $MeV\cdot fm^{-3}$. As for the maximum pressure of $3000$ $MeV\cdot fm^{-3}$, this was selected to ensure the complete prediction of the $M-R$ graph from all EoSs, and in particular the prediction of the potential maximum mass of the Neutron Star (see Fig. \ref{fig:mainNS_MR}).

The solution of TOV equations (see files \texttt{tov\_solver\_NS.py} and \texttt{tov\_solver\_NS\_par.py} in Table \ref{tab:TOV_solve_codes_NS}) for the $21$ 'main' EoSs reveals the $M-R$ curves in Fig. \ref{fig:mainNS_MR}. As shown, the diversity of EoSs is passed through and preserved in the $M-R$ curves. One can see, that some curves reach masses greater than $2.5M_\odot$, while others barely surpass the mass of $2M_\odot$. In addition, the curves differ in the minimum radius they reach. Most curves approach radii at least at 9 km. However, there are also curves, such as the M-R curves of HLPS-3, PS and W, in which the radius does not fall below 10 km. On the contrary, all $M-R$ curves start from high radii at small masses, as expected for compact stars with crust (see Table \ref{tab:QS_vs_NS}).

Finally, we note that the 21 'main' EoSs leave several empty spaces, both at the $\epsilon-P$ and $M-R$ planes. These spaces need to be covered by more curves, so that much more data can be collected, to establish a better connection between the $M-R$ curves and the EoSs from which they were derived. Nevertheless, the diversity of the 21 'main' EoSs, makes them suitable for testing the performance and accuracy, of the regression models we develop (see chapter \ref{Build models Comp}).

\subsection{Polytropic and linear EoSs}\label{TOV_solved_polylinNS}
A more complex representation of the Neutron Star can be achieved, using polytropic EoSs. In this case, the general form of the total EoS can be written as:
\begin{equation}\label{total_EOS_polyNS}
    \epsilon(P) = 
    \begin{cases}
        \epsilon_{crust}(P), &P<P_{crust-core} \\
        \epsilon_{main}(P), &P_{crust-core}\leq P<P_0 \\
        \epsilon_{poly}(P), & P_0\leq P\leq P_{n}
    \end{cases}
\end{equation}
where the crust EoS $\epsilon_{crust}(P)$ consists of the same $4$ EoSs as in Eq. \ref{total_crustEOS} and the core EoS consists of two layers:
\begin{equation}\label{total_coreEOS_polyNS}
    \epsilon_{core}(P) =
    \begin{cases}
        \epsilon_{main}(P), &P_{crust-core}\leq P<P_0 \\
        \epsilon_{poly}(P), & P_0\leq P\leq P_{n}
    \end{cases}
\end{equation}
The first layer starts from the boundary pressure $P_{crust-core}$ (the transition point from the crust to the core of the star) and reaches up to a pressure $P_0$, featuring one of the 'main' EoSs for Neutron Stars. Then, the second layer starts and spans across the region $[P_0,P_n]$, featuring $n$ pressure (or mass density) segments and a polytropic EoS, parametrized with $n$ piecewise polytropes (see section \ref{PolyLin_EOSs}).

It is obvious, that one must determine the values of the pressures $P_{crust-core}$, $P_0$ and $P_n$, as well as the number of pressure segments $n$ in the polytropic area and the value of the parameter $\Gamma$ in each one of them. In our study, we choose the HLPS-2 and HLPS-3 EoSs for the first core layer. The HLPS-2 model is sufficiently shifted to the left (at least for small masses, as shown in Fig. \ref{fig:mainNS_MR}), making it suitable for generating polytropic EoSs with $M-R$ diagrams, which successfully scan the range of radii smaller than $10km$ and masses smaller than $2M_\odot$. On the other hand, the more stiff form of the HLPS-3 model, allows the generation of polytropic EoSs for scanning regions with masses greater than $2M_\odot$ and radii greater than $10km$. Of course, a region of overlap of the polytropic $M-R$ diagrams, derived from the two models, is expected and to some extent desirable.

\begin{figure}[h!]
    \centering
    \includegraphics[width=0.8\linewidth]{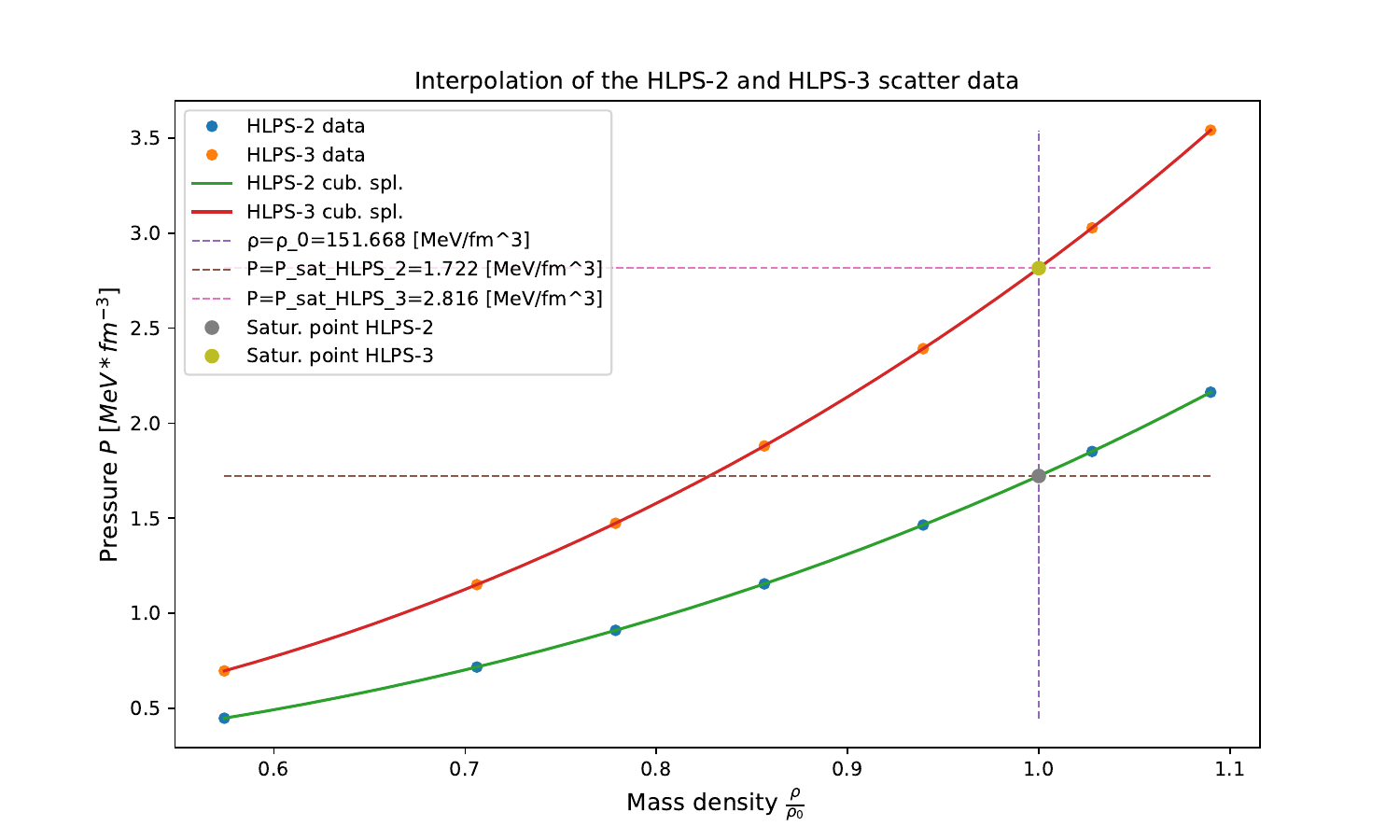}
    \caption{Interpolating via cubic splines the scatter data (pressure vs mass density) of the HLPS-2 and HLPS-3 models (see \cite{hebeler2013equation}) to obtain the value of pressure at saturation density $\rho_0$. The horizontal axis features the ratio $\frac{\rho}{\rho_0}$. See \texttt{StudyPolyNS.ipynb} in Table \ref{tab:TOV_solve_codes_NS}.}, 
    \label{fig:polyNS_interpld}
\end{figure}

Hence, the value of $P_{crust-core}$ is at $0.184$ $MeV\cdot fm^{-3}$ (see section \ref{TOV_solved_mainNS}). Next, we assume that the polytropic layer of the core, starts at the nuclear saturation density: $\rho_0=\rho_{sat}=2.7\cdot 10^{14}$ $g\cdot cm^{-3}$ or $\rho_0 c^2=151.67$ $MeV\cdot fm^{-3}$. Therefore, the pressure $P_0$ in Eqs. \ref{total_EOS_polyNS} and \ref{total_coreEOS_polyNS} equals to the pressure at nuclear saturation ($P_{sat}$). Its value differs for the two models, HLPS-2 and HLPS-3, and has to be calculated separately. We get the scatter data of pressure $P$ and mass density $\rho$ times $c^2$, around the nuclear saturation point (see Table \ref{tab:Scatter_data_saturation}). Then, we evaluate the cubic spline that passes through these points, considering the mass density values as their $x$-coordinates and the pressure values as their $y$-coordinates. At last, we take the output of the cubic spline at nuclear saturation density (see Fig. \ref{fig:polyNS_interpld}). We have: 
\begin{itemize}
    \item HLPS-2: $P_{sat\_HLPS-2}=P_{cub. spl\_HLPS-2}(\rho_0c^2)=1.722$ $MeV\cdot fm^{-3}$

    \item HLPS-3: $P_{sat\_HLPS-3}=P_{cub. spl\_HLPS-3}(\rho_0c^2)=2.816$ $MeV\cdot fm^{-3}$
\end{itemize}

\begin{table}[h]
    \centering
    \begin{tabular}{|c|c|c|c|}
    \hline
    \hline
       $\rho c^2$  & $\frac{\rho}{\rho_0}$ & $P$ (HLPS-2) & $P$ (HLPS-3) \\
    \hline   
        87.07  & 0.57 & 0.4470 & 0.6960 \\
        107.1  & 0.71 & 0.7162 & 1.150 \\
        118.1  & 0.78 & 0.9094 & 1.473 \\
        129.9  & 0.86 & 1.154 & 1.880 \\
        142.5  & 0.94 & 1.464 & 2.392 \\
        155.9  & 1.03 & 1.851 & 3.028 \\
        165.3  & 1.1 & 2.163 & 3.542 \\
    \hline
    \hline     
    \end{tabular}
    \caption{Numerical data of mass density $\rho c^2$ and pressure $P$ for the HLPS-2 and HLPS-3 EoSs, around nuclear saturation density $\rho_0 c^2=151.67$  $MeV\cdot fm^{-3}$ (see \cite{hebeler2013equation}). The units in columns 1, 3 and 4 are in $MeV\cdot fm^{-3}$, while the values in column 2 are dimensionless.}
    \label{tab:Scatter_data_saturation}
\end{table}

The value of the final pressure $P_n$ depends on the value of the mass density $\rho_n$ at the right endpoint of the $n$-th (last) polytropic segment and the values of the parameter $\Gamma$ among all segments. Two parameterizations with different values of $\Gamma$, result in different values of pressure $P_n$ at the same mass density $\rho_n$. On the contrary, two parameterizations with same values of $\Gamma$ and $\rho_n$, result always in the same value of pressure $P_n$, regardless the sequence of $\Gamma$ values. In our study, we consider as reference point the density $\rho_{ref}=7.5\rho_0$, which corresponds to the maximum mass of the star \cite{raithel2016neutron}. At first, we create $n$ segments, evenly spaced in the logarithm of the region $[\rho_0,\rho_{ref}]$. The first polytrope starts at density $\rho_0$ and the last polytrope ends at density $\rho_n=\rho_{ref}$. Then, we increase the value $\rho_n$, based on the value of $\Gamma$ at the first segment, to ensure complete prediction of the $M-R$ curve. So, the $n$-th segment will be longer than the rest. More specifically, we apply the following:
\begin{itemize}
    \item For $\Gamma_1\leq2$: $\rho_n=\rho_{ref}\rightarrow{\rho_n=15\rho_0}$ (soft).
    \item For $2<\Gamma_1\leq3$: $\rho_n=\rho_{ref}\rightarrow{\rho_n=12\rho_0}$ (intermediate).
    \item For $\Gamma_1>3$: $\rho_n=\rho_{ref}\rightarrow{\rho_n=9\rho_0}$ (stiff).
\end{itemize}

\begin{figure}[h]
    \centering
    \includegraphics[width=\linewidth]{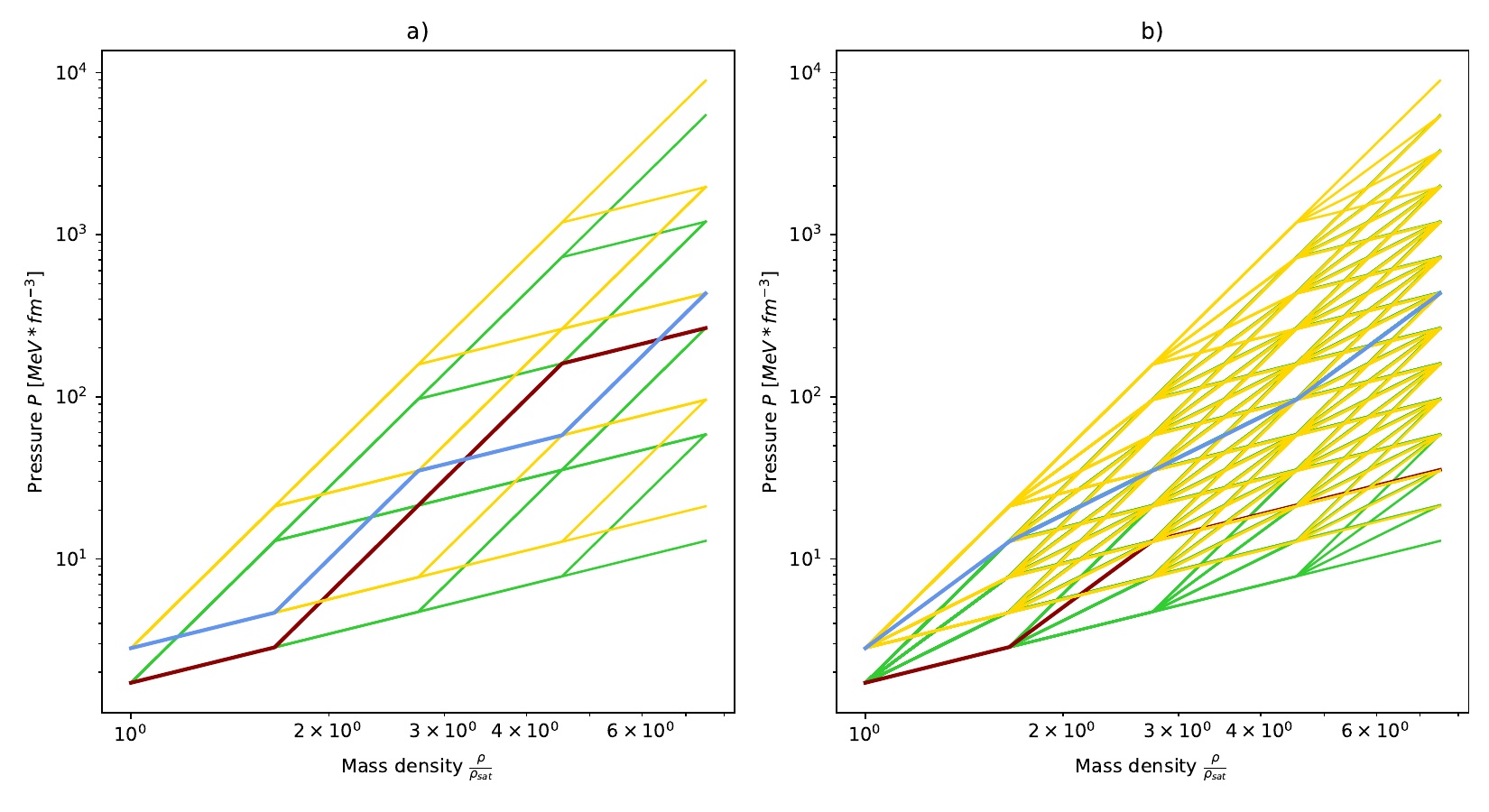}
    \caption{Grids of polytropes for $4$ mass density segments and a) $2$ available choices $\{1,4\}$ or b) $4$ available choices $\{1,2,3,4\}$ for $\Gamma$ values. The green grid corresponds to polytropes that start at the saturation pressure of HLPS-2 ($1.722$ $MeV\cdot fm^{-3}$), while the yellow grid corresponds to polytropes that start at the saturation pressure of HLPS-3 ($2.816$ $MeV\cdot fm^{-3}$). The red line features (from left to right) the sequence $\{\Gamma:1\rightarrow{}4\rightarrow{}4\rightarrow{}1\}$ in left graph and the sequence $\{\Gamma:1\rightarrow{}3\rightarrow{}1\rightarrow{}1\}$ in right graph. The blue line features (from left to right) the sequence $\{\Gamma:1\rightarrow{}4\rightarrow{}1\rightarrow{}4\}$ in left graph and the sequence $\{\Gamma:3\rightarrow{}2\rightarrow{}2\rightarrow{}3\}$ in right graph. See \texttt{StudyPolyNS.ipynb} in Table \ref{tab:TOV_solve_codes_NS}.} 
    \label{fig:polyNS_P_grid}
\end{figure}

As for the values of $\Gamma$ at each segment, one can experiment. For two available choices: $\Gamma\in\{1,4\}$, a grid of polytropes is formed with two branches at each endpoint, as shown in the left graph of Fig \ref{fig:polyNS_P_grid}. On the contrary, for four available choices: $\Gamma\in\{1,2,3,4\}$, a grid of polytropes is formed with four branches at each endpoint, as shown in the right graph of Fig \ref{fig:polyNS_P_grid}. We observe, that a grid with more branches is denser and more detailed in scanning the same area, as it offers more configuration options and thus more polytropic EoS. In particular, using Eq. \ref{number_of_polyEoSs}, for $n=4$ segments and $l=2$ $\Gamma$ choices we get: $f=2^4=16$ mock polytropic EoSs, while for $n=4$ segments and $l=4$ $\Gamma$ choices we get: $f=4^4=256$ mock polytropic EoSs.

Finally, we have to treat the possible violation of causality, as we discussed in section \ref{PolyLin_EOSs}. Assuming the violation happens at a pressure $P_{tr}<P_n$, we keep the polytropic behavior till that pressure and then employ a \textit{Maxwell} transition to linear behavior. We consider the continuity of the EoS, so the term $\Delta\epsilon$ in Eq \ref{Maxwell_construct} vanishes. Furthermore, we fix the slope of the linear EoS at the causality limit: $c_s/c=1$ or $(c_s/c)^{-2}=d\epsilon/dP=1$. In this case, the total EoS of the Neutron Star reads:
\begin{equation}\label{total_EOS_polylinNS}
    \epsilon(P) = 
    \begin{cases}
        \epsilon_{crust}(P), &P<P_{crust-core} \\
        \epsilon_{main}(P), &P_{crust-core}\leq P<P_0 \\
        \epsilon_{poly}(P), & P_0\leq P\leq P_{tr} \\
        \epsilon_{lin}(P), & P_{tr}< P \leq P_n
    \end{cases}
    = 
     \begin{cases}
        \epsilon_{crust}(P), &P<P_{crust-core} \\
        \epsilon_{main}(P), &P_{crust-core}\leq P<P_0 \\
        \epsilon_{poly}(P), & P_0\leq P\leq P_{tr} \\
        \epsilon_{poly}(P_{tr})+P-P_{tr}, & P_{tr}< P \leq P_n
    \end{cases}
\end{equation}
with core EoS:
\begin{equation}\label{total_coreEOS_polylinNS}
    \epsilon_{core}(P) =
    \begin{cases}
        \epsilon_{main}(P), &P_{crust-core}\leq P<P_0 \\
        \epsilon_{poly}(P), & P_0\leq P\leq P_{tr} \\
        \epsilon_{lin}(P), & P_{tr}< P \leq P_n
    \end{cases}
    =
    \begin{cases}
        \epsilon_{main}(P), &P_{crust-core}\leq P<P_0 \\
        \epsilon_{poly}(P), & P_0\leq P\leq P_{tr} \\
        \epsilon_{poly}(P_{tr})+P-P_{tr}, & P_{tr}< P \leq P_n
    \end{cases}
\end{equation}

Choosing the option with four mass density segments and four choices for $\Gamma$ ($\{1,2,3,4\}$), we produce $256$ mock EoSs, for each of the two 'main' EoSs, HLPS-2 and HLPS-3. That is, $512$ mock EoSs in total. In this situation, a parallel process is necessary, solving the TOV equations for many mock EoSs at the same time (see files \texttt{tov\_solver\_polyNS\_par.py} and \texttt{tov\_solver\_polyNS\_par2.py} in Table \ref{tab:TOV_solve_codes_NS}). This way, we obtain the solution data much faster, than solving the TOV equations for each mock EoS separately.

\begin{figure}[h!]
    \centering
    \includegraphics[width=\linewidth, height=20.5cm]{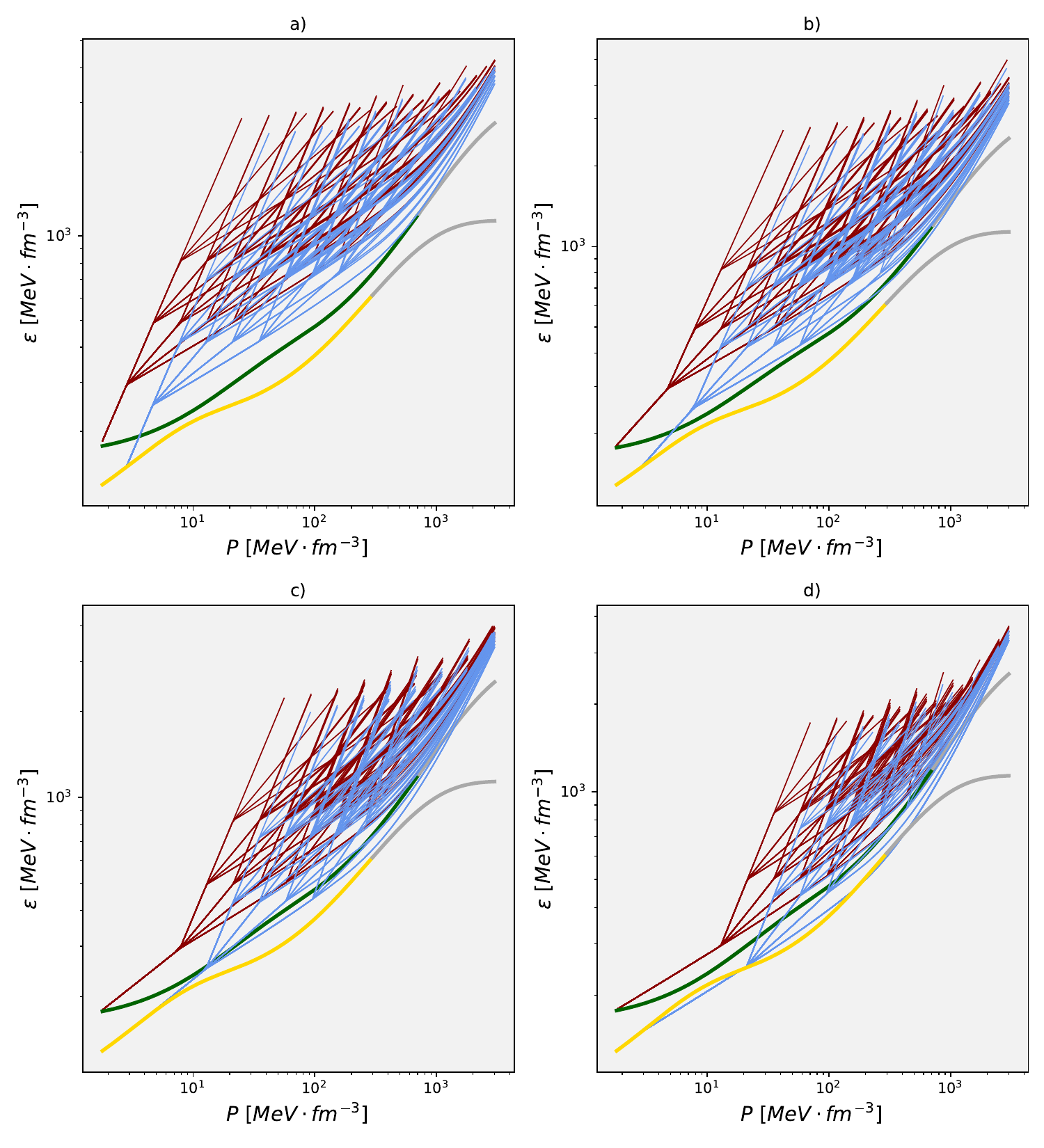}
    \caption{Plots of the $\epsilon-P$ curves of mock EoSs. Both axes are in logarithmic scale (log-log). The red grids correspond to mock EoSs derived from the HLPS-2 'main' EoS (green curve), while the blue grids correspond to mock EoSs derived from the HLPS-3 'main' EoS (yellow curve). Moreover, the grids are separated in subplots, based on the value of $\Gamma$ at the first polytropic segment ($n=1$). We have: a) $\Gamma_1=1$, b) $\Gamma_1=2$, c) $\Gamma_1=3$ and d) $\Gamma_1=4$. The gray ending in the curves of both HLPS-2 and HLPS-3 indicates the violation of causality. Notice that the grids do not feature such ending, since we have fixed the mock EoSs under the causality restrictions. See \texttt{ExoticStarsResults\_1.ipynb} in Table \ref{tab:TOV_handle_data}.} 
    \label{fig:linNS_EOSs_sep}
\end{figure}
\begin{figure}[h!]
    \centering
    \includegraphics[width=\linewidth, height=20.5cm]{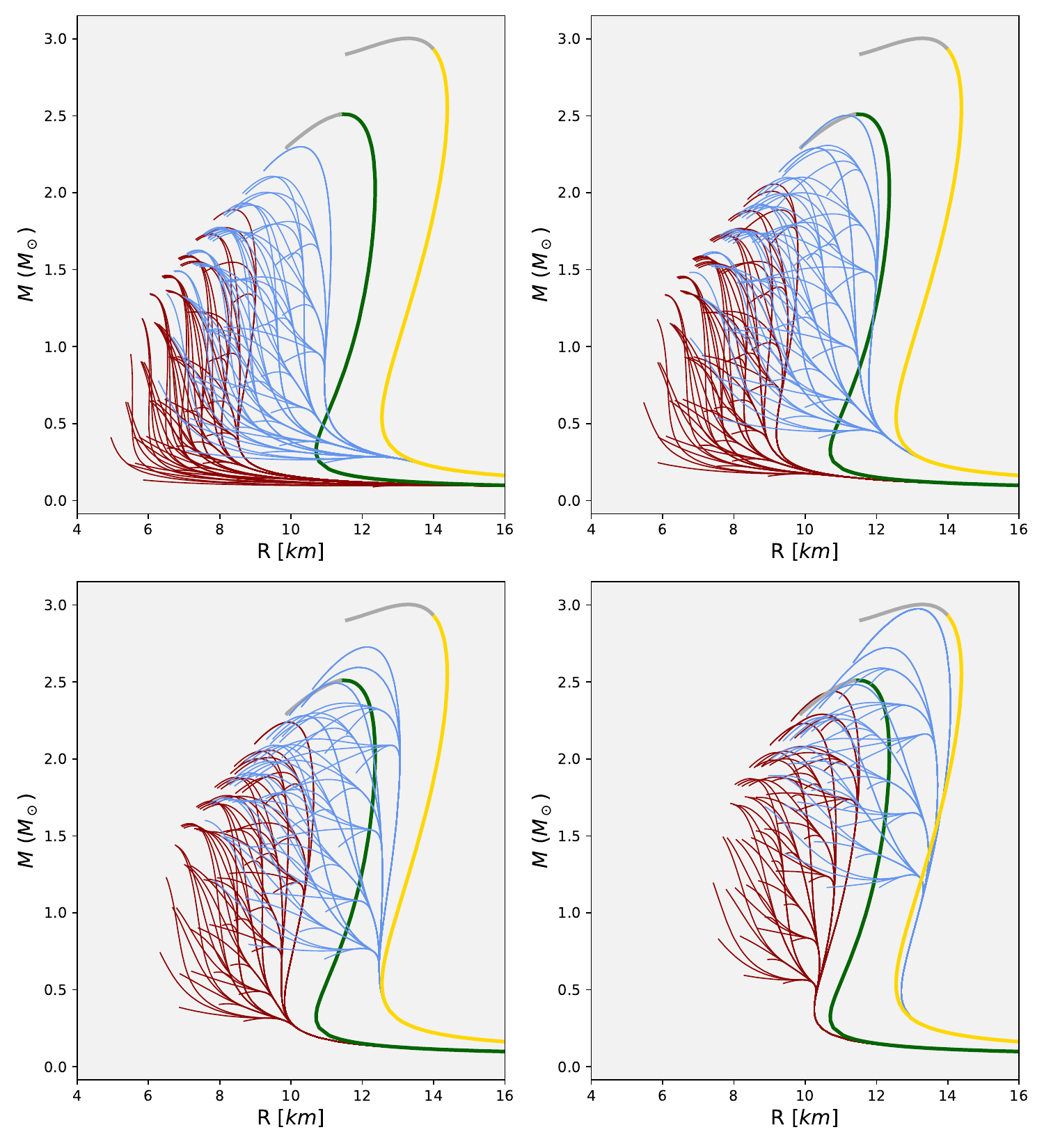}
    \caption{Plots of the $M-R$ curves of mock Neutron Stars EoSs. The red $M-R$ curves correspond to mock EoSs derived from the HLPS-2 'main' EoS, while the blue $M-R$ curves correspond to mock EoSs derived from the HLPS-3 'main' EoS. The $M-R$ curves of HLPS-2 and HLPS-3 EoSs, are also included in each subplot, with green and yellow color, respectively. Moreover, the $M-R$ curves are separated in subplots, based on the value of $\Gamma$ at the first polytropic segment ($n=1$). We have: a) $\Gamma_1=1$, b) $\Gamma_1=2$, c) $\Gamma_1=3$ and d) $\Gamma_1=4$. The gray ending in the $M-R$ curves of both HLPS-2 and HLPS-3 indicates the violation of causality. Notice that the $M-R$ curves of the mock EoSs do not feature such ending, since we have fixed the mock EoSs under the causality restrictions. See \texttt{ExoticStarsResults\_1.ipynb} in Table \ref{tab:TOV_handle_data}.} 
    \label{fig:linNS_MR_sep}
\end{figure}
\clearpage

In Fig. \ref{fig:linNS_EOSs_sep} we present the $\epsilon-P$ curves of the 512 mock EoSs, as produced from the solution data of TOV equations. As one can see, the curves terminate at different pressures. The softer ones, might have a final pressure of few dozens of $MeV\cdot fm^{-3}$, while the stiff ones might terminate at pressures of thousands of $MeV\cdot fm^{-3}$! In any case, we set an upper bound on the final pressure at $3000$ $MeV\cdot fm^{-3}$, to align with the results of the 'main' EoSs in Fig. \ref{fig:mainNS_EoSs}. Furthermore, the parametrization with greater values of $\Gamma$ results in mock EoSs that are closer to the behavior of the 'main' EoS, from which they were derived. The grids for $\Gamma_1=1$ are wider (in logarithmic scale) and most mock EoSs are soft and far from the respective 'main' EoS. As the value of $\Gamma_1$ increases the grids get tighter and more mock EoSs are stiff enough, tending to follow the behavior of HLPS-2 or HLPS-3. This is the effect of the polytropic parameterization. 

A similar situation is expected for the $M-R$ curves. Indeed, one can see in Fig. \ref{fig:linNS_MR_sep} that the $M-R$ curves of the mock EoSs shift to the right and upward, as the value of $\Gamma$ in all segments becomes higher, tending to match the $M-R$ curves of HLPS-2 and HLPS-3. The overlap between the mock EoSs of HLPS-2 and HLPS-3 in some areas, is another significant aspect, as it offers a denser coverage of the $M-R$ plane. Since, the $M-R$ curves are shifted to the right, the overlap area is also shifted to that direction. Hence, the parameterization with small values of $\Gamma$ serves for scanning regions of small masses and radii ($M\leq1.5M_\odot$ and $R\leq9km$). On the contrary, parameterizations with higher values of $\Gamma$ allow acquiring information for greater masses and radii and eventually for $M\geq2M_\odot$ and $R\geq10km$.

\begin{figure}[h!]
    \centering
    \includegraphics[width=\linewidth,height=8cm]{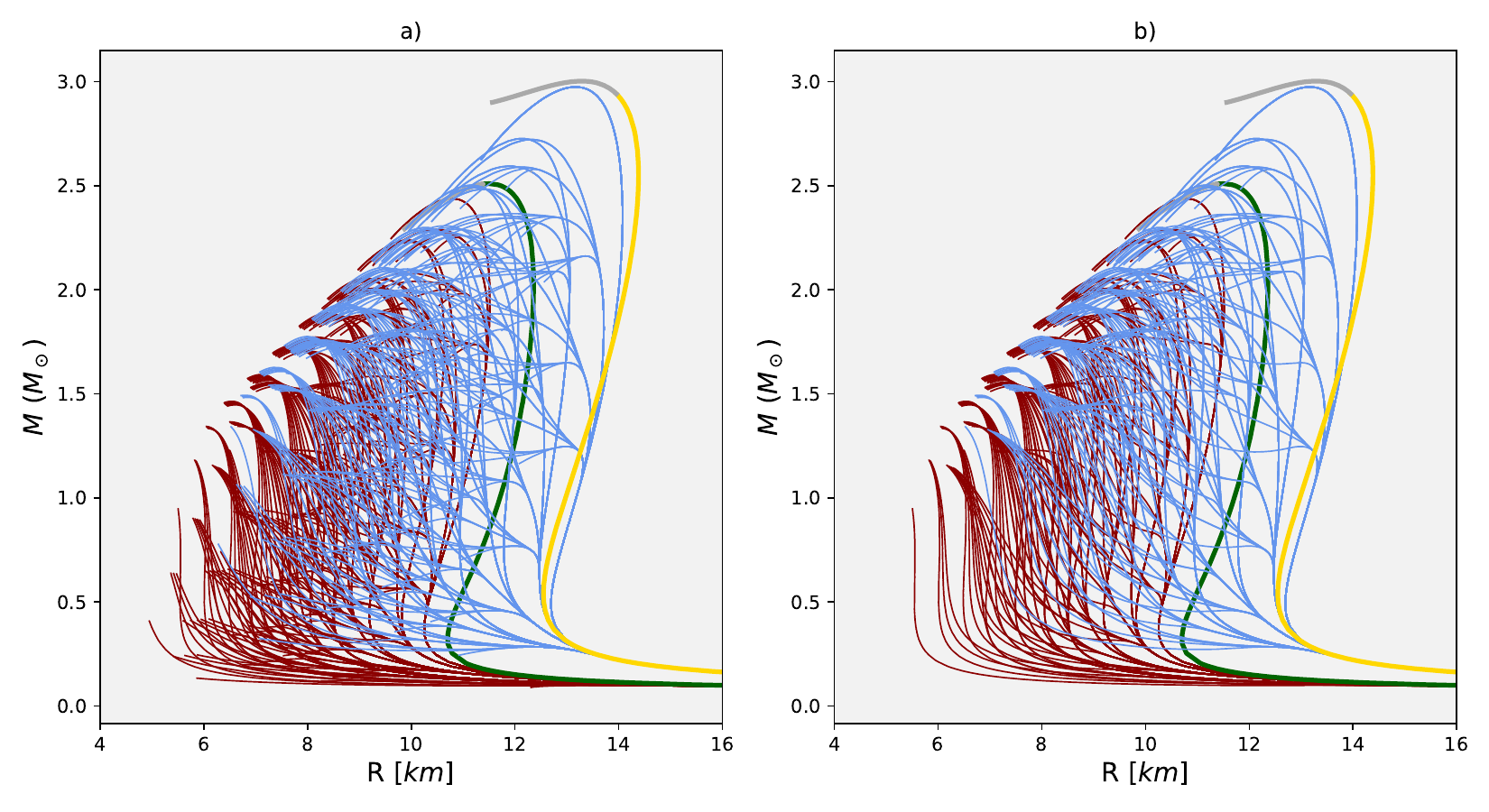}
    \caption{Plots of the $M-R$ curves of mock Neutron Stars EoSs. a) The $M-R$ curves of all $512$ mock EoSs, derived from HLPS-2 and HLPS-3 'main' EoSs, for all $\Gamma$ combinations in $4$ mass density segments. b) The $M-R$ curves of the $304$ out of the $512$ mock EoSs, exceeding the pressure of $850$ $MeV\cdot fm^{-3}$. The $M-R$ curves of HLPS-2 (green) and HLPS-3 (yellow), are also included, with gray endings marking the violation of causality. See \texttt{ExoticStarsResults\_1.ipynb} in Table \ref{tab:TOV_handle_data}.} 
    \label{fig:linNS_MR_all}
\end{figure}

This is exactly what is depicted in the overall picture in the left graph of Fig. \ref{fig:linNS_MR_all}: with the use of the $512$ mock EoSs, the part of $M-R$ space, where the occurrence of Neutron Stars is expected, is covered systematically and in detail. Now, in our analysis we aim to reconstruct values of the energy density $\epsilon$ in the pressure range $[10,800]$ $MeV\cdot fm^{-3}$. Of course, this means that all the mock EoSs, that do not surpass this value of pressure, need to be ignored. For more acceptable results, we add an extra threshold of $50$ $MeV\cdot fm^{-3}$ and choose to neglect all the EoSs that terminate under the pressure of $850$ $MeV\cdot fm^{-3}$. This filtering leaves $304$ mock EoSs to work with, as shown in the right graph of Fig. \ref{fig:linNS_MR_all}, scanning the $M-R$ space equally well, as the $512$ ones. 

\section{Methodology and solutions for Quark Stars EoSs}\label{TOV_solved_QS}

\begin{figure}[h!]
    \centering
    \includegraphics[width=\linewidth,height=13cm]{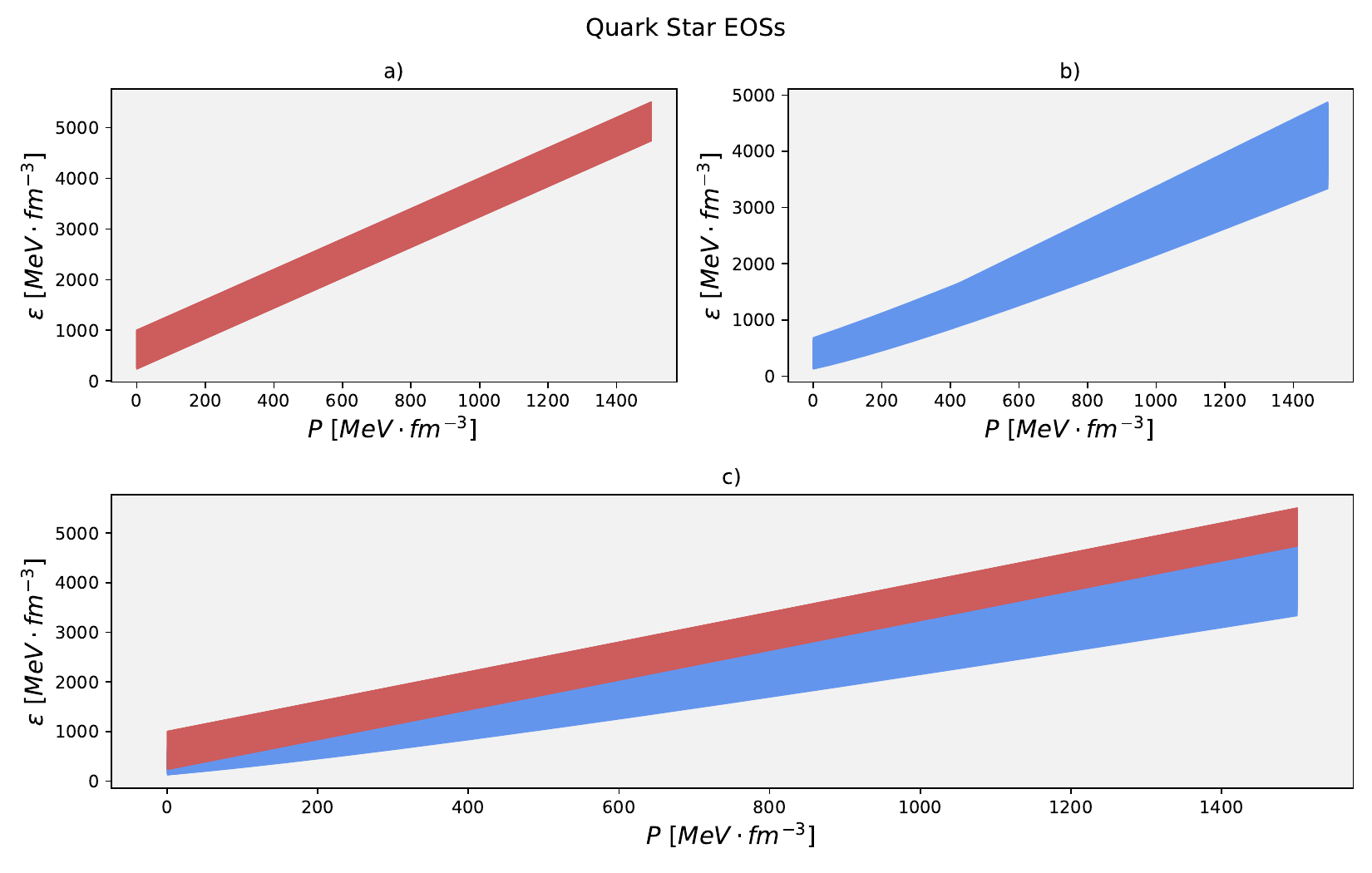}
    \caption{Plots of the $\epsilon-P$ curves of Quark Stars EoSs. a) The $381$ EoSs representing the MIT bag model are depicted with red color. The values of the bag parameter $B$ range in the interval $[60,250]$ $MeV\cdot fm^{-3}$ with a step of $0.5$ $MeV\cdot fm^{-3}$. b) The $510$ EoSs representing the CFL model are depicted with blue color. Both parameters $B$ (bag) and $\Delta$ (gap) get values in the interval $[60,250]$ $MeV\cdot fm^{-3}$ with a step of $5$ $MeV\cdot fm^{-3}$. c) Combined graph of the $\epsilon-P$ curves of both MIT bag and CFL quark matter models. See \texttt{ExoticStarsResults\_1.ipynb} in Table \ref{tab:TOV_handle_data}.} 
    \label{fig:QS_EOS_all}
\end{figure}

The methodology is quite different, when it comes to the solution of the TOV equations for Quark Stars EoSs. Since Quark Stars do not have a crust, only the core needs to be described by an equation of state\footnote{Of course, one can consider the existence of crust in Quark Stars (see section \ref{Selfbound Stars}) and include, also, an equation of state for that layer of the star}. The simplest EoS one can use, is the EoS derived from the MIT bag model. Rewriting Eq. \ref{MIT_EOS_press} in form $\epsilon=\epsilon(P)$, we get:
\begin{equation}\label{MIT_EOS_enrg}
    \epsilon = 3P +4B
\end{equation}
We immediately notice, that this is the equation of a straight line, with fixed slope equal to $3$ and adjustable increment involving the bag parameter $B$ (often denoted as $B_{eff}$). Thus, one can change the value of $B$ and produce many EoSs with different increment, as shown in a) of Fig. \ref{fig:QS_EOS_all}. The only constraint that is applied, relates to the minimum value of the bag constant: $B>57$ $MeV\cdot fm^{-3}$, as we discussed in section \ref{Selfbound Stars}. Demanding dense coverage in the $\epsilon-P$ and $M-R$ planes, as we did with the EoSs of Neutron Stars, we took values of $B$ in the interval $[60,250]$ $MeV\cdot fm^{-3}$ with a step of $0.5$ $MeV\cdot fm^{-3}$. Thus, we resulted in $381$ EoSs representing the MIT bag model. 

Additionally, a more complex approach was made, using the EoS for CFL quark matter. The form of the latter was presented in section \ref{CFL matter}, as follows:
\begin{equation}\label{CFL_EOS_enrg2_again}
    \epsilon = 3P+4B-\frac{9\alpha\mu^2}{(\hbar c)^3\pi^2}
\end{equation}
Note, that $\alpha$ is a function of the mass of strange quark $m_s$ and the gap parameter $\Delta$: $\alpha=\alpha(m_s,\Delta)$. On the other hand, the value of $\mu$ depends directly on the values of the pressure $P$ and bag constant $B$ and indirectly on the values of $m_s$ and $\Delta$ through $\alpha$. That is, the CFL EoS is a three-parameter equation, with the three parameters $m_s$, $B$ and $\Delta$, the values of which affect both the slope and the increment of the $\epsilon-P$ curve, as shown in b) of Fig.  \ref{fig:QS_EOS_all}.

\begin{figure}[h!]
    \centering
    \includegraphics[width=\linewidth, height=9.5cm]{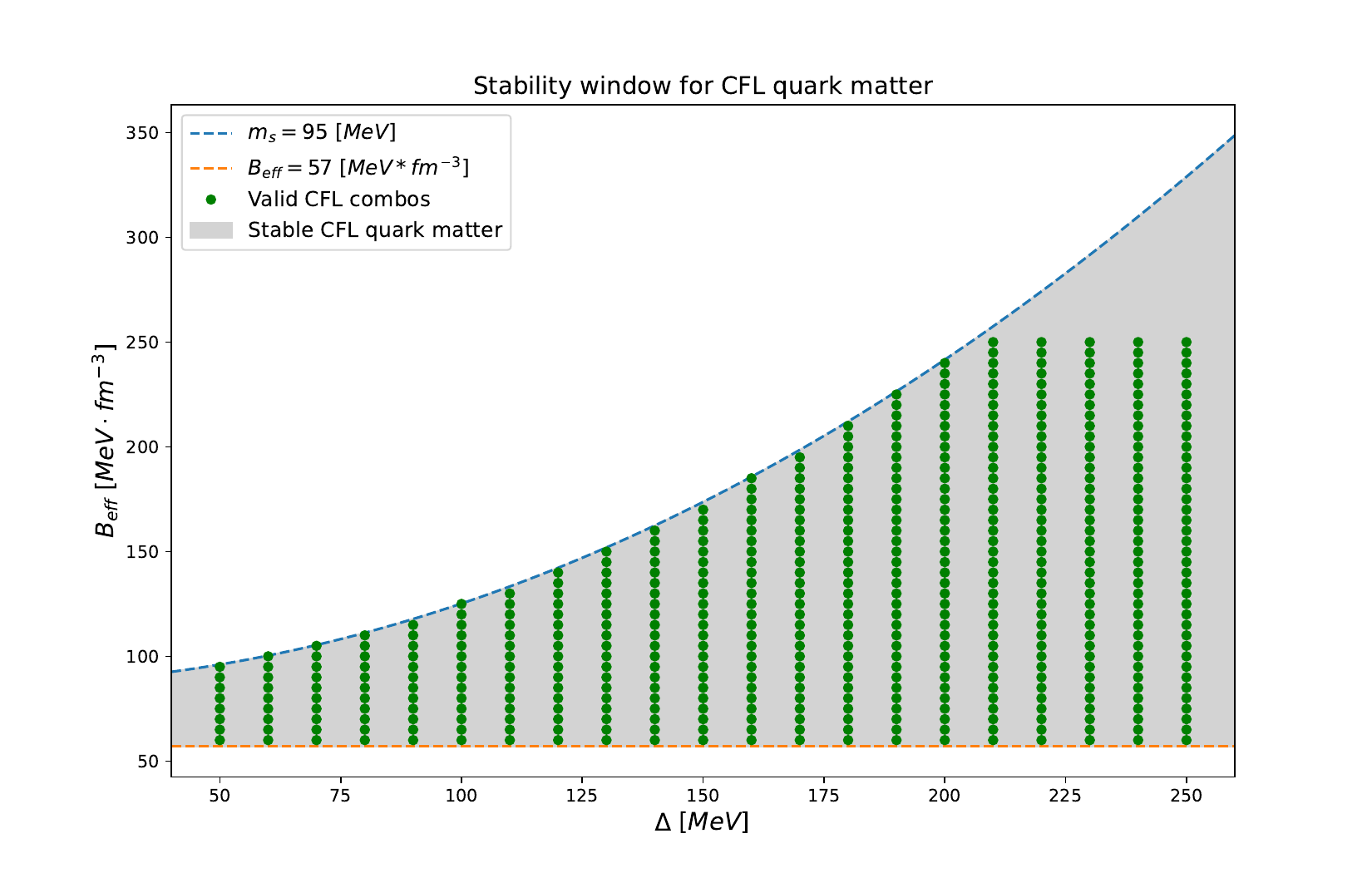}
    \caption{Scanning the stability window region for CFL quark matter with $m_s=95$ $MeV$. See \texttt{ExoticStarsResults\_1.ipynb} in Table \ref{tab:TOV_handle_data}.} 
    \label{fig:valid_cfl_combos}
\end{figure}

We fix the value of $m_s$ at $m_s=95$ $MeV$ (typical value for the mass of strange quark), making Eq. \ref{CFL_EOS_enrg2_again} a two-parameter equation. For $B$ and $\Delta$, we assumed that both  get values in the interval $[60,250]$ $MeV\cdot fm^{-3}$ with a step of $5$ $MeV\cdot fm^{-3}$. Two constraints apply this time for $B$: one for its minimum value (from Eq. \ref{bag_low_constraint}) and one for its maximum value (from Eq. \ref{bag_high_constraint}). For constant value of $m_s$, the last one yields to the equation of a curve, namely the $B_{max}=B_{max}(\Delta)$ curve, as shown with blue color and dashed-line style in Fig. \ref{fig:valid_cfl_combos}. The orange dashed-line curve marks the minimum value of $57$ $MeV\cdot fm^{-3}$. The two curves combined, define a stability region for CFL quark matter in the $B-\Delta$ plane (gray color in Fig. \ref{fig:valid_cfl_combos}). The coordinates of the green points, which scan the stability region, are valid combinations of $B$ and $\Delta$ values for $m_s=95$ $MeV$. Our assumptions, led to $510$ different valid combinations, that is $510$ different EoSs representing the CFL model.

As presented in the top-left graph of Fig. \ref{fig:QS_EOS_all}, the change in the value of $B$ for the EoSs of the MIT bag model, leads to a range of $[240,1000]$ $MeV\cdot fm^{-3}$ for the values of energy density $\epsilon$ at zero pressure. On the contrary, the EoSs of the CFL model feature a narrower range and lower values of energy density at zero pressure (for the same range of values for $B$ and $\Delta$). Furthermore, the top-right graph in Fig. \ref{fig:QS_EOS_all} reveals, that the slope of the CFL $\epsilon-P$ curves shifts to higher values after pressure greater than $500$ $MeV\cdot fm^{-3}$ and the curves become softer. The shift is different among the different combinations of $B$ and $\Delta$. Finally, the $\epsilon-P$ curves of the two quark matter models, overlap as shown in the bottom graph of Fig. \ref{fig:QS_EOS_all}. The overlap is bigger in low pressures and reduces until the final pressure $1500$ $MeV\cdot fm^{-3}$. The value of the latter was chosen to fully predict the $M-R$ curve, similar to the value $3000$ $MeV\cdot fm^{-3}$, which was chosen for Neutron Stars (see sections \ref{TOV_solved_mainNS} and \ref{TOV_solved_polylinNS}).

\begin{figure}[h!]
    \centering
    \includegraphics[width=\linewidth,height=12cm]{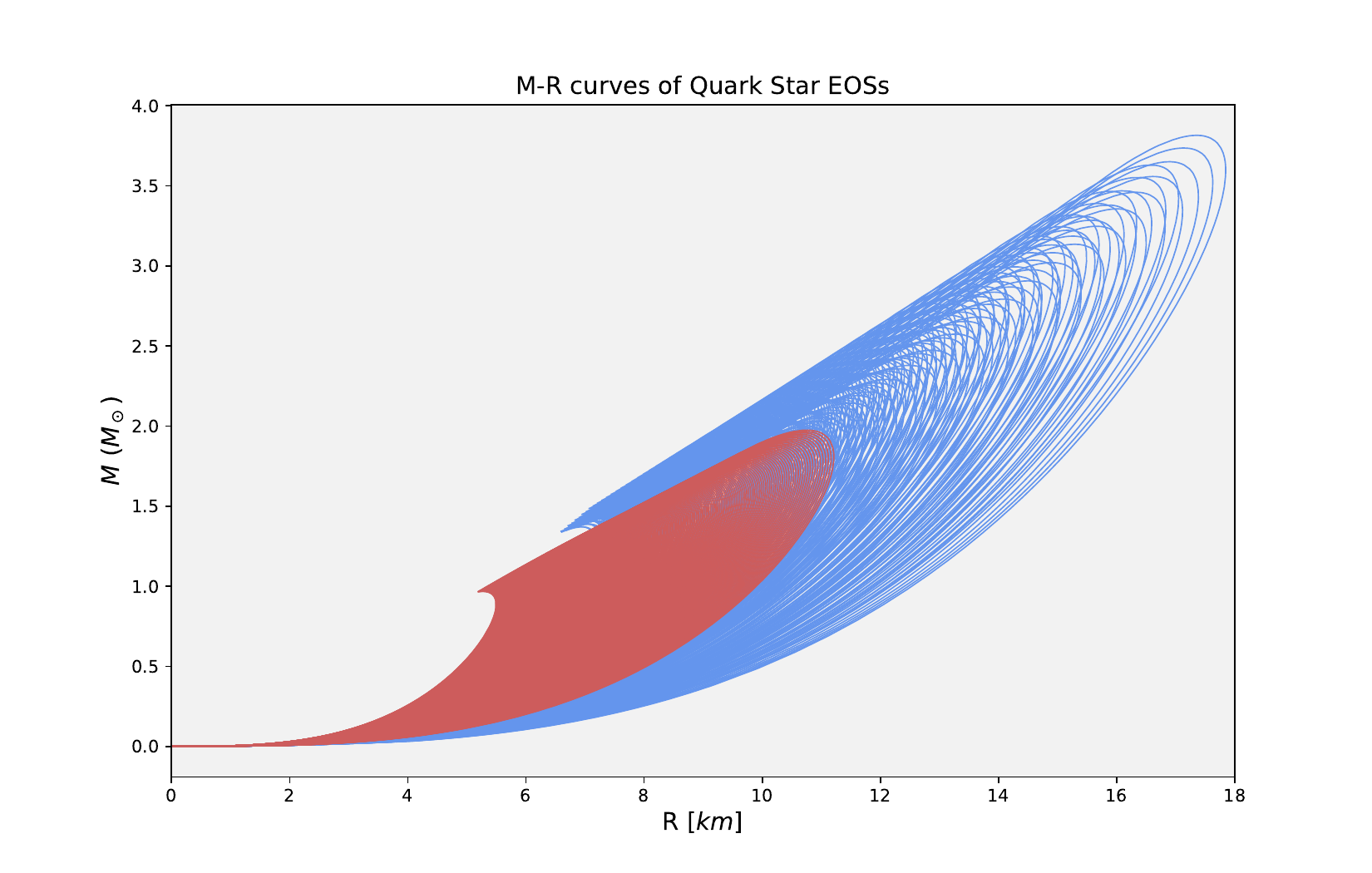}
    \caption{Plots of the $M-R$ curves of Quark Stars EoSs. The $381$ $M-R$ curves of MIT bag model EoSs are depicted with red color, while the $510$ $M-R$ curves of CFL model EoSs are depicted with blue color. In total we get $891$ curves scanning the $M-R$ space. See \texttt{ExoticStarsResults\_1.ipynb} in Table \ref{tab:TOV_handle_data}.} 
    \label{fig:QS_MR_all}
\end{figure}

The resulting $M-R$ curves in Fig. \ref{fig:QS_MR_all} provide some useful insights. All these curves start at very small masses for small radii, as expected for selfbound stars without crust (see Table \ref{tab:QS_vs_NS}). However, the CFL $M-R$ curves (blue) extend over a much wider range in the $M-R$ plane, reaching masses nearly four times the mass of the Sun ($M\sim 4M_\odot$) and radii greater than $16km$ ($R\geq16km$), in some extreme cases. The overlap between the curves of the two models is, also, present and covers a quite large area: $M\leq2M_\odot$ and $R\leq11km$.

With $891$ curves in total, the part of $M-R$ space, where the occurrence of Quark Stars is expected, is scanned systematically and in detail. This amount of EoSs required parallel programming, in order to solve the TOV equations for many EoSs simultaneously and save time (see files \texttt{tov\_solver\_mitQS\_par.py} and \texttt{tov\_solver\_cflQS\_par.py} in Table \ref{tab:TOV_solve_codes_QS}). Moreover, here none of the EoSs has to discarded, since all the $\epsilon-P$ curves reach the final pressure of $1500$ $MeV\cdot fm^{-3}$, as opposed to the polytropic mock EoSs for Neutron Stars. Also, none of them violates causality, meaning no fixing of the slope $d\epsilon/dP$ is needed. Therefore, a wider range of pressure (than $[10,800]$ $MeV\cdot fm^{-3}$) can be selected, in which the values of energy density will be reconstructed, as we will see in the next chapters.

\chapter{Building and Fitting Regression Models}\label{Build models Comp}

We move to the next phase of our study, the analysis. This phase includes several steps, from the collection and preparation of certain solution data to be suitable for regression techniques, to the fine-tuning of machine and deep learning models and the evaluation of their accuracy. These subjects will be addressed thoroughly in this chapter, for better understanding of our work and its results.

\section{Data preparation}\label{Data_preparation}

\subsection{Sampling}\label{Data_sampling}
The collection of appropriate data and their proper processing is essential for effective training of regression models, that serve the purposes of this dissertation: the reconstruction of an EoS from its $M-R$ curve. We start from the sampling of the explanatory data (or features) of regression (we will refer to them as X data often from now on). These include the values of mass $M$ and the respective values of radius $R$, of selected points from $M-R$ curves. The points should be representative of the entire $M-R$ curve. 

\begin{figure}[h!]
    \centering
    \includegraphics[width=\linewidth,height=8.5cm]{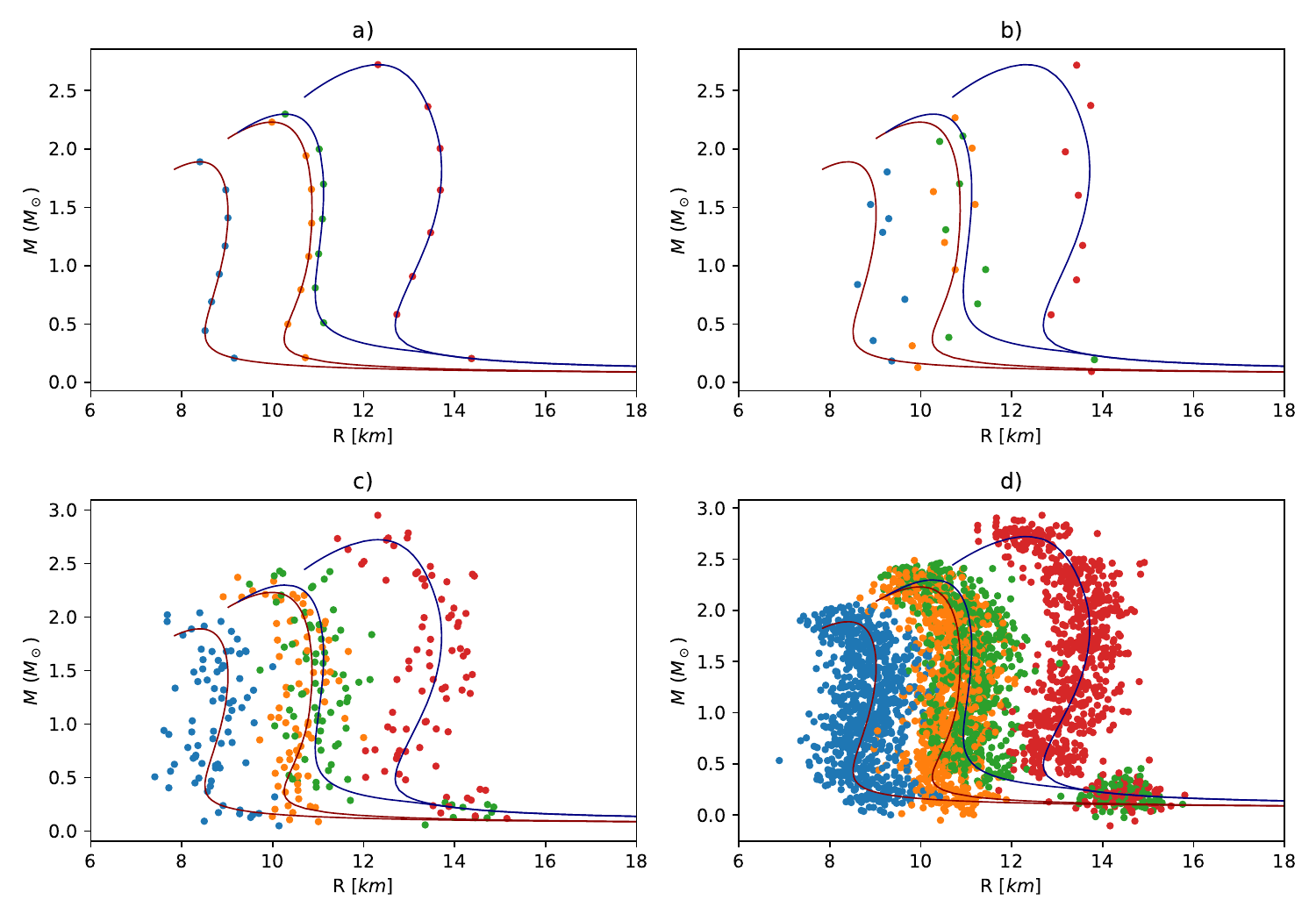}
    \caption{Sampling example of mass and radius data, using $8$ points from each of the M-R curves of the following polytropic EoSs: HLPS-2\_ADDDL (blue), HLPS-2\_DCDCL (orange),  HLPS-3\_ADDDL (green) and HLPS-3\_DCDCL (red). The respective M-R curves are plotted too. The graphs depict: a) the noise-free basic observation of M-R points for each EoS, b) $1$ random M-R observation per EoS, c) $10$ random M-R observations per EoS and d) $100$ random M-R observations per EoS. Each random observation includes additional observational noise: $\Delta M\sim0.1M_\odot$ and $\Delta R\sim0.5km$. See \texttt{ExoticStarsResults\_2.ipynb} in Table \ref{tab:TOV_handle_data}.} 
    \label{fig:sample_MR_8pts_NS}
\end{figure}

\begin{figure}[h!]
    \centering
    \includegraphics[width=\linewidth,height=8.5cm]{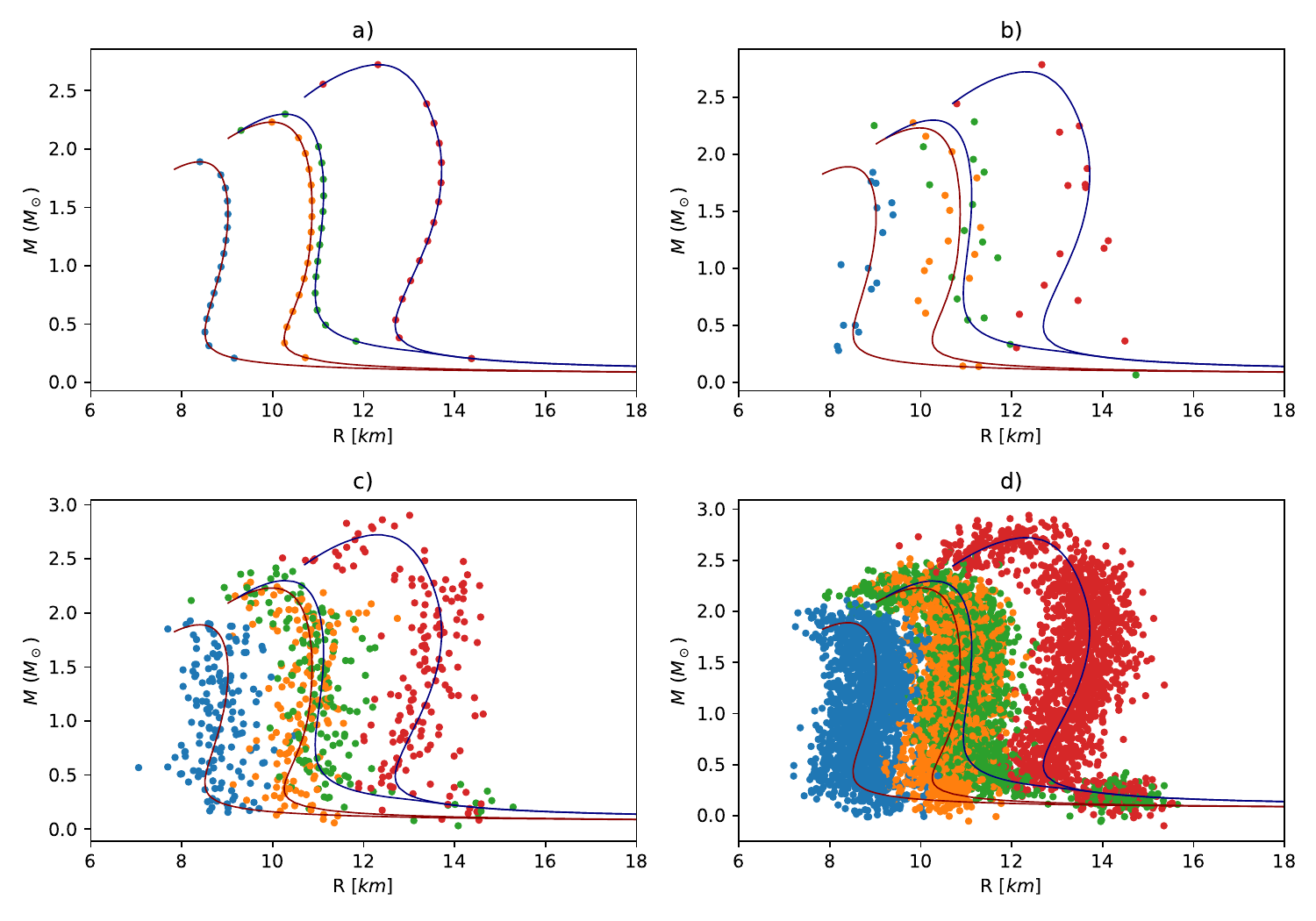}
    \caption{Sampling example of mass and radius data, using $16$ points from each of the M-R curves of the following polytropic EoSs: HLPS-2\_ADDDL (blue), HLPS-2\_DCDCL (orange),  HLPS-3\_ADDDL (green) and HLPS-3\_DCDCL (red). The respective M-R curves are plotted too. The graphs depict: a) the noise-free basic observation of M-R points for each EoS, b) $1$ random M-R observation per EoS, c) $10$ random M-R observations per EoS and d) $100$ random M-R observations per EoS. Each random observation includes additional observational noise: $\Delta M\sim0.1M_\odot$ and $\Delta R\sim0.5km$. See \texttt{ExoticStarsResults\_2.ipynb} in Table \ref{tab:TOV_handle_data}.} 
    \label{fig:sample_MR_16pts_NS}
\end{figure}

\begin{figure}[h!]
    \centering
    \includegraphics[width=\linewidth,height=8.5cm]{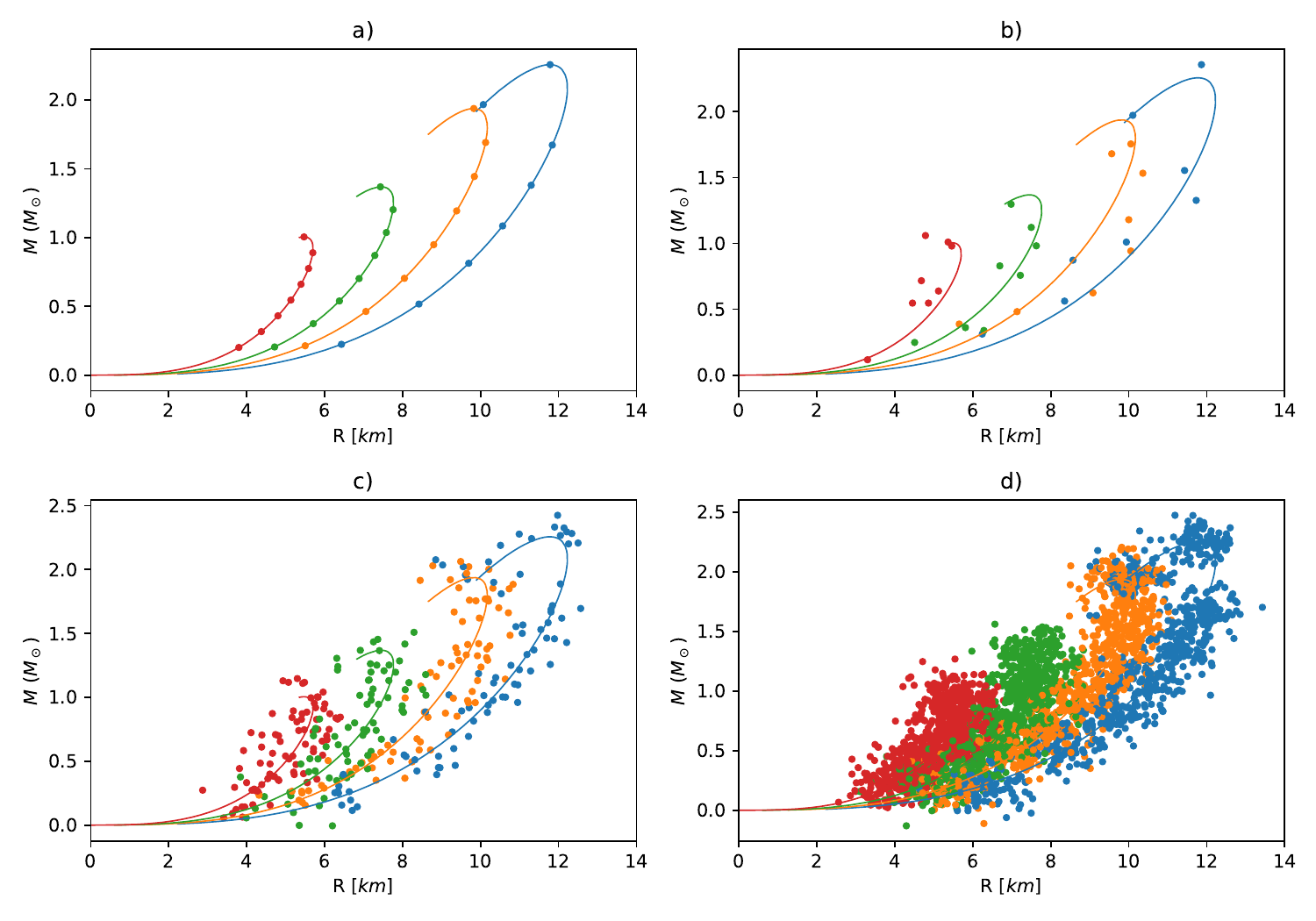}
    \caption{Sampling example of mass and radius data, using $8$ points from each of the M-R curves of the following quark matter EoSs: CFL-50 (blue), CFL-250 (orange), MITbag-131 (green) and MITbag-345 (red). The respective M-R curves are plotted too. The graphs depict: a) the noise-free basic observation of M-R points for each EoS, b) $1$ random M-R observation per EoS derived, c) $10$ random M-R observations per EoS and d) $100$ random M-R observations per EoS. Each random observation includes additional observational noise: $\Delta M\sim0.1M_\odot$ and $\Delta R\sim0.5km$. See \texttt{ExoticStarsResults\_2.ipynb} in Table \ref{tab:TOV_handle_data}.} 
    \label{fig:sample_MR_8pts_QS}
\end{figure}

\begin{figure}[h!]
    \centering
    \includegraphics[width=\linewidth,height=8.5cm]{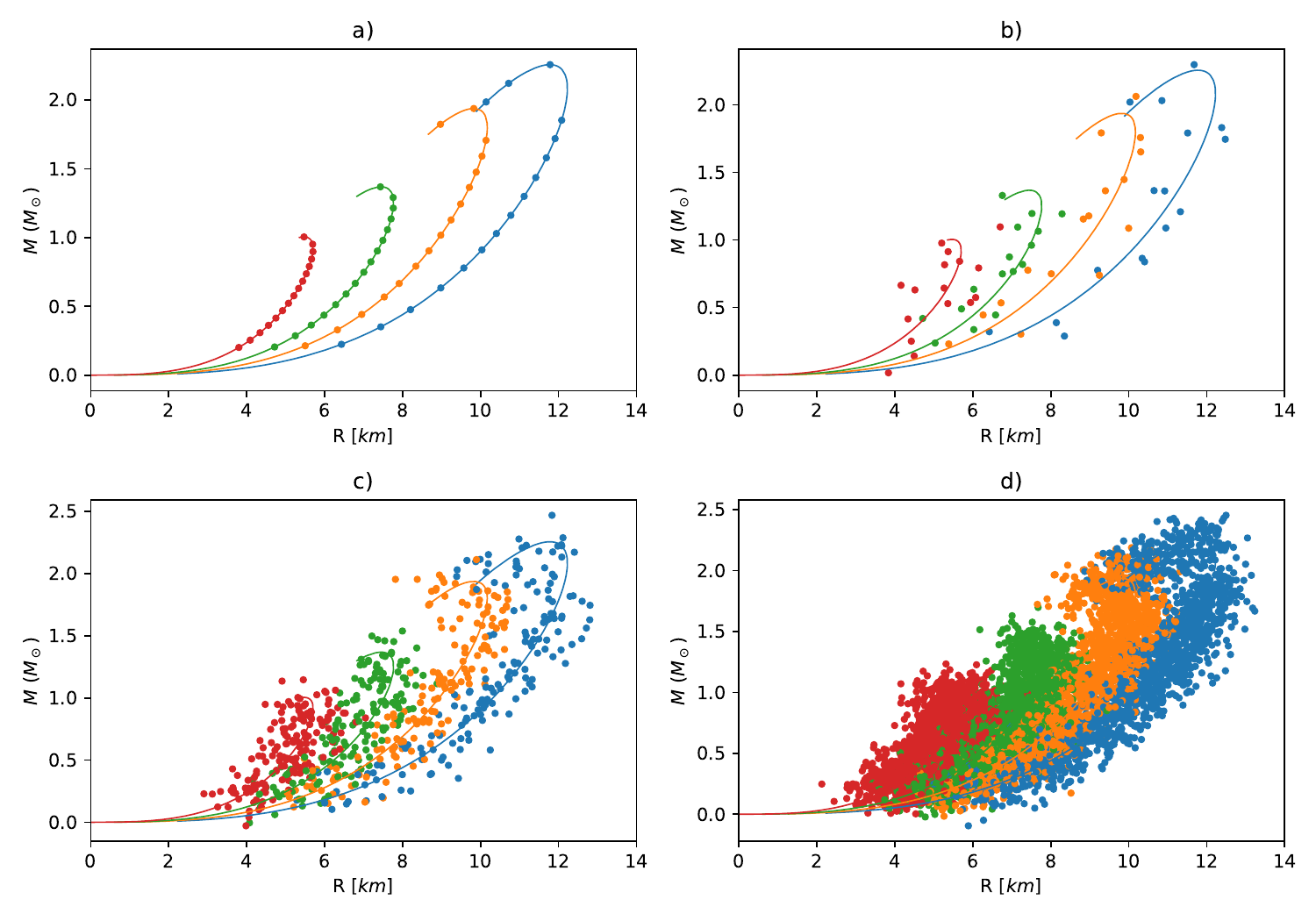}
    \caption{Sampling example of mass and radius data, using $16$ points from each of the M-R curves of the following quark matter EoSs: CFL-50 (blue), CFL-250 (orange), MITbag-131 (green) and MITbag-345 (red). The respective M-R curves are plotted too. The graphs depict: a) the noise-free basic observation of M-R points for each EoS, b) $1$ random M-R observation per EoS derived, c) $10$ random M-R observations per EoS and d) $100$ random M-R observations per EoS. Each random observation includes additional observational noise: $\Delta M\sim0.1M_\odot$ and $\Delta R\sim0.5km$. See \texttt{ExoticStarsResults\_2.ipynb} in Table \ref{tab:TOV_handle_data}.} 
    \label{fig:sample_MR_16pts_QS}
\end{figure}

In this regard, we follow two options: scanning every available $M-R$ curve, with either $8$ points (as shown in Figs. \ref{fig:sample_MR_8pts_NS} and \ref{fig:sample_MR_8pts_QS}) or $16$ points (as shown in Figs. \ref{fig:sample_MR_16pts_NS} and \ref{fig:sample_MR_16pts_QS}). We set a lower threshold at $0.2M_\odot$ and take the corresponding number of points in the interval $[0.2M_\odot,M_{max}]$, with equally spaced mass values. The resulting points form the basic observation of each $M-R$ curve (with 16 X data or 32 X data respectively). This way, both scanning options offer an efficient representation of the $M-R$ curves. Of course, with the 16-points option, one captures more details, but at the cost of heavier datasets and possible longer training times of the regression models. 

Another significant aspect is the errors involved in the observation of a Compact Star event. We take these errors into account by introducing artificial noise in the $M-R$ observations. More specifically, we take the mass values $M_i$ and radius values $R_i$ in the basic $M-R$ curve observation (without noise), and add a normally distributed error in each one of them. For the mass values, we assume an observational noise with standard deviation: $\Delta M_i=0.1M_\odot$, while for the radius values we assume an observational noise with standard deviation: $\Delta R_i=0.5km$.

Thus, we can create as many random observations per EoS (or per $M-R$ curve) as we wish. Creating $1$ random observation is merely enough to depict the effect of noise, as shown in b) graphs of Figs. \ref{fig:sample_MR_8pts_NS}, \ref{fig:sample_MR_16pts_NS}, \ref{fig:sample_MR_8pts_QS}, \ref{fig:sample_MR_16pts_QS}. In contrast, the $10$ random observations offer a more decent sampling and representation of the shape of the $M-R$ curve, as can be seen in the c) graphs of the same figures. Finally, with $100$ observations per EoS on can obtain the most detailed representation of the $M-R$ curve, without much difference between the options of $8$ and $16$ points, while maintaining the size of the datasets at normal levels. A larger number of random observations per EoS would only provide negligible improvement and would result in heavier datasets and training times.


\clearpage

At this point, let's discuss about the naming and the notation of our available EoSs. For the mock polytropic EoSs, we use the notation:
\begin{itemize}\label{notation_polylinNS}
    \item \textbf{HLPS-X\_$\Gamma_1\Gamma_2\Gamma_3\Gamma_4$L}
\end{itemize}
where X takes either the value $2$ or $3$, based on the "main" EoS, that was used to produce the mock polytropic EoS (see section \ref{TOV_solved_polylinNS}). The values of the $\Gamma_i$ come next (notice that we have $4$ such values, one for each of the mass density segments). Instead of using the numerical values, we choose to include them in a coded format. Since, we considered $4$ possible choices for each $\Gamma_i$: $[1,2,3,4]$, we assign to each of them a letter, starting from letter $A$. This leads to the following codification:
\begin{itemize}
    \item A: $\Gamma_i=1$
    \item B: $\Gamma_i=2$
    \item C: $\Gamma_i=3$
    \item D: $\Gamma_i=4$
\end{itemize}
The suffix L indicates that the possible violation of causality has been taken into account and handled accordingly, by including a linear part to the EoS (see Eqs. \ref{total_EOS_polylinNS} and \ref{total_coreEOS_polylinNS}). For example, the notation "HLPS-3\_DCDCL" corresponds to a mock polytropic EoS, that was derived from the HLPS-3 "main" EoS and features the sequence $\{\Gamma:4\rightarrow{3}\rightarrow{4}\rightarrow{3}\}$ among 4 polytropic segments. It is also very likely, that there is a transition to linear behavior after a pressure $P_{tr}$, where the causality limit has been breached.

The case is quite simpler, for the notation of quark matter EoSs. More specifically, we use the notation:
\begin{itemize}\label{notation_QS}
    \item \textbf{QS\_model-N ($B$, $\Delta$)}
\end{itemize}
where the QS\_model is either the "MITbag" or the "CFL" model. The number N occurs from the grouping of the EoSs during the parallel solving process (see see files \texttt{tov\_solver\_mitQS\_par.py} and \texttt{tov\_solver\_cflQS\_par.py} in Table \ref{tab:TOV_solve_codes_QS}) and its actually irrelevant. What matters most, is the values of the parameters $B$ and $\Delta$ (see section \ref{TOV_solved_QS}). The latter are included inside the parenthesis. Of course, for the case of the MIT bag model, $\Delta$ parameter is absent and its value is denoted as "-". Thus, the notation "MITbag-131 (125, -)" points to an EoS of the MIT bag model, with $B=125$ $MeV\cdot fm^{-3}$, while the notation "CFL-50 (70, 120)" points to an EoS of the CFL model, with $B=70$ $MeV\cdot fm^{-3}$ and $\Delta=120$ $MeV$.

Moving on, we need to sample the response data (or targets) of regression (we will refer to them as Y data often from now on). These will have to carry information about the EoSs from which the $M-R$ curves were derived, via the solution of TOV equations. We are limited in two options. The first one, involves collecting the values of slope $d\epsilon/dP$ of the $\epsilon-P$ curve, or equivalently the values of local speed of sound $c_s$, since $c_s/c=(d\epsilon/dP)^{-1/2}$. This option raises some major issues. 

\begin{figure}[h!]
    \centering
    \includegraphics[width=\linewidth,height=9.5cm]{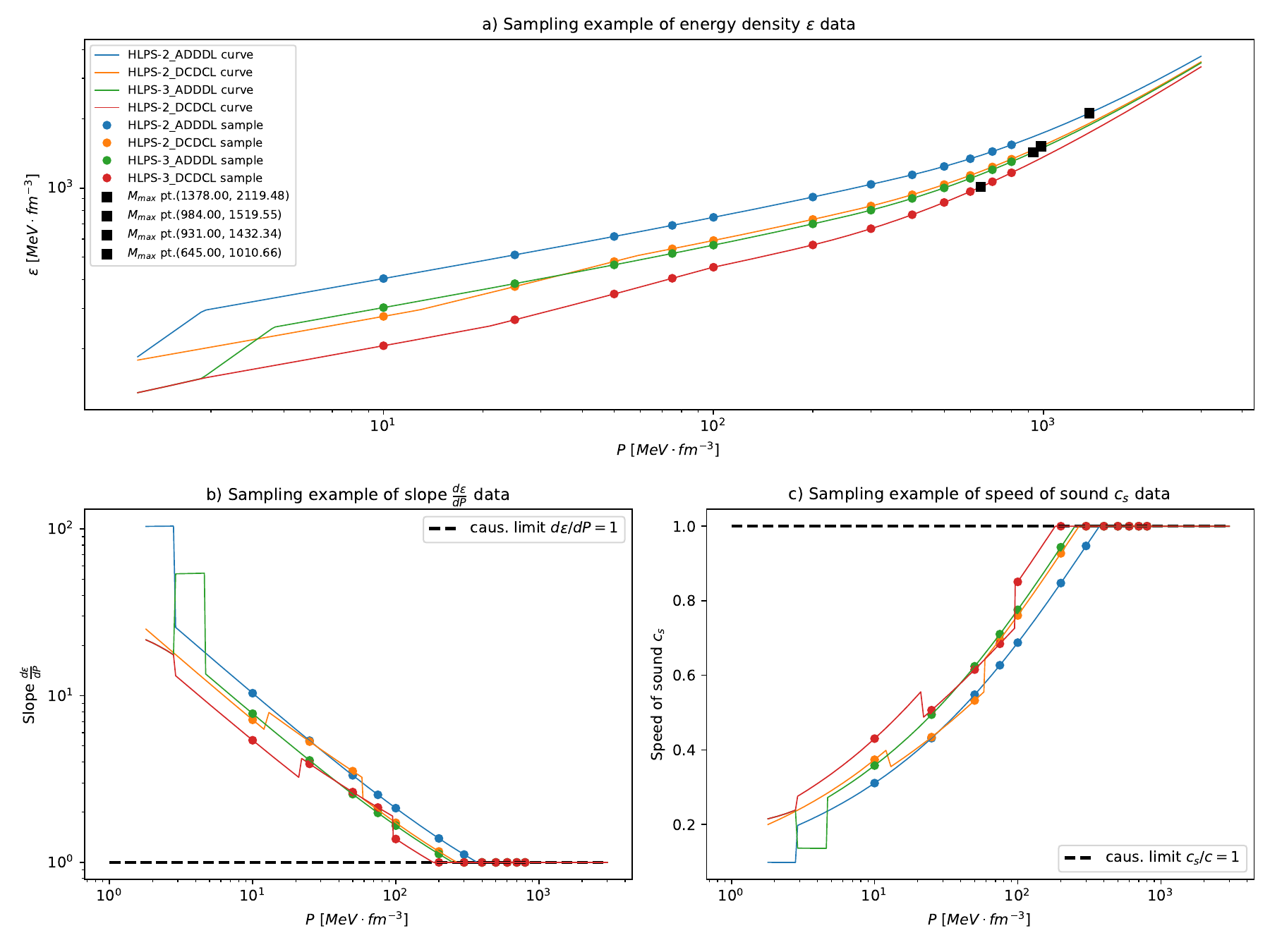}
    \caption{Sampling example of data from Neutron Star EoSs. The values of a) energy density $\epsilon$, b) slope $d\epsilon/dP$ and c) speed of sound $c_s$, at $12$ values of pressure $P$: \{10, 25, 50, 75, 100, 200, 300, 400, 500, 600, 700, 800\} $MeV\cdot fm^{-3}$ are collected and displayed as points. The maximum mass points: $M_{max}$ pt. $(P_{M_{max}}, \epsilon_{M_{max}})$, are also included and displayed as black squares in graph a), along with the causality limit in graphs b) and c). See \texttt{ExoticStarsResults\_2.ipynb} in Table \ref{tab:TOV_handle_data}.} 
    \label{fig:sample_EOS_NS}
\end{figure}

\begin{figure}[h!]
    \centering
    \includegraphics[width=\linewidth,height=9.5cm]{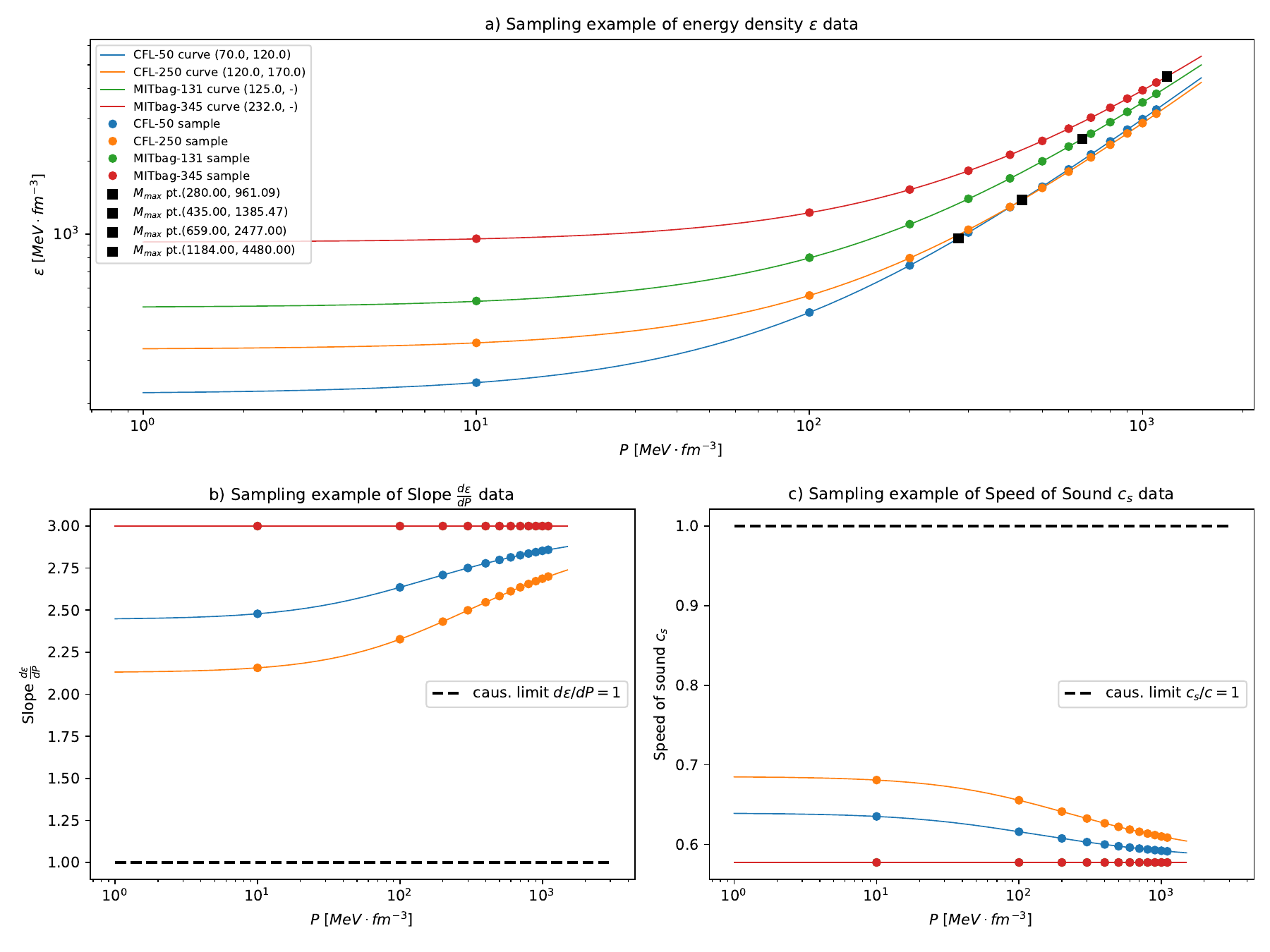}
    \caption{Sampling example of data from Quark Star EoSs. The values of a) energy density $\epsilon$, b) slope $d\epsilon/dP$ and c) speed of sound $c_s$, at $12$ values of pressure $P$: \{10, 100, 200, 300, 400, 500, 600, 700, 800, 900, 1000, 1100\} $MeV\cdot fm^{-3}$ are collected and displayed as points. The maximum mass points: $M_{max}$ pt. $(P_{M_{max}}, \epsilon_{M_{max}})$, are also included and displayed as black squares in graph a), along with the causality limit in graphs b) and c). See \texttt{ExoticStarsResults\_2.ipynb} in Table \ref{tab:TOV_handle_data}.} 
    \label{fig:sample_EOS_QS}
\end{figure}

For the mock polytropic EoSs of Neutron Stars, the slope $d\epsilon/dP$ and the speed of sound $c_s$ exhibit discontinuities at some values of pressure, as shown in graphs b) and c) of Fig. \ref{fig:sample_EOS_NS}. This is due to the change in the value of the $\Gamma$ parameter, between the polytropic segments. Furthermore, the slope $d\epsilon/dP$ might be fixed at the causality limit, leading to the collection of the same value $d\epsilon/dP=1$ or $c_s/c=1$ at different values of pressure, and consequently to possible confusion during the training of the regression models. On the other hand, the slope and the speed of sound of the parametric EoSs of Quark Stars, do not violate causality and depict a smooth and continuous behavior, as shown in graphs b) and c) of Fig. \ref{fig:sample_EOS_QS}. However, the constant value of the slope and speed of sound, at $d\epsilon/dP=3$ or $c_s/c=1/3$, respectively, for all the MIT bag model EoSs, makes them indistinguishable. Therefore, the reconstruction of the $d\epsilon/dP-P$ or $c_s-P$ curves, from $M-R$ data, is proved extremely difficult and quite useless in obtaining the original EoS.

\clearpage

 Then, we rely on the second option: the collection of the values of energy density $\epsilon$ at $12$ values of pressure. For the polytropic Neutron Stars EoSs, we choose these pressure values to be: \{10, 25, 50, 75, 100, 200, 300, 400, 500, 600, 700, 800\} $MeV\cdot fm^{-3}$, as shown in graph a) of Fig. \ref{fig:sample_EOS_NS}. Notice, that the sampling is denser in the region [10,100] $MeV\cdot fm^{-3}$. In this region, the hadronic EoSs turn stiffer (see Figs. \ref{fig:mainNS_EoSs} and \ref{fig:linNS_EOSs_sep}) and more points need to be added for detailed capture of the EoS's behavior. In the other region, [100, 800] $MeV\cdot fm^{-3}$, the pressure points are equally spaced. The final value: $P_{12}=800$ $MeV\cdot fm^{-3}$, is arbitrarily selected, to reach sufficiently big values of mass density and exceed (in most cases) the maximum mass point. Following the same approach for Quark Stars EoSs, we choose the $12$ pressure values to be: \{10, 100, 200, 300, 400, 500, 600, 700, 800, 900, 1000, 1100\} $MeV\cdot fm^{-3}$, as shown in graph a) of Fig. \ref{fig:sample_EOS_QS}. The sampling is, this time, uniformly distributed in pressure axis, since the quark matter EoSs are linear (or almost linear) and do not exhibit sudden changes in their behavior (see Fig. \ref{fig:QS_EOS_all}). Thus, no denser sampling is needed at some pressure regions. This allows, to increase the final sampling pressure $P_{12}$, from 800 to 1100 $MeV\cdot fm^{-3}$, and be able to reconstruct the quark matter EoSs in a larger pressure region.

\subsection{Shuffling}\label{Data_shuffling}

Following the methodology of section \ref{Data_sampling}, one results in a dataset with $12$ columns for the Y data and $16$ or $32$ columns for the X data, based on the number of $M-R$ points. We choose the first half of the X data columns to be occupied by mass values ($M_i$) and the second half by radius values ($R_i$). Each row of the dataset represents a random observation. Thus, for the $304$ polytropic hadronic EoSs and $100$ random observations per EoS, one shall obtain a dataset with $30.400$ rows. Similarly, for the $891$ quark matter EoSs ($381$ MIT bag and $510$ CFL), one shall obtain a dataset with $89.100$ rows.

\begin{figure}[h!]
    \centering
    \includegraphics[width=\linewidth,height=9.5cm]{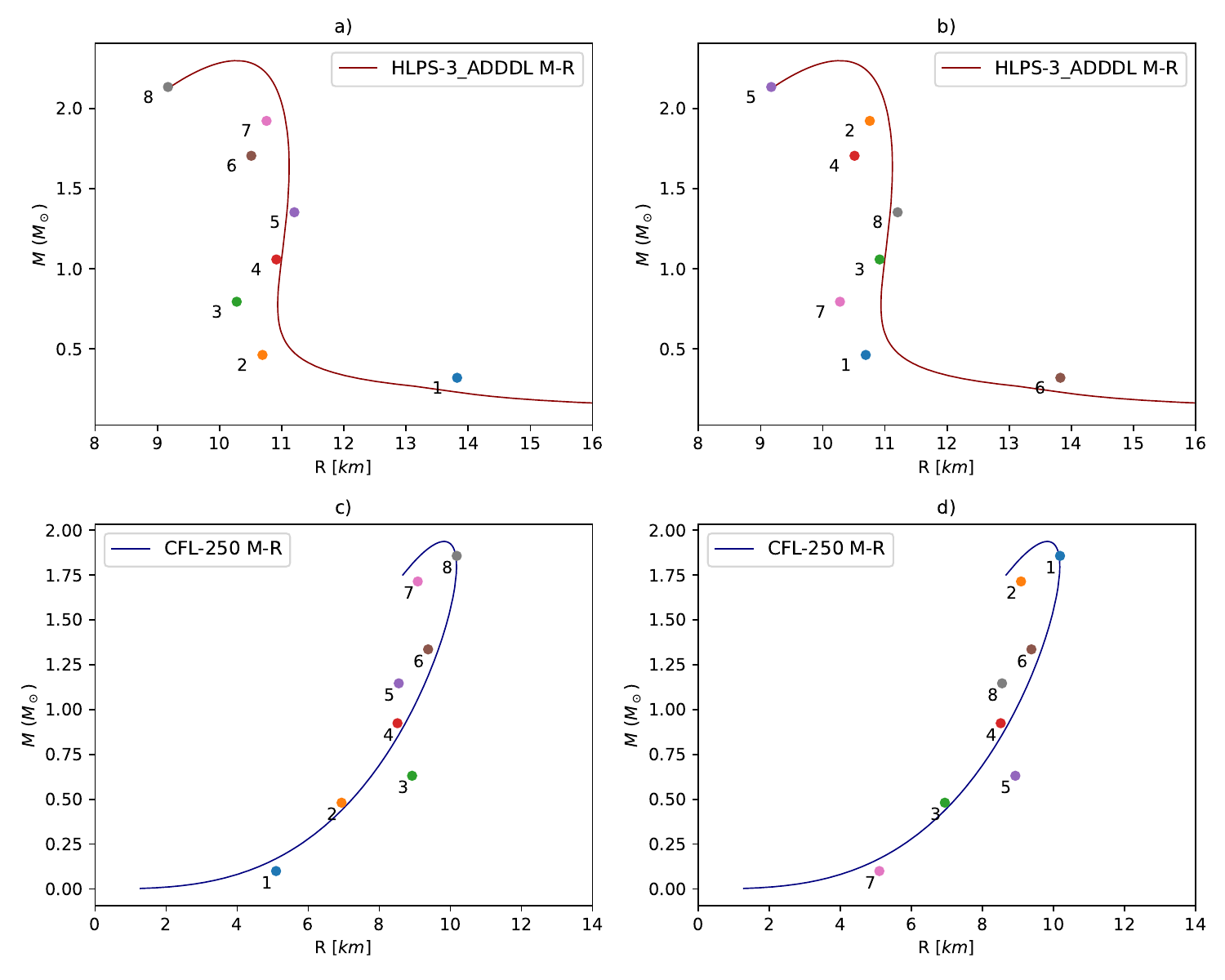}
    \caption{Shuffling example of M-R data from a) and b) a Neutron Star M-R curve, c) and d) a Quark Star M-R curve. The numbers indicate the order in which the points will be recorded in the final dataset. Thus, without shuffling, the points are recorded in ascending order of mass, as shown in a) and c). On the contrary, after shuffling, the points are recorded in random order, as shown in b) and d). Notice, that the points include observational noise and that the shuffling affects only the order of recording and not their coordinates. See \texttt{ExoticStarsResults\_2.ipynb} in Table \ref{tab:TOV_handle_data}.} 
    \label{fig:shuffle_MR}
\end{figure}

For better training of regression models and more generalized and reliable results, we introduce two levels of data shuffling. The first level involves the shuffling of the available EoSs models. For Neutron Stars, we want to include mock EoSs, derived from both the 'main' EoS of HLPS-2 and the 'main' EoS HLPS-3, in both the training and testing parts of the dataset. Likewise, for Quark Stars, we want the inclusion of both types of EoSs, MIT bag and CFL, in both the training and test datasets. So, we record the EoSs in random order on the final dataset. However, we demand no-splitting of the random observations of a single EoS model. That is, all $100$ rows (or rather $100$ random observations) of an EoS, having the same values of energy column-wise, must belong strictly to one of the two sets, either the training or the test set. Cases, where even $90$ rows of an EoS belong to the training set and only $10$ rows belong to the test set, are prohibited and need to be fixed. This way, we exclude data leakage and the regression model is not biased, yielding reliable results, when it comes to evaluating it on EoSs from the test set or other foreign EoSs, since these EoSs were not included in the training process (their behavior is unknown to the regression model).

To further improve the accuracy of our models, we perform a second shuffling, this time on the $M-R$ data. As shown in graphs a) and c) of Fig. \ref{fig:shuffle_MR}, without shuffling, the $M-R$ curve is scanned from bottom to top, i.e. in ascending mass order. With this approach, possible linear correlations might occur between the columns of X data, rendering some features depended on others. An efficient and precise regression model, requires all the features to be independent. Here, comes the row-wise shuffling to cure this dependency. The $M-R$ points are now recorded in the dataset in random order, as shown in graphs b) and d) of Fig. \ref{fig:shuffle_MR}. The order differs even from row to row, making each observation unique, in the way it scans the $M-R$ curve. An important comment needs to be made, regarding the pairing of mass and radius values. Since the points ($M_i$, $R_i$) are of interest and not the values of mass ($M_i$) and radius ($R_i$) separately, one has to be cautious on the recording after shuffling. In particular, the shuffling order should be the same for the mass and radius values in a single row. For example, if the shuffle appoints mass $M_3$ to the column where mass $M_7$ was originally (before the shuffle), then radius $R_3$ must also be appointed to the original column of radius $R_7$. Otherwise, the information of point ($M_3$, $R_3$), and all points ($M_i$, $R_i$) in general, will be lost.

To verify the success of the row-wise shuffling we employ the Pearson's correlation coefficient. The point estimation of the latter, for $n$ observations, reads \cite{rencher2012methods}:
\begin{equation}\label{Pearson_coeff}
    r_{x_jx_k} = \frac{s_{jk}}{\sqrt{s_{jj}s_{kk}}}=\frac{\sum\limits_{i=1}^n x_{ij}x_{ik}-n\bar{x}_j\bar{x}_k}{\sqrt{(\sum\limits_{i=1}^nx_{ij}^2-n\bar{x}_j^2)(\sum\limits_{i=1}^nx_{ik}^2-n\bar{x}_k^2})}
\end{equation}
where $s_{jk}$ is the unbiased estimator of the \textit{covariance} $\sigma_{jk}$ between the $j$-th and the $k$-th variables \cite{rencher2012methods}:
\begin{equation}\label{covariance_XjXk}
    s_{jk} = \frac{1}{n-1}\left(\sum\limits_{i=1}^n x_{ij}x_{ik}-n\bar{x}_j\bar{x}_k\right).
\end{equation}
Similarly, $s_{jj}=s_j^2$ is the unbiased estimator of the \textit{variance} $\sigma_{jj}$ of the $j$-th variable (\textit{covariance} between the $j$-th variable and itself) \cite{rencher2012methods}:
\begin{equation}\label{variance_Xj}
    s_{jj} = \frac{1}{n-1}\left(\sum\limits_{i=1}^n x_{ij}^2-n\bar{x}_j^2\right).
\end{equation}
The quantities $\bar{x}_j$ and $\bar{x}_k$ are the \textit{sample mean values} of the $j$-th and the $k$-th variable, respectively.

Pearson's coefficient is particularly useful in detecting linear correlations between the columns of a sample. By definition, $r_{x_jx_k}\in [-1,1]$, that is, it normalizes the \textit{covariance} $\sigma_{jk}$ with the product of standard deviations of $x_j$ and $x_k$. When $|r_{x_j x_k}|=1$, the values in the corresponding columns $j$ and $k$ are linearly correlated: $x_j=ax_k+b$. Moreover, cases where $0.9\leq|r_{x_j x_k}|<1$ indicate strong, but not absolute, linear correlation between the variables $x_j$ and $x_k$. In contrast, results of type $0<|r_{x_j x_k}|\leq0.8$ imply a case of less strong linear correlation, which becomes even weaker as we approach zero. Finally, the case $r_{x_jx_k}=0$ corresponds to linearly uncorrelated variables $x_j$ and $x_k$, but not necessarily independent \cite{benesty2009pearson}.

\begin{table}[h]
\centering
\tiny
\begin{tabular}{l*{16}{c}}
\toprule
\textbf{i$\backslash$j} & M\_1 & M\_2 & M\_3 & M\_4 & M\_5 & M\_6 & M\_7 & M\_8 & R\_1 & R\_2 & R\_3 & R\_4 & R\_5 & R\_6 & R\_7 & R\_8 \\
\midrule
M\_1 & \colorpearson{1.000} & \colorpearson{0.001} & \colorpearson{0.002} & \colorpearson{0.002} & \colorpearson{-0.003} & \colorpearson{0.001} & \colorpearson{-0.002} & \colorpearson{-0.001} & \colorpearson{-0.019} & \colorpearson{-0.009} & \colorpearson{-0.005} & \colorpearson{-0.003} & \colorpearson{-0.004} & \colorpearson{-0.004} & \colorpearson{-0.000} & \colorpearson{-0.001} \\
M\_2 & \colorpearson{0.001} & \colorpearson{1.000} & \colorpearson{0.430} & \colorpearson{0.481} & \colorpearson{0.501} & \colorpearson{0.513} & \colorpearson{0.521} & \colorpearson{0.526} & \colorpearson{0.278} & \colorpearson{0.340} & \colorpearson{0.401} & \colorpearson{0.431} & \colorpearson{0.448} & \colorpearson{0.469} & \colorpearson{0.477} & \colorpearson{0.491} \\
M\_3 & \colorpearson{0.002} & \colorpearson{0.430} & \colorpearson{1.000} & \colorpearson{0.716} & \colorpearson{0.752} & \colorpearson{0.768} & \colorpearson{0.777} & \colorpearson{0.784} & \colorpearson{0.406} & \colorpearson{0.494} & \colorpearson{0.585} & \colorpearson{0.634} & \colorpearson{0.661} & \colorpearson{0.688} & \colorpearson{0.705} & \colorpearson{0.727} \\
M\_4 & \colorpearson{0.002} & \colorpearson{0.481} & \colorpearson{0.716} & \colorpearson{1.000} & \colorpearson{0.837} & \colorpearson{0.854} & \colorpearson{0.866} & \colorpearson{0.874} & \colorpearson{0.456} & \colorpearson{0.556} & \colorpearson{0.657} & \colorpearson{0.708} & \colorpearson{0.737} & \colorpearson{0.766} & \colorpearson{0.783} & \colorpearson{0.805} \\
M\_5 & \colorpearson{-0.003} & \colorpearson{0.501} & \colorpearson{0.752} & \colorpearson{0.837} & \colorpearson{1.000} & \colorpearson{0.896} & \colorpearson{0.907} & \colorpearson{0.914} & \colorpearson{0.482} & \colorpearson{0.584} & \colorpearson{0.686} & \colorpearson{0.741} & \colorpearson{0.772} & \colorpearson{0.802} & \colorpearson{0.820} & \colorpearson{0.844} \\
M\_6 & \colorpearson{0.001} & \colorpearson{0.513} & \colorpearson{0.768} & \colorpearson{0.854} & \colorpearson{0.896} & \colorpearson{1.000} & \colorpearson{0.927} & \colorpearson{0.935} & \colorpearson{0.494} & \colorpearson{0.597} & \colorpearson{0.702} & \colorpearson{0.757} & \colorpearson{0.790} & \colorpearson{0.820} & \colorpearson{0.839} & \colorpearson{0.863} \\
M\_7 & \colorpearson{-0.002} & \colorpearson{0.521} & \colorpearson{0.777} & \colorpearson{0.866} & \colorpearson{0.907} & \colorpearson{0.927} & \colorpearson{1.000} & \colorpearson{0.948} & \colorpearson{0.502} & \colorpearson{0.604} & \colorpearson{0.712} & \colorpearson{0.768} & \colorpearson{0.801} & \colorpearson{0.832} & \colorpearson{0.851} & \colorpearson{0.875} \\
M\_8 & \colorpearson{-0.001} & \colorpearson{0.526} & \colorpearson{0.784} & \colorpearson{0.874} & \colorpearson{0.914} & \colorpearson{0.935} & \colorpearson{0.948} & \colorpearson{1.000} & \colorpearson{0.505} & \colorpearson{0.610} & \colorpearson{0.717} & \colorpearson{0.774} & \colorpearson{0.807} & \colorpearson{0.838} & \colorpearson{0.856} & \colorpearson{0.882} \\
R\_1 & \colorpearson{-0.019} & \colorpearson{0.278} & \colorpearson{0.406} & \colorpearson{0.456} & \colorpearson{0.482} & \colorpearson{0.494} & \colorpearson{0.502} & \colorpearson{0.505} & \colorpearson{1.000} & \colorpearson{0.867} & \colorpearson{0.780} & \colorpearson{0.721} & \colorpearson{0.683} & \colorpearson{0.645} & \colorpearson{0.609} & \colorpearson{0.569} \\
R\_2 & \colorpearson{-0.009} & \colorpearson{0.340} & \colorpearson{0.494} & \colorpearson{0.556} & \colorpearson{0.584} & \colorpearson{0.597} & \colorpearson{0.604} & \colorpearson{0.610} & \colorpearson{0.867} & \colorpearson{1.000} & \colorpearson{0.893} & \colorpearson{0.849} & \colorpearson{0.812} & \colorpearson{0.775} & \colorpearson{0.735} & \colorpearson{0.688} \\
R\_3 & \colorpearson{-0.005} & \colorpearson{0.401} & \colorpearson{0.585} & \colorpearson{0.657} & \colorpearson{0.686} & \colorpearson{0.702} & \colorpearson{0.712} & \colorpearson{0.717} & \colorpearson{0.780} & \colorpearson{0.893} & \colorpearson{1.000} & \colorpearson{0.921} & \colorpearson{0.898} & \colorpearson{0.869} & \colorpearson{0.835} & \colorpearson{0.790} \\
R\_4 & \colorpearson{-0.003} & \colorpearson{0.431} & \colorpearson{0.634} & \colorpearson{0.708} & \colorpearson{0.741} & \colorpearson{0.757} & \colorpearson{0.768} & \colorpearson{0.774} & \colorpearson{0.721} & \colorpearson{0.849} & \colorpearson{0.921} & \colorpearson{1.000} & \colorpearson{0.933} & \colorpearson{0.915} & \colorpearson{0.884} & \colorpearson{0.843} \\
R\_5 & \colorpearson{-0.004} & \colorpearson{0.448} & \colorpearson{0.661} & \colorpearson{0.737} & \colorpearson{0.772} & \colorpearson{0.790} & \colorpearson{0.801} & \colorpearson{0.807} & \colorpearson{0.683} & \colorpearson{0.812} & \colorpearson{0.898} & \colorpearson{0.933} & \colorpearson{1.000} & \colorpearson{0.936} & \colorpearson{0.912} & \colorpearson{0.875} \\
R\_6 & \colorpearson{-0.004} & \colorpearson{0.469} & \colorpearson{0.688} & \colorpearson{0.766} & \colorpearson{0.802} & \colorpearson{0.820} & \colorpearson{0.832} & \colorpearson{0.838} & \colorpearson{0.645} & \colorpearson{0.775} & \colorpearson{0.869} & \colorpearson{0.915} & \colorpearson{0.936} & \colorpearson{1.000} & \colorpearson{0.934} & \colorpearson{0.905} \\
R\_7 & \colorpearson{-0.000} & \colorpearson{0.477} & \colorpearson{0.705} & \colorpearson{0.783} & \colorpearson{0.820} & \colorpearson{0.839} & \colorpearson{0.851} & \colorpearson{0.856} & \colorpearson{0.609} & \colorpearson{0.735} & \colorpearson{0.835} & \colorpearson{0.884} & \colorpearson{0.912} & \colorpearson{0.934} & \colorpearson{1.000} & \colorpearson{0.915} \\
R\_8 & \colorpearson{-0.001} & \colorpearson{0.491} & \colorpearson{0.727} & \colorpearson{0.805} & \colorpearson{0.844} & \colorpearson{0.863} & \colorpearson{0.875} & \colorpearson{0.882} & \colorpearson{0.569} & \colorpearson{0.688} & \colorpearson{0.790} & \colorpearson{0.843} & \colorpearson{0.875} & \colorpearson{0.905} & \colorpearson{0.915} & \colorpearson{1.000} \\
\bottomrule
\end{tabular}
\caption{Correlation matrix for Neutron Stars regression data with 16 features ($8$ M-R points), before row-wise shuffling. Correlations with $|\rho|>0.8$ are highlighted with \textcolor{red}{red}, while correlations with $|\rho|\leq0.8$ are highlighted with \textcolor{green!60!black}{green}. The elements of the main diagonal, where $\rho=1$, are shown in \textbf{black}. See \texttt{assessing\_regression\_data.ipynb} in Table \ref{tab:train_test_ML}.}
\label{tab:corr_mat_before_rwsh}
\end{table}

\begin{table}[h]
\centering
\tiny
\begin{tabular}{l*{16}{c}}
\toprule
\textbf{i$\backslash$j} & M\_1 & M\_2 & M\_3 & M\_4 & M\_5 & M\_6 & M\_7 & M\_8 & R\_1 & R\_2 & R\_3 & R\_4 & R\_5 & R\_6 & R\_7 & R\_8 \\
\midrule
M\_1 & \colorpearson{1.000} & \colorpearson{0.013} & \colorpearson{0.006} & \colorpearson{0.010} & \colorpearson{0.007} & \colorpearson{0.003} & \colorpearson{0.022} & \colorpearson{0.016} & \colorpearson{-0.015} & \colorpearson{0.287} & \colorpearson{0.288} & \colorpearson{0.291} & \colorpearson{0.286} & \colorpearson{0.284} & \colorpearson{0.287} & \colorpearson{0.288} \\
M\_2 & \colorpearson{0.013} & \colorpearson{1.000} & \colorpearson{0.002} & \colorpearson{0.010} & \colorpearson{0.025} & \colorpearson{0.010} & \colorpearson{0.013} & \colorpearson{0.004} & \colorpearson{0.295} & \colorpearson{-0.005} & \colorpearson{0.294} & \colorpearson{0.293} & \colorpearson{0.288} & \colorpearson{0.298} & \colorpearson{0.292} & \colorpearson{0.290} \\
M\_3 & \colorpearson{0.006} & \colorpearson{0.002} & \colorpearson{1.000} & \colorpearson{0.020} & \colorpearson{0.005} & \colorpearson{0.013} & \colorpearson{0.009} & \colorpearson{0.012} & \colorpearson{0.290} & \colorpearson{0.287} & \colorpearson{-0.008} & \colorpearson{0.286} & \colorpearson{0.293} & \colorpearson{0.287} & \colorpearson{0.291} & \colorpearson{0.293} \\
M\_4 & \colorpearson{0.010} & \colorpearson{0.010} & \colorpearson{0.020} & \colorpearson{1.000} & \colorpearson{0.011} & \colorpearson{0.019} & \colorpearson{0.030} & \colorpearson{0.018} & \colorpearson{0.306} & \colorpearson{0.308} & \colorpearson{0.300} & \colorpearson{0.008} & \colorpearson{0.305} & \colorpearson{0.302} & \colorpearson{0.307} & \colorpearson{0.305} \\
M\_5 & \colorpearson{0.007} & \colorpearson{0.025} & \colorpearson{0.005} & \colorpearson{0.011} & \colorpearson{1.000} & \colorpearson{0.021} & \colorpearson{0.012} & \colorpearson{0.011} & \colorpearson{0.292} & \colorpearson{0.294} & \colorpearson{0.294} & \colorpearson{0.298} & \colorpearson{-0.001} & \colorpearson{0.294} & \colorpearson{0.292} & \colorpearson{0.295} \\
M\_6 & \colorpearson{0.003} & \colorpearson{0.010} & \colorpearson{0.013} & \colorpearson{0.019} & \colorpearson{0.021} & \colorpearson{1.000} & \colorpearson{0.024} & \colorpearson{0.019} & \colorpearson{0.298} & \colorpearson{0.299} & \colorpearson{0.300} & \colorpearson{0.291} & \colorpearson{0.298} & \colorpearson{0.001} & \colorpearson{0.297} & \colorpearson{0.300} \\
M\_7 & \colorpearson{0.022} & \colorpearson{0.013} & \colorpearson{0.009} & \colorpearson{0.030} & \colorpearson{0.012} & \colorpearson{0.024} & \colorpearson{1.000} & \colorpearson{0.015} & \colorpearson{0.297} & \colorpearson{0.303} & \colorpearson{0.304} & \colorpearson{0.303} & \colorpearson{0.300} & \colorpearson{0.301} & \colorpearson{0.013} & \colorpearson{0.304} \\
M\_8 & \colorpearson{0.016} & \colorpearson{0.004} & \colorpearson{0.012} & \colorpearson{0.018} & \colorpearson{0.011} & \colorpearson{0.019} & \colorpearson{0.015} & \colorpearson{1.000} & \colorpearson{0.299} & \colorpearson{0.296} & \colorpearson{0.301} & \colorpearson{0.300} & \colorpearson{0.304} & \colorpearson{0.299} & \colorpearson{0.305} & \colorpearson{0.003} \\
R\_1 & \colorpearson{-0.015} & \colorpearson{0.295} & \colorpearson{0.290} & \colorpearson{0.306} & \colorpearson{0.292} & \colorpearson{0.298} & \colorpearson{0.297} & \colorpearson{0.299} & \colorpearson{1.000} &\colorpearson{0.677} & \colorpearson{0.678} & \colorpearson{0.675} & \colorpearson{0.684} & \colorpearson{0.680} & \colorpearson{0.683} & \colorpearson{0.679} \\
R\_2 & \colorpearson{0.287} & \colorpearson{-0.005} & \colorpearson{0.287} & \colorpearson{0.308} & \colorpearson{0.294} & \colorpearson{0.299} & \colorpearson{0.303} & \colorpearson{0.296} & \colorpearson{0.677} & \colorpearson{1.000} & \colorpearson{0.679} & \colorpearson{0.675} & \colorpearson{0.681} & \colorpearson{0.678} & \colorpearson{0.684} & \colorpearson{0.681} \\
R\_3 & \colorpearson{0.288} & \colorpearson{0.294} & \colorpearson{-0.008} & \colorpearson{0.300} & \colorpearson{0.294} & \colorpearson{0.300} & \colorpearson{0.304} & \colorpearson{0.301} & \colorpearson{0.678} & \colorpearson{0.679} & \colorpearson{1.000} & \colorpearson{0.682} & \colorpearson{0.682} & \colorpearson{0.684} & \colorpearson{0.684} & \colorpearson{0.682} \\
R\_4 & \colorpearson{0.291} & \colorpearson{0.293} & \colorpearson{0.286} & \colorpearson{0.008} & \colorpearson{0.298} & \colorpearson{0.291} & \colorpearson{0.303} & \colorpearson{0.300} & \colorpearson{0.675} & \colorpearson{0.675} & \colorpearson{0.682} & \colorpearson{1.000} & \colorpearson{0.678} & \colorpearson{0.684} & \colorpearson{0.681} & \colorpearson{0.680} \\
R\_5 & \colorpearson{0.286} & \colorpearson{0.288} & \colorpearson{0.293} & \colorpearson{0.305} & \colorpearson{-0.001} & \colorpearson{0.298} & \colorpearson{0.300} & \colorpearson{0.304} & \colorpearson{0.684} & \colorpearson{0.681} & \colorpearson{0.682} & \colorpearson{0.678} & \colorpearson{1.000} & \colorpearson{0.684} & \colorpearson{0.688} & \colorpearson{0.682} \\
R\_6 & \colorpearson{0.284} & \colorpearson{0.298} & \colorpearson{0.287} & \colorpearson{0.302} & \colorpearson{0.294} & \colorpearson{0.001} & \colorpearson{0.301} & \colorpearson{0.299} & \colorpearson{0.680} & \colorpearson{0.678} & \colorpearson{0.684} & \colorpearson{0.684} & \colorpearson{0.684} & \colorpearson{1.000} & \colorpearson{0.685} & \colorpearson{0.679} \\
R\_7 & \colorpearson{0.287} & \colorpearson{0.292} & \colorpearson{0.291} & \colorpearson{0.307} & \colorpearson{0.292} & \colorpearson{0.297} & \colorpearson{0.013} & \colorpearson{0.305} & \colorpearson{0.683} & \colorpearson{0.684} & \colorpearson{0.684} & \colorpearson{0.681} & \colorpearson{0.688} & \colorpearson{0.685} & \colorpearson{1.000} & \colorpearson{0.682} \\
R\_8 & \colorpearson{0.288} & \colorpearson{0.290} & \colorpearson{0.293} & \colorpearson{0.305} & \colorpearson{0.295} & \colorpearson{0.300} & \colorpearson{0.304} & \colorpearson{0.003} & \colorpearson{0.679} & \colorpearson{0.681} & \colorpearson{0.682} & \colorpearson{0.680} & \colorpearson{0.682} & \colorpearson{0.679} & \colorpearson{0.682} & \colorpearson{1.000}\\
\bottomrule
\end{tabular}
\caption{Correlation matrix for Neutron Stars regression data with 16 features ($8$ M-R points), after row-wise shuffling. Correlations with $|\rho|>0.8$ are highlighted with \textcolor{red}{red}, while correlations with $|\rho|\leq0.8$ are highlighted with \textcolor{green!60!black}{green}. The elements of the main diagonal, where $\rho=1$, are shown in \textbf{black}. See \texttt{assessing\_regression\_data.ipynb} in Table \ref{tab:train_test_ML}.}
\label{tab:corr_mat_after_rwsh}
\end{table}

In Tables \ref{tab:corr_mat_before_rwsh} and \ref{tab:corr_mat_after_rwsh} we present the \textit{correlation matrices} of a Neutron Stars' regression dataset with $16$ features ($8$ M-R points), before and after row-wise shuffling. As we can see in Table \ref{tab:corr_mat_before_rwsh}, there are several evidences of strong linear correlation. First, the values of masses higher than $M_\odot$ ($M_4,\dots,M_8$) are most likely to be connected to each other, a correlation that becomes stronger ($r_{M_jM_k}>0.9$) as we get to the final masses in sampling: $M_7$ and $M_8$. The same pattern appears between the values of radii. Almost all radii, exhibit strong linear correlations with each other and this connection is reinforced for the radii corresponding to masses $M_5,\dots,M_8$. Again, we might meet extreme cases of $r_{R_jR_k}>0.9$. The radius $R_1$ is an exception, since it is highly correlated only to radius $R_2$. Finally, values of radii $R_5,\dots,R_8$ seem to be linearly correlated to values of masses $M_4,\dots,M_8$ with $r_{M_jR_k}>0.8$. 

Hence, linear correlations pose a significant problem in our analysis, as almost half of the variables are not independent to each other. Row-wise shuffling becomes a vitally important step. Indeed, Table \ref{tab:corr_mat_after_rwsh} shows that, after this shuffling, there is not a single case where: $|r_{X_jX_k}|>0.7$. The strong correlations have been weakened and the feature variables are no more related. We obtain the same or similar results for the other datasets used in our analysis, for Quark stars and/or with more features (see \texttt{assessing\_regression\_data.ipynb} in Table \ref{tab:train_test_ML}).

\section{Fine-tuning}\label{Fine_tune}
Having the regression data prepared, we are ready to feed them in our models for fitting. For the machine learning algorithms, the first $80\%$ of the dataset is used as training dataset and the remaining $20\%$ as test dataset, to evaluate the performance of the model. As last step before training, the feature data are scaled, by applying the \texttt{StandardScaler()}. This scaler standardizes features by removing the mean and scaling to unit variance (see \href{https://scikit-learn.org/stable/modules/generated/sklearn.preprocessing.StandardScaler.html}{documentation}). Centering and scaling happens separately on each feature variable by calculating the relevant statistics on the samples in the training set. Mean and standard deviation are then stored to be utilized on later data using the \texttt{transform()} method. Standardization of a dataset is a common requirement for many machine learning estimators: they might perform poorly if the individual features do not more or less look like standard normally distributed data. For example, many elements included in the objective function of a learning algorithm, consider that all features are centered around 0 and have variance in the same order. If a feature has a variance that is orders of magnitude larger than others, it could prevail over the objective function and render the estimator unable to learn from other features properly, as expected. The scaler brings all features to the same (unit) scale and allows the model to learn from all of them equally.

In section \ref{CrossVal_GridSrch}, we denoted the importance of \textbf{Cross-Validation} and \textbf{Grid Search} techniques. Subsequently, for every model we choose to perform a 5-fold cross-validation combined with a grid search over certain hyperparameters. The respective values of these hyperparameter are presented in Tables \ref{tab:grid_dtree}, \ref{tab:grid_rf}, \ref{tab:grid_gradboost} and \ref{tab:grid_xgboost}, both for Neutron Stars (NS) and Quark Stars (QS). For each combination of hyperparameters, the model is trained and validated via cross-validation, in order to get the mean estimation of the loss from 5-folds. The final fine-tuned model is obtained by comparing the mean losses over all combinations. The best combination of hyperparameters is the one, for which the model exhibits the minimum mean loss. Since every fold corresponds to a different training process, the model is trained $5\times\text{combos}$ times before is fully optimized. For example, a \textit{Decision Tree} model with $216$ different combinations is trained $5x216=1080$ times during the learning process.

\begin{table}[h]
    \centering
    \begin{tabular}{|p{4cm}|p{3cm}|p{3cm}|p{6cm}|}
    \hline
    \hline
    \textbf{Hyperparameters}  &  \textbf{Values NS} & \textbf{Values QS} & \textbf{Brief Description}\\
    \hline
    \hline
    \vspace{0.005cm} max\_depth   & \vspace{0.005cm} [None, 5, 10, \textcolor{green!60!black}{20}] & \vspace{0.005cm} [None, 5, 10, \textcolor{green!60!black}{20}] & Limits tree depth; prevents overfitting\\
    \hline
    \vspace{0.005cm} min\_samples\_split  & \vspace{0.005cm} [\textcolor{green!60!black}{2}, 5, 10] & \vspace{0.005cm} [\textcolor{blue}{2}, \textcolor{red}{5}, 10] & Minimum samples to split an internal node\\
    \hline
    \vspace{0.005cm} min\_samples\_leaf   & \vspace{0.005cm} [1, 2, \textcolor{green!60!black}{5}] & \vspace{0.005cm} [1, 2, \textcolor{green!60!black}{5}] & Minimum samples at a leaf node; helps in generalization\\
    \hline
    \vspace{0.005cm} max\_features   & \vspace{0.005cm} [\textcolor{green!60!black}{None}, 'sqrt', 'log2'] & \vspace{0.005cm} [\textcolor{red}{None}, 'sqrt', \textcolor{blue}{'log2'}] & Number of features considered at each split\\
    \hline
    \vspace{0.005cm} criterion  & \vspace{0.005cm} ['squared\_error', \textcolor{green!60!black}{'friedman\_mse'}] & \vspace{0.005cm} ['squared\_error', \textcolor{green!60!black}{'friedman\_mse'}] & 'squared\_error' (default), 'friedman\_mse'; may help with variance \\
    \hline 
    \hline
    \end{tabular}
    \caption{Grid of hyperparameters' values for the tuning of \textit{Decision Tree} models. Total combinations: $4\times3\times3\times3\times2 = 216$. The resulted values for $16$ features are highlighted with \textcolor{red}{red}, while the resulted values for $32$ features are highlighted with \textcolor{blue}{blue}. If the values are same for $16$ and $32$ features, they are shown with \textcolor{green!60!black}{green}. See \texttt{train\_test\_dtree\_regress.ipynb} in Table \ref{tab:train_test_ML}.}
    \label{tab:grid_dtree}
\end{table}

\begin{table}[h]
    \centering
    \begin{tabular}{|p{4cm}|p{3cm}|p{3cm}|p{6cm}|}
    \hline
    \hline
    \textbf{Hyperparameters}  &  \textbf{Values NS} & \textbf{Values QS} & \textbf{Brief Description}\\
    \hline
    \hline
    \vspace{0.005cm} n\_estimators  & \vspace{0.005cm} [25, \textcolor{green!60!black}{50}] & \vspace{0.005cm} [25, \textcolor{green!60!black}{50}] & Number of trees in the forest\\
    \hline
    \vspace{0.005cm} max\_depth   & \vspace{0.005cm} [\textcolor{green!60!black}{None}, 10, 20] & \vspace{0.005cm} [\textcolor{green!60!black}{None}, 10, 20] & Maximum depth of each tree\\
    \hline
    \vspace{0.005cm} min\_samples\_split  & \vspace{0.005cm} [\textcolor{green!60!black}{20}, 40] & \vspace{0.005cm} [\textcolor{green!60!black}{20}, 40] & Minimum samples to split a tree node\\
    \hline
    \vspace{0.005cm} min\_samples\_leaf   & \vspace{0.005cm} [\textcolor{green!60!black}{10}, 12, 14] & \vspace{0.005cm} [\textcolor{green!60!black}{10}, 12, 14] & Minimum samples at a leaf node; helps reduce overfitting\\
    \hline
    \vspace{0.005cm} max\_features   & \vspace{0.005cm} [\textcolor{green!60!black}{None}, 'sqrt', 'log2'] & \vspace{0.005cm} [\textcolor{green!60!black}{None}, 'sqrt', 'log2'] & Number of features to consider per split\\
    \hline
    \vspace{0.005cm} criterion  & \vspace{0.005cm} [\textcolor{green!60!black}{'squared\_error'}] & \vspace{0.005cm} [\textcolor{green!60!black}{'squared\_error'}] &  Default criterion \\
    \hline 
    \vspace{0.005cm} random\_state  & \vspace{0.005cm} Fixed int \textcolor{green!60!black}{45} & \vspace{0.005cm} Fixed int \textcolor{green!60!black}{45} &  Reproducibility\\
    \hline 
    \hline
    \end{tabular}
    \caption{Grid of hyperparameters' values for the tuning of \textit{Random Forest} models. Total combinations: $2\times3\times2\times3\times3\times1 = 108$. The resulted values for $16$ features are highlighted with \textcolor{red}{red}, while the resulted values for $32$ features are highlighted with \textcolor{blue}{blue}. If the values are same for $16$ and $32$ features, they are shown with \textcolor{green!60!black}{green}. See \texttt{train\_test\_rf\_regress.ipynb} in Table \ref{tab:train_test_ML}.}
    \label{tab:grid_rf}
\end{table}

\begin{table}[h]
    \centering
    \begin{tabular}{|p{4cm}|p{3cm}|p{3cm}|p{6cm}|}
    \hline
    \hline
    \textbf{Hyperparameters}  &  \textbf{Values NS} & \textbf{Values QS} & \textbf{Brief Description}\\
    \hline
    \hline
    \vspace{0.005cm} n\_estimators  & \vspace{0.005cm} [50, \textcolor{green!60!black}{100}] & \vspace{0.005cm} [50, \textcolor{green!60!black}{100}] & Total boosting stages (trees)\\
    \hline
    \vspace{0.005cm} learning\_rate  & \vspace{0.005cm} [0.01, \textcolor{green!60!black}{0.05}] & \vspace{0.005cm} [0.01, \textcolor{green!60!black}{0.05}] & Shrinks each tree’s impact; lower + more trees = often better\\
    \hline
    \vspace{0.005cm} max\_depth   & \vspace{0.005cm} [3, \textcolor{green!60!black}{5}] & \vspace{0.005cm} [3, \textcolor{green!60!black}{5}] & Maximum depth of each tree\\
    \hline
    \vspace{0.005cm} min\_samples\_split  & \vspace{0.005cm} [\textcolor{blue}{2}, \textcolor{red}{5}] & \vspace{0.005cm} [2, \textcolor{green!60!black}{5}] & Minimum samples to split a tree node\\
    \hline
    \vspace{0.005cm} min\_samples\_leaf   & \vspace{0.005cm} [1, \textcolor{green!60!black}{2}] & \vspace{0.005cm} [1, \textcolor{green!60!black}{2}] & Minimum samples at a leaf node; helps reduce overfitting\\
    \hline
    \vspace{0.005cm} max\_features   & \vspace{0.005cm} [\textcolor{green!60!black}{'sqrt'}, 'log2'] & \vspace{0.005cm} [\textcolor{green!60!black}{'sqrt'}, 'log2'] & Feature subset per tree/split\\
    \hline
    \vspace{0.005cm} subsample  & \vspace{0.005cm} [\textcolor{green!60!black}{1.0}] & \vspace{0.005cm} [\textcolor{green!60!black}{1.0}] &  Fraction of samples used per tree; $<1.0\xrightarrow{}$ stochastic boosting\\
    \hline 
    \vspace{0.005cm} loss  & \vspace{0.005cm} [\textcolor{green!60!black}{'squared\_error'}] & \vspace{0.005cm} [\textcolor{green!60!black}{'squared\_error'}] &  Aligning with criterion in \textit{Random Forest}\\
    \hline
    \vspace{0.005cm} random\_state  & \vspace{0.005cm} Fixed int \textcolor{green!60!black}{45} & \vspace{0.005cm} Fixed int \textcolor{green!60!black}{45} &  Reproducibility\\
    \hline 
    \hline
    \end{tabular}
    \caption{Grid of hyperparameters' values for the tuning of \textit{Gradient Boosting} models. Total combinations: $2\times2\times2\times2\times2\times2\times1\times1\times1 = 64$. The resulted values for $16$ features are highlighted with \textcolor{red}{red}, while the resulted values for $32$ features are highlighted with \textcolor{blue}{blue}. If the values are same for $16$ and $32$ features, they are shown with \textcolor{green!60!black}{green}. See \texttt{train\_test\_gradboost\_regress.ipynb} in Table \ref{tab:train_test_ML}.}
    \label{tab:grid_gradboost}
\end{table}
\clearpage

\begin{table}[h]
    \centering
    \begin{tabular}{|p{4cm}|p{3cm}|p{3cm}|p{6cm}|}
    \hline
    \hline
    \textbf{Hyperparameters}  &  \textbf{Values NS} & \textbf{Values QS} & \textbf{Brief Description}\\
    \hline
    \hline
    \vspace{0.005cm} n\_estimators  & \vspace{0.005cm} [50, \textcolor{green!60!black}{100}] & \vspace{0.005cm} [50, \textcolor{green!60!black}{100}] & Total boosting stages (trees)\\
    \hline
    \vspace{0.005cm} learning\_rate  & \vspace{0.005cm} [0.05, \textcolor{green!60!black}{0.1}] & \vspace{0.005cm} [0.05, \textcolor{green!60!black}{0.1}] & Shrinks each tree’s impact; lower + more trees = often better\\
    \hline
    \vspace{0.005cm} max\_depth   & \vspace{0.005cm} [3, 5, \textcolor{green!60!black}{7}] & \vspace{0.005cm} [3, 5, \textcolor{green!60!black}{7}] & Maximum depth of each tree: increasing this value will make the model more complex and more likely to overfit\\
    \hline
    \vspace{0.005cm} subsample  & \vspace{0.005cm} [0.7, \textcolor{green!60!black}{1.0}] & \vspace{0.005cm} [\textcolor{green!60!black}{0.7}, 1.0] &  Fraction of samples used per tree; $<1.0\xrightarrow{}$ stochastic boosting, occurs once in every boosting iteration, prior to growing trees to prevent overfitting\\
    \hline 
    \vspace{0.005cm} colsample\_bytree  & \vspace{0.005cm} [\textcolor{red}{0.7}, \textcolor{blue}{1.0}] & \vspace{0.005cm} [0.7, \textcolor{green!60!black}{1.0}] &  The subsample ratio of columns when constructing each tree. Subsampling occurs once for every tree constructed.\\
    \hline
    \vspace{0.005cm} reg\_alpha  & \vspace{0.005cm} [\textcolor{green!60!black}{0.1}] & \vspace{0.005cm} [\textcolor{green!60!black}{0.1}] &  $L_1$ regularization term on weights. Increasing this value will make model more conservative, range: $[0,\infty]$.\\
    \hline 
    \vspace{0.005cm} reg\_lambda  & \vspace{0.005cm} [1.0, \textcolor{green!60!black}{5.0}] & \vspace{0.005cm} [\textcolor{green!60!black}{1.0}, 5.0] &  $L_2$ regularization term on weights. Increasing this value will make model more conservative, range: $[0,\infty]$.\\
    \hline 
    \hline
    \end{tabular}
    \caption{Grid of hyperparameters' values for the tuning of \textit{XGBoost} models. Total combinations: $2\times2\times3\times2\times2\times1\times2 = 96$. The resulted values for $16$ features are highlighted with \textcolor{red}{red}, while the resulted values for $32$ features are highlighted with \textcolor{blue}{blue}. If the values are same for $16$ and $32$ features, they are shown with \textcolor{green!60!black}{green}. See \texttt{train\_test\_xgboost\_regress.ipynb} in Table \ref{tab:train_test_ML}.}
    \label{tab:grid_xgboost}
\end{table}

Now, as depicted in Tables \ref{tab:grid_dtree}, \ref{tab:grid_rf}, \ref{tab:grid_gradboost} and \ref{tab:grid_xgboost}, all machine learning models tend to absorb as much information from the data as possible and become as specific to the problem as possible, but without being over-fitted (as we will see from the results of the metrics in the next chapter). Indeed, the best estimators are those with the maximum given number of trees and these trees are grown to the maximum given depth. As for the number of samples per split and per leaf, these vary based on the algorithm and the number of features to achieve the best performance. Moreover, the use of all features and the bigger learning rate seem to be the best options, in most cases, for accurate learning. The subsampling in \textit{XGBoost} (Table \ref{tab:grid_xgboost}) is different for Neutron and Quark Stars, while the column subsampling by tree (\texttt{colsample\_bytree}) differs for Neutron Stars data, based on the number of features.

In general, we see that the models result in the same best combinations for both number of features ($16$ or $32$) with minor deviations. \textit{Decision Trees} (Table \ref{tab:grid_dtree}) exhibit two such deviations, in the \texttt{min\_samples\_split} and \texttt{max\_features} for Quark Stars. \textit{Gradient Boosting} (Table \ref{tab:grid_gradboost}) and \textit{XGBoost} (Table \ref{tab:grid_xgboost}) exhibit one such deviation, in \texttt{min\_samples\_split} and \texttt{colsample\_bytree} for Neutron Stars, as discussed previously. \textit{Random Forest} (Table \ref{tab:grid_rf}) shows not such deviations. This low amount of deviations might be a sign of similar performance of all algorithms, regardless of the number of M-R points used to scan the $M-R$ curves. Besides, the best combinations seem to be the same for Neutron Stars and Quark Stars, with any differences probably due to the different size of the datasets (see subsection \ref{Data_shuffling}).

The \texttt{criterion} or \texttt{loss} hyperparameter, was chosen to be either 'squared\_error', or 'friedman\_mse', to align with our selection of the general losses: \texttt{MSE} and \texttt{MSLE} (see section \ref{CrossVal_GridSrch}). We have to mention that both cross-validation and grid search processes, use the \texttt{MSLE} for the mean loss estimation and the fine-tuning, respectively, following the methodology of \cite{fujimoto2018methodology}. We also, fixed the \texttt{random\_state}, whenever possible, at $45$, to provide easy reproducibility of our models. However, we advise the reader to take the optimizations above with caution: a new learning procedure with the use of our codes in Table \ref{tab:train_test_ML}, might lead to slightly different optimizations, since all these algorithms are statistical and/or stochastic learning procedures. 

We will close this chapter, by describing the architecture of the \textit{Deep Neural Network} (\textit{DNN}) models we developed for the purposes of our analysis. The entire structure of our neural networks is presented in Table \ref{tab:DNN_structure}. The input layer is the only point of differentiation in our \textit{DNN} models: it includes 16 units for the case of 8 M-R points or 32 units for the case of 16 M-R points. The rest structure is the same for all cases and for both Neutron and Quark Stars. It includes 3 hidden layers with decreasing number of units, by a factor of 2, as we move to the output. All hidden layers are activated with the ReLU function for effectiveness in computations (see subsection \ref{NNs_activ_func}). 

\begin{table}[h]
    \centering
    \begin{tabular}{|p{5cm}|p{6cm}|p{5cm}|}
    \hline
    \hline
    \textbf{Layers}  &  \textbf{Number of units (neurons)} &  \textbf{Activation function}\\
    \hline
    \hline
    Dense\_0 (Input) & 16 or 32, based on features & - \\
    \hline
    Dense\_1 (Hidden Layer 1) & 128 & ReLU \\
    Batch\_Normalization & 128 & - \\
    Dropout & drops $50\%$ of neurons & - \\
    \hline
    Dense\_2 (Hidden Layer 2) & 64 & ReLU \\
    Batch\_Normalization & 64 & - \\
    Dropout & drops $50\%$ of neurons & - \\
    \hline
    Dense\_3 (Hidden Layer 3) & 32 & ReLU \\
    Batch\_Normalization & 32 & - \\
    \hline
    Dense\_4 (Output) & 12 & - \\
    \hline
    \hline
    \end{tabular}
    \caption{The structure of the Deep Neural Networks (\textit{DNN}) models we built for the purposes of this dissertation. Each \textit{DNN} features 3 hidden layers and incorporates the techniques of batch normalization and dropout to prevent overfitting. We selected the name \textit{DNN-3} for these models. See \texttt{train\_test\_dnn3\_regress.ipynb} in Table \ref{tab:train_test_DL}.}
    \label{tab:DNN_structure}
\end{table}

This formation was chosen, in order to avoid too much overspecialization on our data, between the hidden layers. We also, applied \textit{batch normalization} (see subsection \ref{DNN_regulirize}) between all hidden layers and between the last hidden layer and the output layer, for regularization. Besides, we incorporated \textit{dropout} procedures of order $50\%$ (see subsection \ref{DNN_regulirize}), between the hidden layers, towards the same direction: the reduce of overfitting. That is, between Hidden Layer 1 and Hidden Layer 2, the algorithm eliminates (randomly for each batch of observations) half of the neurons and the weights of the rest neurons are scaled up by a factor of 2. Notice, that there is not a dropout process between the last hidden layer (Hidden Layer 3) and the output, as we wish all information from this hidden layer to be used in the final regression, which makes the predictions of the target variables. Finally, the output layer contains 12 units and comes without activation, since we aim to predict (without interference from activations) 12 values of energy density $\epsilon$. 

When it comes to fitting and evaluating the performance of our \textit{DNN} models, we followed the same partition of the entire dataset as in machine learning, with an additional step for validation. The first $80\%$ of the dataset was again reserved as training dataset and the last $20\%$. Then, the first $80\%$ of the training dataset was used as the final training dataset and the last $20\%$ of the training dataset was used as validation set. This way, we could evaluate the stability of training of our \textit{DNN} models, by comparing their performance on the final training data and the validation data, per epoch. The batch size is selected to be at 128 and the total number of epochs was set at 1000 for each model. We selected the loss function MSLE to be optimized, for ease of comparison with the machine learning models. The optimizer we applied, was the \texttt{Adam} optimizer (see \href{https://www.tensorflow.org/api_docs/python/tf/keras/optimizers/Adam}{documentation}) with a learning rate of $0.001$ for backpropagation (see subsection \ref{NNs_backpropag}). The results of all regression models are presented in the next chapter.

\chapter{Final results and discussion}\label{Results Comp}

Coming to the end of our study, in this chapter, we show the basic results of our analysis. We start with the general results of metrics and learning curves. We also make a comment about the fitting time of all models. Then we proceed to more specialized results, regarding the reconstruction of the 21 "main" equations of state for hadronic stars and 20 equations of state for quark stars (10 MIT bag, 10 CFL). We close this chapter with a discussion over these results, with separate references on the performance of each regression model.

\section{Metrics, learning curves and fitting time}\label{General_results}

\begin{figure}[h]
    \centering
    \includegraphics[width=0.75\linewidth]{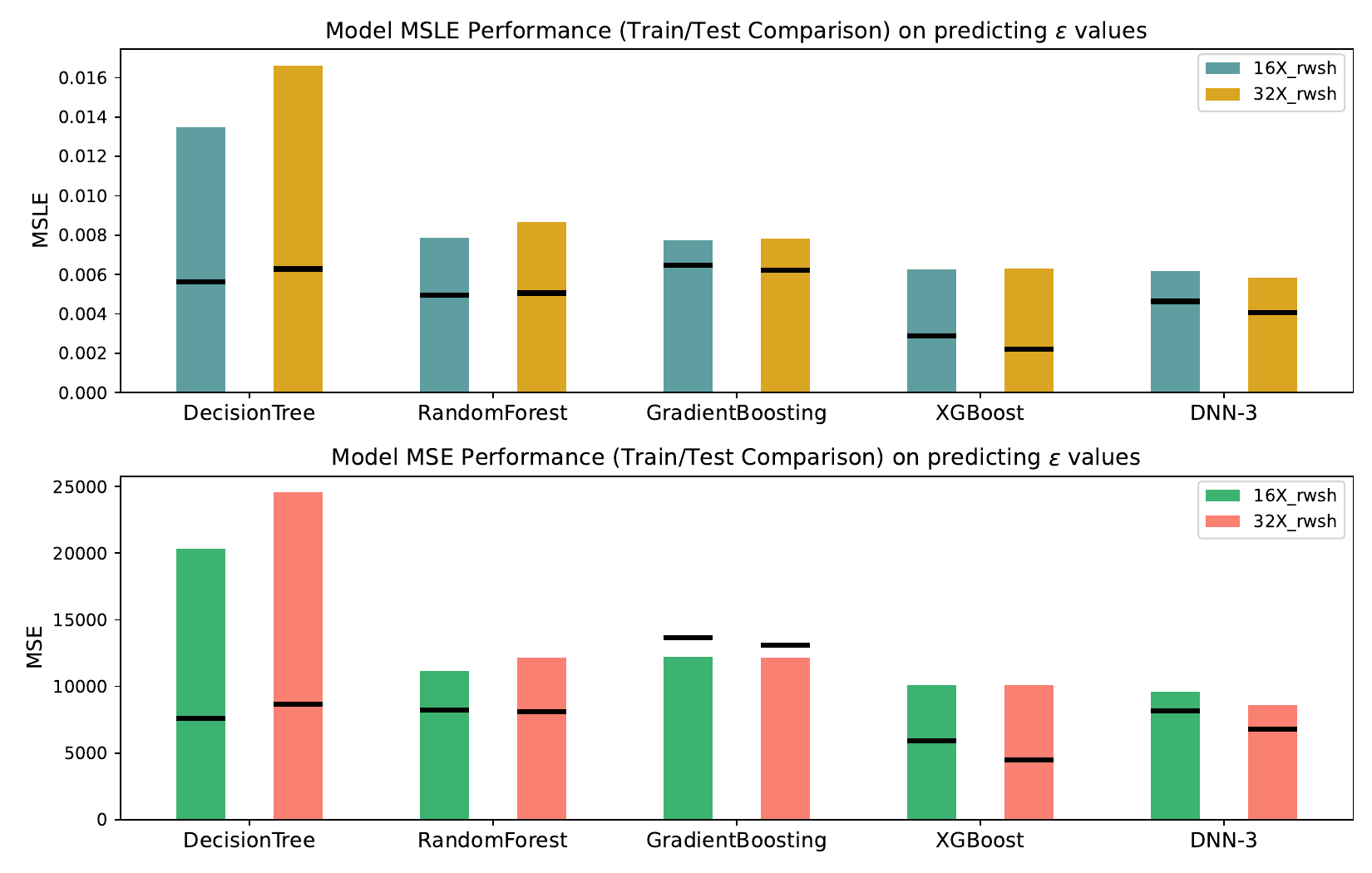}
    \caption{Metrics results for regression on Neutron Stars data. \textit{Top}: MSLE results. \textit{Bottom}: MSE results. The results are presented in grouped bar plots. There are as many groups as the different algorithms used. Each group contains two bar plots, one for each different number of features (\textit{left bar}: 16 features, \textit{right bar}: 32 features). The black lines in each bar, depict the respective training result, i.e. the performance of the model on the training dataset itself after fitting. See \texttt{metrics\_learning\_curves\_final\_results.ipynb} in Table \ref{tab:final_results}.}
    \label{fig:metrics_results_NS}
\end{figure}

After fitting our regression models, we evaluated their performance on the training dataset itself and the test dataset, to obtain the general accuracy and check for possible overfitting. Figure \ref{fig:metrics_results_NS} contains an overview of the results for regression on Neutron Stars data. At first glance, one can notice the significant weaker performance of the \textit{Decision Tree} models. Indeed, \textit{Decision Tree} models exhibit an error that is twice the error of the other models. Moreover, they have low accuracy, since the training value (black line) is less than the half of the corresponding test value (height of bar), for both metrics MSE and MSLE. Then, we observe a continuous improvement on performance, as we move to the right of Fig. \ref{fig:metrics_results_NS}. 

This improvement, although welcoming, is actually small. That is, \textit{Decision Tree} excluded, all other models seem to perform in a similar way. The \textit{Gradient Boosting} models have similar accuracy with the \textit{Random Forest} ones. Moreover, \textit{XGBoost} models have similar accuracy with the \textit{DNN-3} models. Based exclusively on the test results of the metrics, we can say that the \textit{XGBoost} models are the best machine learning models in our study, since they antagonize fairly the much complex \textit{DNN-3} models. 

Now, an important notice, concerning all models, is the fact that their accuracy is similar,  regardless of the number of features used in the input. As can be seen in Fig. \ref{fig:metrics_results_NS}, there are even cases, like the \textit{Gradient Boosting} and the \textit{XGBoost} ones, where the results are identical for 16 and 32 features (8 and 16 M-R points). This would imply, that a possible reconstruction of an EoS could be done we equal accuracy, either using 8 or 16 M-R points from its respective $M-R$ curve on these models. Another important notice, is that the behavior of the two metrics MSLE and MSE is the same for all models. The only minor deviation is observed for the \textit{Gradient Boosting} algorithm, where the MSE results indicate that the models perform better on the test set rather than the training set, while the MSLE results indicate the opposite. However, the training and test results are close enough for both metrics. Hence, an equivalent depiction of the performance of our models can be provided, using whichever of the two metrics, MSE and MSLE, we desire.

\begin{figure}[h!]
    \centering
    \includegraphics[width=0.75\linewidth]{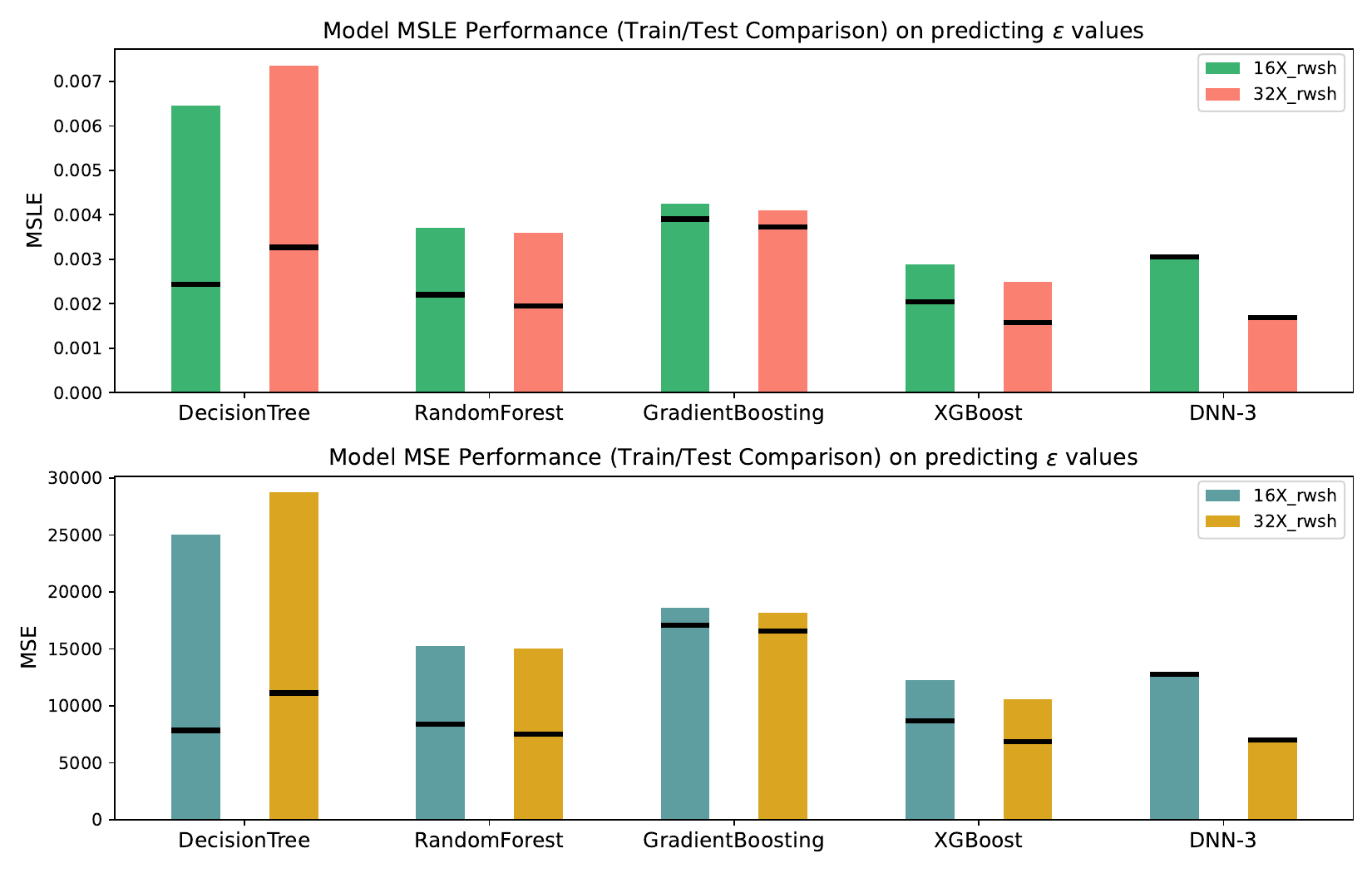}
    \caption{Metrics results for regression on Quark Stars data. \textit{Top}: MSLE results. \textit{Bottom}: MSE results. The results are presented in grouped bar plots. There are as many groups as the different algorithms used. Each group contains two bar plots, one for each different number of features (\textit{left bar}: 16 features, \textit{right bar}: 32 features). The black lines in each bar, depict the respective training result, i.e. the performance of the model on the training dataset itself after fitting. See \texttt{metrics\_learning\_curves\_final\_results.ipynb} in Table \ref{tab:final_results}.}
    \label{fig:metrics_results_QS}
\end{figure}

The same can be said for the performance of our regression models on Quark Stars data, shown in Fig. \ref{fig:metrics_results_QS}. Things are pretty much similar to the Neutron Stars' case. All other algorithms perform significantly better than the \textit{Decision Tree} one. Furthermore, the accuracy of \textit{Random Forest} models is comparable to the accuracy of \textit{Gradient Boosting} ones (this time though \textit{Random Forest} performs better on test set than \textit{Gradient Boosting}). And the \textit{XGBoost} models exhibit comparable accuracy to the \textit{DNN-3} models, even being better on test set, for 16 features. 

A quite interesting notice, is the improvement on predictions when larger datasets are used and 32 features variables are used instead of 16, in comparison with the Neutron star's case. It seems that the larger dataset for Quark Stars (89100 rows vs 30400 rows for Neutron Stars) and the much simplest nature of Quark Star EoSs (linear or almost linear behavior), allow the algorithms \textit{Gradient Boosting}, \textit{XGBoost} and \textit{DNN-3}, to analyze more data and learn more effectively the complex connections between the $M-R$ curves and the EoSs, from which they were derived. Thus, the metrics results for the test set are closer to the ones of the training set, than in Fig. \ref{fig:metrics_results_NS}, for these three types of algorithms. In practice, the best improvement is found in \textit{DNN-3} models, having identical values for test and train results in Fig. \ref{fig:metrics_results_QS} (the black lines are exactly on the top of the bars). The same models exhibit, also, the biggest improvement when the number of features is increased. Indeed, the use of 32 feature variables in the \textit{DNN-3} models for Quark Stars, results in approximately half the error of the use of 16 feature variables.

\begin{figure}[h]
    \centering
    \includegraphics[width=0.75\linewidth]{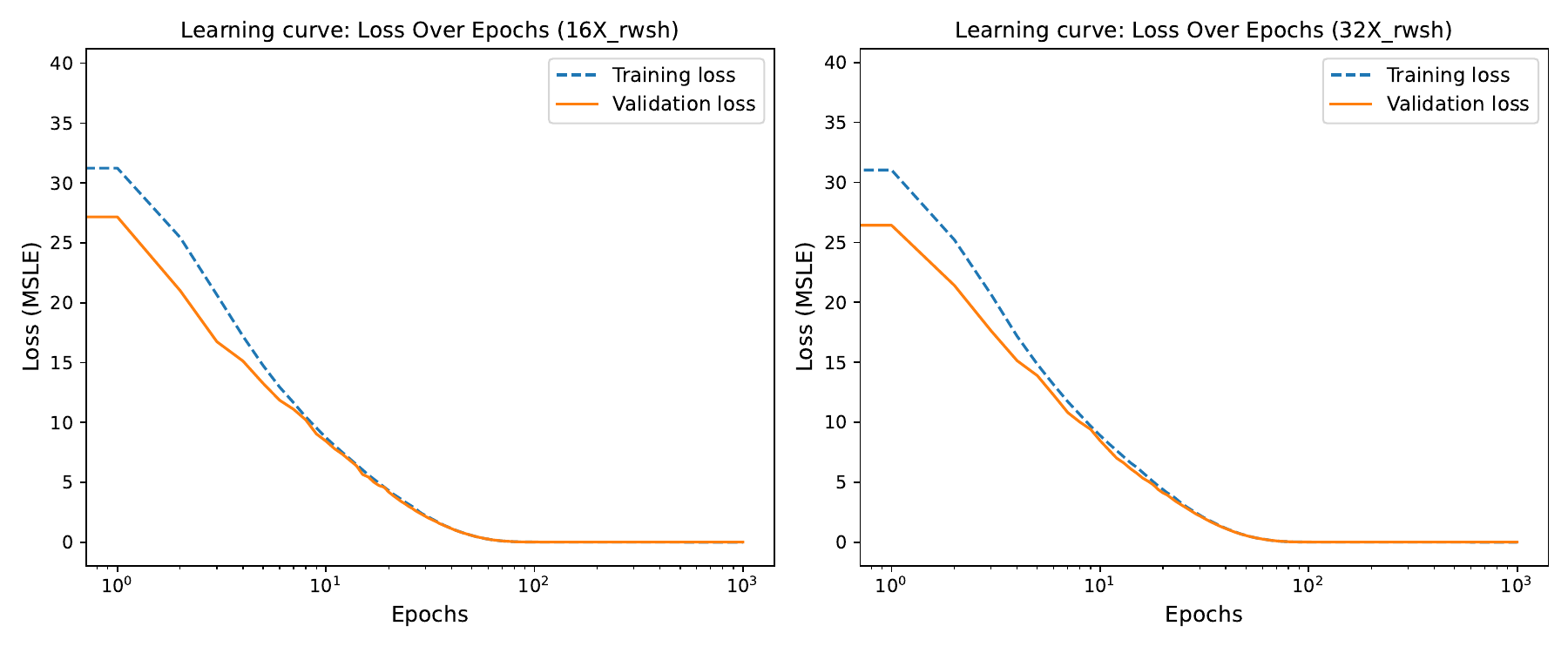}
    \caption{Learning curves of \textit{DNN-3} models trained and validated on Neutron Stars data. The axes are in semi=log scale. See \texttt{metrics\_learning\_curves\_final\_results.ipynb} in Table \ref{tab:final_results}.}
    \label{fig:learning_curve_NS_semilog}
\end{figure}

\begin{figure}[h]
    \centering
    \includegraphics[width=0.75\linewidth]{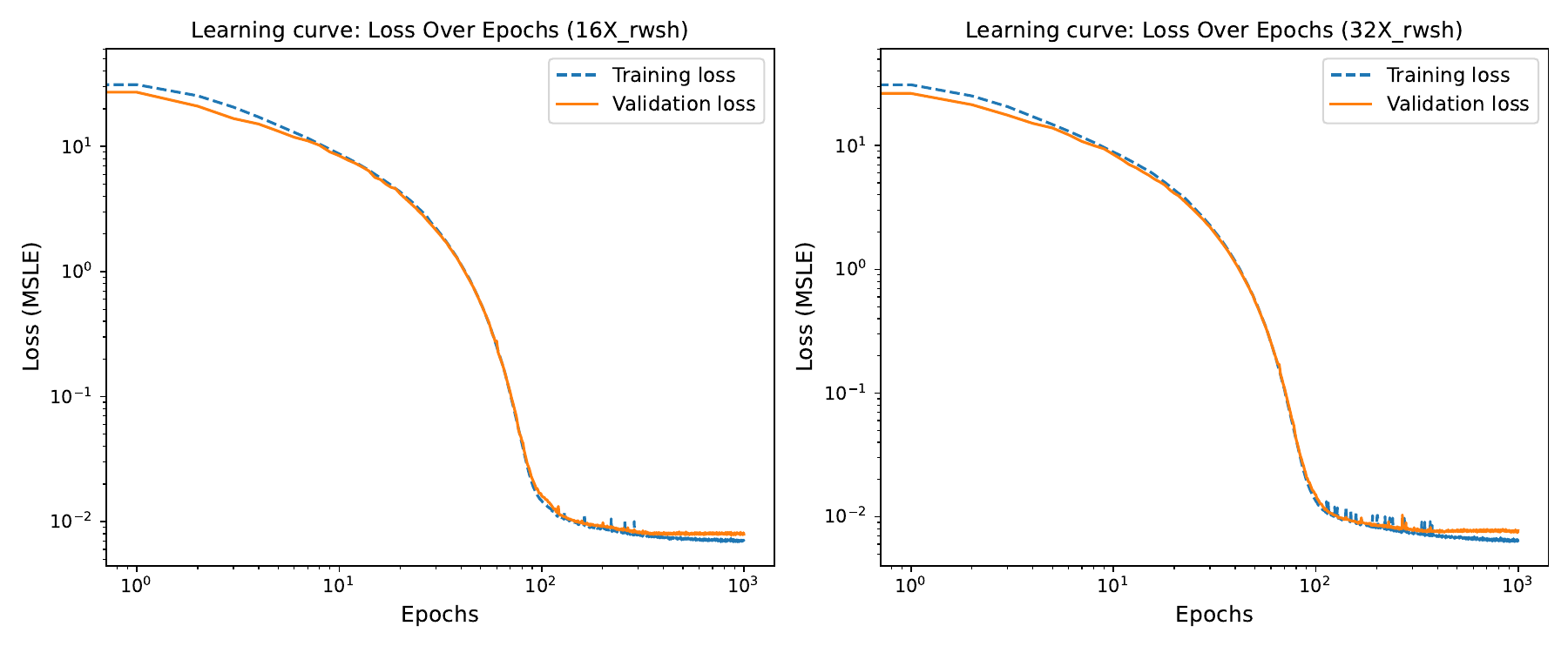}
    \caption{Learning curves of \textit{DNN-3} models trained and validated on Neutron Stars data. The axes are in log=log scale. See \texttt{metrics\_learning\_curves\_final\_results.ipynb} in Table \ref{tab:final_results}.}
    \label{fig:learning_curve_NS_loglog}
\end{figure}

\begin{figure}[h]
    \centering
    \includegraphics[width=0.75\linewidth]{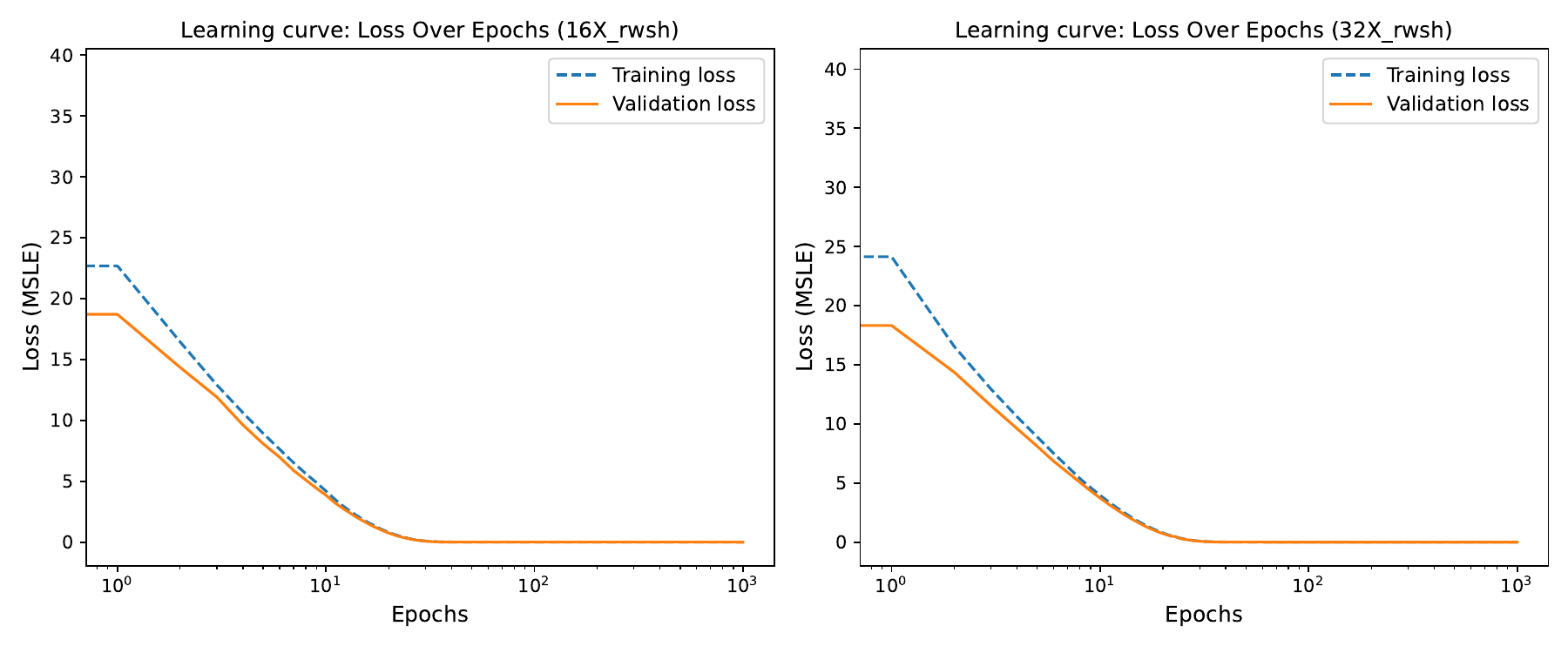}
    \caption{Learning curves of \textit{DNN-3} models trained and validated on Quark Stars data. The axes are in semi=log scale. See \texttt{metrics\_learning\_curves\_final\_results.ipynb} in Table \ref{tab:final_results}.}
    \label{fig:learning_curve_QS_semilog}
\end{figure}

\begin{figure}[h!]
    \centering
    \includegraphics[width=0.75\linewidth]{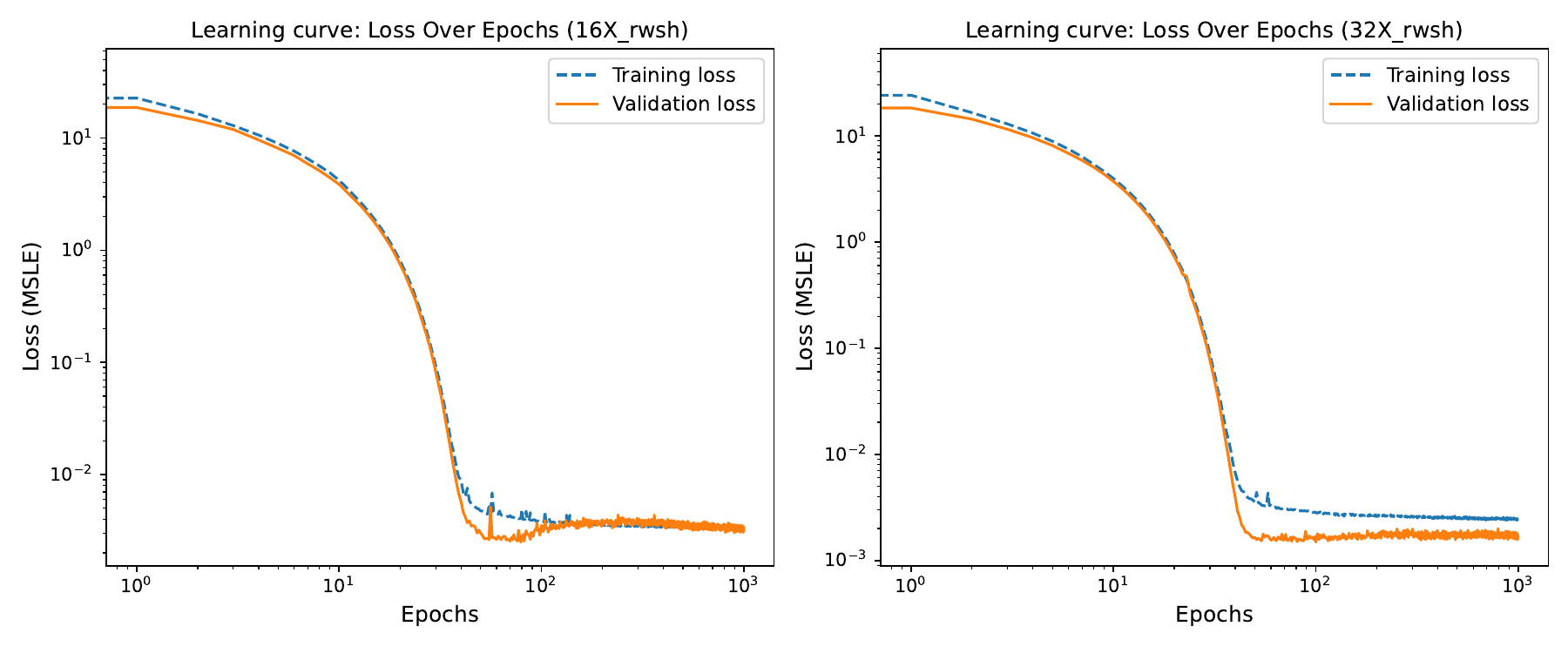}
    \caption{Learning curves of \textit{DNN-3} models trained and validated on Quark Stars data. The axes are in log=log scale. See \texttt{metrics\_learning\_curves\_final\_results.ipynb} in Table \ref{tab:final_results}.}
    \label{fig:learning_curve_QS_loglog}
\end{figure}

This is why, we have to study the fit of these models further. Fortunately, the \textit{TensorFlow} framework allows us to capture the loss function history during training and make the learning curve of a \textit{DNN} model, as shown in \texttt{train\_test\_dnn3\_regress.ipynb} notebook in Table \ref{tab:train_test_DL} and \texttt{metrics\_learning\_curves\_final\_results.ipynb} in Table \ref{tab:final_results}. We present these curves in Figs. \ref{fig:learning_curve_NS_semilog}, \ref{fig:learning_curve_NS_loglog}, \ref{fig:learning_curve_QS_semilog} and \ref{fig:learning_curve_QS_loglog}. The existence of the validation set provides insights of the stability of training, through direct comparison of the training loss and the validation loss. 

The scaling of the axes is, also, important for demonstrating and capturing as much details as possible. The x-axis requires, always, log scale, due to the large number of epochs (1000). In contrast, we are free to choose between linear or log scale on the y-axis. The first choice is suitable when the loss maintains the same order of magnitude (or it changes by at most one order of magnitude), through epochs. In our analysis, this option results in showing a great training process, where both the MSLE training loss and validation loss start from a few dozen and become very small at approximately 600 epochs, as shown in Figs. \ref{fig:learning_curve_NS_semilog} and \ref{fig:learning_curve_QS_semilog}. Additionally, the validation loss seems to be constantly below the training loss, implying the \textit{DNN-3} models would perform better (or at least the same) on foreign data, rather than (as) the given training dataset.  

However, the semi-log option fails to capture the final order of magnitude of the losses, as well as any deviations between the validation and the training in the epochs interval where it appears to coincide (600-1000). This is where, a log-log figure comes in handy. Indeed, from Figs. \ref{fig:learning_curve_NS_loglog} and \ref{fig:learning_curve_QS_loglog}, one can obtain the order of $10^{-3}$ as the final order of magnitude of the losses. Hence, the losses become 4 orders of magnitude smaller during fitting. We can, also, observe that for Neutron Stars the training loss ultimately falls behind w the validation loss and that the validation loss becomes continuously bigger after 400 epochs, indicating the model begins to loose its accuracy (see Fig. \ref{fig:learning_curve_NS_loglog}).  On the contrary, for Quark Stars the validation loss, is always smaller than the training loss (see Fig. \ref{fig:learning_curve_QS_loglog}), ensuring better performance on foreign Quark Stars data. Besides, the significant decrease of the losses happens sooner for Quark Stars (at epochs $\approx 20$), than Neutron Stars (at epochs $\approx 40$), again perhaps due to simpler form and larger dataset of Quark Stars EoSs. Moreover, the noises of the losses have to be treated with caution in the log-log scale. In our case, a noise at the initial epochs ($<10$) is significantly larger (and might need to be cured), than a noise at later epochs, although the latter may appear to be bigger, due to the log-log scaling. Last, but not least, all Figs. \ref{fig:learning_curve_NS_semilog}, \ref{fig:learning_curve_NS_loglog}, \ref{fig:learning_curve_QS_semilog} and \ref{fig:learning_curve_QS_loglog} confirm the similar (almost identical) effectiveness of the \textit{DNN-3} models, either 16 (\textit{left} graphs) or 32 (\textit{right} graphs) feature variables are given as input. 

\begin{table}[h]
    \centering
    \begin{tabular}{|c|c|c|c|c|}
    \hline
    \hline  
    \textbf{Algorithm} & \textbf{16 feats. NS} & \textbf{32  feats. NS} & \textbf{16 feats. QS} & \textbf{32 feats. QS} \\
    \hline
    \hline
    Decision Tree  & $\approx 20$ sec & $\approx 30$ sec & $\approx 70$ sec & $\approx 2$ min\\
    \hline
    Random Forest  & $\approx 3$ min $20$ sec & $\approx 6$ min & $\approx 14$ min & $\approx 25$ min\\
    \hline
    Gradient Boosting  & $\approx 14$ min $15$ sec & $\approx 17$ min $30$ sec & $\approx 47$ min $30$ sec & $\approx 61$ min $30$ sec \\
    \hline
    XGBoost  & $\approx 3$ min & $\approx 5$ min $30$ sec & $\approx 5$ min $30$ sec & $\approx 10$ min $30$ sec \\
    \hline
    DNN-3  & $\approx 4$ min $40$ sec & $\approx 5$ min & $\approx 12$ min & $\approx 12$ min $30$ sec\\
    \hline
    \hline     
    \end{tabular}
    \caption{Total fitting times of all regression models. For the machine learning algorithms the times refer to the combined application of 5-fold cross-validation and grid search. Columns 2 and 3 represent fitting times on Neutron Stars data, with 24200 rows of training data ($80\%\times 30400$) and with 16 or 32 feature variables, respectively. Columns 4 and 5 represent fitting times on Quarks Stars data, with 71300 rows of training data ($80\%\times 89100$) and with 16 or 32 feature variables.}
    \label{tab:fit_times}
\end{table}

A study, though, of the effectiveness of statistical learning procedures would be incomplete, without a reference on the fitting times. As we discussed in section \ref{Fine_tune}, for machine learning algorithms we employed the techniques of cross-validation (5-fold) and grid search. The combination of these two methods, resulted in multiple trainings of each model, before the final evaluation. The large amount of trainings can lead to extremely long fitting times, rendering the whole analysis process computationally infeasible. In order to reduce the total time it takes the models to reach final evaluation, we parallelized the grid search procedure. All computations were executed on an \href{https://www.intel.com/content/www/us/en/products/sku/236849/intel-core-ultra-9-processor-185h-24m-cache-up-to-5-10-ghz/specifications.html}{Intel Ultra 9 185H} CPU, utilizing 18 of the total 22 available threads. 

The resulted fitting times are presented in Table \ref{tab:fit_times}. As expected, the faster algorithm is \textit{Decision Tree}, due to its much simpler structure. In contrast, the slower algorithm is \textit{Gradient Boosting}, exceeding the threshold of 1 hour in the most extreme case of Quark Stars data with 32 feature variables. \textit{Random Forest} models being the generalization of \textit{Decision Tree} models, exhibit $10\times$ to $12\times$ longer fitting times than the latter. Furthermore, \textit{Random Forest} models exhibit poor performance scaling on more complex data with more feature variables, as doubling the number of features leads to a doubling of the fitting time. The case is the same for the \textit{XGBoost} models. However, \textit{XGBoost} models have a much better scaling on larger data: offering double fitting times for approximately 3 times larger training datasets (71300 for Quark Stars to 24200 for Neutron Stars). Finally, the \textit{DNN-3} models, provide the best escalation on more complex data, having identical fitting times for the two different numbers of feature variables. In other words, it seems that the fitting time of \textit{DNN-3} models is only affected from the size of datasets. We have to denote though, that that the training dataset used in the training of the \textit{DNN-3} models is smaller (19400 rows for Neutron Stars and 57000 rows for Quark Stars), than the training dataset used in ML algorithms, due to the additional splitting step to obtain the validation set (see section \ref{Fine_tune}). Yet, with less training information, the \textit{DNN-3} models perform better than most ML algorithms and with comparable fitting times, confirming the superiority of deep learning over machine learning techniques.

\section{Reconstructing Neutron Stars' EoSs}\label{recontstruct_NS}
Below, we present our reconstruction attempt of the 21 "main" equations of state for hadronic stars:

\begin{figure}[h]
    \centering
    \includegraphics[width=0.99\linewidth]{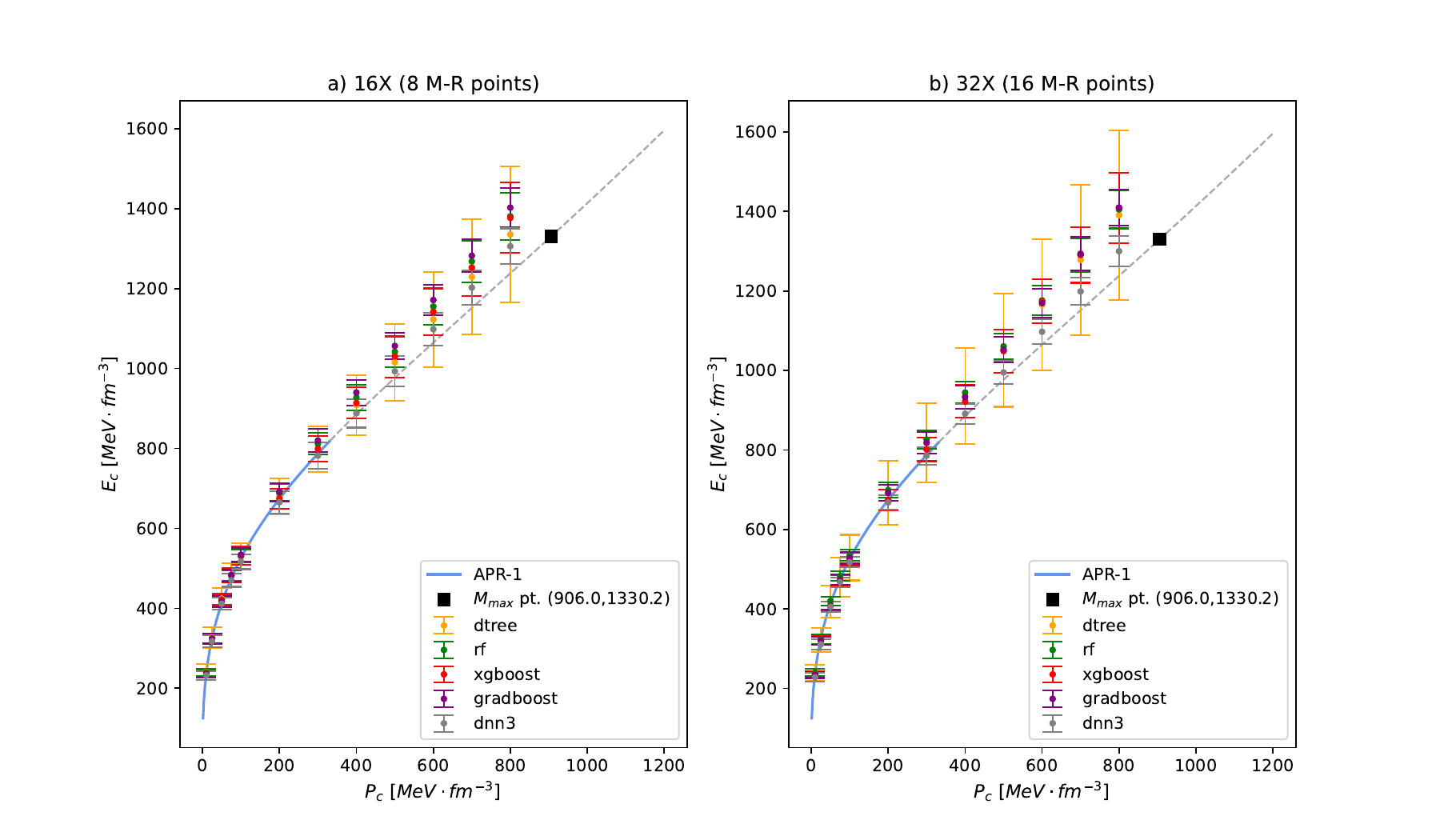}
    \caption{Reconstructing the \textit{APR-1} EoS}
    \label{fig:APR-1_EOS_predict}
\end{figure}

\begin{figure}[h!]
    \centering
    \includegraphics[width=0.99\linewidth]{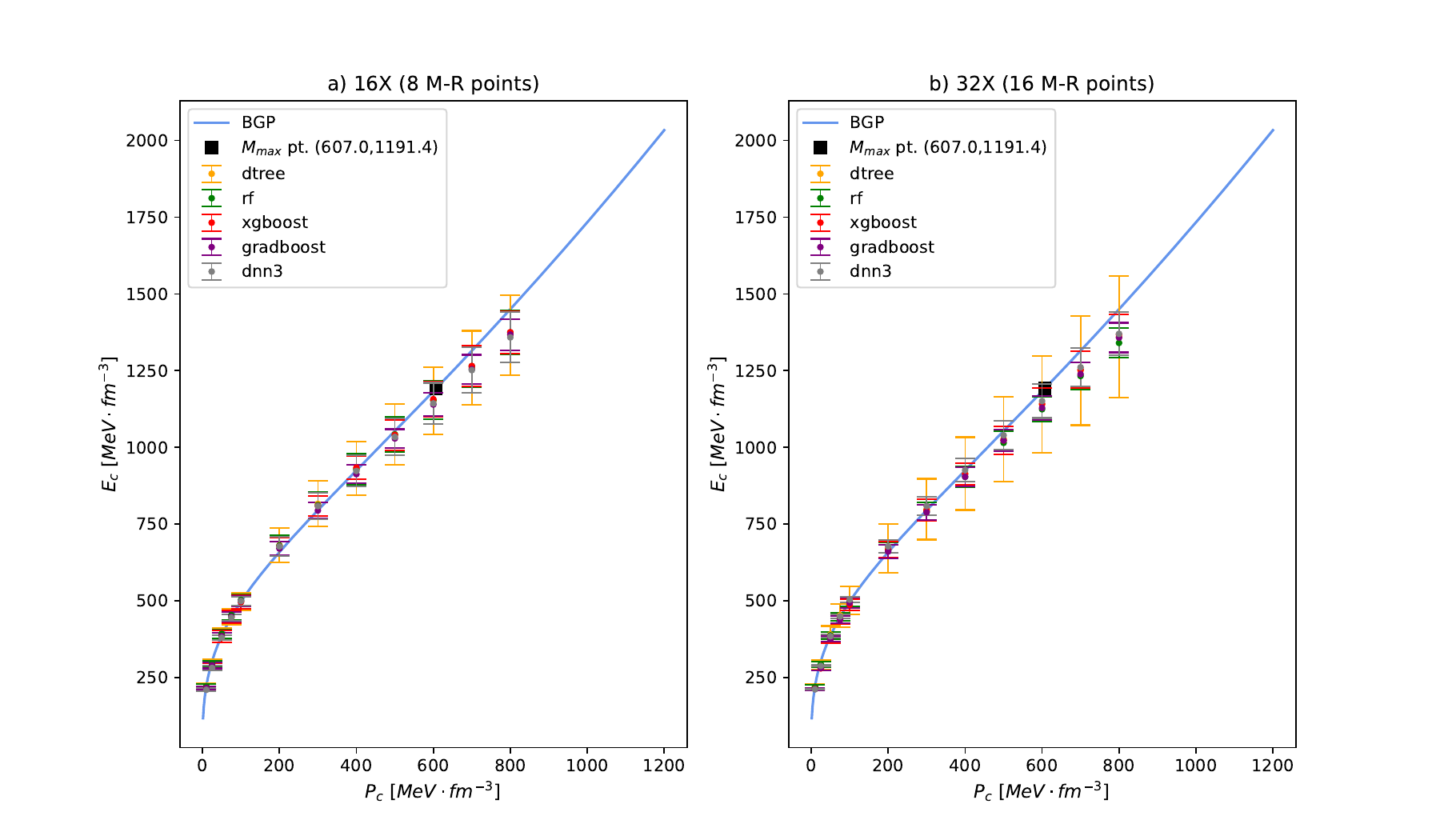}
    \caption{Reconstructing the \textit{BGP} EoS}
    \label{fig:BGP_EOS_predict}
\end{figure}

\begin{figure}[h]
    \centering
    \includegraphics[width=\linewidth]{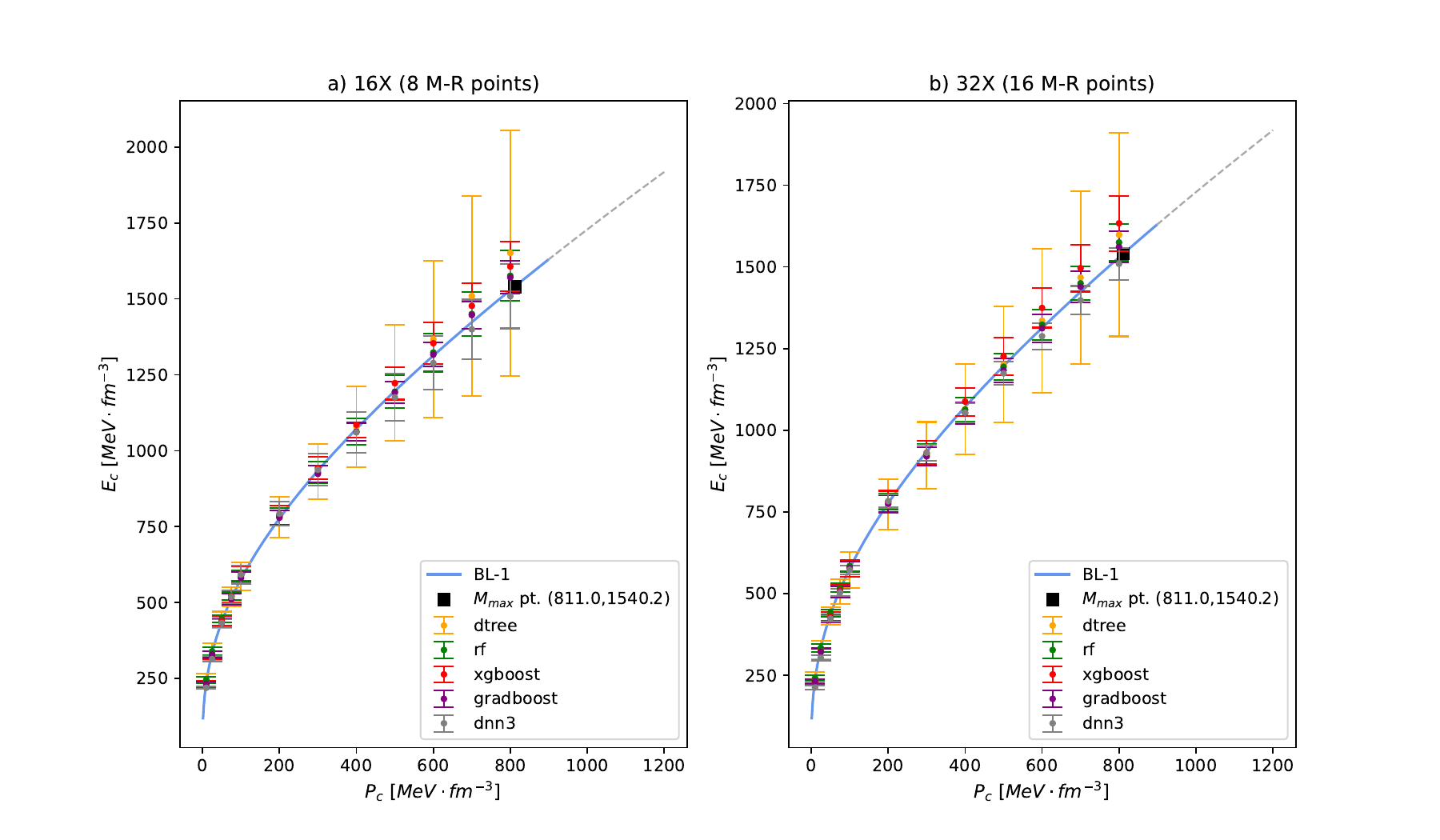}
    \caption{Reconstructing the \textit{BL-1} EoS}
    \label{fig:BL-1_EOS_predict}
\end{figure}

\begin{figure}[h!]
    \centering
    \includegraphics[width=\linewidth]{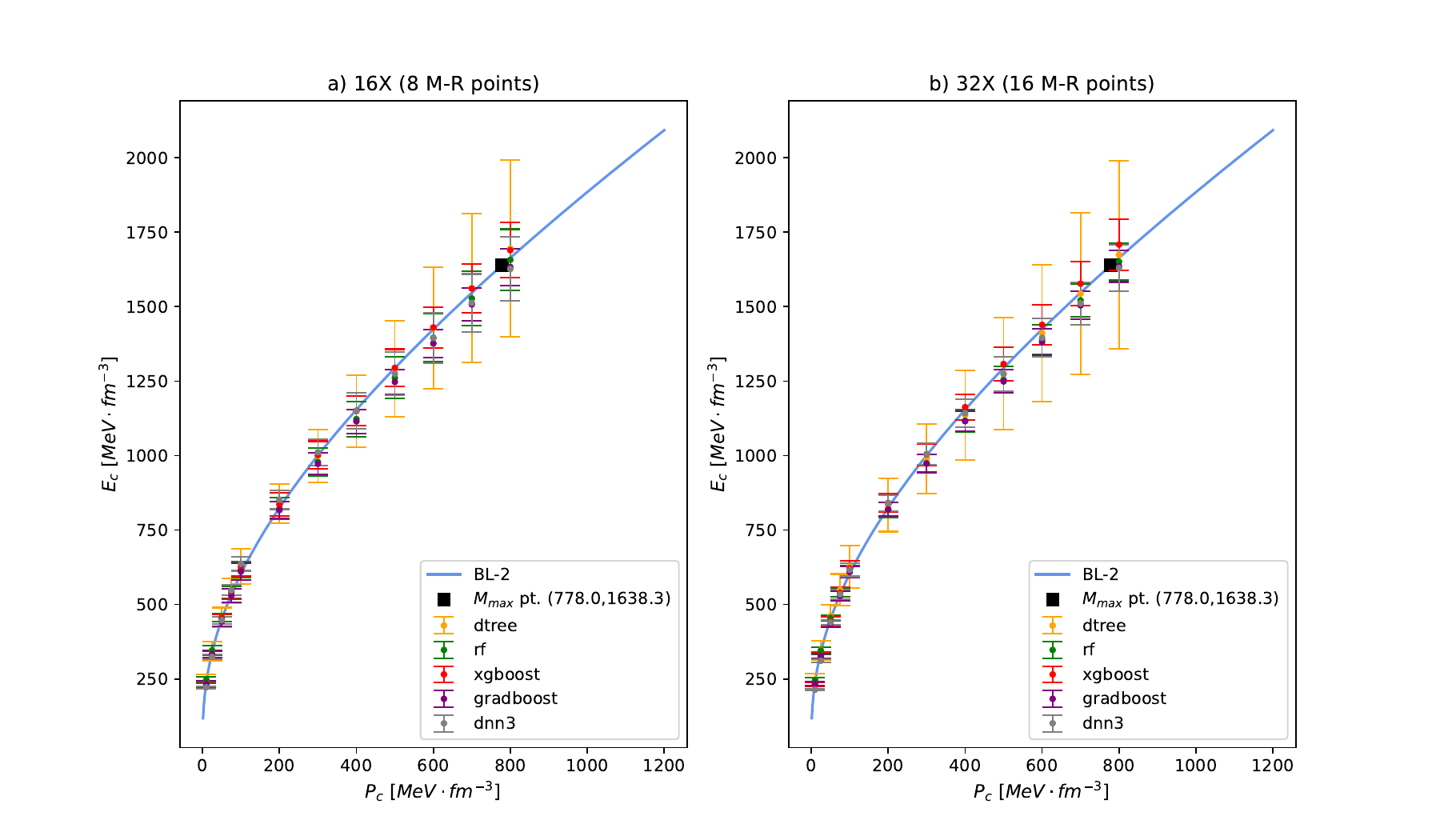}
    \caption{Reconstructing the \textit{BL-2} EoS}
    \label{fig:BL-2_EOS_predict}
\end{figure}

\begin{figure}[h]
    \centering
    \includegraphics[width=\linewidth]{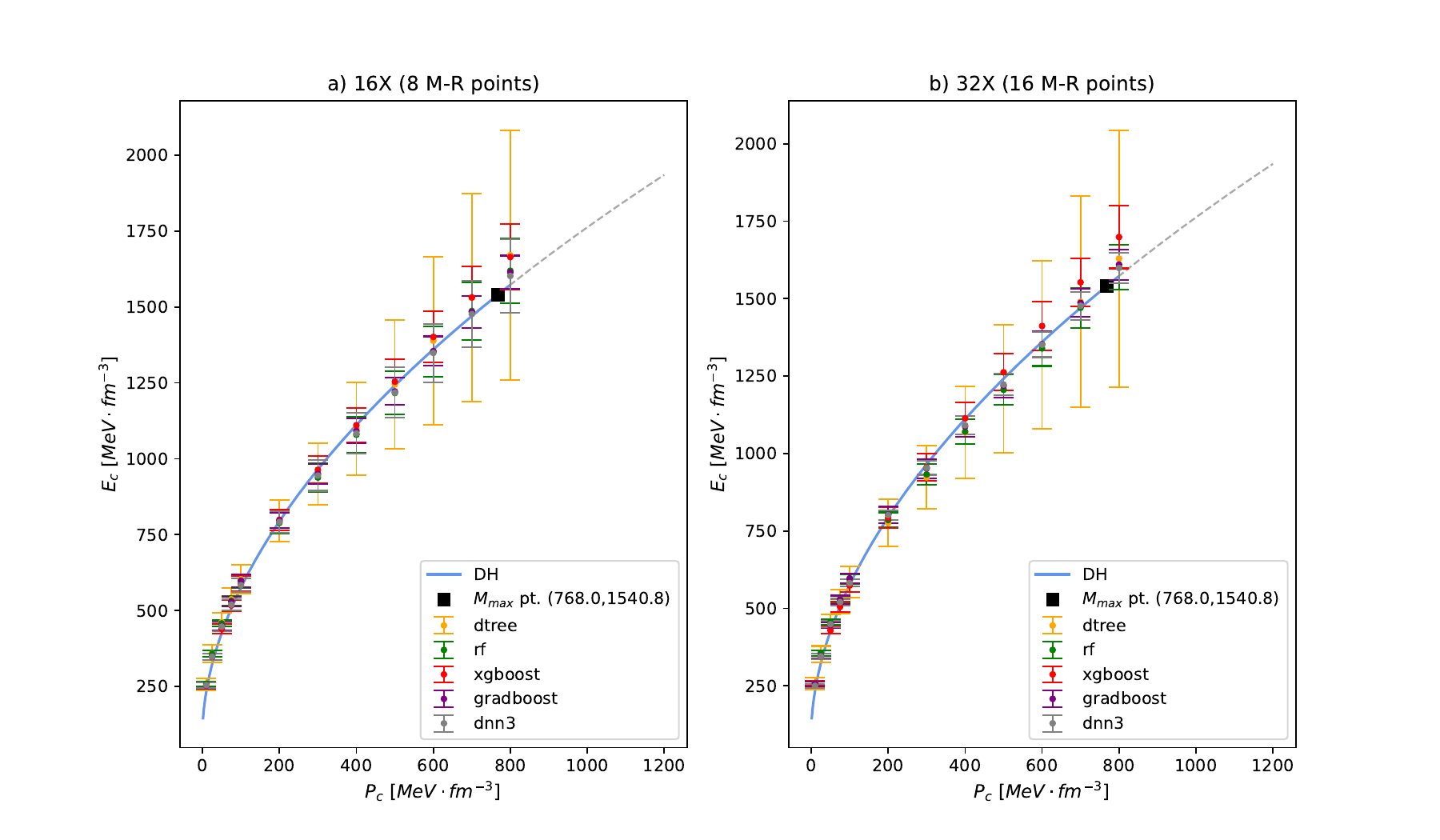}
    \caption{Reconstructing the \textit{DH} EoS}
    \label{fig:DH_EOS_predict}
\end{figure}

\begin{figure}[h!]
    \centering
    \includegraphics[width=\linewidth]{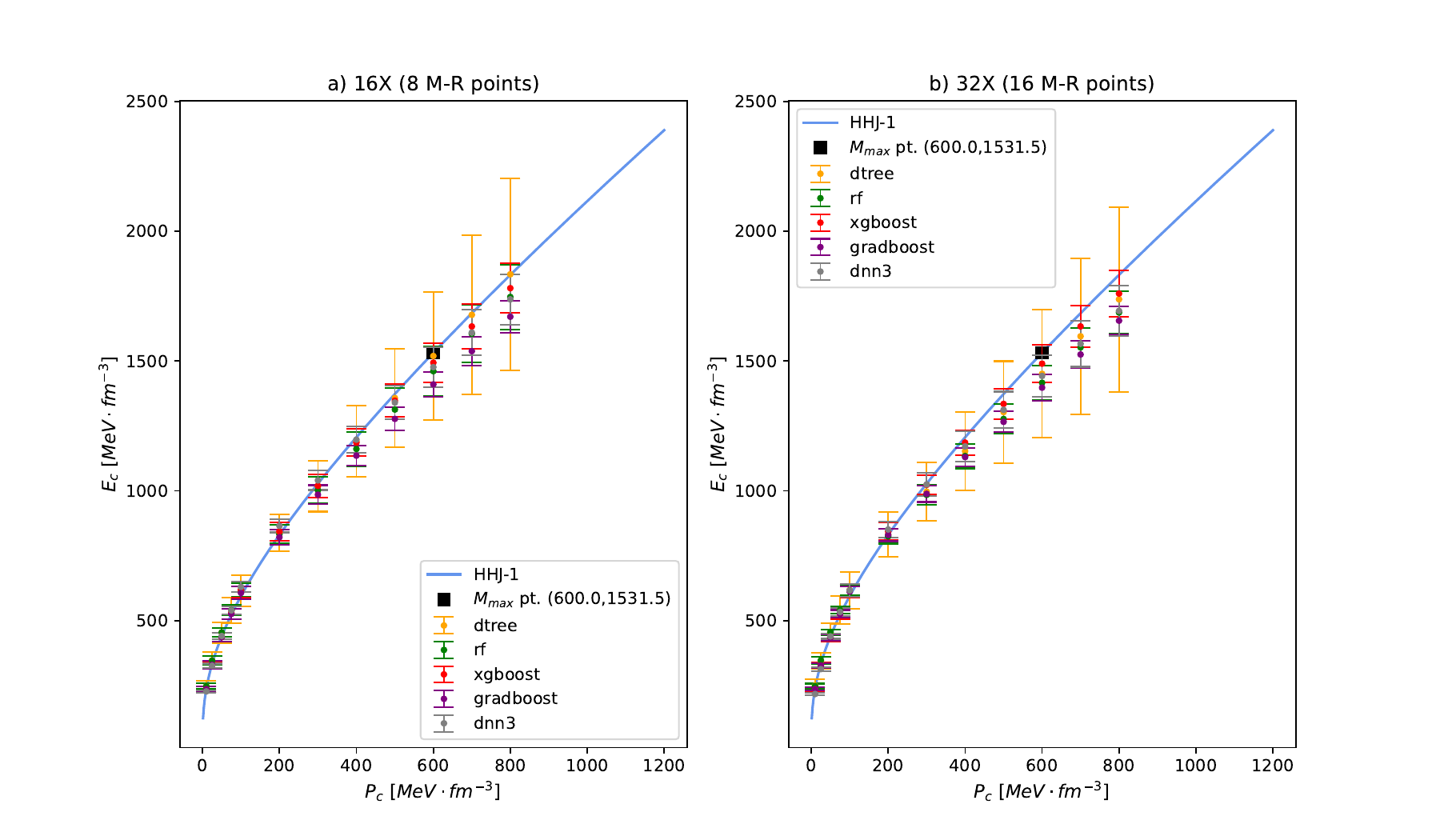}
    \caption{Reconstructing the \textit{HHJ-1} EoS}
    \label{fig:HHJ-1_EOS_predict}
\end{figure}

\begin{figure}[h]
    \centering
    \includegraphics[width=\linewidth]{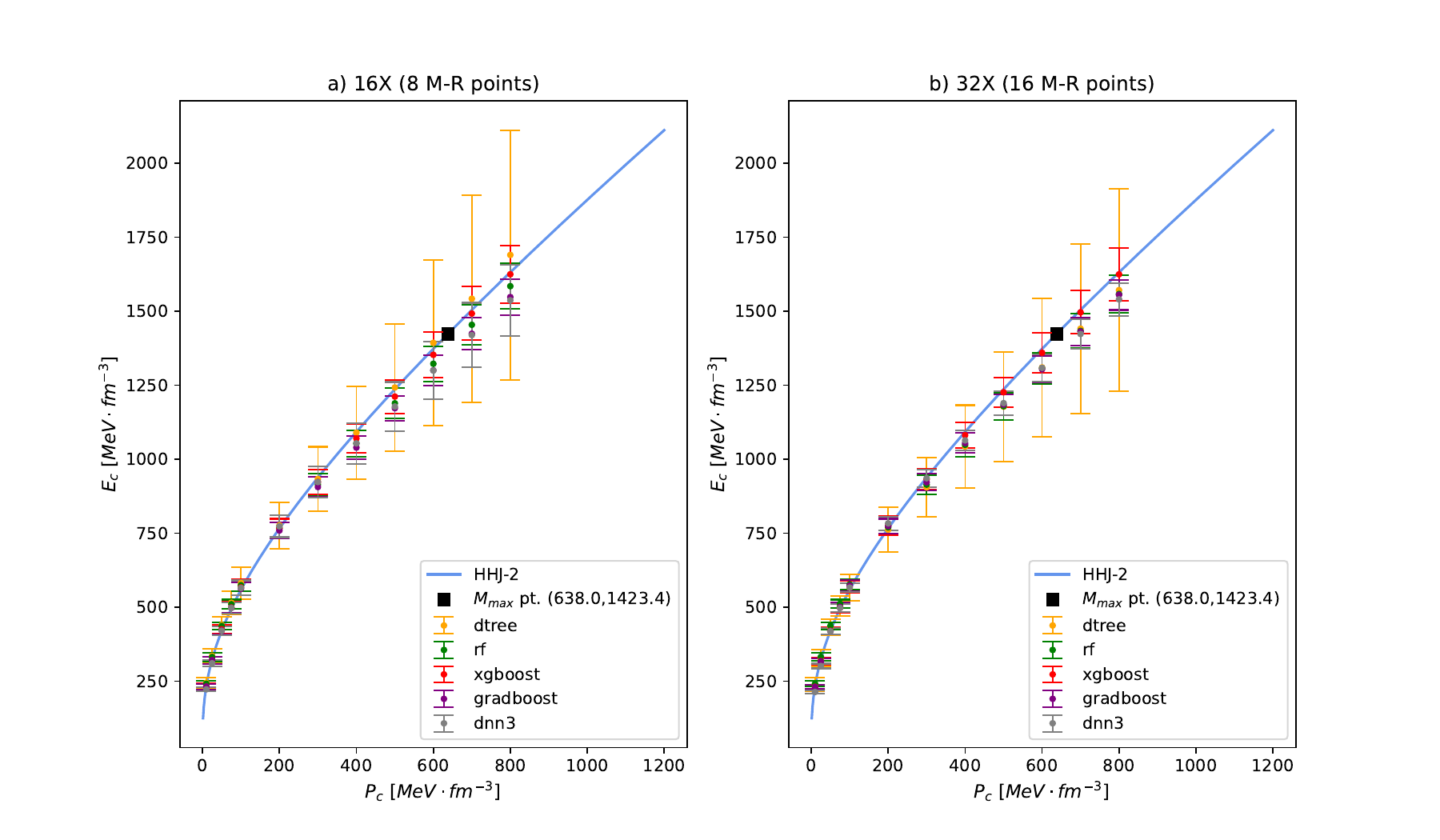}
    \caption{Reconstructing the \textit{HHJ-2} EoS}
    \label{fig:HHJ-2_EOS_predict}
\end{figure}

\begin{figure}[h!]
    \centering
    \includegraphics[width=\linewidth]{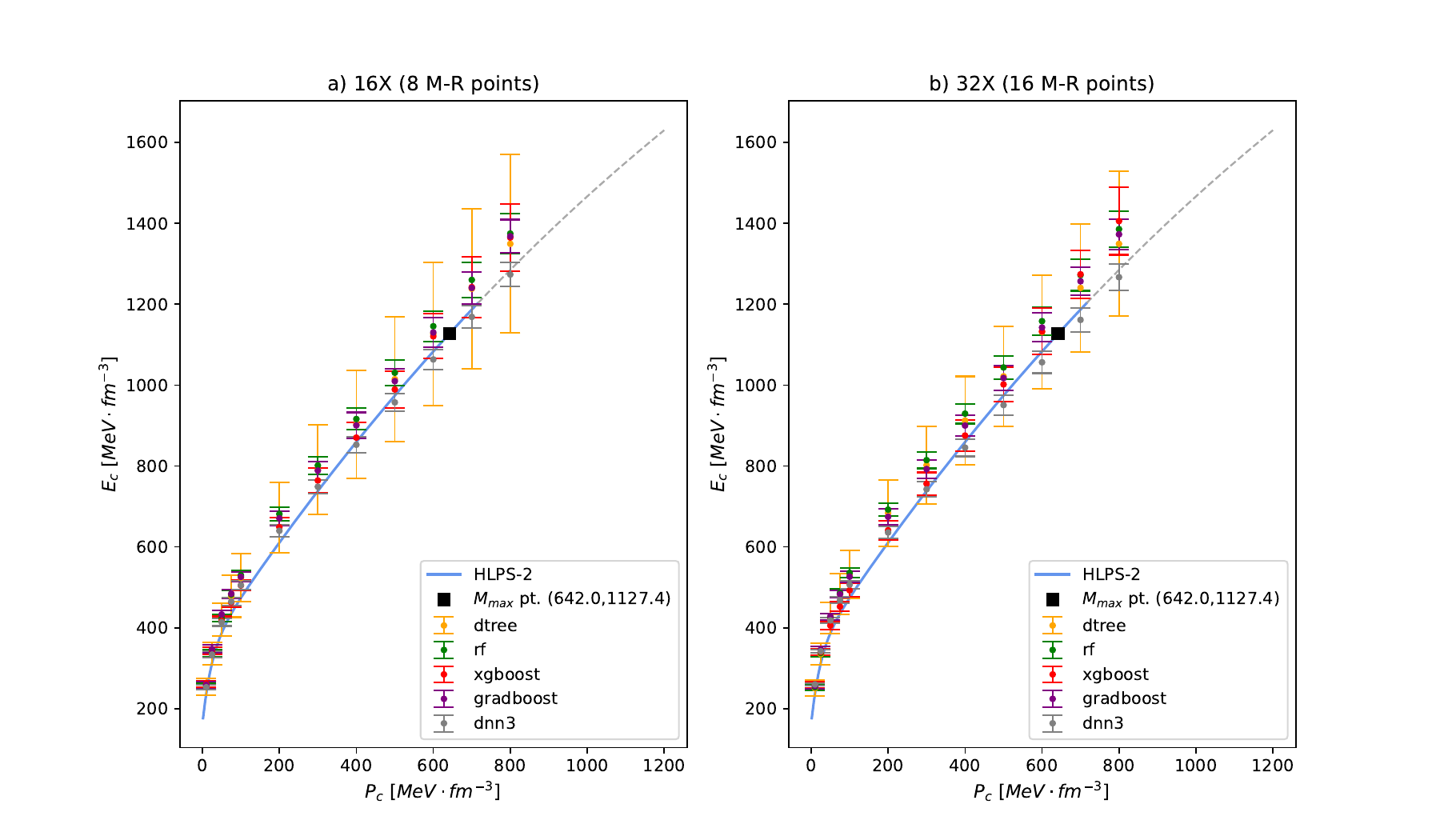}
    \caption{Reconstructing the \textit{HLPS-2} EoS}
    \label{fig:HLPS-2_EOS_predict}
\end{figure}

\begin{figure}[h]
    \centering
    \includegraphics[width=\linewidth]{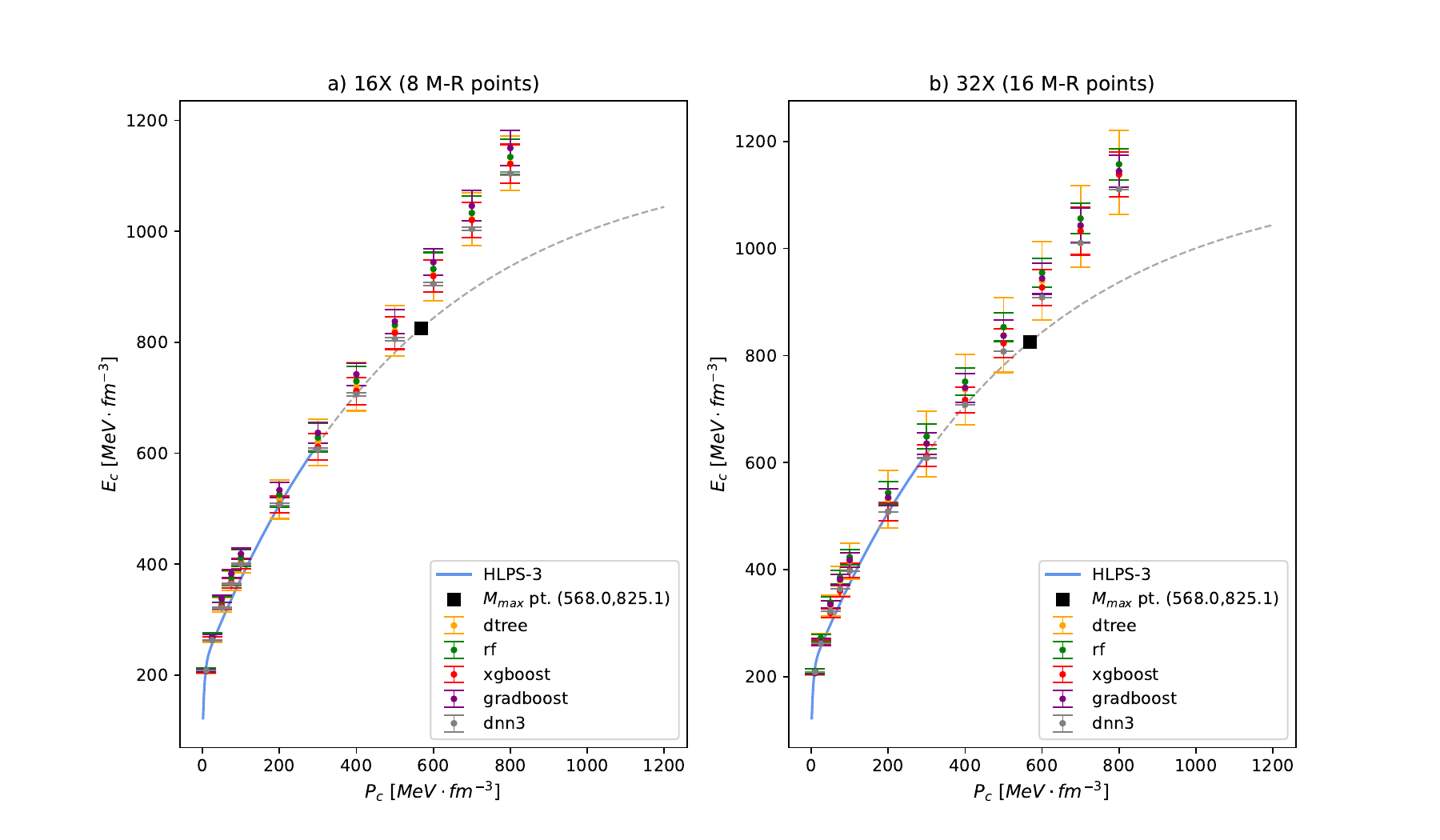}
    \caption{Reconstructing the \textit{HLPS-3} EoS}
    \label{fig:HLPS-3_EOS_predict}
\end{figure}

\begin{figure}[h!]
    \centering
    \includegraphics[width=\linewidth]{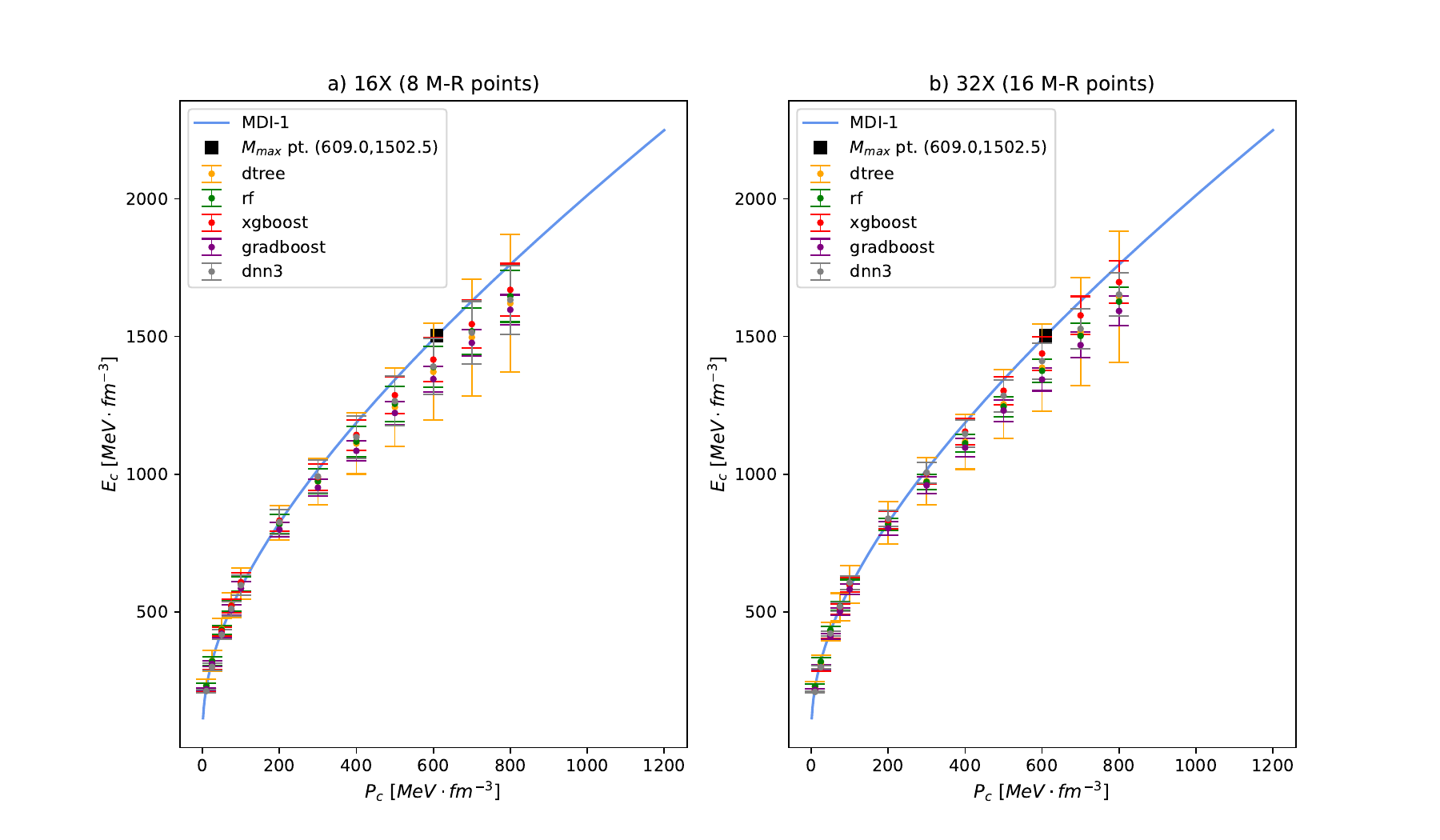}
    \caption{Reconstructing the \textit{MDI-1} EoS}
    \label{fig:MDI-1_EOS_predict}
\end{figure}

\begin{figure}[h]
    \centering
    \includegraphics[width=\linewidth]{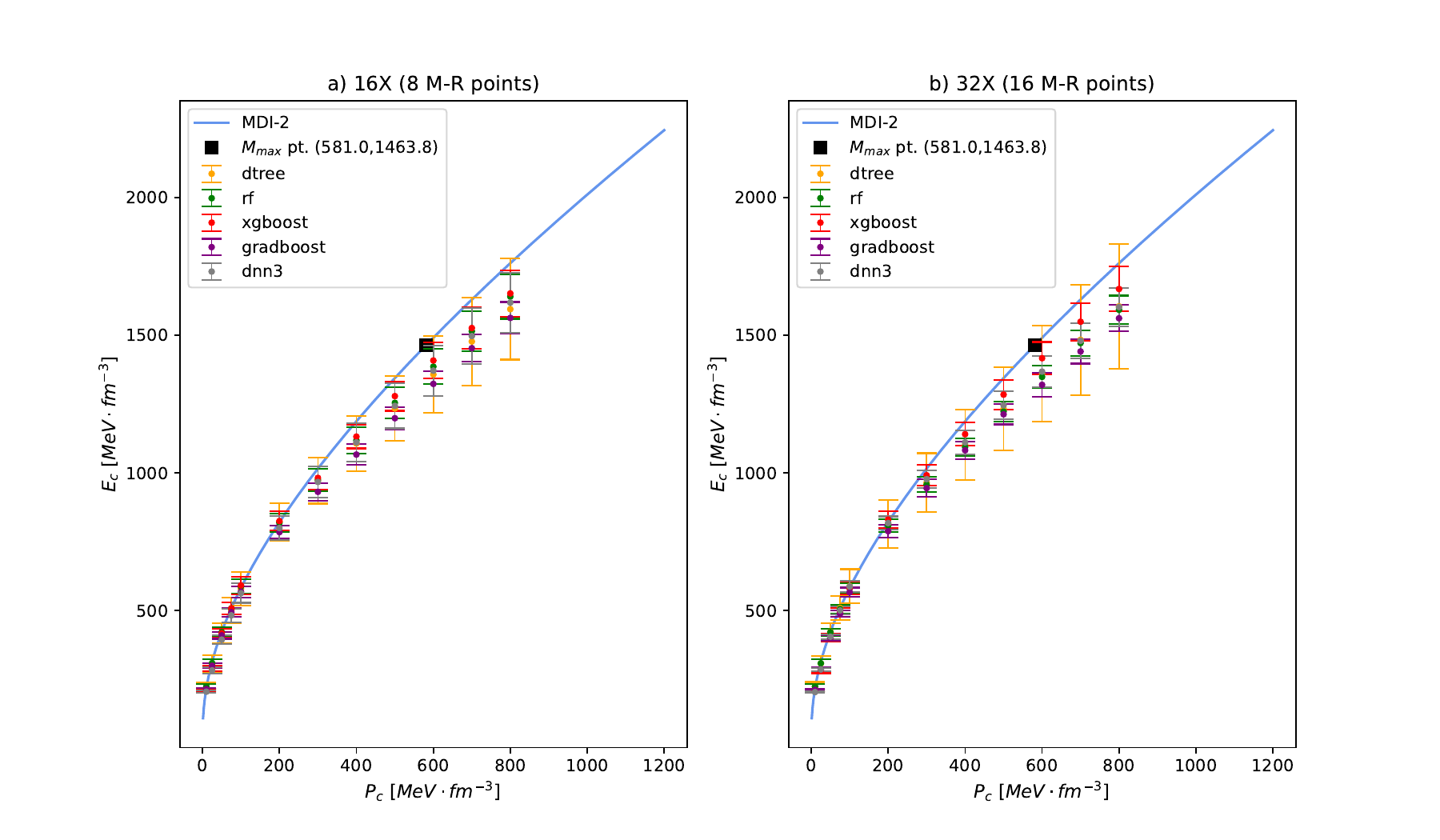}
    \caption{Reconstructing the \textit{MDI-2} EoS}
    \label{fig:MDI-2_EOS_predict}
\end{figure}

\begin{figure}[h!]
    \centering
    \includegraphics[width=\linewidth]{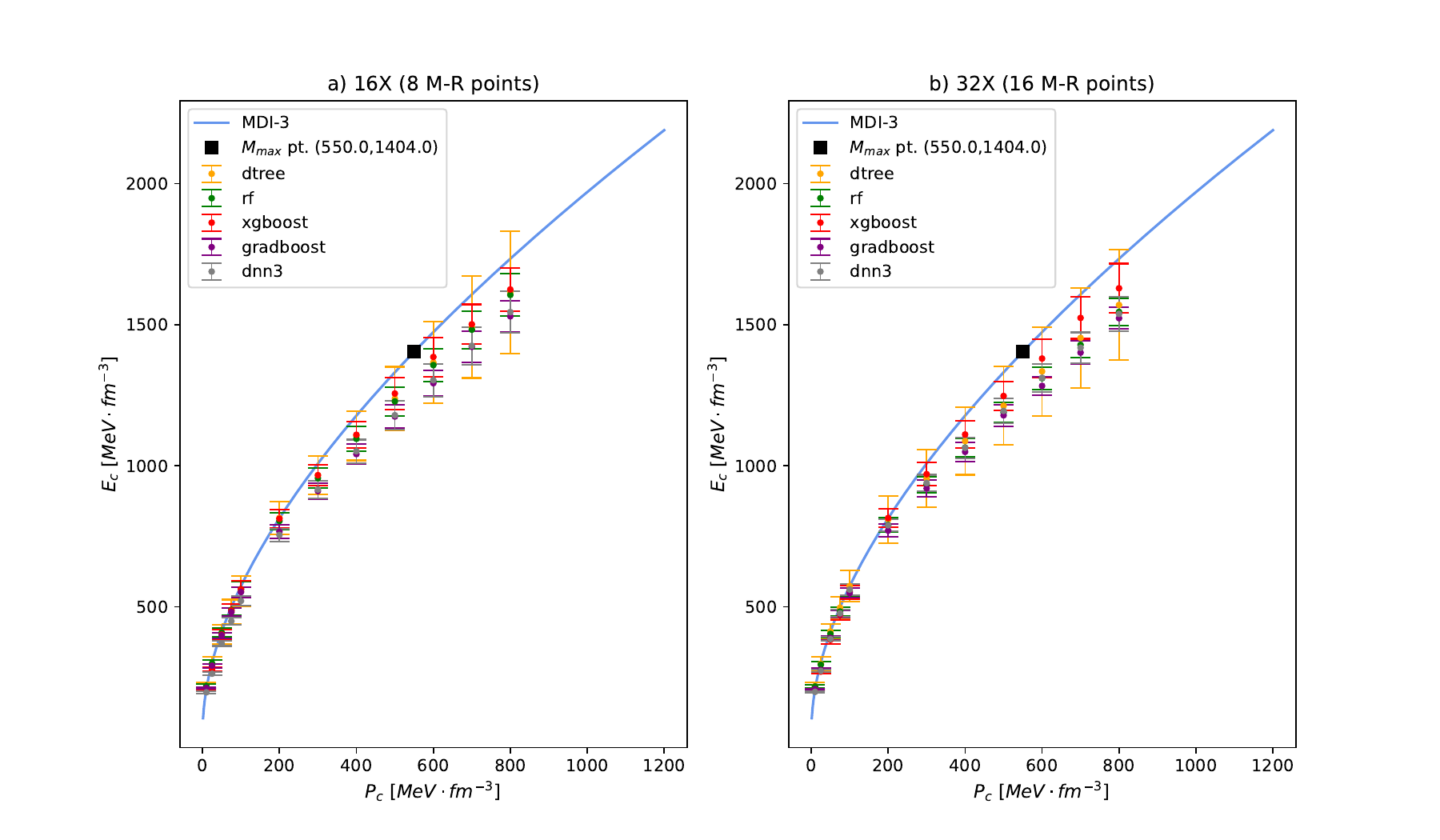}
    \caption{Reconstructing the \textit{MDI-3} EoS}
    \label{fig:MDI-3_EOS_predict}
\end{figure}

\begin{figure}[h]
    \centering
    \includegraphics[width=\linewidth]{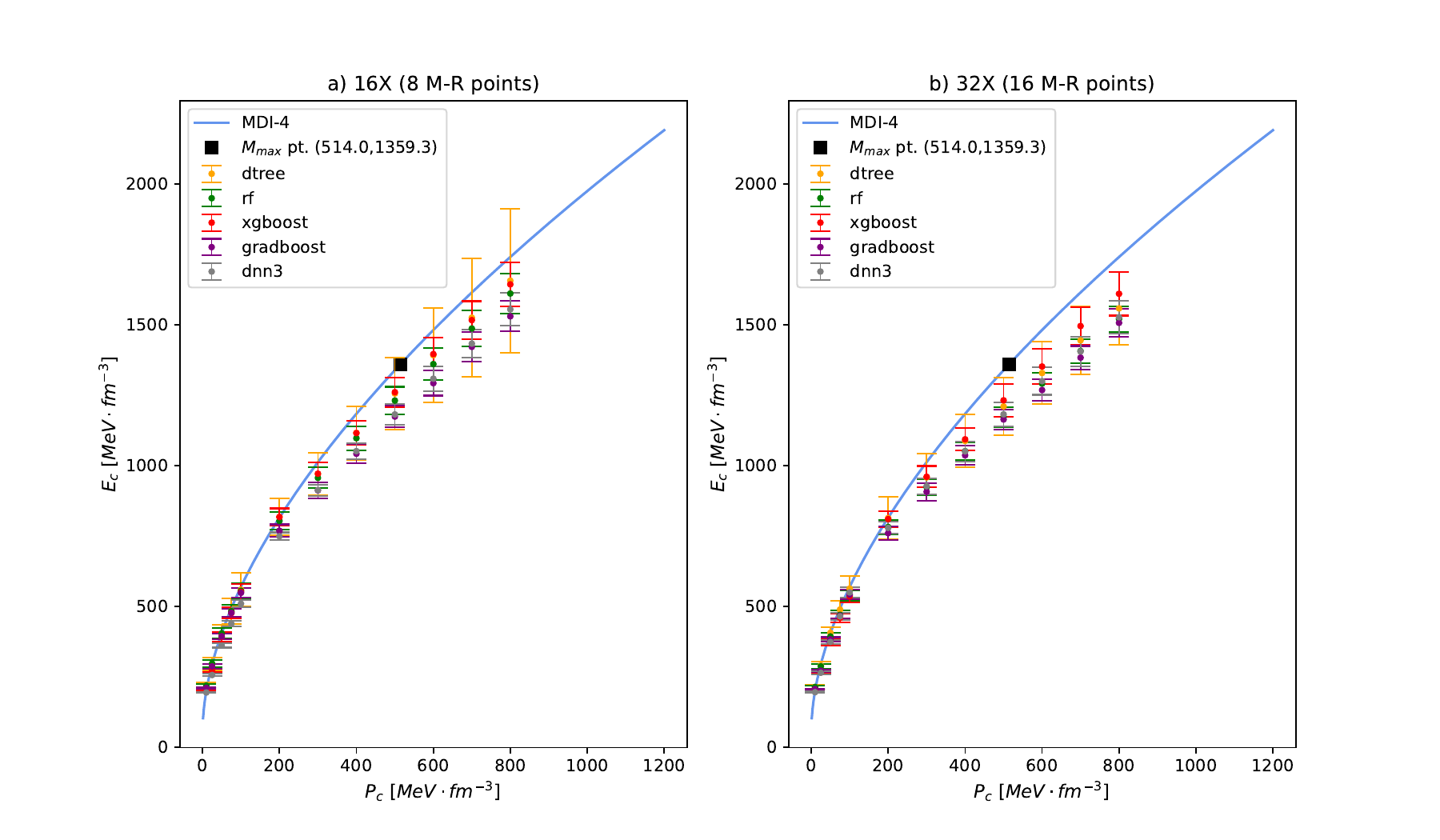}
    \caption{Reconstructing the \textit{MDI-4} EoS}
    \label{fig:MDI-4_EOS_predict}
\end{figure}

\begin{figure}[h!]
    \centering
    \includegraphics[width=\linewidth]{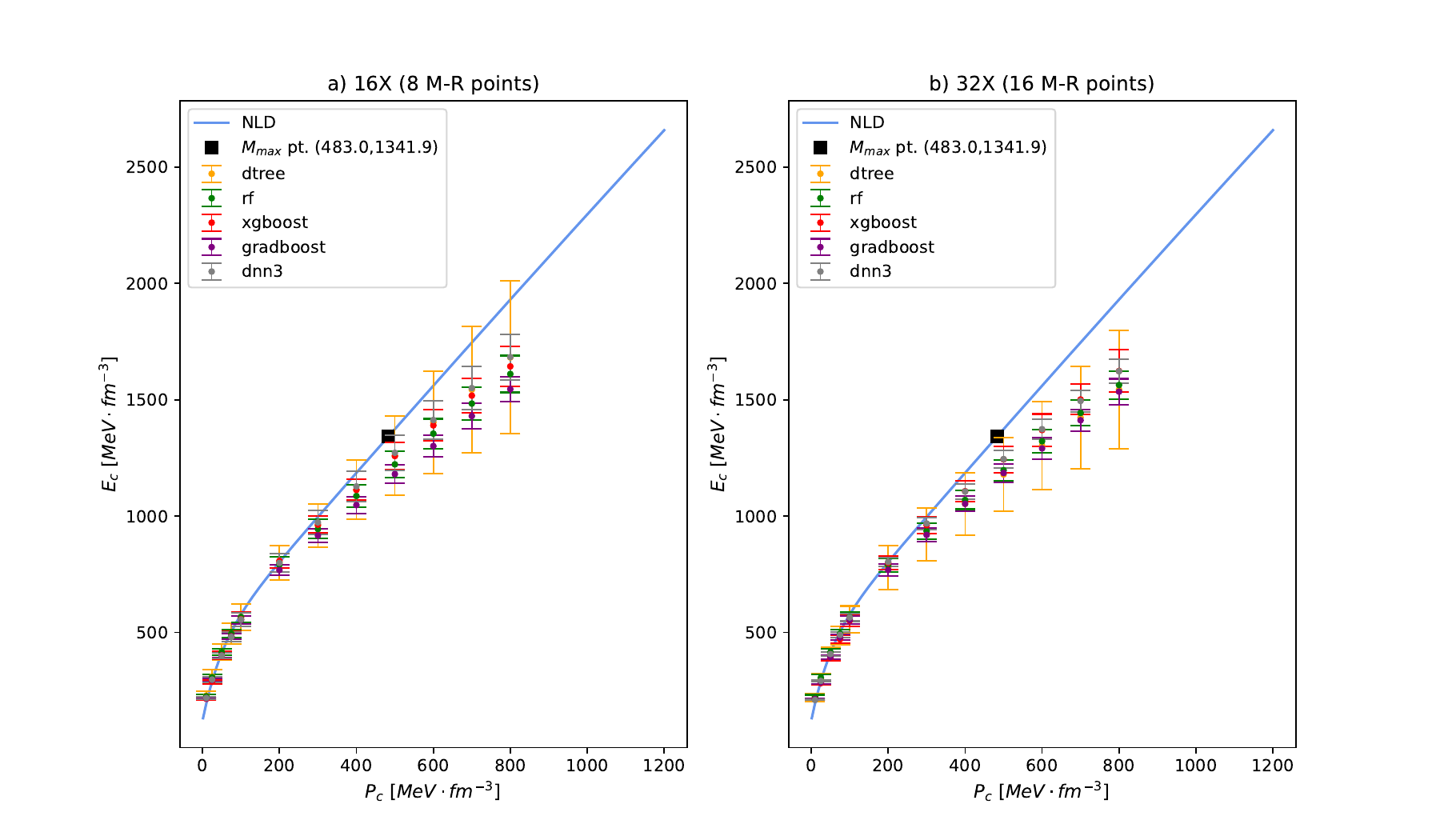}
    \caption{Reconstructing the \textit{NLD} EoS}
    \label{fig:NLD_EOS_predict}
\end{figure}

\begin{figure}[h]
    \centering
    \includegraphics[width=\linewidth]{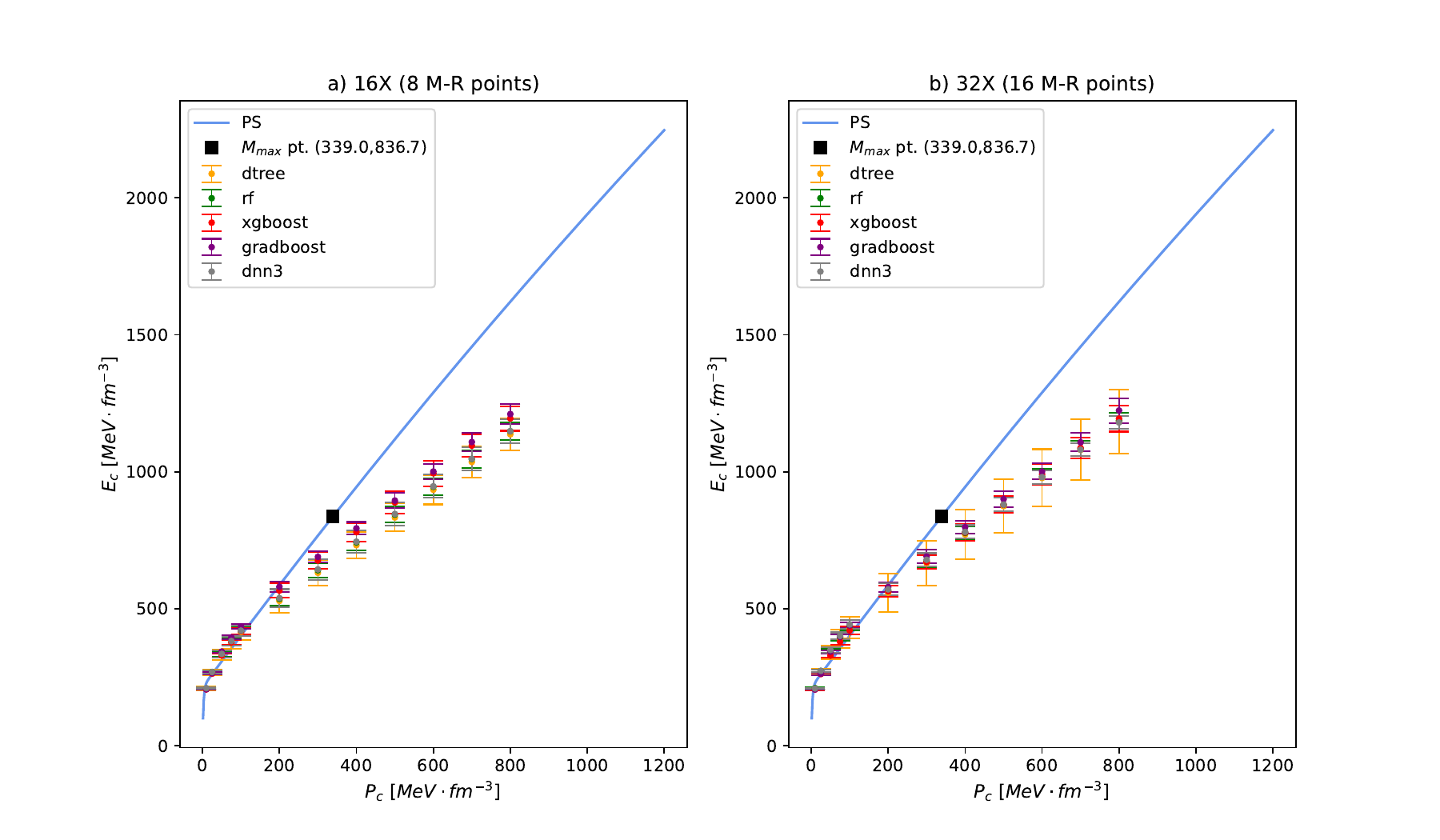}
    \caption{Reconstructing the \textit{PS} EoS}
    \label{fig:PS_EOS_predict}
\end{figure}

\begin{figure}[h!]
    \centering
    \includegraphics[width=\linewidth]{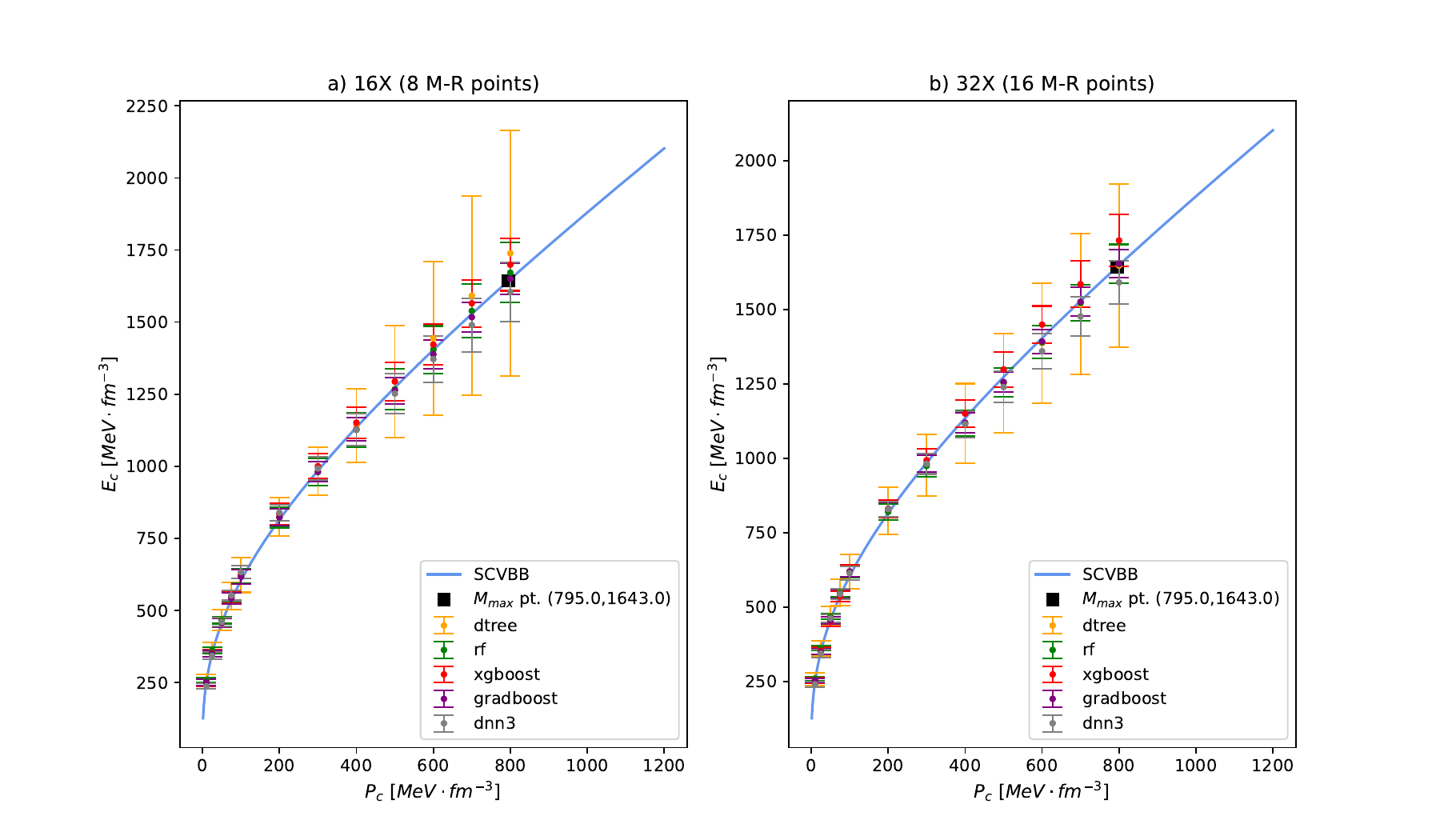}
    \caption{Reconstructing the \textit{SCVBB} EoS}
    \label{fig:SCVBB_EOS_predict}
\end{figure}

\begin{figure}[h]
    \centering
    \includegraphics[width=\linewidth]{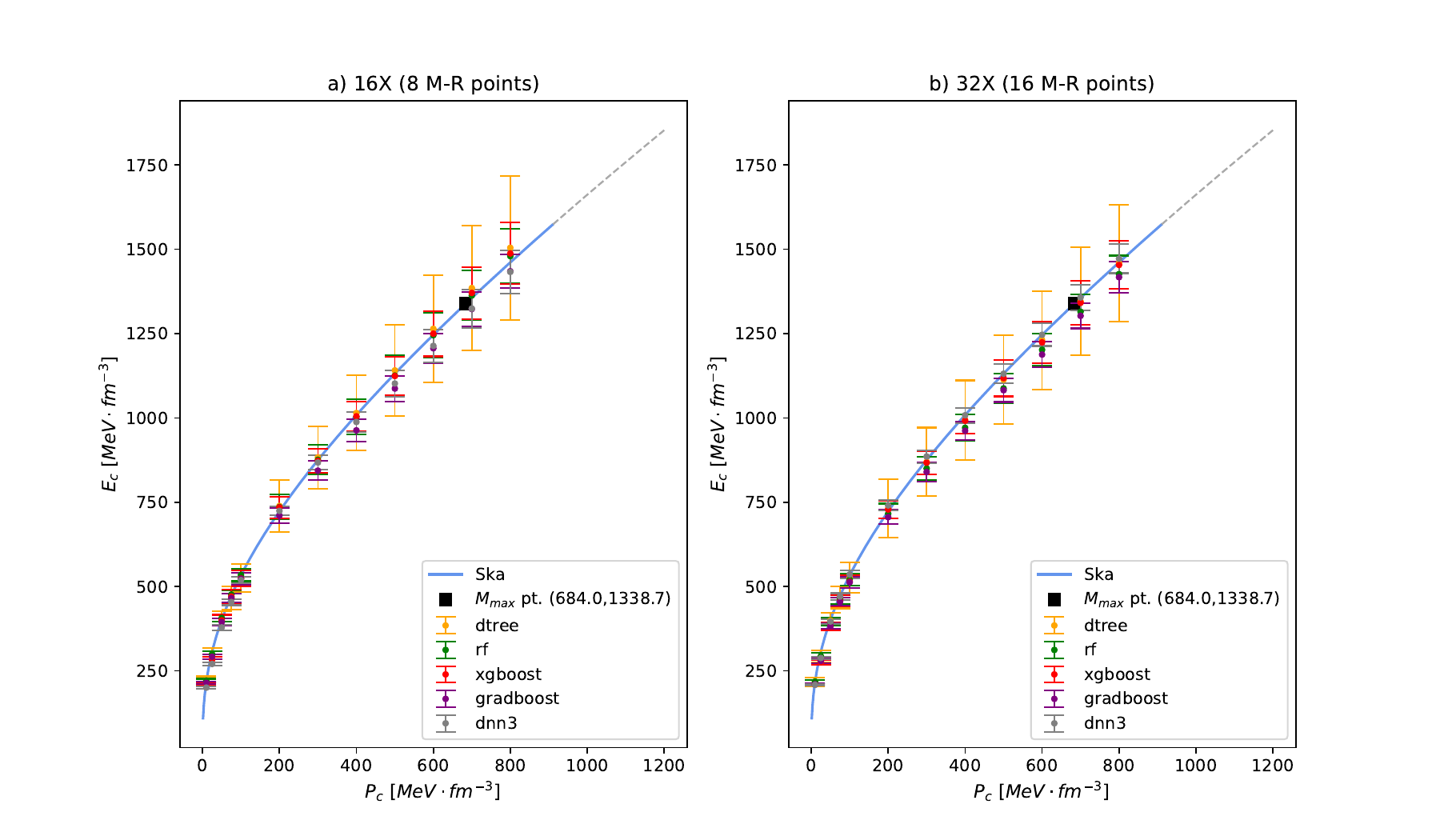}
    \caption{Reconstructing the \textit{Ska} EoS}
    \label{fig:Ska_EOS_predict}
\end{figure}

\begin{figure}[h!]
    \centering
    \includegraphics[width=\linewidth]{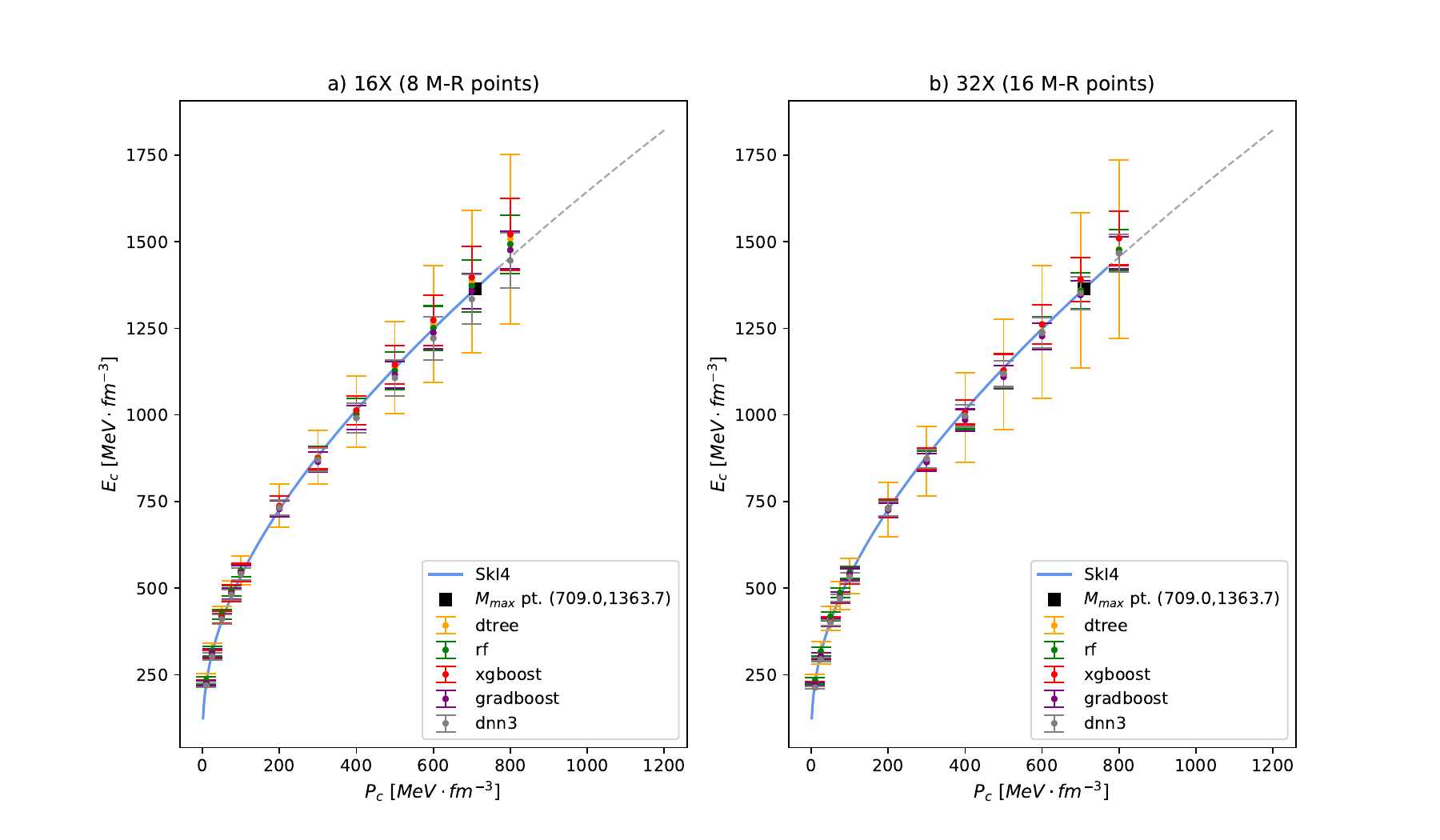}
    \caption{Reconstructing the \textit{SkI4} EoS}
    \label{fig:SkI4_EOS_predict}
\end{figure}

\begin{figure}[h]
    \centering
    \includegraphics[width=\linewidth]{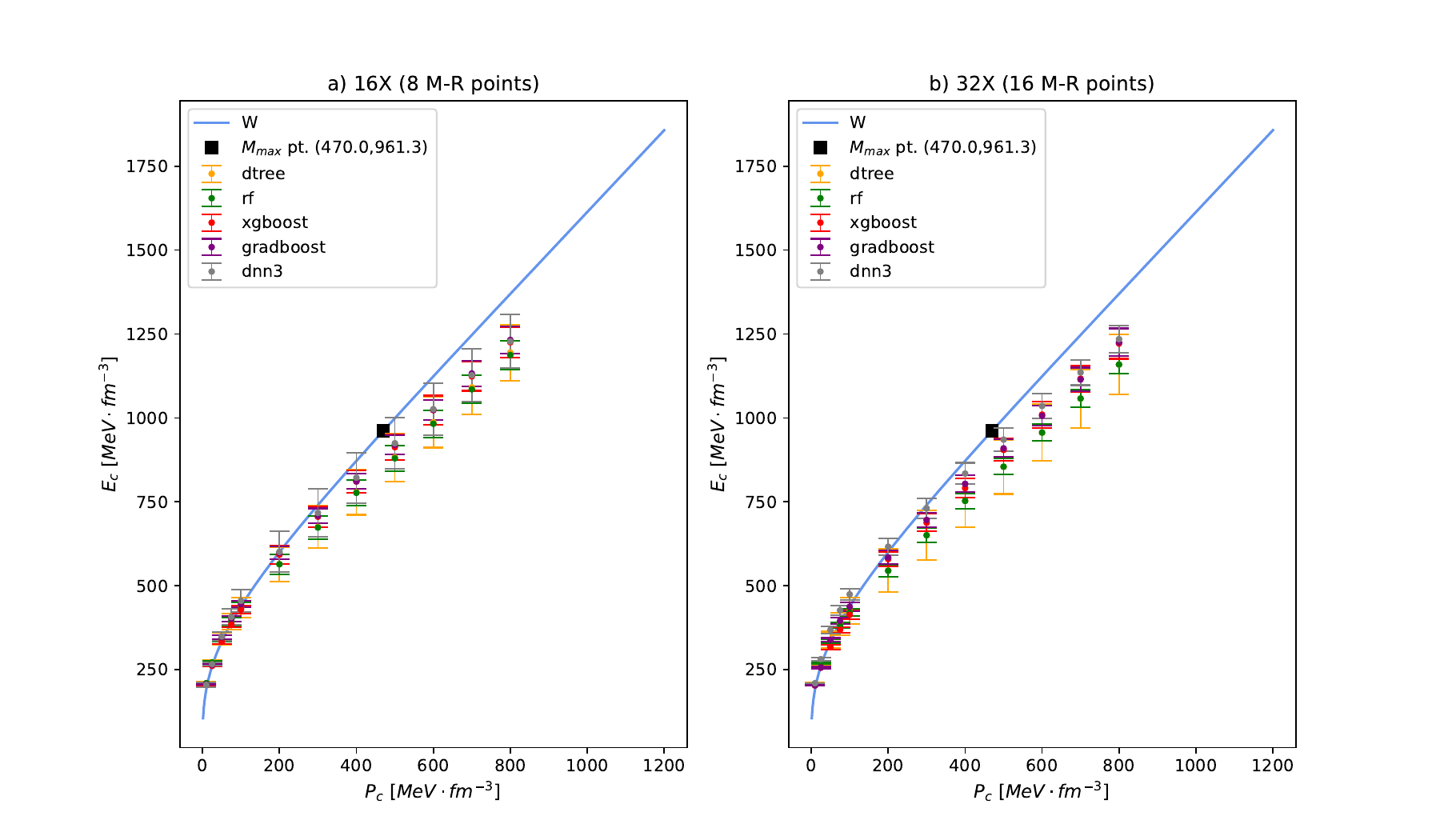}
    \caption{Reconstructing the \textit{W} EoS}
    \label{fig:W_EOS_predict}
\end{figure}

\begin{figure}[h!]
    \centering
    \includegraphics[width=\linewidth]{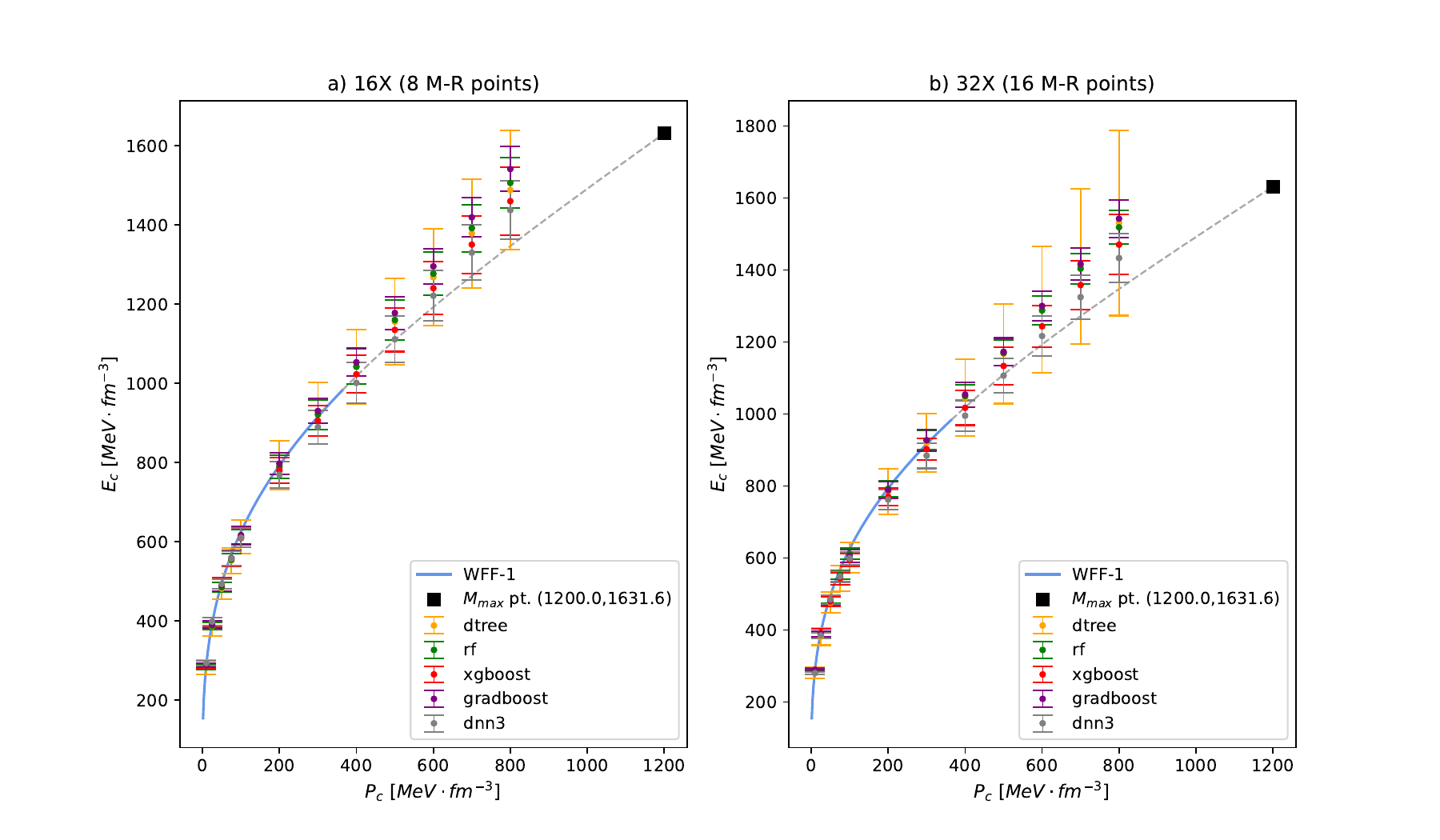}
    \caption{Reconstructing the \textit{WFF-1} EoS}
    \label{fig:WFF-1_EOS_predict}
\end{figure}
\FloatBarrier

\begin{figure}[h]
    \centering
    \includegraphics[width=0.97\linewidth]{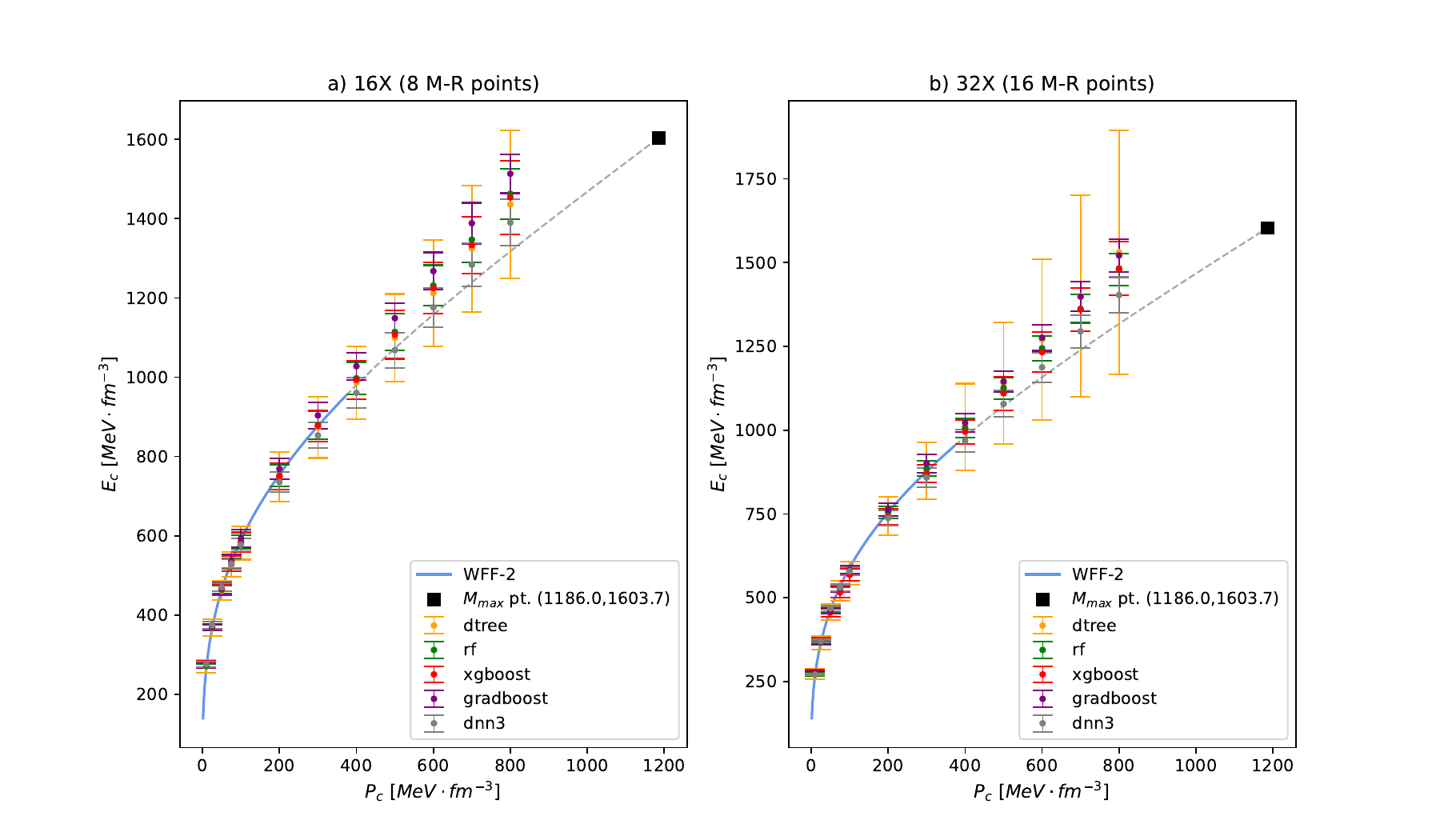}
    \caption{Reconstructing the \textit{WFF-2} EoS}
    \label{fig:WFF-2_EOS_predict}
\end{figure}

\section{Reconstructing Quark Stars' EoSs}\label{recontstruct_QS}
Below, we present the reconstruction of 20 equations of state for quark stars (10 MIT bag, 10 CFL):

\begin{figure}[h!]
    \centering
    \includegraphics[width=0.97\linewidth]{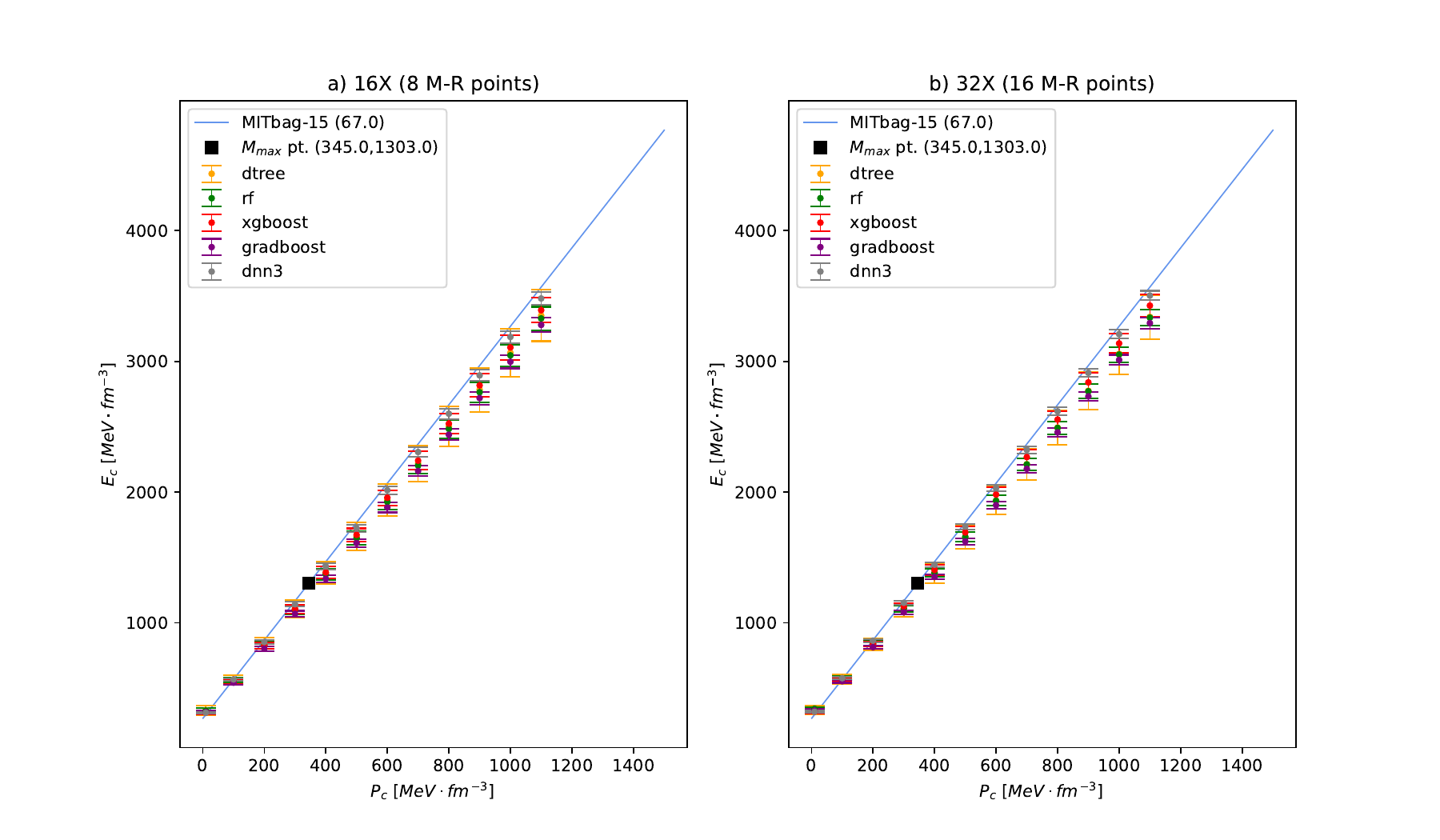}
    \caption{Reconstructing the \textit{MITbag-15} EoS ($B=67$ $MeV\cdot fm^{-3}$)}
    \label{fig:MITbag-15_EOS_predict}
\end{figure}

\begin{figure}[h]
    \centering
    \includegraphics[width=\linewidth]{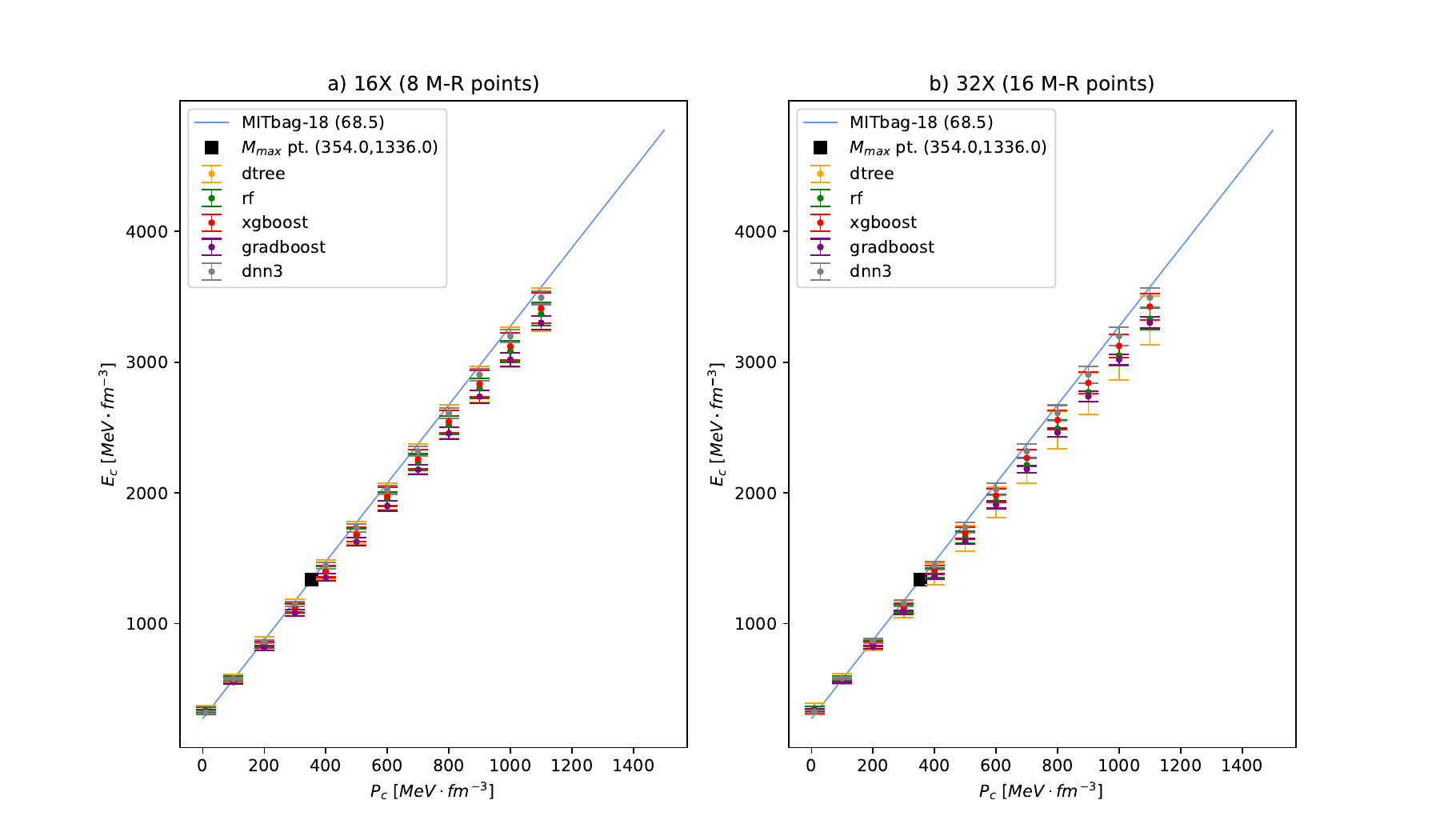}
    \caption{Reconstructing the \textit{MITbag-18} EoS ($B=68.5$ $MeV\cdot fm^{-3}$)}
    \label{fig:MITbag-18_EOS_predict}
\end{figure}

\begin{figure}[h!]
    \centering
    \includegraphics[width=\linewidth]{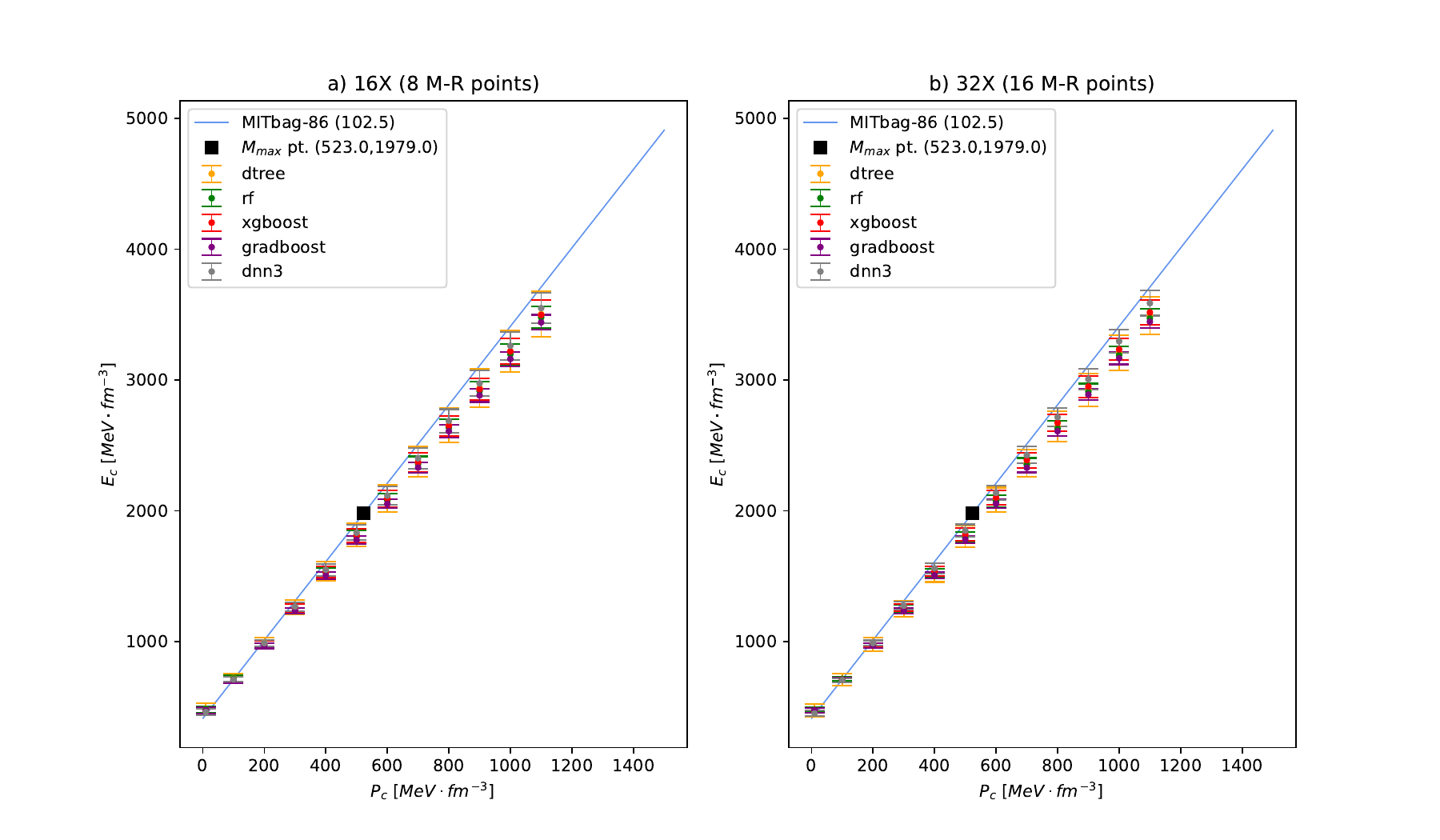}
    \caption{Reconstructing the \textit{MITbag-86} EoS ($B=102.5$ $MeV\cdot fm^{-3}$)}
    \label{fig:MITbag-86_EOS_predict}
\end{figure}

\begin{figure}[h]
    \centering
    \includegraphics[width=\linewidth]{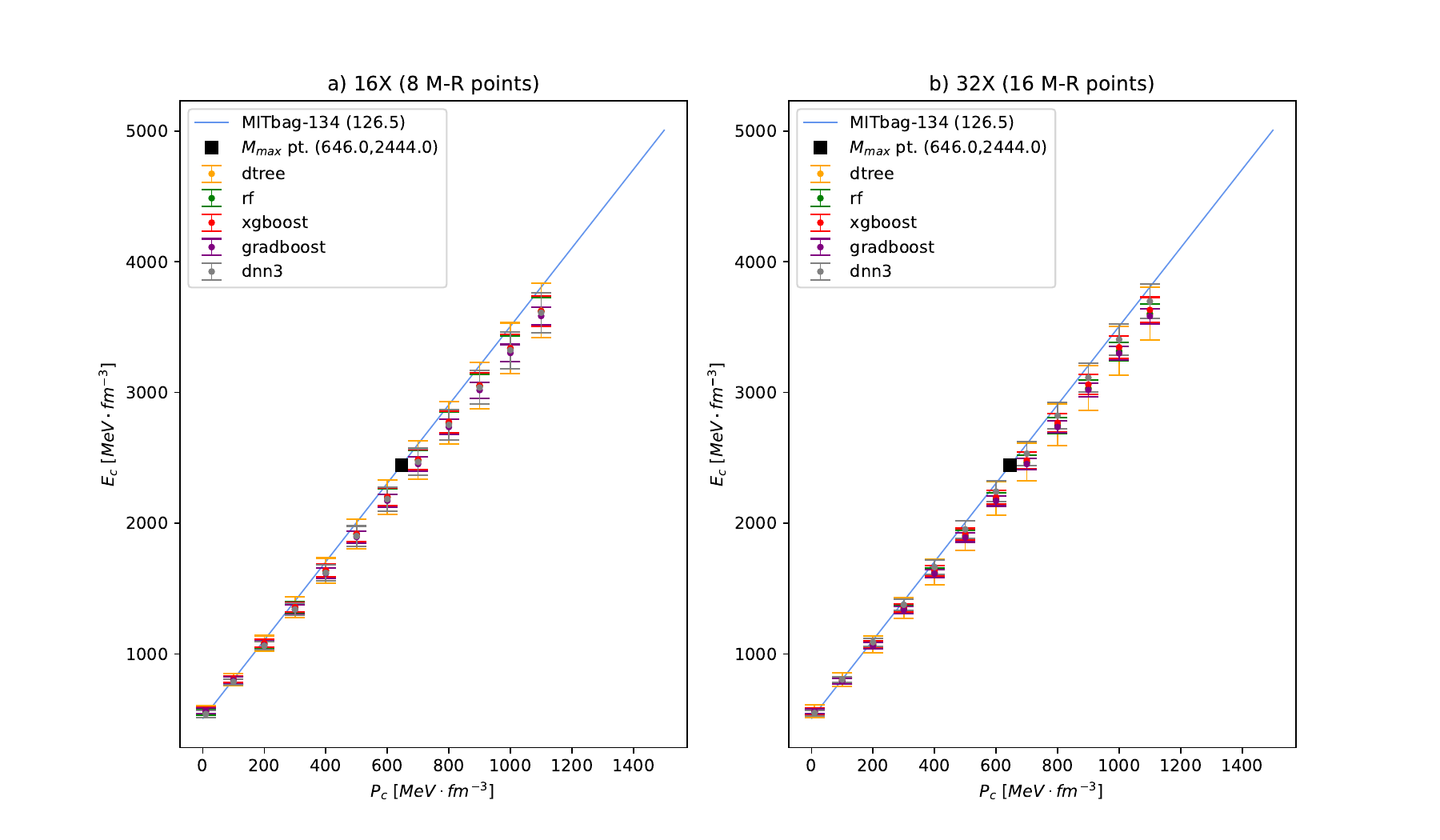}
    \caption{Reconstructing the \textit{MITbag-134} EoS ($B=126.5$ $MeV\cdot fm^{-3}$)}
    \label{fig:MITbag-134_EOS_predict}
\end{figure}

\begin{figure}[h!]
    \centering
    \includegraphics[width=\linewidth]{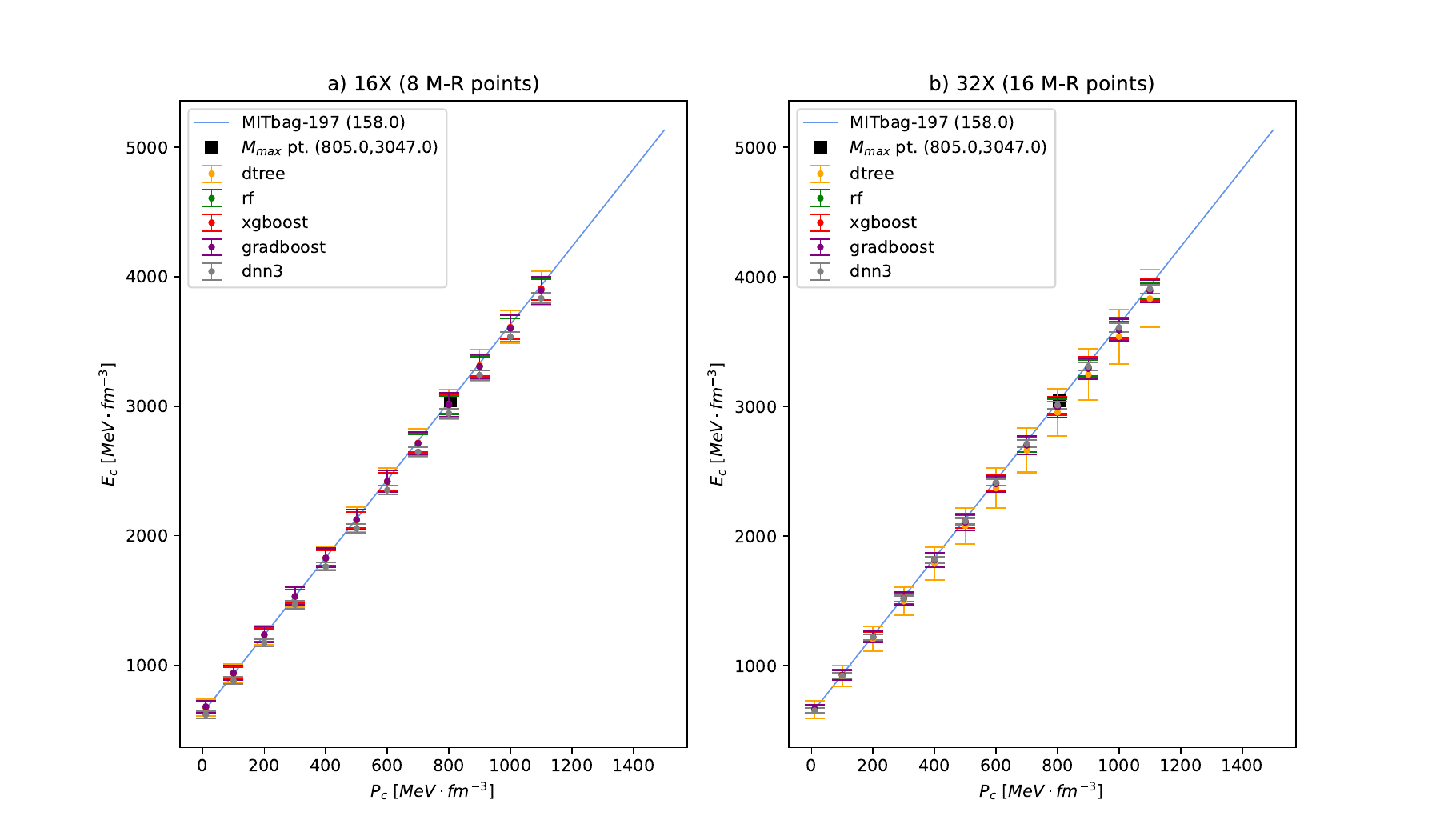}
    \caption{Reconstructing the \textit{MITbag-197} EoS ($B=158$ $MeV\cdot fm^{-3}$)}
    \label{fig:MITbag-197_EOS_predict}
\end{figure}

\begin{figure}[h]
    \centering
    \includegraphics[width=\linewidth]{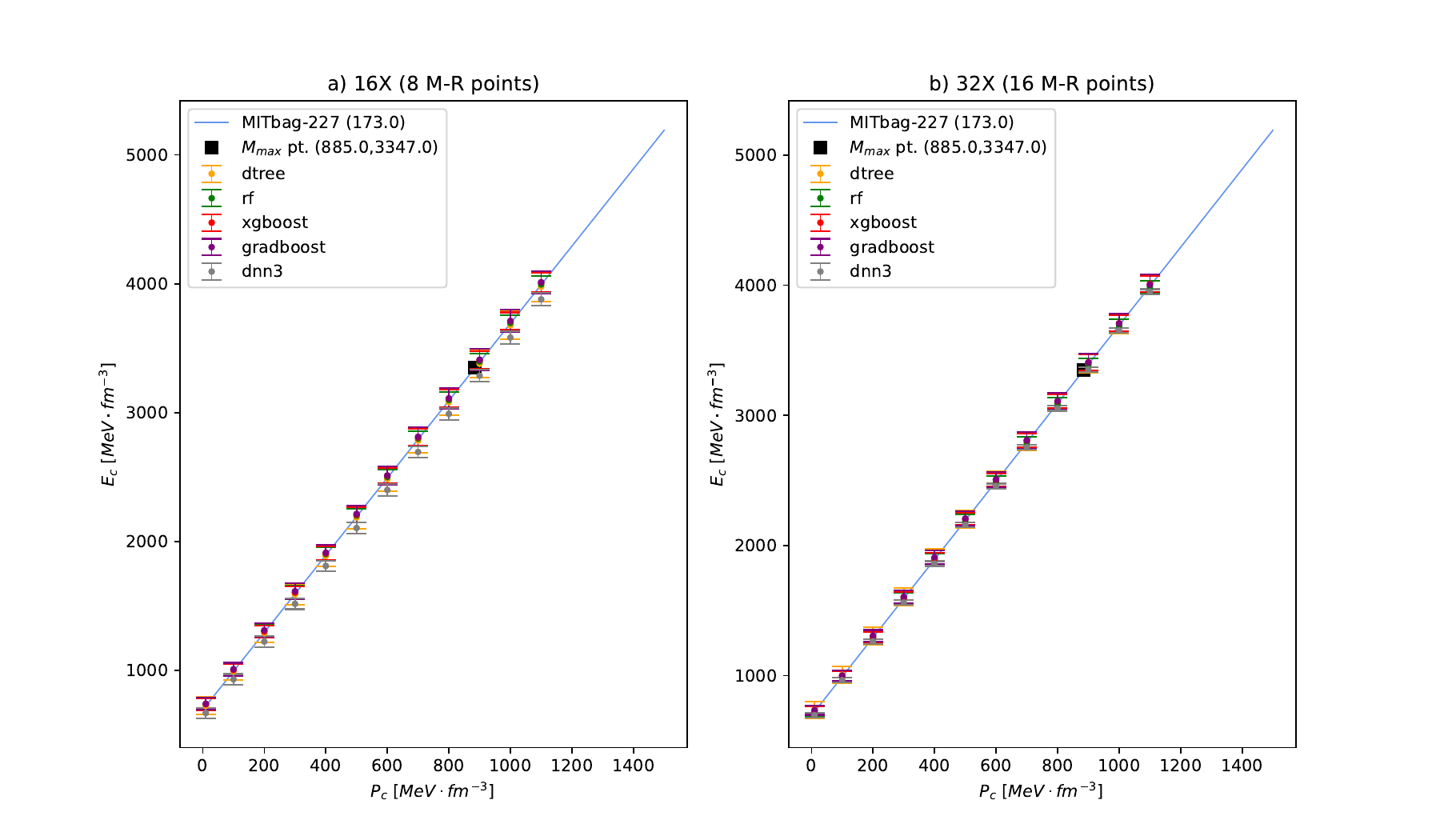}
    \caption{Reconstructing the \textit{MITbag-227} EoS ($B=173$ $MeV\cdot fm^{-3}$)}
    \label{fig:MITbag-227_EOS_predict}
\end{figure}

\begin{figure}[h!]
    \centering
    \includegraphics[width=\linewidth]{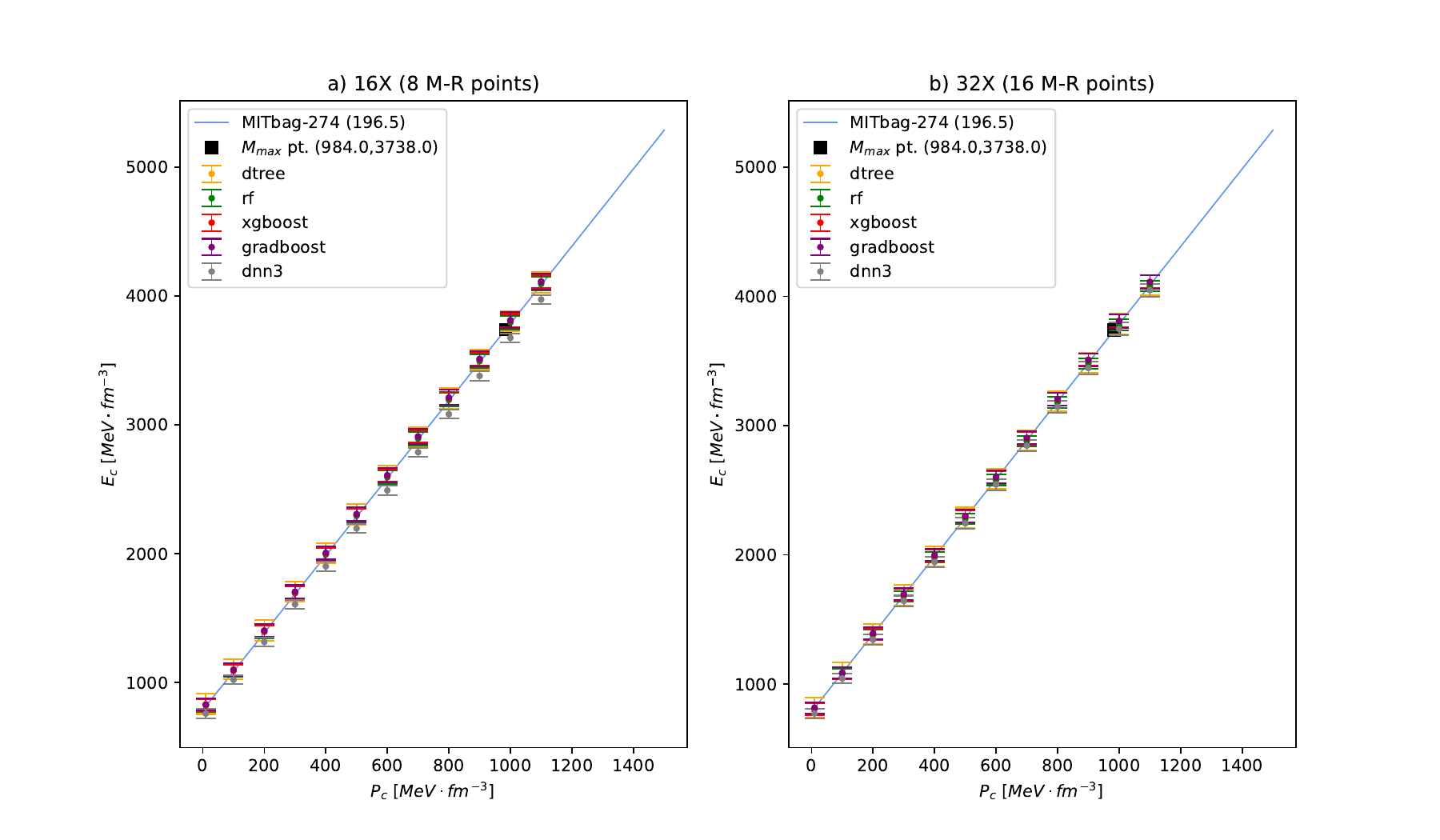}
    \caption{Reconstructing the \textit{MITbag-274} EoS ($B=196.5$ $MeV\cdot fm^{-3}$)}
    \label{fig:MITbag-274_EOS_predict}
\end{figure}

\begin{figure}[h]
    \centering
    \includegraphics[width=\linewidth]{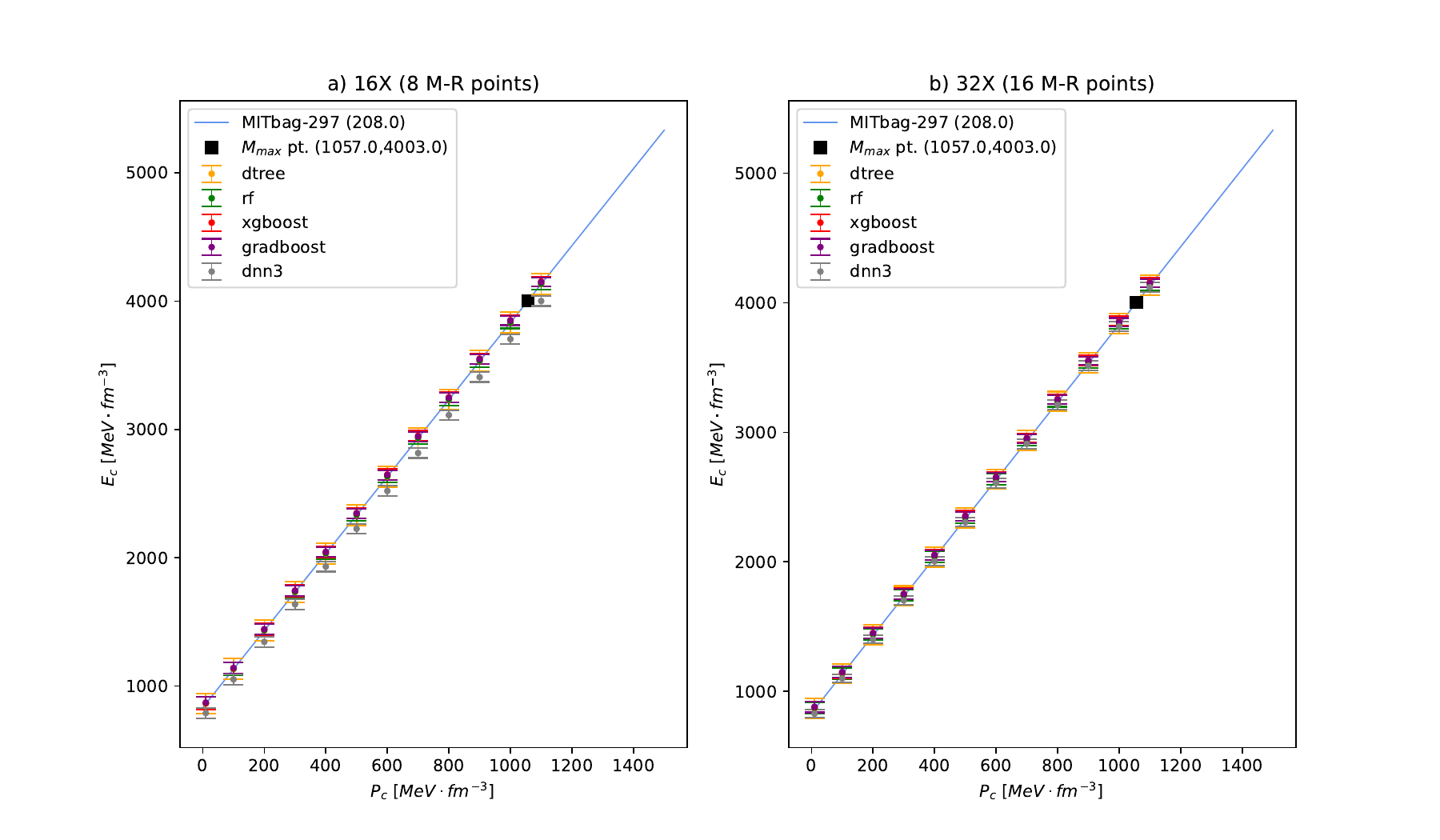}
    \caption{Reconstructing the \textit{MITbag-297} EoS ($B=208$ $MeV\cdot fm^{-3}$)}
    \label{fig:MITbag-297_EOS_predict}
\end{figure}

\begin{figure}[h!]
    \centering
    \includegraphics[width=\linewidth]{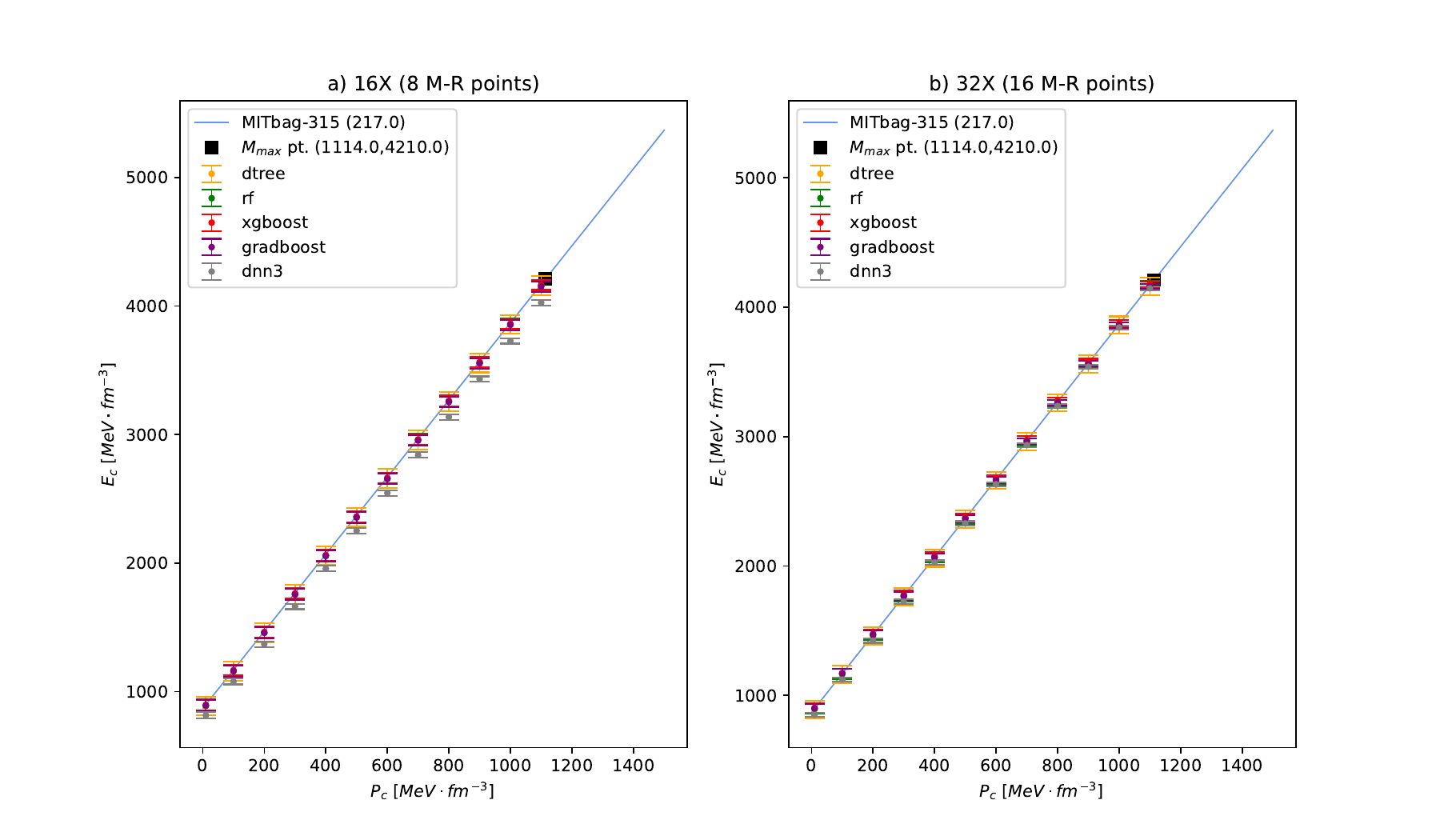}
    \caption{Reconstructing the \textit{MITbag-315} EoS ($B=217$ $MeV\cdot fm^{-3}$)}
    \label{fig:MITbag-315_EOS_predict}
\end{figure}

\begin{figure}[h]
    \centering
    \includegraphics[width=\linewidth]{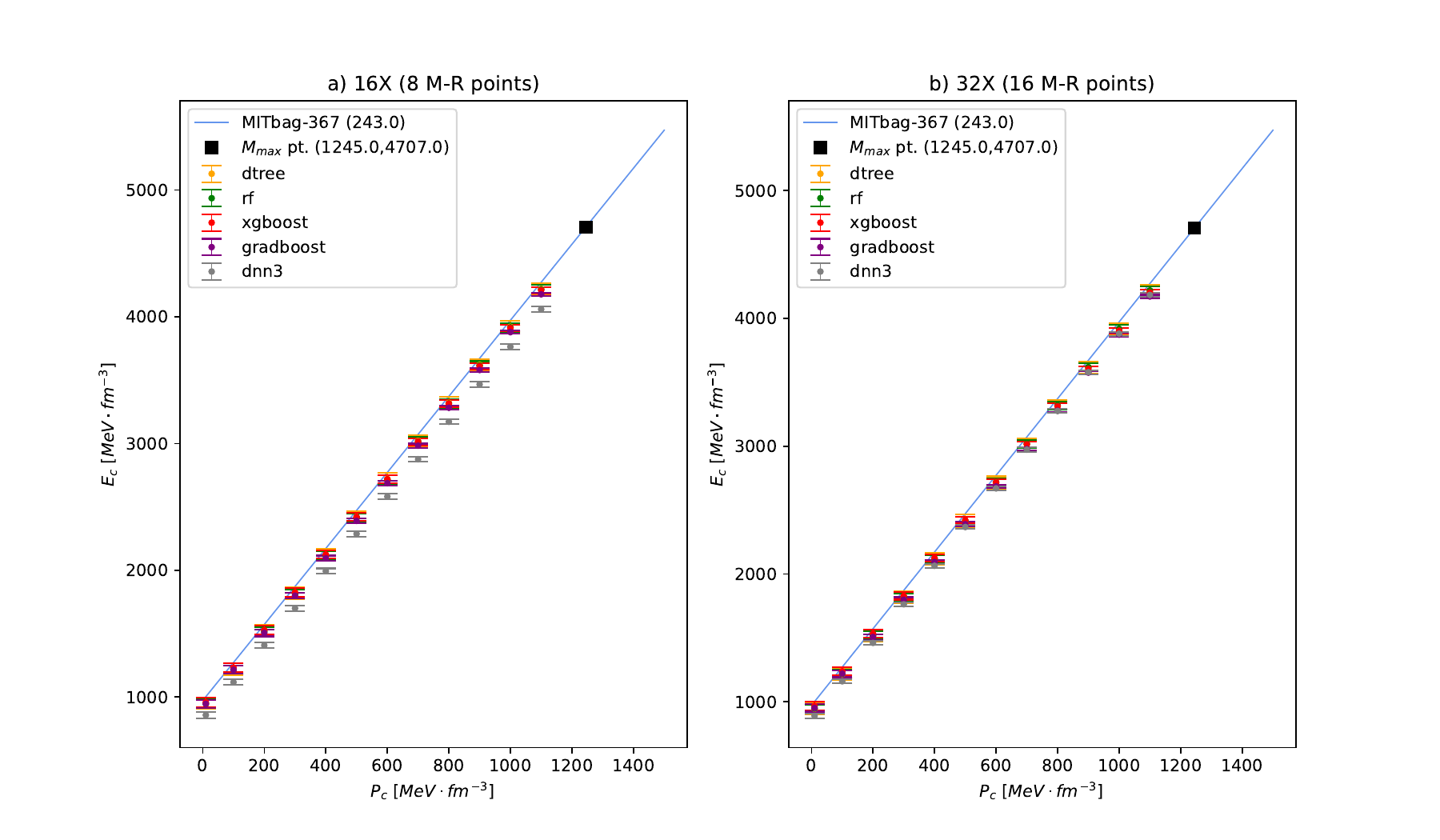}
    \caption{Reconstructing the \textit{MITbag-367} EoS ($B=243$ $MeV\cdot fm^{-3}$)}
    \label{fig:MITbag-367_EOS_predict}
\end{figure}

\begin{figure}[h!]
    \centering
    \includegraphics[width=\linewidth]{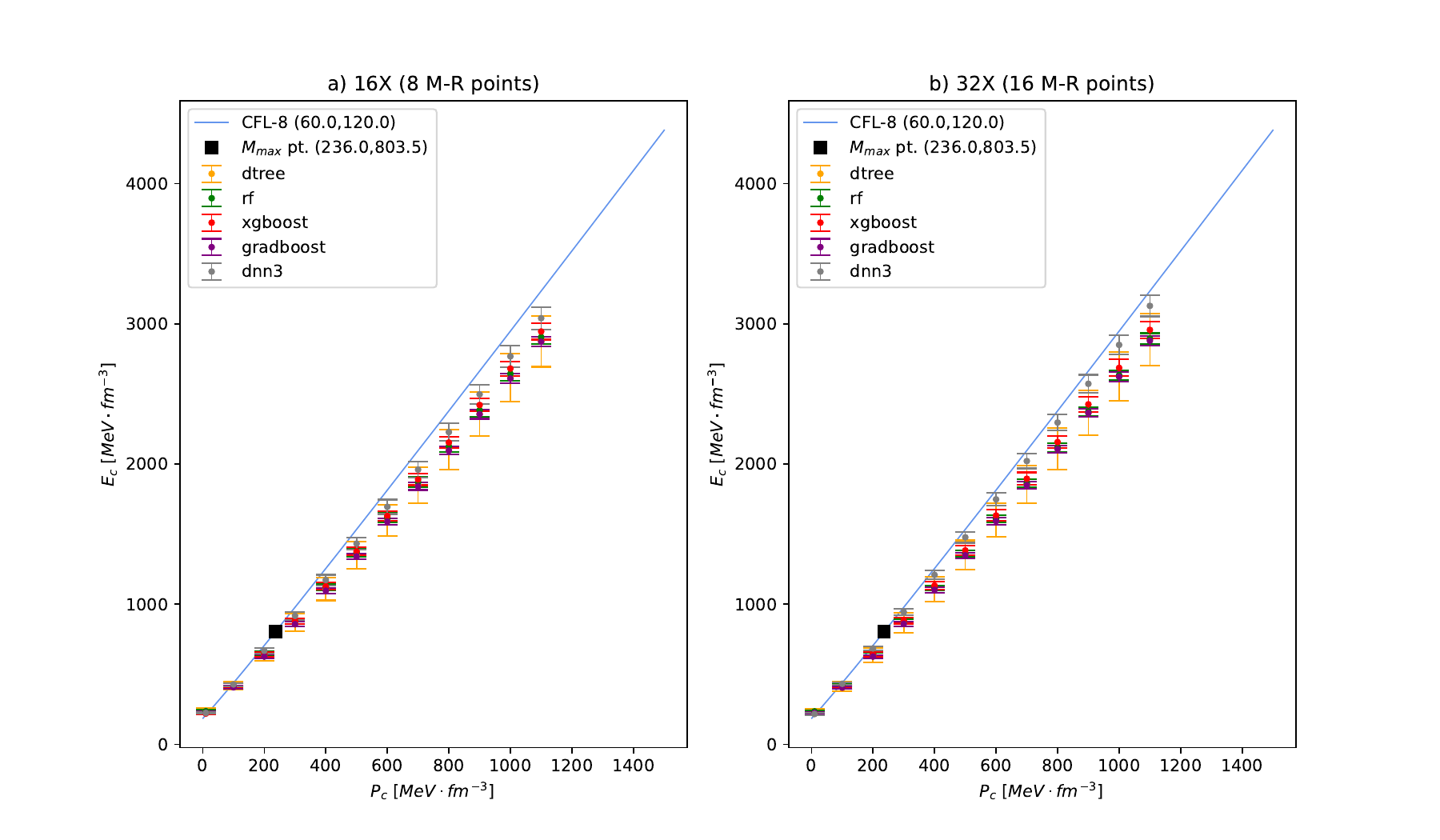}
    \caption{Reconstructing the \textit{CFL-8} EoS ($B=60$ $MeV\cdot fm^{-3}$, $\Delta=120$ $MeV$)}
    \label{fig:CFL-8_EOS_predict}
\end{figure}

\begin{figure}[h]
    \centering
    \includegraphics[width=\linewidth]{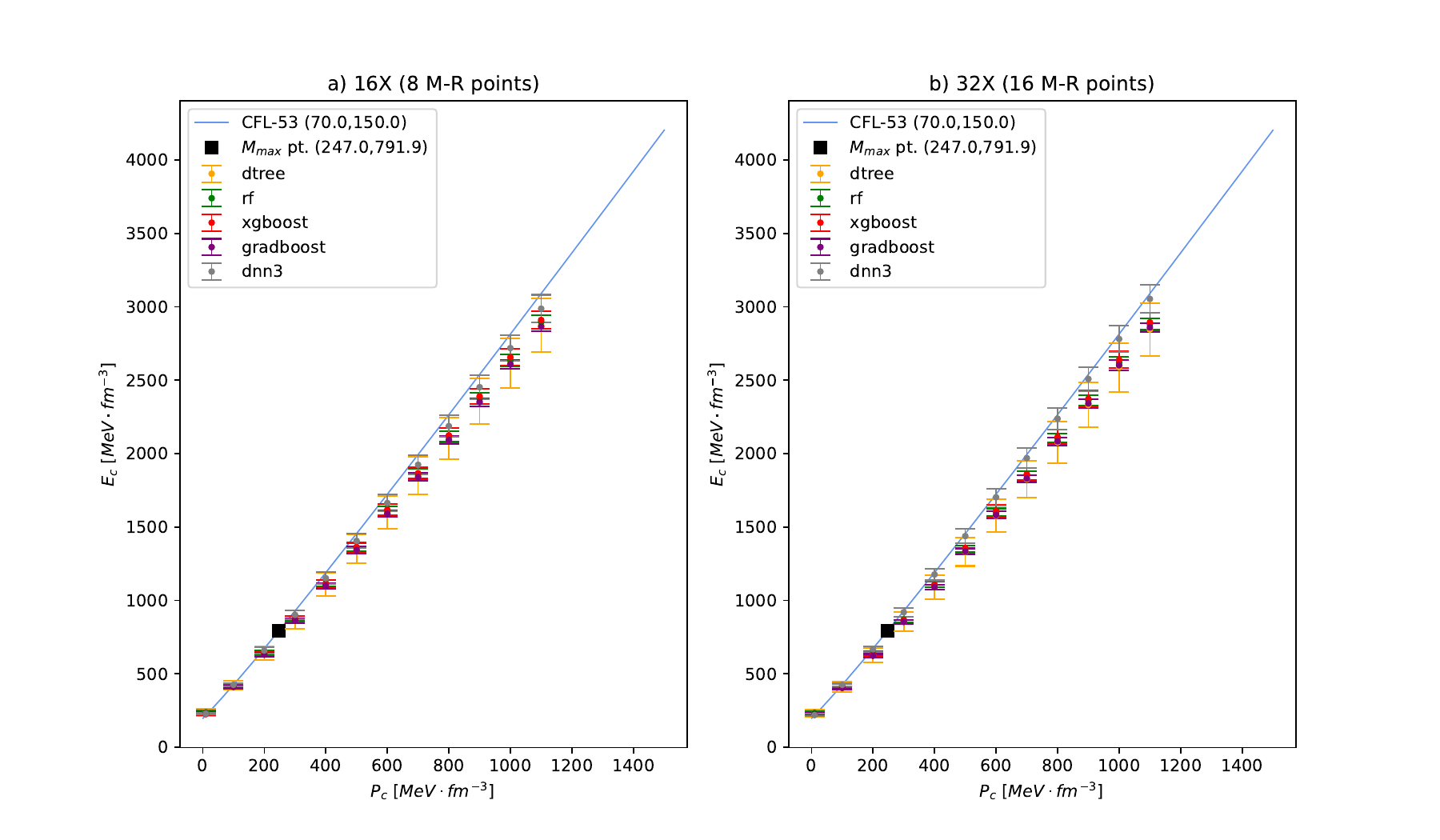}
    \caption{Reconstructing the \textit{CFL-53} EoS ($B=70$ $MeV\cdot fm^{-3}$, $\Delta=150$ $MeV$)}
    \label{fig:CFL-53_EOS_predict}
\end{figure}

\begin{figure}[h!]
    \centering
    \includegraphics[width=\linewidth]{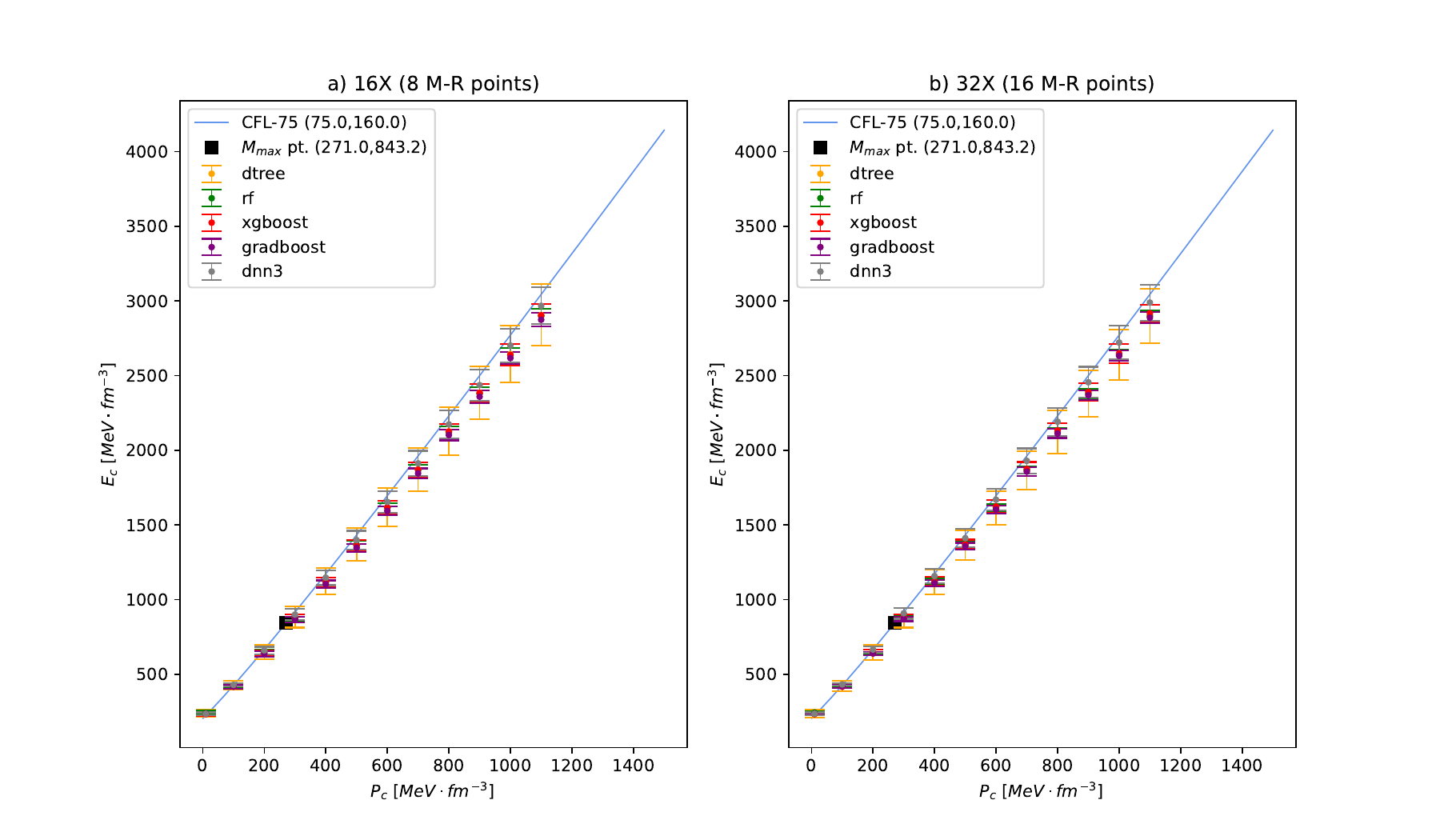}
    \caption{Reconstructing the \textit{CFL-75} EoS ($B=75$ $MeV\cdot fm^{-3}$, $\Delta=160$ $MeV$)}
    \label{fig:CFL-75_EOS_predict}
\end{figure}

\begin{figure}[h]
    \centering
    \includegraphics[width=\linewidth]{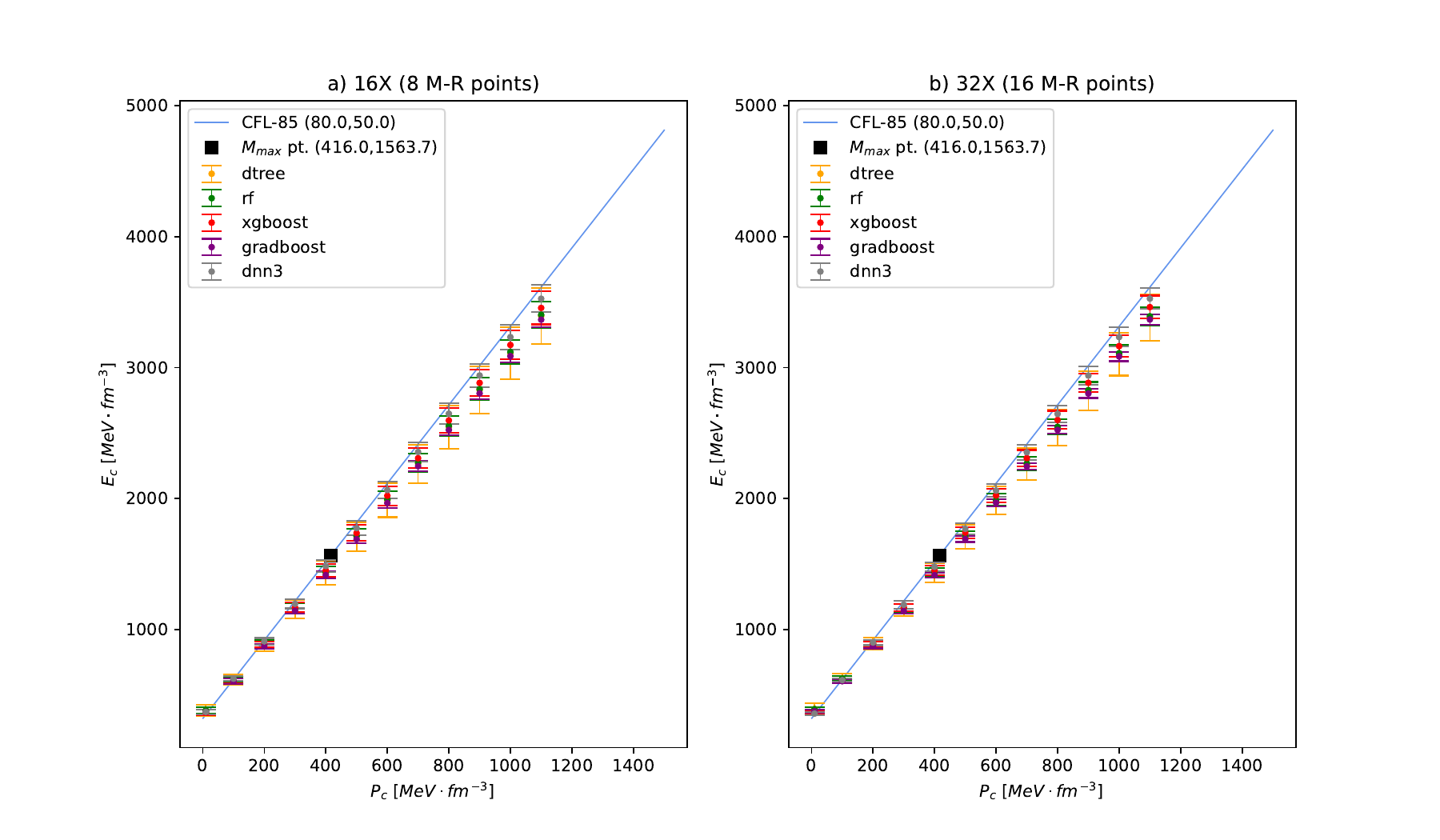}
    \caption{Reconstructing the \textit{CFL-85} EoS ($B=80$ $MeV\cdot fm^{-3}$, $\Delta=50$ $MeV$)}
    \label{fig:CFL-85_EOS_predict}
\end{figure}

\begin{figure}[h!]
    \centering
    \includegraphics[width=\linewidth]{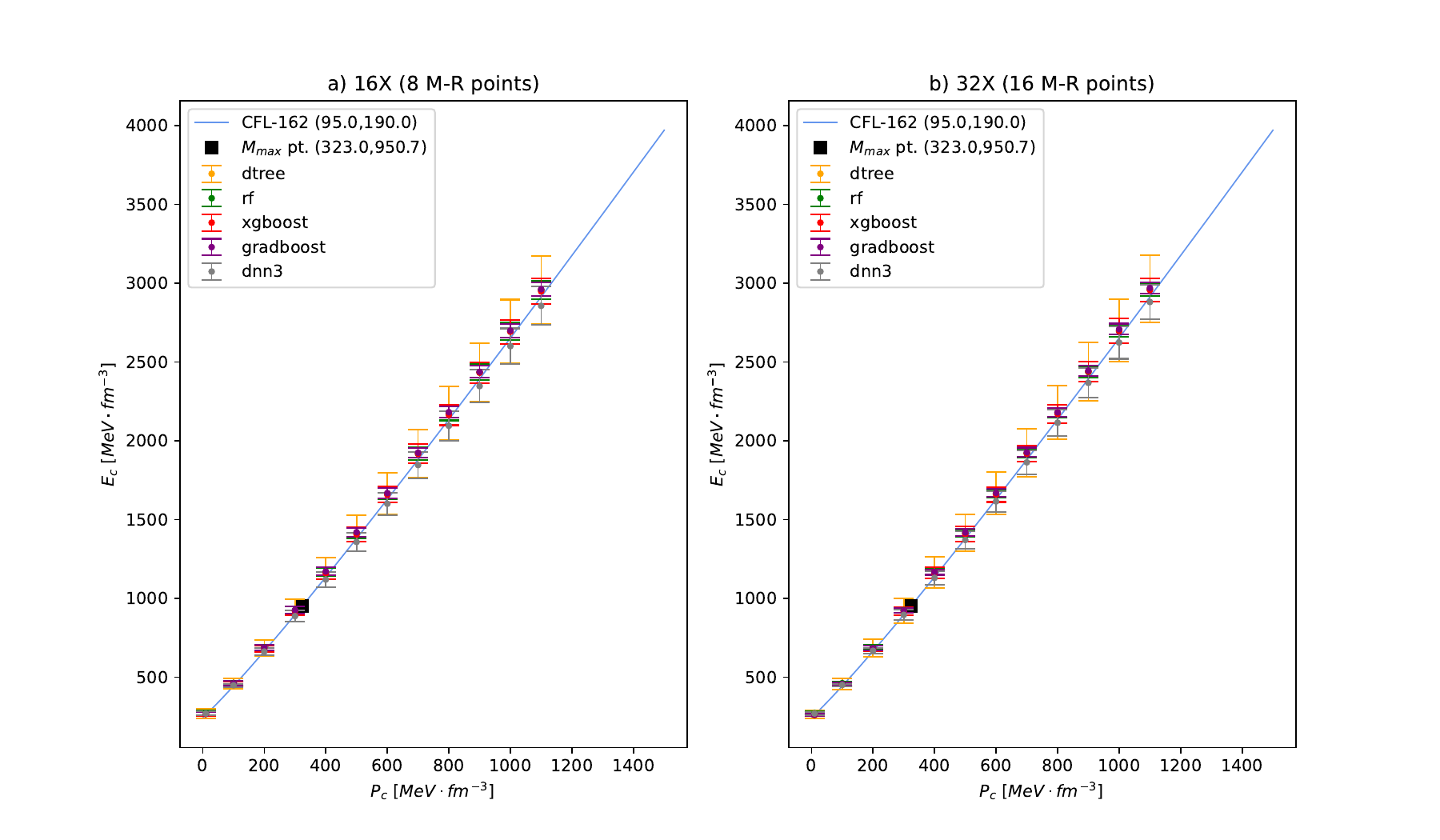}
    \caption{Reconstructing the \textit{CFL-162} EoS ($B=95$ $MeV\cdot fm^{-3}$, $\Delta=190$ $MeV$)}
    \label{fig:CFL-162_EOS_predict}
\end{figure}

\begin{figure}[h]
    \centering
    \includegraphics[width=\linewidth]{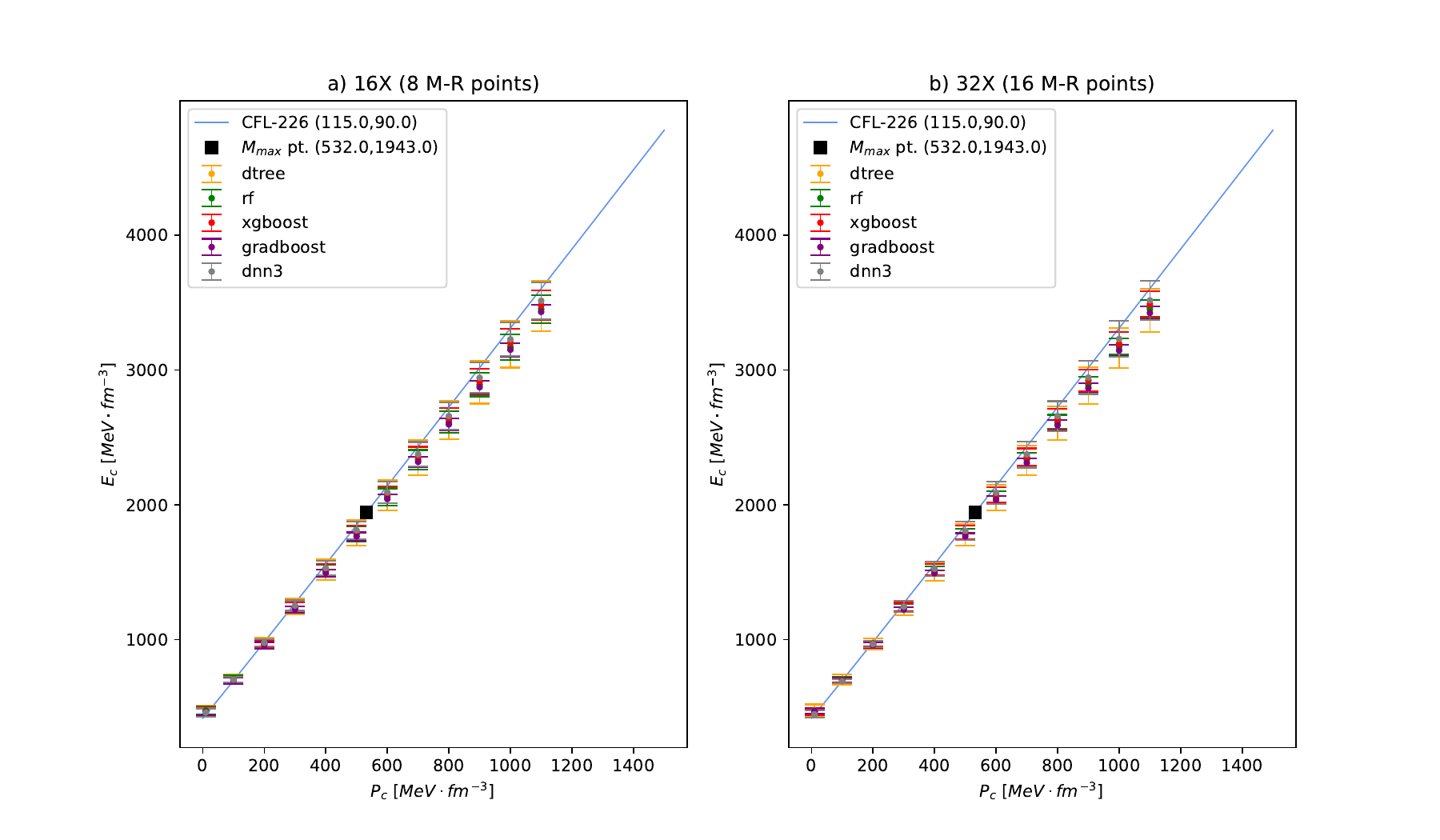}
    \caption{Reconstructing the \textit{CFL-226} EoS ($B=115$ $MeV\cdot fm^{-3}$, $\Delta=90$ $MeV$)}
    \label{fig:CFL-226_EOS_predict}
\end{figure}

\begin{figure}[h!]
    \centering
    \includegraphics[width=\linewidth]{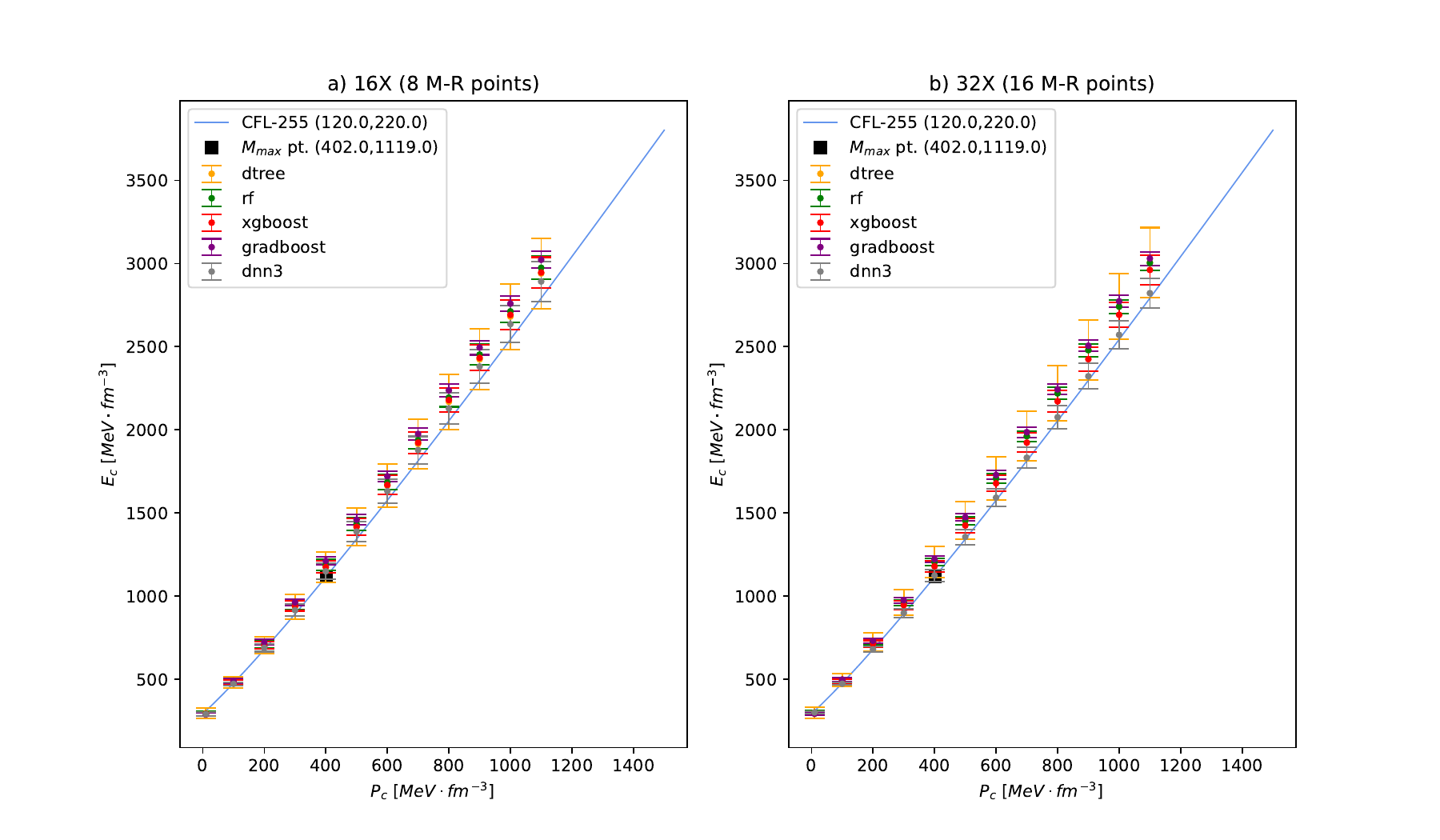}
    \caption{Reconstructing the \textit{CFL-255} EoS ($B=120$ $MeV\cdot fm^{-3}$, $\Delta=220$ $MeV$)}
    \label{fig:CFL-255_EOS_predict}
\end{figure}

\begin{figure}[h]
    \centering
    \includegraphics[width=\linewidth]{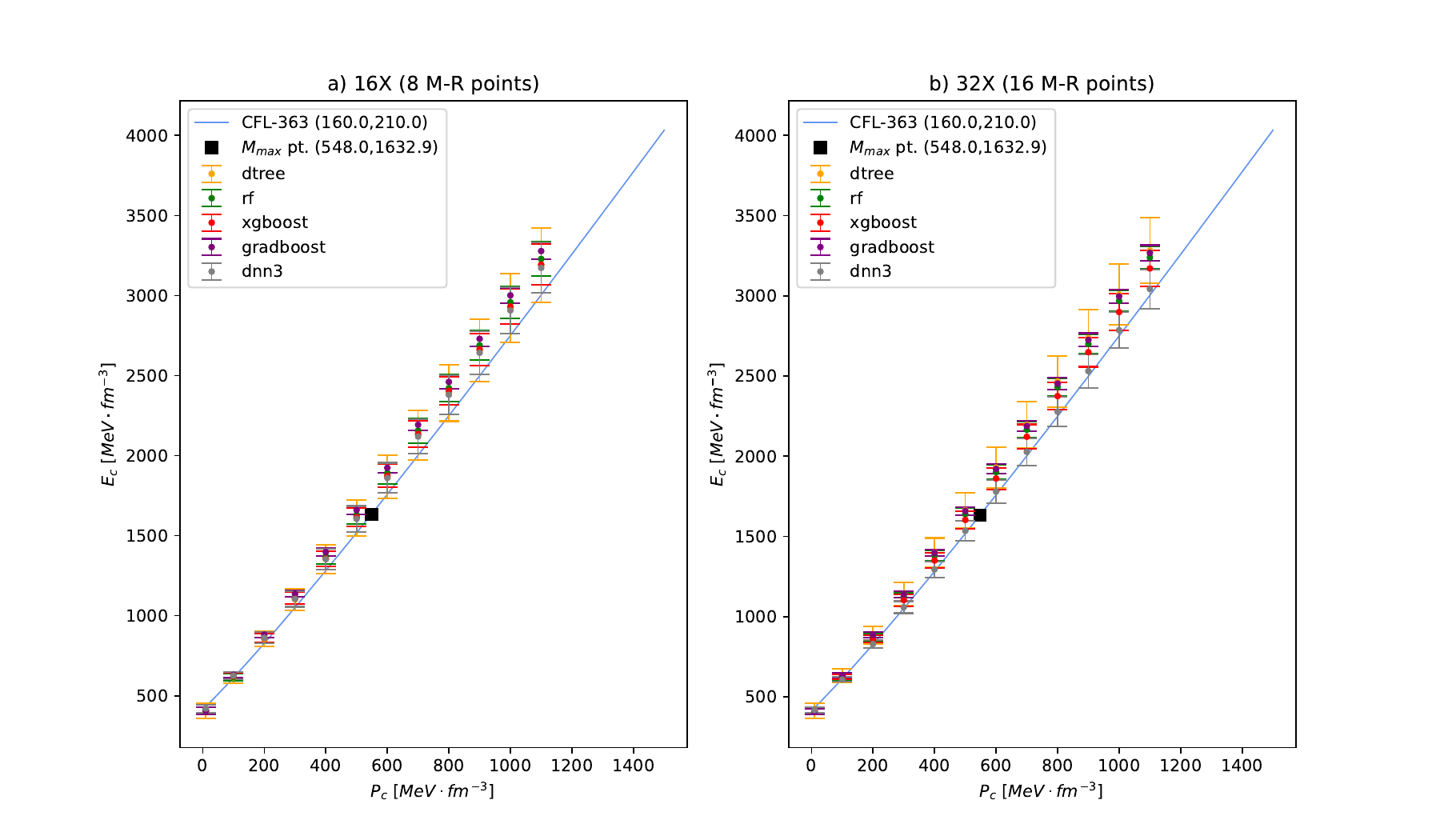}
    \caption{Reconstructing the \textit{CFL-363} EoS ($B=160$ $MeV\cdot fm^{-3}$, $\Delta=210$ $MeV$)}
    \label{fig:CFL-363_EOS_predict}
\end{figure}

\begin{figure}[h!]
    \centering
    \includegraphics[width=\linewidth]{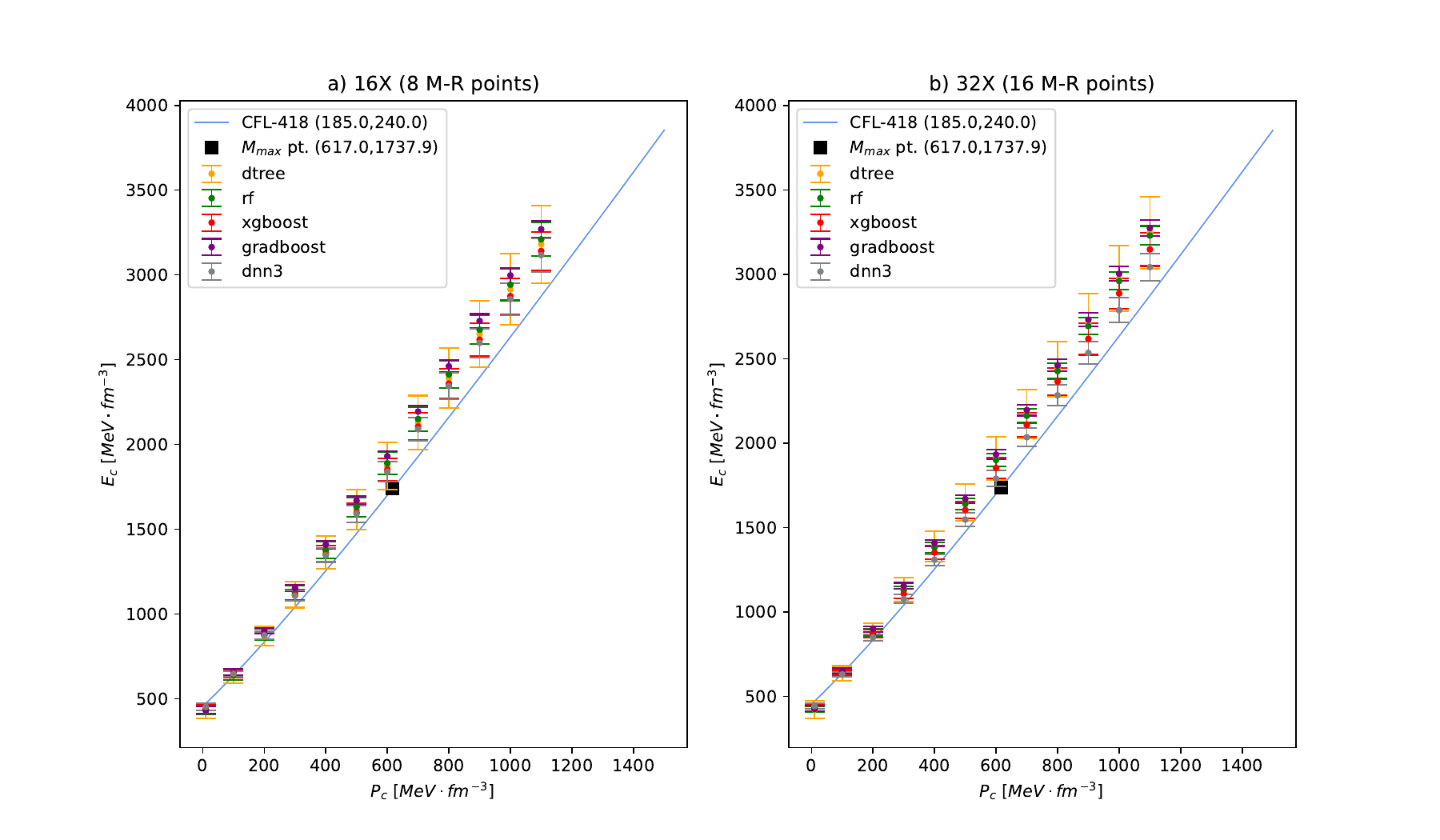}
    \caption{Reconstructing the \textit{CFL-418} EoS ($B=185$ $MeV\cdot fm^{-3}$, $\Delta=240$ $MeV$)}
    \label{fig:CFL-418_EOS_predict}
\end{figure}

\FloatBarrier

\begin{figure}[h]
    \centering
    \includegraphics[width=\linewidth]{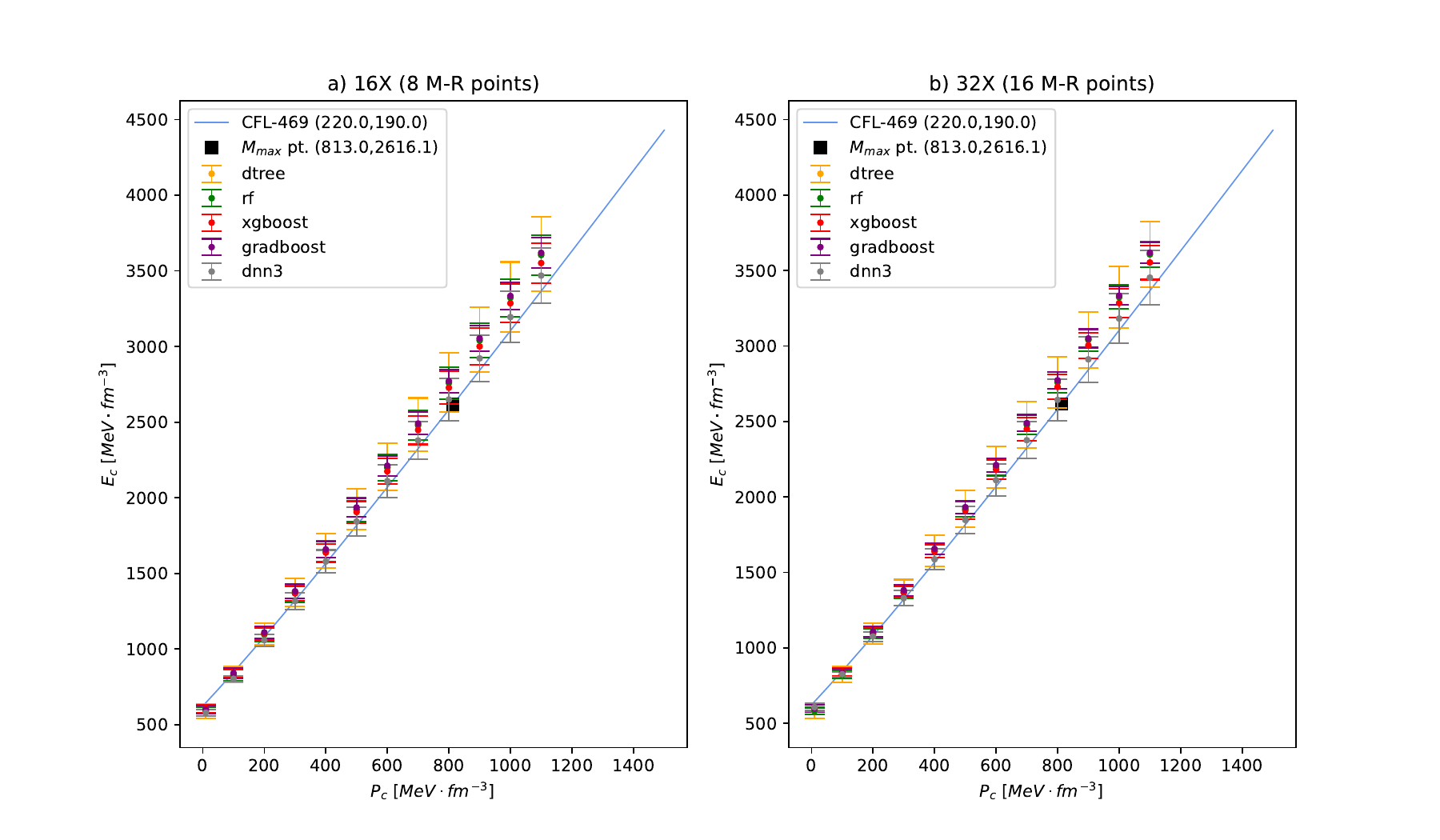}
    \caption{Reconstructing the \textit{CFL-469} EoS ($B=220$ $MeV\cdot fm^{-3}$, $\Delta=190$ $MeV$)}
    \label{fig:CFL-469_EOS_predict}
\end{figure}

\section{Discussion}

We shall begin the discussion, by describing the methodology we followed to obtain the figures in sections \ref{recontstruct_NS} and \ref{recontstruct_QS}. For each EoS to be reconstructed, we took 100 random observations of $M-R$ points, from its respective $M-R$ curve. We also shuffled these $M-R$ points, differently per observation. This way we created samples of data similar to the ones used for fitting the regression models. Then, we computed the predictions of energy density at the selected values of pressure (see subsection \ref{Data_sampling}), by feeding each $M-R$ observation into our regression models. Thus, for each model, we resulted in 100 predictions of the energy density, per pressure value. We calculated the mean value and the standard deviation of the 100 predictions and made error bars, in order to assess the accuracy and variance of the respective model in reconstructing the EoS. In Figs. \ref{fig:APR-1_EOS_predict} - \ref{fig:CFL-469_EOS_predict}, the error bars of the predictions from each regression model are depicted, along with the respective original EoS. The circle in the center of the error bars, corresponds to the mean prediction, while the total length of the error bars is twice the value of the predictions' standard deviation.

The first notice, like in general results of section \ref{General_results}, is the significantly poorer performance of the \textit{Decision Tree} models compared to all other algorithms. \textit{Decision Tree} models exhibit low accuracy, since the mean prediction deviates highly from the original EoS and also have high variance. On the contrary, the \textit{DNN-3} models have the minimum variance on predictions among all algorithms. However, the \textit{DNN-3} model is not always the best algorithm for reconstruction: for a certain EoS, its mean predictions might deviate more from the original EoS, than the mean predictions of some machine learning models. As for the \textit{Random Forest}, \textit{Gradient Boosting} and \textit{XGBoost} models, these perform the same on average, as we denoted in section \ref{General_results}, as well.

Now, in left graphs of Figs. \ref{fig:APR-1_EOS_predict} - \ref{fig:CFL-469_EOS_predict} the reconstruction is made using 8 points from the $M-R$ curves, while in right graphs of Figs. \ref{fig:APR-1_EOS_predict} - \ref{fig:CFL-469_EOS_predict} the reconstruction is made using 16 points from the $M-R$ curves. As it seems, in most cases, we obtain similar results, regardless the number of points. In practice, the 16 M-R points offer little improvement and, in some cases, they might actually confuse the algorithms, leading them to perform worse, than using 8 M-R points. This means, we can reconstruct an EoS effectively with less information from the M-R curve.

Moving on, we notice the perfect (or almost perfect) reconstruction of every EoS at lower pressures ($<100$ $MeV\cdot fm^{-1}$). Furthermore, we denote the gradual but evident increase in variance as we move into high pressures for all algorithms. This confirms our discussion in the early chapters \ref{NS Theory} and \ref{QS Theory}. The equation of state is well-defined in low mass densities (lower than nuclear saturation density) and their formula becomes unknown as we reach extremely high mass densities, due to the uncertainty in the composition of matter at these densities. Nevertheless, we get reliable results till the maximum mass point (black square in figures): the mean predictions from all regression models are very close or coincide with the original EoS till that point. In other words, since the maximum mass point marks the transition between stable and unstable Compact Stars configurations, we can claim that we reconstruct the stable part of the EoS with good accuracy.

In more specific results for Neutron Stars, the violation of causality might be a factor that affects the performance of the regression models. The reader has to remember that all algorithms are trained on data that do not violate causality. Hence, our models start to exhibit bigger variance when entered regions of causality violation and the maximum mass point lies inside this region, like in Figs. \ref{fig:APR-1_EOS_predict}, \ref{fig:HLPS-3_EOS_predict}, \ref{fig:WFF-1_EOS_predict} and \ref{fig:WFF-2_EOS_predict} of the EoSs \textit{APR-1}, \textit{HLPS-3}, \textit{WFF-1} and \textit{WFF-2}. On the other hand, the good reconstruction of many hadronic EoSs shows, that the behavior of these EoSs can indeed be reproduced by multimodal parameterization. In other words, the polytropic parameterization is an effective technique in capturing the details of an EoS in all pressures. Of course, there might be exceptions, like in Figs. \ref{fig:MDI-1_EOS_predict}-\ref{fig:PS_EOS_predict} and \ref{fig:W_EOS_predict} of the EoSs \textit{MDI-i} ($i\in\{1,2,3,4\}$), \textit{NLD}, \textit{PS} and \textit{W}. Besides, it is remarkable to see, that there is no perfect algorithm in reconstructing hadronic EoSs. Algorithms that perform great with some EoSs, might perform poorly with other EoSs.

Finally, we arrive at the reconstruction of quark matter EoSs. This reconstruction exhibits much smaller variances compared to the reconstruction of hadronic EoSs. This is due to the simpler and better-defined form of Quark Stars EoSs. The latter, is also the reason, why we get good results even after the maximum mass points. That is, we can predict effectively a region the unstable part of the EoSs. Moreover, we observe the evident superiority of the \textit{DNN-3} models: these models have noticeably smaller variance and there are closer to the original EoS than any other algorithm, in most cases. Moreover, the use of 16 M-R points, instead of 8 M-R points. offers significant improvement on the performance. However, cases of moderate results may occur, where the predictions do not follow exactly the straight line of the original EoS, like in Figs. \ref{fig:MITbag-15_EOS_predict}, \ref{fig:MITbag-86_EOS_predict}, \ref{fig:CFL-8_EOS_predict} and \ref{fig:CFL-418_EOS_predict}, for example. We speculate this might be due to the overlap area of the \textit{MITbag} and \textit{CFL} curves. When an $M-R$ curve from that area is provided, the algorithms cannot clearly distinguish which type of EoS to predict (\textit{MITbag} or \textit{CFL}) and yield poorer results. This is something we should be concerned about in future implementations.

\phantomsection
\chapter*{Conclusions - Epilogue}\label{Conc - Eplg}
\addcontentsline{toc}{part}{Conclusions - Epilogue}
This dissertation includes an extensive theoretical part and an extensive computational part, underlining our systematic approach to the problem of EoS reconstruction. In theoretical part, we covered the necessary physics governing Compact Stars and presented the mathematical formalism behind the operation of machine and deep learning algorithms, along with tips for better optimization. In computational part, we tried to encapsulate as much as possible from the theoretical part into our codes. Our first concern, was the production of large amount of data, for dense and detailed coverage of the $M-R$ space. We achieved that, through generation of mock EoSs. We saw that the polytropic parameterization of hadronic EoSs requires more steps and much more caution to align with physics of Neutron Stars, compared to the parameterization of Quark Stars EoSs, which only required a careful scanning of the stability window of strange quark matter. Our next concern, was the quality of our data. Here, we introduced the sampling and shuffling techniques, which helped in producing sufficiently uncorrelated data with no leakage. Then, we moved on feeding these data on regression algorithms, to obtain the much desired connection between the EoS and its M-R curve. Our primary goal was to get models with good optimization and good generalization to foreign data. We can claim that we have achieved this to a significant degree, based on the results we presented in the last chapter. In addition, The use of five different algorithms provides important insights:
\begin{itemize}
    \item \textit{Decision Tree} models have the lower reliability and the higher variance in predictions, as expected by definition. We advise the reader to process the results of these models with great caution and avoid the use of them to make reconstruction attempts of EoSs. The only good use of \textit{Decision Tree} models, would be to observe the performance improvement between these models and the \textit{Random Forest} ones.

    \item \textit{Random Forest}, \textit{Gradient Boosting} and \textit{XGBoost} models, perform the same on average. In practice, though, the \textit{XGBoost} models are the fastest in training and offer, among the three, the better generalization to more complex data (more features, more targets, etc.) and larger datasets. On the contrary, \textit{Gradient Boosting} models, are the slower ones, in total. Hence, we advise the reader to avoid the use of \textit{Gradient Boosting}, when very large datasets (with $>$100000 rows) or datasets with many features and target variables, are employed in training. Though, both \textit{XGBoost} and  \textit{Gradient Boosting} models needed a wrapper to adapt in multi-output regression. 
    
    \item \textit{Random Forest} models can be seen as "comfort" models, when all other machine learning algorithms fail. Their natively support for multi-output regression and the satisfyingly short training times are more than welcome properties in the problem of EoS reconstruction.

    \item \textit{Deep Neural Networks} offer the lower variance in predictions of energy density and the best generalization to more complex data, confirming the power of deep learning. They are, also, the most customizable. The selection of three hidden layers seems to be quite effective and appropriate and yields, in practice, better results than machine learning algorithms. One can easily experiment with larger datasets and more hidden layers, as well as different activations and regularization techniques. However, the vast variety of customization choices might ultimately lead to confusion and carelessly choosing a specific architecture for the neural network can make the results worse.
\end{itemize}

We could not leave the reader, without some suggestions for future improvements. Regarding the sampling, one could use and predict more values of energy density. A denser sampling in the area around the maximum mass point might be the answer in reconstructing effectively the unstable part of an EoS, besides the stable part. The accuracy of EoS reconstruction, using less than 8 M-R points from the entire M-R curve, or from a part of it, is also a matter worth investigating. In addition, more independent features could be sought to help capture more details of the M-R curves and EoSs. On the other hand, regarding the regression algorithms, one could experiment with more hyperparameters in machine learning or denser neural networks, keeping in mind computational effectiveness and improvement of accuracy. The addition and evaluation of other machine learning or deep learning algorithms, like PINNs (Physics Informed Neural Networks), could also be examined. Finally, we propose the combined prediction of the maximum mass point, along with the reconstruction of the EoS, for the user of the algorithm to know the transition between stable and unstable Compact Stars configurations. 

All the aforementioned, indicate there is still much to do. Of course, our work is not perfect and leaves several things open for improvement. However, it is a step forward in addressing such a difficult problem in Theoretical Physics, as the reconstruction of an EoS from its M-R curve. We hope that our implementation will provide a fertile ground for future efforts and for the even greater exploitation of computational science and data analysis, especially through machine and deep learning, in this field of Physics.


\appendixtitleformat

\part{Appendices}

\appendix

\chapter{Achieving hydrostatic equilibrium in compact stars}\label{Hydrostatic Equilibrium}

\section{Relativistic framework}\label{relat frame}
Treating hydrostatic equilibrium in neutron stars (and massive stars in general) with a Newtonian approach raises several weaknesses \cite{kanakis2019constraints,kourmpetis2024nuclear}. A remarkable one, is the failure to produce predictions about the potential maximum mass of a neutron star. Laplace also states that the escape velocity $\sqrt{GM/R}$ may ultimately exceed the speed of light. An approach, within the framework of \textbf{General Relativity}, is therefore more preferable, as it overcomes these limitations, and provides additional constraints on compactness. Moreover, general relativity dictates the existence of an upper bound for the mass density within the star, whenever a measurement of its mass is performed, consequently leading to an upper bound for the ultimate energy density of cold, static matter in the universe.

The starting point is the full \textbf{Einstein}'s field equations \cite{schaffner2020compact}:
\begin{equation}\label{full_Ein_eqs}
    G_{\mu\nu} = R_{\mu\nu}-\frac{1}{2}g_{\mu\nu}R = 8\pi G\cdot T_{\mu\nu}
\end{equation}
where $G_{\mu\nu}$ is the Einstein tensor, $ R_{\mu\nu}$ is the Ricci tensor, $R$ is the Ricci scalar (or scalar curvature), $g_{\mu\nu}$ is the metric tensor of spacetime, $G$ is the Newtonian constant of gravitation and $T_{\mu\nu}$ is the energy-momentum tensor. The assumption $c=1$ for the speed of light is also applied, otherwise the constant factor that multiplies the $T_{\mu\nu}$ tensor would be $\frac{8\pi G}{c^4}$.

The star is treated as a sphere with radius $R$, which must be larger than the \textbf{Schwarzschild} radius: $R_s=2GM$. Equation \ref{full_Ein_eqs} needs to be solved for both the exterior ($r>R$) and the interior ($r\leq R$) of the star, and the two solutions must match each other at the radius $R$ of the star.
The solution of the star's exterior, as suggested by the \textit{Birkhoff’s theorem}, is the Schwarzschild metric. We then seek the solution of \ref{full_Ein_eqs} for the interior of the star. The preferable ansatz for the metric, considering the static and spherical symmetric configuration of the star, is the following:
\begin{equation}\label{metric_int}
    ds^2=-e^{2\alpha(r)}\cdot dt^2+e^{2\beta(r)}\cdot dr^2 + r^2\cdot (d\theta^2+\sin^2{\theta}d\phi^2)=-e^{2\alpha(r)}\cdot dt^2+e^{2\beta(r)}\cdot dr^2 + r^2\cdot d\Omega^2
\end{equation}
where $d\Omega$ is the differential solid angle  and $\alpha(r)$, $\beta(r)$ are metric functions. 

Now, one has to determine
the elements of the Einstein tensor $G_{\mu\nu}$ in the left-hand side of \ref{full_Ein_eqs} and results in the expressions below:

\begin{equation}\label{G00_exp}
    G_{00}=\frac{1}{r^2}e^{2(\alpha-\beta)}\cdot (2r\beta^{'}-1+e^{2\beta})
\end{equation}
\begin{equation}\label{G11_exp}
    G_{11}=\frac{1}{r^2}\cdot(2r\alpha^{'}+1-e^{2\beta})
\end{equation}
\begin{equation}\label{G22_exp}
    G_{22}=r^2 e^{-2\beta}\cdot\left[\alpha^{''}+(\alpha^{'})^2-\alpha^{'}\beta^{'}+\frac{1}{r}(\alpha^{'}-\beta^{'})\right]
\end{equation}
\begin{equation}\label{G33_exp}
    G_{33}=\sin^2\theta\cdot G_{22}
\end{equation}
since the Ricci scalar $R$, calculated from the components of the Ricci tensor, is equal to \cite{schaffner2020compact}:
\begin{equation}\label{Ricci_scalar}
    R = 2e^{-2\beta}\left[-\alpha^{''}-(\alpha^{'})^2+\alpha^{'}\beta^{'}+\frac{2}{r}(\beta^{'}-\alpha^{'})-\frac{1}{r^2}\right]+\frac{2}{r^2}
\end{equation}
Let's denote that the primes ($'$) indicate first derivative and the double primes ($''$) second derivative with respect to $r$. 

As for the right-hand side of \ref{full_Ein_eqs}, the energy-momentum tensor $T_{\mu\nu}$ needs to be specified. Modeling the matter in the interior of the star as a perfect fluid:
\begin{equation}\label{Tmn_perf_fluid}
    T_{\mu\nu} = (\epsilon+P)u_{\mu}u_{\nu}+P\cdot g_{\mu\nu}
\end{equation}
where $\epsilon$ is the energy density and $P$ is the pressure, and fixing the four-vector to be at rest with respect to the matter:
\begin{equation}\label{4vec_rest}
    u^{\mu}=(1,0,0,0)  
\end{equation}
we get the following expression for the $T_{\mu\nu}$ tensor:
\begin{equation}\label{Tmn_explicit}
    T_{\mu\nu} = 
    \begin{pmatrix}
    e^{2\alpha}\cdot \epsilon & 0 & 0 & 0 \\
    0 & e^{2\beta}\cdot P & 0 & 0 \\
    0 & 0 & r^2 \cdot P & 0 \\
    0 & 0 & 0 & r^2 \sin^2\theta\cdot P
    \end{pmatrix}
\end{equation}
Notice that, in this case, the metric coefficients squared (see \ref{metric_int}) appear in the components of $T_{\mu\nu}$.
Substituting the expressions \ref{G00_exp}-\ref{G33_exp} for the Einstein tensor and the expression \ref{Tmn_explicit} for the energy-momentum tensor, into the full Einstein field equations \ref{full_Ein_eqs}, will give us three independent differential equations:
\begin{equation}\label{ind_diff_eq1}
    \frac{1}{r^2}e^{-2\beta}\cdot (2r\beta^{'}-1+e^{2\beta})=
    8\pi G\cdot \epsilon
\end{equation}
\begin{equation}\label{ind_diff_eq2}
    \frac{1}{r^2}e^{-2\beta}\cdot(2r\alpha^{'}+1-e^{2\beta})=
     8\pi G\cdot P
\end{equation}
\begin{equation}\label{ind_diff_eq3}
    e^{-2\beta}\cdot\left[\alpha^{''}+(\alpha^{'})^2-\alpha^{'}\beta^{'}+\frac{1}{r}(\alpha^{'}-\beta^{'})\right]=
    8\pi G\cdot P
\end{equation}
that seek solution.

\section{The TOV equations}\label{TOV Theory}
We can define the metric function $\beta(r)$, including a mass function $m_r(r)$, as shown below \cite{schaffner2020compact}:
\begin{equation}\label{beta_metric}
    e^{2\beta(r)}=\left[1-\frac{2Gm_r(r)}{r}\right]^{-1}
\end{equation}
This form, allows an easy recovery of the exterior Schwarzschild metric, thus the expression on the right-hand side is also denoted as the Schwarzschild factor. In fact, when the radius of the star is reached: $r=R$, the metric component is equal to that of the Schwarzschild metric with the total mass of the star $m_r(r = R) = M$.

Substituting \ref{beta_metric} in \ref{ind_diff_eq1}, results in the differential equation:
\begin{equation}\label{TOV_mass}
    \frac{dm_r(r)}{dr}=4\pi r^2\epsilon(r)
\end{equation}
which is the first of the two \textbf{Tolman-Oppenheimer-Volkoff} (TOV) equations we need in our study. To get, the second equation, one has to start by rewriting \ref{ind_diff_eq2}, using again equation \ref{beta_metric} for $\beta(r)$ metric:
\begin{equation}\label{ind_diff_eq2_alt}
    \frac{d\alpha(r)}{dr}=\frac{Gm_r(r)}{r^2}\left(1+\frac{4\pi r^3P(r)}{m_r(r)}\right)\left(1-\frac{2Gm_r(r)}{r}\right)^{-1}
\end{equation}
With equation \ref{ind_diff_eq2_alt}, the metric function $\alpha(r)$ is fixed in terms of the mass function $m_r(r)$ and the pressure $P(r)$. For the complete elimination of $\alpha(r)$, it is better to use the energy-momentum conservation equation, rather than the latter differential equation \ref{ind_diff_eq3}:
\begin{equation}\label{energ_moment_cons}
    \nabla_{\mu}T^{\mu\nu}=0
\end{equation}
In our case, the only relevant component is for $\mu=\nu=1$, resulting to the following differential equation for pressure:
\begin{equation}\label{P_diff_eq}
    \frac{dP(r)}{dr}=-(\epsilon(r)+P(r))\frac{d\alpha(r)}{dr}
\end{equation}
The second TOV equation is then obtained by combining equations \ref{ind_diff_eq2_alt} and \ref{P_diff_eq}:
\begin{equation}\label{TOV_pressure}
    \frac{dP(r)}{dr}=-\frac{Gm_r(r)\epsilon(r)}{r^2}\left(1+\frac{P(r)}{\epsilon(r)}\right)\left(1+\frac{4\pi r^3P(r)}{m_r(r)}\right)\left(1-\frac{2Gm_r(r)}{r}\right)^{-1}
\end{equation}
Notice that the first term on the right-hand side is the Newtonian term for the hydrostatic equilibrium of a compact star:
\begin{equation}\label{Newton_equilib}
    \frac{dP(r)}{dr} = -\frac{Gm_r(r)\rho(r)}{r^2}
\end{equation}
with $\rho(r)$ the mass density. The other three terms are corrections from General Relativity. Specifically, we have \cite{schaffner2020compact}:
\begin{itemize}
    \item $\left(1+\frac{P(r)}{\epsilon(r)}\right)$: modifies the mass density $\rho(r)$ and takes into account that gravity couples to the energy density $\epsilon(r)$ and the pressure $P(r)$ of matter

    \item $\left(1+\frac{4\pi r^3P(r)}{m_r(r)}\right)$: modifies the mass function $m_r(r)$ and adds another correction term from the pressure of matter

    \item $\left(1-\frac{2Gm_r(r)}{r}\right)^{-1}$: modifies the radius and takes into account the warpage of spacetime that is described by the Schwarzschild factor
\end{itemize}

In the above form of the TOV equations we assumed $c=1$. An alternative form can be found in literature, where the mass density $\rho(r)$ is used instead of the energy density $\epsilon(r)$, with $\rho(r)=\epsilon(r)/c^2$:
\begin{equation}\label{TOV_press_mass_2}
\begin{aligned}
    \frac{dP(r)}{dr}&=-\frac{Gm_r(r)\rho(r)}{r^2}\left(1+\frac{P(r)}{c^2\rho(r)}\right)\left(1+\frac{4\pi r^3P(r)}{c^2m_r(r)}\right)\left(1-\frac{2Gm_r(r)}{c^2r}\right)^{-1} \\
    \frac{dm_r(r)}{dr}&=4\pi r^2\rho(r)
\end{aligned}    
\end{equation}
and $c\neq1$.

\chapter{Python Codes}\label{Python Codes}
In this Appendix we present tables summarizing the Python codes we developed for this dissertation. The tables contain links to the codes in \texttt{GitHub}.

\begin{table}[htbp]
    \centering
    \begin{tabular}{|p{7cm}|p{10cm}|}
\hline
\hline
\textbf{Filename} & \textbf{Brief Description} \\
\hline
         \href{https://github.com/istergak/MSc-Computational-Physics-AUTH/blob/main/Thesis%20-%20ML%20and%20ANNs%20regression%20models%20for%20Exotic%20Star's%20EOSs/Part%201%20-%20Solving%20the%20TOV%20equations%20for%20Hadronic%20and%20Quark%20Stars/eos_lib_NS.py}{eos\_lib\_NS.py} & Defining (numerically and symbolically) and storaging the 'main' EOSs for the core and the crust EOSs of Neutron Stars in lists. \\

\hline
         \href{https://github.com/istergak/MSc-Computational-Physics-AUTH/blob/main/Thesis%20-%20ML%20and%20ANNs%20regression%20models%20for%20Exotic%20Star's%20EOSs/Part%201%20-%20Solving%20the%20TOV%20equations%20for%20Hadronic%20and%20Quark%20Stars/tov_solver_NS.py}{tov\_solver\_NS.py} & Solving the TOV equations serially for a single 'main' core EOS of a Neutron Star included in \texttt{eos\_lib\_NS.py} module. The crust EOSs are always included. \\
\hline
         \href{https://github.com/istergak/MSc-Computational-Physics-AUTH/blob/main/Thesis%20-%20ML%20and%20ANNs%20regression%20models%20for%20Exotic%20Star's%20EOSs/Part%201%20-%20Solving%20the%20TOV%20equations%20for%20Hadronic%20and%20Quark%20Stars/tov_solver_NS_par.py}{tov\_solver\_NS\_par.py} & Solving the TOV equations in parallel for a selected number of 'main' models as the core EOS of a Neutron Star included in \texttt{eos\_lib\_NS.py} module. Each model is distributed to a single thread for solution. The crust EOSs are always included. \\
\hline
        \href{https://github.com/istergak/MSc-Computational-Physics-AUTH/blob/main/Thesis%20-%20ML%20and%20ANNs%20regression%20models%20for%20Exotic%20Star's%20EOSs/Part%201%20-%20Solving%20the%20TOV%20equations%20for%20Hadronic%20and%20Quark%20Stars/StudyPolyNS.ipynb}{StudyPolyNS.ipynb} & Studying the general methodology of parametrizing an EOS, using piecewise polytropes. The pressure values of the HLPS-2 and HLPS-3 'main' EOSs at the nuclear saturation density are being determined. \\
\hline
       \href{https://github.com/istergak/MSc-Computational-Physics-AUTH/blob/main/Thesis%20-%20ML%20and%20ANNs%20regression%20models%20for%20Exotic%20Star's%20EOSs/Part%201%20-%20Solving%20the%20TOV%20equations%20for%20Hadronic%20and%20Quark%20Stars/tov_solver_polyNS_par.py}{tov\_solver\_polyNS\_par.py} & Solving the TOV equations in parallel for a selected number of polytropic mock EOSs (combined with a 'main' EOS model: either HLPS-2 or HLPS-3) as the core EOSs of a Neutron Star. Each mock EOS is distributed to a single thread for solution. The crust EOSs are always included. \\
\hline
       \href{https://github.com/istergak/MSc-Computational-Physics-AUTH/blob/main/Thesis%20-%20ML%20and%20ANNs%20regression%20models%20for%20Exotic%20Star's%20EOSs/Part%201%20-%20Solving%20the%20TOV%20equations%20for%20Hadronic%20and%20Quark%20Stars/tov_solver_polyNS_par2.py}{tov\_solver\_polyNS\_par2.py} & Same as \texttt{tov\_solver\_polyNS\_par.py} module, but corrections in the polytropic part of the mock EOSs are being made to avoid violation of causality. The corrections involve the replace of the polytropic part of the mock EOS that violates causality with a linear part, with fixed slope that does not violate causality. \\
\hline
\hline
    \end{tabular}
    \caption{Python codes for solving the TOV equations for Neutron Stars}
    \label{tab:TOV_solve_codes_NS}
\end{table}

\begin{table}[htbp]
    \centering
    \begin{tabular}{|p{7cm}|p{10cm}|}
\hline
\hline
\textbf{Filename} & \textbf{Brief Description} \\
\hline
       \href{https://github.com/istergak/MSc-Computational-Physics-AUTH/blob/main/Thesis%20-%20ML%20and%20ANNs%20regression%20models%20for%20Exotic%20Star's%20EOSs/Part%201%20-%20Solving%20the%20TOV%20equations%20for%20Hadronic%20and%20Quark%20Stars/eos_lib_QS.py}{eos\_lib\_QS.py} & Defining (numerically and symbolically) and storaging CFL EOSs of Quark Stars in lists. \\
\hline
      \href{https://github.com/istergak/MSc-Computational-Physics-AUTH/blob/main/Thesis%20-%20ML%20and%20ANNs%20regression%20models%20for%20Exotic%20Star's%20EOSs/Part%201%20-%20Solving%20the%20TOV%20equations%20for%20Hadronic%20and%20Quark%20Stars/tov_solver_cflQS.py}{tov\_solver\_cflQS.py} & Solving the TOV equations serially for a single CFL EOS of a Quark Star included in \texttt{eos\_lib\_QS.py} module. No crust EOSs are included. \\
\hline
      \href{https://github.com/istergak/MSc-Computational-Physics-AUTH/blob/main/Thesis%20-%20ML%20and%20ANNs%20regression%20models%20for%20Exotic%20Star's%20EOSs/Part%201%20-%20Solving%20the%20TOV%20equations%20for%20Hadronic%20and%20Quark%20Stars/tov_solver_cflQS_par.py}{tov\_solver\_cflQS\_par.py} & Solving the TOV equations in parallel for a selected number of CFL EOSs of a Quark Star. No crust EOSs are included. The user can determine the ranges of the $B_{eff}$ and $\Delta$ parameters and generate arbitrarily the preferred CFL models. Each model is distributed to a single thread for solution.
\\
\hline
      \href{https://github.com/istergak/MSc-Computational-Physics-AUTH/blob/main/Thesis%20-%20ML%20and%20ANNs%20regression%20models%20for%20Exotic%20Star's%20EOSs/Part%201%20-%20Solving%20the%20TOV%20equations%20for%20Hadronic%20and%20Quark%20Stars/tov_solver_mitQS_par.py}{tov\_solver\_mitQS\_par.py} & Solving the TOV equations in parallel for a selected number of MIT bag EOSs of a Quark Star. No crust EOSs are included. The user can determine the range of the $B_{eff}$ parameter and generate arbitrarily the preferred MIT bag models. Each model is distributed to a single thread for solution.
\\
\hline
\hline
    \end{tabular}
    \caption{Python codes for solving the TOV equations for Quark Stars}
    \label{tab:TOV_solve_codes_QS}
\end{table}

\begin{table}[htbp]
    \centering
    \begin{tabular}{|p{7cm}|p{10cm}|}
\hline
\hline
\textbf{Filename} & \textbf{Brief Description} \\
\hline
    \href{https://github.com/istergak/MSc-Computational-Physics-AUTH/blob/main/Thesis%20-%20ML%20and%20ANNs%20regression%20models%20for%20Exotic%20Star's%20EOSs/Part%202%20-%20Handling%20the%20TOV%20equations%20solution%20data%20for%20Exotic%20Stars/ExoticStarsDataHandling.py}{ExoticStarsDataHandling.py} &  Module containing functions and classes for: a) validating the parameters of polytropic Neutron Stars EOSs ($\Gamma$ parameter) and CFL Quark Stars EOSs ($B_{eff}$ and $\Delta$ parameters), b) plotting $E_c-P_c$, $c^2_s-P_c$ and $M-R$ curves of the respective EOSs and c) sampling and shuffling data for regression purposes.   \\
\hline
      \href{https://github.com/istergak/MSc-Computational-Physics-AUTH/blob/main/Thesis%20-%20ML%20and%20ANNs%20regression%20models%20for%20Exotic%20Star's%20EOSs/Part%202%20-%20Handling%20the%20TOV%20equations%20solution%20data%20for%20Exotic%20Stars/ExoticStarsDataHandling2.py}{ExoticStarsDataHandling2.py} & A different version of \texttt{ExoticStarsDataHandling.py} module, containing major or minor adjustments in some classes. \\
\hline
      \href{https://github.com/istergak/MSc-Computational-Physics-AUTH/blob/main/Thesis%20-%20ML%20and%20ANNs%20regression%20models%20for%20Exotic%20Star's%20EOSs/Part%202%20-%20Handling%20the%20TOV%20equations%20solution%20data%20for%20Exotic%20Stars/plot_curves_NS.py}{plot\_curves\_NS.py} & Module that plots the $E_c-P_c$ and $M-R$ curves of the respective main Neutron Star EOSs. Needs to be executed from a terminal. \\
\hline
     \href{https://github.com/istergak/MSc-Computational-Physics-AUTH/blob/main/Thesis%20-%20ML%20and%20ANNs%20regression%20models%20for%20Exotic%20Star's%20EOSs/Part%202%20-%20Handling%20the%20TOV%20equations%20solution%20data%20for%20Exotic%20Stars/ExoticStarsResults_1.ipynb}{ExoticStarsResults\_1.ipynb} & Using \texttt{ExoticStarsDataHandling.py} module to plot curves. \\
\hline
     \href{https://github.com/istergak/MSc-Computational-Physics-AUTH/blob/main/Thesis%20-%20ML%20and%20ANNs%20regression%20models%20for%20Exotic%20Star's%20EOSs/Part%202%20-%20Handling%20the%20TOV%20equations%20solution%20data%20for%20Exotic%20Stars/ExoticStarsResults_2.ipynb}{ExoticStarsResults\_2.ipynb} & Using \texttt{ExoticStarsDataHandling.py} module to sample and shuffle data for regression purposes. \\     
\hline
\hline
    \end{tabular}
    \caption{Python codes for handling the data from the solution of TOV equations}
    \label{tab:TOV_handle_data}
\end{table}

\begin{table}[htbp]
    \centering
    \begin{tabular}{|p{7cm}|p{10cm}|}
\hline
\hline
\textbf{Filename} & \textbf{Brief Description} \\
\hline
    \href{https://github.com/istergak/MSc-Computational-Physics-AUTH/blob/main/Thesis%20-%20ML%20and%20ANNs%20regression%20models%20for%20Exotic%20Star's%20EOSs/Part%203%20-%20Training%20and%20testing%20ML%20regression%20algorithms/data_analysis_ES_ML.py}{data\_analysis\_ES\_ML.py} &  Module containing functions and classes for a) assessing linear correlations in regression data, b) training and testing machine learning regression models, c) storaging the fitting results in .pkl files and loading the .pkl files of trained models, and d) storaging summary results in .csv files, loading and presenting the summary results in \textbf{PrettyTable} and \textbf{bar plots} forms   \\
\hline
      \href{https://github.com/istergak/MSc-Computational-Physics-AUTH/blob/main/Thesis%20-%20ML%20and%20ANNs%20regression%20models%20for%20Exotic%20Star's%20EOSs/Part%203%20-%20Training%20and%20testing%20ML%20regression%20algorithms/assessing_regression_data.ipynb}{assessing\_regression\_data.ipynb} & Using \texttt{data\_analysis\_ES\_ML.py} module for assessing linear correlations in our regression data. \\
\hline
     \href{https://github.com/istergak/MSc-Computational-Physics-AUTH/blob/main/Thesis%20-%20ML%20and%20ANNs%20regression%20models%20for%20Exotic%20Star's%20EOSs/Part%203%20-%20Training%20and%20testing%20ML%20regression%20algorithms/train_test_dtree_regress.ipynb}{train\_test\_dtree\_regress.ipynb} & Using \texttt{data\_analysis\_ES\_ML.py} module for fitting \textit{Decision Tree} models on our regression data and storaging the results in .pkl files. \\ 
\hline
     \href{https://github.com/istergak/MSc-Computational-Physics-AUTH/blob/main/Thesis%20-%20ML%20and%20ANNs%20regression%20models%20for%20Exotic%20Star's%20EOSs/Part%203%20-%20Training%20and%20testing%20ML%20regression%20algorithms/train_test_rf_regress.ipynb}{train\_test\_rf\_regress.ipynb} & Using \texttt{data\_analysis\_ES\_ML.py} module for fitting \textit{Random Forest} models on our regression data and storaging the results in .pkl files. \\
\hline
     \href{https://github.com/istergak/MSc-Computational-Physics-AUTH/blob/main/Thesis%20-%20ML%20and%20ANNs%20regression%20models%20for%20Exotic%20Star's%20EOSs/Part%203%20-%20Training%20and%20testing%20ML%20regression%20algorithms/train_test_gradboost_regress.ipynb}{train\_test\_gradboost\_regress.ipynb} & Using \texttt{data\_analysis\_ES\_ML.py} module for fitting \textit{Gradient Boosting} models on our regression data and storaging the results in .pkl files. \\
\hline
     \href{https://github.com/istergak/MSc-Computational-Physics-AUTH/blob/main/Thesis%20-%20ML%20and%20ANNs%20regression%20models%20for%20Exotic%20Star's%20EOSs/Part%203%20-%20Training%20and%20testing%20ML%20regression%20algorithms/train_test_xgboost_regress.ipynb}{train\_test\_xgboost\_regress.ipynb} & Using \texttt{data\_analysis\_ES\_ML.py} module for fitting \textit{XGBoost} models on our regression data and storaging the results in .pkl files. \\
\hline     
    \href{https://github.com/istergak/MSc-Computational-Physics-AUTH/blob/main/Thesis%20-%20ML%20and%20ANNs%20regression%20models%20for%20Exotic%20Star's%20EOSs/Part%203%20-%20Training%20and%20testing%20ML%20regression%20algorithms/assessing_summary_ml_reg.ipynb}{assessing\_summary\_ml\_reg.ipynb} & Using \texttt{data\_analysis\_ES\_ML.py} module for loading the .pkl files of trained models, storaging summary results in .csv files, loading and presenting the summary results in \textbf{PrettyTable} and \textbf{bar plots} forms. \\     
\hline
\hline
    \end{tabular}
    \caption{Python codes for fitting and assessing machine learning regression models}
    \label{tab:train_test_ML}
\end{table}

\begin{table}[htbp]
    \centering
    \begin{tabular}{|p{7cm}|p{10cm}|}
\hline
\hline
\textbf{Filename} & \textbf{Brief Description} \\
\hline
    \href{https://github.com/istergak/MSc-Computational-Physics-AUTH/blob/main/Thesis%20-%20ML%20and%20ANNs%20regression%20models%20for%20Exotic%20Star's%20EOSs/Part%204%20-%20Training%20and%20testing%20ANN%20regression%20models/data_analysis_ES_ANNs.py}{data\_analysis\_ES\_ANNs.py} &  Module containing functions and classes for a) assessing linear correlations in regression data, b) building and fitting deep learning regression models and c) storaging the fitting results in .pkl files and loading the .pkl files of trained models.   \\
\hline
      \href{https://github.com/istergak/MSc-Computational-Physics-AUTH/blob/main/Thesis%20-%20ML%20and%20ANNs%20regression%20models%20for%20Exotic%20Star's%20EOSs/Part%204%20-%20Training%20and%20testing%20ANN%20regression%20models/activation_functions.ipynb}{activation\_functions.ipynb} & Defining and plotting the activation functions \textit{sigmoid} and \textit{ReLU}, along with their first derivatives. \\
\hline
     \href{https://github.com/istergak/MSc-Computational-Physics-AUTH/blob/main/Thesis%20-%20ML%20and%20ANNs%20regression%20models%20for%20Exotic%20Star's%20EOSs/Part%204%20-%20Training%20and%20testing%20ANN%20regression%20models/train_test_dnn3_regress.ipynb}{train\_test\_dnn3\_regress.ipynb} & Using \texttt{data\_analysis\_ES\_ANNs.py} module for building and fitting \textit{Deep Neural Network} models (with 3 hidden layers) on our regression data and storaging the results in .pkl files. \\ 
\hline
\hline
    \end{tabular}
    \caption{Python codes for fitting and assessing deep learning regression models}
    \label{tab:train_test_DL}
\end{table}

\FloatBarrier

\begin{table}[h!]
    \centering
    \begin{tabular}{|p{8cm}|p{9cm}|}
\hline
\hline
\textbf{Filename} & \textbf{Brief Description} \\
\hline
    \href{https://github.com/istergak/MSc-Computational-Physics-AUTH/blob/main/Thesis%20-%20ML%20and%20ANNs%20regression%20models%20for%20Exotic%20Star's%20EOSs/Part%205%20-%20Final%20Results/metrics_learning%20curves_final_results.ipynb}{metrics\_learning\_curves\_final\_results.ipynb} &  Obtaining and illustrating the final results of a) metrics MSLE and MSE and b) learning curves of DNN models. \\
\hline
      \href{https://github.com/istergak/MSc-Computational-Physics-AUTH/blob/main/Thesis%20-%20ML%20and%20ANNs%20regression%20models%20for%20Exotic%20Star's%20EOSs/Part%205%20-%20Final%20Results/reconstruct_EOS_NS.ipynb}{reconstruct\_EOS\_NS.ipynb} & Reconstructing the 21 "main" EoSs and 6 mock poly-linear EoSs of Neutron Stars. \\
\hline
     \href{https://github.com/istergak/MSc-Computational-Physics-AUTH/blob/main/Thesis%20-%20ML%20and%20ANNs%20regression%20models%20for%20Exotic%20Star's%20EOSs/Part%205%20-%20Final%20Results/reconstruct_EOS_QS.ipynb}{reconstruct\_EOS\_QS.ipynb} & Reconstructing 10 MIT bag and 10 CFL EoSs of Quark Stars. \\ 
\hline
\hline
    \end{tabular}
    \caption{Python codes for final results and reconstruction of Compact Star EoSs}
    \label{tab:final_results}
\end{table}

The whole repository on \texttt{GitHub} can be found on the following link: \href{https://github.com/istergak/MSc-Computational-Physics-AUTH/tree/main/Thesis%20-%20ML%20and%20ANNs%20regression%20models%20for%20Exotic%20Star's%20EOSs}{Thesis - ML and ANNs regression models for Exotic Star's EoSs}.


\end{document}